\documentclass[a4paper,fdp-mod]{w-art-mod}

\makeatletter
\def\input@path{{styles/}{chapters/}{pictures/}}
\makeatother
\usepackage[
    hyperrefon,      
    fancylayoutoff,  
    symbolson,       
    booktoarticleon, 
    colortableson,   
    compressciteoff  
    ]{jo}

\IFHYPER{

\hypersetup{%
  pdfpagemode = UseOutlines,
  colorlinks  = true,
}

\hypersetup{%
  pdftitle    = {Quantum Field Theories Coupled to Supergravity:
                 AdS/CFT and Local Couplings},
  pdfsubject  = {Thesis (Dissertation, Ludwig-Maximilians-Universit\"at, M\"unchen)},
  pdfauthor   = {Johannes Gro\ss{}e},
  pdfkeywords = {Yang-Mills theory, AdS/CFT correspondence, QCD, 
                 chiral symmetry breaking, meson spectrum, Higgsbranch,
                 heavy-light mesons, space-time dependent couplings, 
                 c-theorem, a-theorem},
  pdfcreator  = {pdflatex},
  pdfproducer = {LaTeX with hyperref},
  }

}

\newif\ifBeautifulForm     
\BeautifulFormfalse        
\def\finalize#1{}          %
\renewcommand\marginpar[2][]{}
\newcommand{\naive}{na\-\"{\i}ve}
\newcommand{\role}{r\^{o}le}
\newcommand{\emt}{en\-er\-gy-mo\-men\-tum tensor}
\begin{document}

  \title[QFTs Coupled to SUGRA: AdS/CFT and Local Couplings]
    {Quantum Field Theories Coupled to Supergravity:
     AdS/CFT Correspondence and Local Couplings}
  \author[J. Gro\ss{}e]{%
    Johannes Gro\ss{}e\footnote{%
    E-mail: \href{mailto:jgrosse@mppmu.mpg.de}
                {\textsf{jgrosse@mppmu.mpg.de}}}}
  \address{%
    Max-Planck-Institut f\"{u}r Physik,\\
    F\"{o}hringer Ring 6, D-80805 M\"{u}nchen, Germany \\[2ex]
    Arnold-Sommerfeld-Center for Theoretical Physics,\\
    Department f\"{u}r Physik, \\
    Ludwig-Maximilians-Universit\"{a}t M\"{u}nchen,\\
    Theresienstra\ss{}e 37, D-80333 M\"{u}nchen, Germany
  }  

  \begin{abstract}
    This article is based on my PhD thesis \cite{JG2006} and covers
    the following topics: Holographic meson spectra in a dilaton flow
    background, the mixed Coulomb--Higgs branch in terms
    of instantons on D7 branes, and a dual description of heavy-light
    mesons. 

    Moreover, in a second part the conformal anomaly of four
    dimensional supersymmetric quantum field theories coupled to
    classical $\cN=1$ supergravity is explored in a superfield
    formulation.  The complete basis for the anomaly and consistency
    conditions, which arise from cohomological considerations, are
    given.  Possible implications for an extension of Zamolodchikov's
    $c$\NB-theorem to four dimensional supersymmetric quantum field
    theories are discussed.
  \end{abstract}
  \maketitle


  \tableofcontents

  \begin{savequote}[\savequotewidth]
  Art is born of the observation and investigation of nature.
  \qauthor{Cicero}
\end{savequote}

\Chapter{Introduction}

An important goal of theoretical physics is the \emph{algorithmic
  compression} of nature to a set of fundamental laws. This means that
a minimal description is sought that encodes a maximum of information
about our universe.  At the current state of knowledge, this
description is in terms of the \emph{standard model} of elementary
particles and Einstein gravity, as well as initial conditions and
parameters.  Although many models used in other areas of physics are
not derived from those fundamental theories, in principle such a
derivation should nevertheless be possible.

The standard model\marginpar{standard model} is a quantum field
theory that describes electromagnetism, the weak and the strong force,
organised by the principle of gauge invariance.  The latter arises
from making the formulation manifestly Lorentz invariant which
requires the introduction of extra non-physical degrees of freedom.
Consequently there are many representations of the same physical
state, which are related by so-called gauge transformations.  Gauge
transformations can be identified with Lie groups having space-time
dependent parameters and form the internal symmetry group of the
standard model, the group $\gr{U}(1) \times\gr{SU}(2)
\times\gr{SU}(3)$, corresponding to quantum electrodynamics
(\acro{QED}) describing photons, the weak interaction, whose gauge
fields are the W and Z bosons responsible for the $\beta$ decay, and
quantum chromodynamics (\acro{QCD}), the theory of the strong force,
which describes the constituents of hadrons like the proton and the
neutron.

We shall first have a closer look at \acro{QED},
\marginpar{\acro{QED} and renormalisation}
which is a remarkably successful theory,
confirmed to an incredible accuracy of up to $10^{-11}$ over
the past decades.  Since a rigorous treatment of interacting quantum
field theories is difficult, an important reason for this success is
the possibility to treat \acro{QED} perturbatively.  In
\emph{perturbation theory} a theory is effectively split into a
solvable part; e.g.~a free theory, and the remainder that renders the
theory unsolvable; e.g.~the interaction terms. Assuming that the
solutions of the free theory are only slightly modified by the
presence of the additional interaction terms allows an expansion in
the coupling constant.  However this expansion is not a true series
expansion since the coupling constants themselves need to be modified
during the expansion by a procedure called \emph{renormalisation} to
absorb infinite contributions arising from the interplay of the
quantisation procedure and perturbation theory. Theories allowing to
absorb these infinities in a finite number of parameters are called
\emph{renormalisable} and can be treated perturbatively in a well
defined manner.

There are basically two points where this strategy can fail and
interestingly both have a connection to \emph{string theory} as will
be seen later.  

The\marginpar{non-renormalisable theories}
first problem arises when trying to tackle non-renormalisable 
theories like gravity. Each order of perturbation theory then produces
a growing number of coupling constants that destroy the predictive
power of the theory. This can either be interpreted as there being
something wrong with the quantisation procedure assuming that
gravity has some miraculous ultraviolet (\acro{UV}) behaviour that
is merely poorly understood or that Einstein gravity is just
an effective field theory that breaks down when leaving its
regime of validity (at the order of the Planck mass 
$m_P \approx 10^{19} \text{ GeV}$) and a more fundamental theory is required. 

In the spirit of the introductory remarks at the beginning, such
a ``more fundamental'' theory, from which also the standard model
of elementary particles should be derived, is a natural goal,
which unfortunately seems to be currently out of reach.
However there exists at least a candidate theory that consistently
quantises gravity and at the 
same time incorporates gauge theories similar to the standard model,
namely \emph{superstring theory}.
Entertainingly this extremely remarkable feature was not what led to its
discovery and it is also not the feature central to this thesis, 
which shall be explicated in the followings.

The\marginpar{strong coupling}
second problem of perturbation theory arises from the phe\-nom\-e\-non of
\emph{running gauge couplings}, a result---though not a 
consequence---of renormalisation. It
is the statement that the strength of the interaction and thus
the validity of perturbation theory depends on the energy scale.
While the electroweak force has small coupling constants at 
low energies, which become large when going to higher energies,
the opposite is true for \acro{QCD}, which is 
\emph{asymptotically free}.
For small energies \acro{QCD} exhibits a phase transition, 
the \emph{confinement}, that effectively screens the theory's
fundamental particles, the quarks and gluons, 
from the dynamics by creating bound states of vanishing colour charge:
hadrons. 
In that sense \acro{QCD} is an accelerator theory that can only
be observed at high energies, although there is very strong evidence
from lattice calculations that \acro{QCD} is also the correct theory
for low energies where ordinary perturbation theory is not applicable
and the dominating degrees of freedom are better recast in an 
effective field theory. However a better understanding of the 
low-energy dynamics of \acro{QCD} and \emph{confinement} 
is still sought after.

Before the break-through of \acro{QCD} there was another 
candidate theory for the strong interaction, which could 
reproduce certain relations in the spectra of low energy
hadron physics: string theory. 

String theory\marginpar{string theory} describes particles as
oscillation modes of strings that propagate through space-time,
joining and splitting along their way, thus sweeping out a
two-dimensional surface, the \emph{world-sheet}.  The action of a
string is that of an idealised soap film; i.e.\ proportional to the
area of the world sheet.  Another interesting feature of the low
energy dynamics of hadrons is the formation of flux tubes between
quarks, which are also string like and even though nowadays perfectly
understandable from a pure \acro{QCD} point of view seemed
to hint at a connection between string theory and hadron physics.  As
will be seen later this connection does indeed exist in the form of
the \emph{\tHooft\marginpar{\tHooft\ expansion} large $N_c$
  expansion} \cite{tHooft:1973jz}, which was born in an attempt to
find a small parameter for perturbative calculations in the strong
coupling regime.  The basic idea is to look at $\gr{SU}(N_c)$
Yang--Mills theories, where $N_c$ is the number of
colours,\footnote{For $N_c=3$ this describes the pure glue part of
  \acro{QCD}.} and perform an expansion in $\frac{1}{N_c}$. This
implies at leading order the \emph{\tHooft\ limit} $N_c\to\infty$, where
additionally $\la := g_{YM}^2 N_c$ is kept fixed, with $g_{YM}$ the
Yang--Mills coupling constant. This particular choice is motivated by
keeping the strong coupling scale $\Lambda_{QCD}$ constant in a
perturbative calculation of the $\beta$ function.  In a double line
notation, the diagrams associated to each order in $\frac{1}{N_c}$ can
be seen to give rise to a topological expansion, which can be
interpreted as a triangulation of two dimensional manifolds, the
string world sheets in a genus expansion.  While this triangulation is
not understood in detail---see \cite{Gopakumar:2003ns} for recent
approaches to this important point---there is nevertheless a map
between a particular gauge theory and string theory in a certain
background.

This\marginpar{\adscft} map, tested by a large number of highly non-trivial checks, 
is \emph{Maldacena}'s conjecture \cite{Maldacena:1998re} of \adscftcorr. 
In its boldest form, it is the statement that $\cN=4$ super-Yang--Mills
(\acro{SYM}) theory, which is a \emph{conformal field theory} (\acro{CFT})
is \emph{dual} to (quantised) type \acro{IIB} string theory
on $\AdS_5 \times \mf{S}^5$. By ``dual'' the existence of a map
is meant that identifies correlation functions of both theories,
thus rendering them actually two different pictures of the same
theory. The details will be reviewed in Chapter~\ref{ch:adscft-overview}.
For now it is sufficient to remark that string theory in that particular
background is still ill-understood, but that there are limits in which 
things are better under control. In the string loop expansion, each
hole in the world sheet comes with a factor of $g_s$, while in a similar
gauge theory Feynman diagram each hole corresponds to a closed loop and
is therefore accompanied by a factor of $g_{YM}^2$. This \naive\ analysis
allows to identify $g_{YM}^2=g_s$, which therefore go to zero simultaneously
in the \tHooft\ limit, demonstrating that the $\frac{1}{N_c}$ expansion
corresponds to a genus expansion of the string world sheet.

From the construction of the $\AdS_5 \times \mf{S}^5$ background in
type \acro{IIB} supergravity (\acro{SUGRA}) theory, which is the small
curvature, low energy limit of type \acro{IIB} superstring theory, it is
possible to derive the relation $\pfrac{L}{\ell_s}^4 \sim \lambda$,
where $L$ is the respective curvature radius of the anti-de~Sitter
space ($\AdS_5$) and the five-sphere ($\mf{S}^5$), and
$\ell_s=\sqrt{\alpha'}$ is the string length.

Therefore, the limit of small curvature $L \gg \ell_s$, where type
\acro{IIB} supergravity on $\AdS_5 \times \mf{S}^5$ is a good
approximation of the corresponding string theory, is dual to taking
$\la$ large in the field theory.
Because $\la$ takes over the \role\ of the coupling constant in the
large $N_c$ limit, with $\la\ll1$ the perturbative regime, the duality
relates said supergravity theory to \emph{strongly coupled} $\cN=4$
\acro{SYM} theory in the large $N_c$ limit.  Since the discovery of the actual mapping
prescription between correlators on both sides of the correspondence
\cite{Witten:1998qj,Gubser:1998bc}, a plethora of non-trivial checks
have been performed \cite{Freedman:1999gp,Lee:1998bx,Nojiri:1998dh,Nojiri:1998yx,Nojiri:2000kh}, that did not
only extend the correspondence to less symmetric regimes but also
provided overwhelming evidence that the conjecture actually holds
true.

This\marginpar{\acro{QFT} coupled to \acro{SUGRA}} thesis is devoted
to studying the coupling between supergravity (\acro{SUGRA}) theories
and quantum field theories. Although the idea was revived by the
discovery of \adscft\ duality, where this coupling is realised
holographically, that is between a four and a five dimensional theory,
it has also been considered earlier in the context of space-time
dependent coupling constants \cite{Bogolyubov:1980nc, Becchi:1974,
  Epstein:1975gp}.

In the first part of this thesis several aspects of \adscftcorr\ 
will be discussed, while the second part uses the idea of
space-time dependent couplings to analyse the conformal anomaly
in super-Yang--Mills theories coupled to minimal supergravity.

Since at a first glance these two subjects seem rather unrelated,
I would like to linger on a bit on the question of what
the two topics have in common before continuing the introduction 
to those two parts.

The idea of space-time dependent couplings\marginpar{space-time
  dependent couplings} is to promote coupling constants to (external)
fields.  Generically the coupling takes the form $\int
d^4x\,\mathcal{J} \cO$, where $\mathcal{J}$ acts as a source for the
operator $\cO$.  A particularly important example for such a
source/operator pair is the metric and the \emt, which couple
according to
\begin{align*}
  S  \mapsto S + \int d^4x \, g^{mn} T_{mn},
\end{align*}
such that allowing coordinate dependence $g^{mn}=g^{mn}(x)$ amounts to coupling the
quantum field theory to a (classical) gravity background---or a
supergravity background for supersymmetric quantum field theories.
Invariance of the action under diffeomorphisms $\delta g^{mn} =
\mathcal{L}_v g^{mn}$ implies $\nabla^m T_{mn}=0$, while from Weyl
invariance ($\delta g^{mn}=2\sigma g^{mn}$) one may conclude
$T_m{}^m=0$. When quantum effects destroy the Weyl symmetry of a
classical theory, the trace of the energy-momentum tensor does not
vanish anymore. It is said to have an \emph{anomaly}: the Weyl or
trace anomaly, which is a standard example of a quantum anomaly. More
will be said about it below.

For\marginpar{\adscft\ mapping of correlation functions} now let us
have a look at the coupling of quantum field field theories to
supergravity from the \adscft\ point of view.  In the \adscftcorr, the
prescription for the calculation of \acro{CFT} correlators in terms of
\acro{SUGRA} fields is given by
\begin{align*}
  \vev{ \exp \int d^4x \, \phi^{(0)} \cO }_\text{CFT} &=
     \exp \bigl\{ - S_\text{SUGRA}[ \phi ] \bigr\} 
     \biggr|_{\phi({\p\AdS}) = \phi^{(0)}, }
\end{align*}
where the right hand side is the generating functional of the
classical supergravity theory, which is evaluated with its fields
$\phi$ determined by their equations of motion and their boundary
values $\phi^{(0)}$ that appear as sources for field theory operators in the
\acro{CFT}. 


\finalize{\begin{sloppypar}}
Much\marginpar{\AdS/\acro{QCD}?} 
of the excitement about the \adscft\ duality came from the
prospect of gaining insight into the strong coupling regime of
Yang--Mills (\acro{YM}) theories and \acro{QCD}. Both $\cN=4$ \acro{SYM} and type
\acro{IIB} \acro{SUGRA} are (almost) entirely determined by their
large symmetry group, namely $\gr{SU}(2,2|4)$. 
For the mapping of operators on both
sides, this is a beautiful feature, but non-supersymmetric \acro{YM}
has a much smaller field content and the problem arises how to get rid
of the extra fields. Furthermore to describe \acro{QCD} quarks are
needed but $\cN=4$ \acro{SYM} contains only one hypermultiplet whose
gauge field forces its adjoint representation on all other fields.
\finalize{\end{sloppypar}}

The conformal group $\gr{SO}(2,4)$ of the \acro{CFT} corresponds to
the isometry group of $\AdS_5$. Similarly the
$\gr{SU}(4)_R$ group is matched by the $\gr{SO}(6)$ isometry group of
the $\mf{S}^5$. Therefore a less supersymmetric \acro{CFT} will be
dual to a \acro{SUGRA} on $\AdS_5\times \mf{M}^5$, where $\mf{M}^5$ is a
suitable less symmetric manifold. Unfortunately the operator map
relies heavily upon the field theory operators being uniquely
determined by their transformational behaviour under the global
symmetry groups, such that reducing the symmetry implies making the
correspondence less precise.  This is especially true when also giving
up the conformal symmetry in order to obtain discrete mass spectra.

Therefore\marginpar{deformed \adscft}
the strategy employed in this thesis will be to describe
theories that are very symmetric in the \acro{UV} but are relevantly
deformed and flow to a less symmetric, phenomenologically more
interesting non-conformal infrared (\acro{IR}) theory.  This allows to still use
the established \adscftcorr\ while at the same time capturing
interesting \acro{IR} physics.

Such a renormalisation group (\acro{RG}) flow is represented by a
supergravity solution that approaches an \AdS\ geometry only towards
the boundary, it is \emph{asymptotically} \AdS.  The interior of the
deformed space corresponds to the field theoretic \acro{IR}.  The
interpretation of the radial direction of the (deformed) \AdS\ space
as the energy scale can be easily seen from considering dilations of
the boundary theory.  Since the boundary theory is conformal such a
dilation should leave the action invariant. To achieve the same in the
\acro{SUGRA} theory, the radial direction has to transform as an
energy to cancel in the metric the transformation of the coordinates
parallel to the boundary. The interpretation of the radial direction
as the renormalisation scale was introduced in
\cite{Henningson:1998gx, Akhmedov:1998vf} and has been used for a
number of checks of the \adscft\ duality, for example calculation of
the ratio of the conformal anomaly at the fixed points of holographic
\acro{RG} flows \cite{Freedman:1999gp,Karch:1999pv}, which coincides
with field theory predictions.

An\marginpar{quarks} important step towards a holographic description of \acro{QCD} is
the introduction of fundamental fields into the correspondence. The
first realisation of such a theory was a string theory in an $\AdS_5
\times \coset{\mf{S}^5}{\mathds{Z}_2}$ background where a number of D7
branes wrapped the $\mathds{Z}_2$ orientifold plane with geometry
$\AdS_5 \times \mf{S}^3$ \cite{Fayyazuddin:1998fb,Aharony:1998xz},
which is dual to an $\cN=2$ $\gr{Sp}(N_c)$ gauge theory. As was
realised by \cite{Karch:2002sh}, a similar scenario of probe D7\NB-branes
wrapping a contractible $\mf{S}^3$ in $\AdS_5\times \mf{S}^5$ leads to a consistent description
of an $\cN=2$ $\gr{SU}(N_c)$ theory, since a contractible $\mf{S}^3$
does not give rise to a tadpole requiring cancellation, nor to an unstable
tachyonic mode due to the Breitenlohner--Freedman bound
\cite{Breitenlohner:1982jf}.  (Further extensions of \adscft\ using D7
branes to include quarks have been presented in
\cite{Bertolini:2001qa,Grana:2001xn,Kruczenski:2003be,Myers:2006qr,
  Sakai:2003wu,Erdmenger:2005bj,Kirsch:2005uy,Arean:2006pk}.\footnote{Related
  models involving other brane setups may be found in
  \cite{Nunez:2003cf,Wang:2003yc,Casero:2006pt,Edelstein:2006kw,Casero:2005se,
    Hirayama:2006jn,Janik:2005zt,Canoura:2005uz,Benvenuti:2005qb}.})
The full string picture is that of a D3\NB-brane stack, whose near
horizon geometry gives rise to an $\AdS_5 \times \mf{S}^5$ space,
probed by parallel D7\NB-branes\marginpar{probe D7\NB-branes} 
wrapping and completely filling an
$\AdS_5 \times \mf{S}^3$ geometry. The strings connecting the two
stacks give rise to an $\cN=2$ hypermultiplet in the fundamental
representation. The resulting field theory is conformal as long as the
two brane stacks coincide. In this case the setup preserves an
$\gr{SO}(4) \times \gr{SO}(2)$ subgroup of the original $\gr{SO}(6)$
isometry, which is dual to an $\gr{SU}(2)_L \times \gr{SU}(2)_R \times
\gr{U}(1)_R$ subgroup of the $\gr{SU}(4)_R$.

Separating the two stacks introduces a quark mass and breaks conformal
symmetry as well as the $\gr{SO}(2) \simeq \gr{U}(1)_R$ symmetry.
Consequently the induced geometry on the D7\NB-branes becomes only
asymptotically $\AdS_5$.  At the same time, the $\mf{S}^3$ starts to
slip of the internal $\mf{S}^5$ when approaching the interior of the
$\AdS_5$ and shrinks to zero size. At that point the quarks decouple
from the \acro{IR} dynamics and the D7\NB-brane seems to end from a
five dimensional point of view.  By solving the Dirac--Born--Infeld
(\acro{DBI}) equations of motion  for the fluctuations of the D7
branes about their embedding the meson spectrum can be determined
\cite{Kruczenski:2003be}.  The setup is reviewed in more detail in
Chapter~\ref{ch:flavour}.

In\marginpar{deformed background geometry} 
Chapter~\ref{ch:dilatondriven}, I discuss how to combine the ideas
laid out above, that is to consider probe D7\NB-branes in background
geometries that only approach $\AdS_5 \times \mf{S}^5$ asymptotically.
The specific geometry under consideration is that of a dilaton flow by
Kehagias--Sfetsos and Gubser \cite{Kehagias:1999tr,Gubser:1999pk},
which preserves an $\gr{SO}(1,3) \times
\gr{SO}(6)$ isometry while breaking conformal invariance and
supersymmetry, thereby allowing chiral symmetry breaking by the
formation of a bilinear quark condensate.

In the framework of \adscftcorr\ all supergravity fields encode two
field theoretic quantities, a source and a vacuum expectation value
(\acro{VEV}). The embedding of a probe D7\NB-brane is determined by a
scalar field arising from the pullback of the ambient metric to the
world volume of the brane.  Solving the equation of motion for this
scalar field $\Phi$ yields the following \acro{UV} behaviour,
\begin{align*}
  \Phi \sim m_q + \frac{\langle \bpsi\psi \rangle}{\rho^2},
\end{align*}
where $\rho$ is the radial coordinate of the \AdS\ space, whose boundary
is approached for $\rho\to\infty$. 

Extending the solution to the interior of the space, it turns out that
generic combinations of the quark mass $m_q$ and the chiral condensate
$\langle \bpsi\psi \rangle$ do not produce solutions that have a
reasonable interpretation as a field theoretic flow; i.e.~are
expressible as a function of the energy scale $\rho$. I demonstrate
that this requirement is sufficient to completely fix the condensate
as a function of the quark mass.  In the limit of vanishing quark mass
there is a non-vanishing bilinear quark \acro{VEV} indicating that the
background is indeed a holographic description of spontaneous chiral
symmetry breaking.

I then determine the mass of the lowest scalar, pseudoscalar and
vector meson by calculating the fluctuations about the embedding
solutions.  Since the equations of motion for the D7 embedding in the
deformed background could only be solved numerically, the same holds
true for the fluctuations about these vacuum solutions.  Still the
spectrum is well understood because it approaches the analytic
solutions of the supersymmetric case in the limit of large quark mass.
This is to be expected since for larger quark mass, the corresponding
mesons decouple from the dynamics at high energies where supersymmetry
is restored.  I show that in the limit of vanishing quark mass, where
chiral symmetry is broken spontaneously, the pseudoscalar meson
becomes massless and is therefore a Goldstone boson for the axial
symmetry. For small quark mass $m_q$, the mass of the Goldstone mode
essentially behaves like $\sqrt{\smash[b]{m_q}}$ in accordance with
predictions from effective field theory.  

Moreover I discuss the spectrum of highly radially excited mesons
(as opposed to excitations on the $\mf{S}^3$, which are not in mutually 
same representations of $\gr{SU}(2)_L\times\gr{SU}(2)_R$).
It is explained why in this holographic setup (as in many others \cite{Schreiber:2004ie})
the field theoretic expectation \cite{Glozman:2004gk,Shifman:2005zn}
of chiral symmetry restoration cannot be met. The reason is 
the infrared being probed more densely in the limit of large 
radial excitations, which also has an interesting effect on the 
heavy-light spectra discussed below.

In\marginpar{non-trivial gauge background} 
Chapter~\ref{ch:higgsbranch} instead of considering a non-trivial geometry,
I discuss the effects of a non-trivial gauge field configuration on 
the brane. 
The spectrum of $N_f\ll N_c$ coincident D7\NB-branes is described by a
non-Abelian \acro{DBI} action plus Wess--Zumino term $C_{4} \wedge F
\wedge F$.  Both scalar and vector fields on the brane are now matrix
valued.  Assuming that the branes are coincident one may diagonalise
and obtain effectively $N_f$ copies of the spectrum of a single
brane---unless there is a contribution from the Wess--Zumino term.
This requires to choose a background configuration with 
non-trivial second Chern class; i.e.~an instanton solution,
which I demonstrate to indeed minimise the D7\NB-brane action. 

The string connecting the D7 and D3\NB-branes
 separated by a distance
$(2\pi\alpha')m_q$ introduces
a massive $\cN=2$ hypermultiplet in the fundamental representation,
which contributes the term $\tilde{Q}_i (m_q + \Phi_3 ) Q^i$ to 
the superpotential. $\tilde{Q}_i$ and $Q^i$ form the 
fundamental hypermultiplet and $\Phi_3$ is the chiral field that
is part of the adjoint $\cN=2$ gauge multiplet. The scalar component of
$\Phi_3$ is an $N_c\times N_c$ matrix. If some of its elements acquire
a \acro{VEV} such that $m_q + \Phi_3$ is zero, then the corresponding
components of the fundamental field may also get a \acro{VEV}
and the theory is on the mixed Coulomb--Higgs branch. 
I show that this Higgs \acro{VEV} corresponds to the instanton size
of above background  and calculate the spectrum 
of scalar and vector mesons as a function of the Higgs \acro{VEV}.
In the limit of vanishing Higgs \acro{VEV} I reproduce the analytic
spectrum of the $\gr{SU}(N_c)$ gauge theory. 
Not surprisingly there is a sense in which the spectrum of 
an infinitely large Higgs \acro{VEV} is equivalent since it belongs
to an $\gr{SU}(N_c-1)$ gauge theory.  I show that this equivalence
holds only up to a non-trivial rearrangement of the spectrum by a 
singular gauge transformation.\footnote{%
After the submission of the thesis, 
the results obtained in Chapter~\ref{ch:higgsbranch} 
have been generalized and extended by \cite{Arean:2007nh} beyond the
approximation employed here; see \cite{Arean:2006vg}
for related earlier works on defect conformal field theories.
}

In\marginpar{heavy-light mesons}
Chapter~\ref{ch:heavylight} mesons consisting of a light 
and a heavy quark are discussed. 
A \naive\ approach would be to use the non-Abelian \acro{DBI}
action, where the diagonal elements of the matrix valued 
scalar field now encode a mass and bilinear condensate
for each of the corresponding $N_f$ quarks. Off-diagonal elements
of the embedding solution would contain mass-mixing terms and 
mixed condensates, which one could set to zero for phenomenological
reasons.
Fluctuations about these embeddings would correspond to the ordinary
same-quark meson for the diagonal elements and to heavy-light mesons
for the off-diagonal entries. 
However the latter are not small with respect to the corresponding
light quark and expansion of the \acro{DBI} action to quadratic order 
is not possible anymore. This step however is crucial to obtain 
an eigenvalue equation for the meson mass.

The approach chosen here is to find an effective description for
heavy-light mesons from the Polyakov action of the string stretched
between \emph{two} D7\NB-branes with different separation from the D3
branes corresponding to two different quark masses.  The separation
is assumed to be large (that is only one quark is heavy, the light
quark is taken massless), such that a semi-classical description of
this long string is possible.  I take the ansatz of a rigid string
spanned in the direction of the separation of the two branes. The
string is not allowed to oscillate or bend but only to move along the
world volume of the D7s.  Then integration over the string length
can be carried out to obtain an effective point-particle-like action.
Its equation of motion is a generalisation of the Klein--Gordon
equation which can be quantised.  I evaluate the resulting eigenvalue
equation for the undeformed \AdS\ background as well as dilaton
deformed backgrounds by Gubser and Kehagias--Sfetsos \cite{Gubser:1999pk,Kehagias:1999tr} and
Constable--Myers \cite{Constable:1999ch}. 

The heavy-light meson spectrum for both deformed geometries approximates
the \AdS\ heavy-light spectrum for large quark mass. This behaviour
is expected because a large quark mass corresponds to the string probing larger
parts of the space-time that are approximately \AdS. 
At the same time, it can be observed that highly excited mesons converge
more slowly to their \AdS\ values. Again this is in accordance with previous
results of Section~\ref{sec:highexcite}, where it has been demonstrated
that highly excited mesons probe the \acro{IR} region of the space time
more densely, where the deviation from the \AdS\ geometry is large.

These\marginpar{B physics} 
heavy-light spectra can be used to determine the mass
of the B~meson by using the results of Chapter~\ref{ch:dilatondriven} as well as the 
experimental values of the Rho and Upsilon meson mass 
to fix the confinement scale and heavy quark mass. 
The prediction for the B~mesons is 20\% above the experimental
value. Since the B~mesons are far in the supersymmetric regime
of this holographic model while at the same time the field
theory is strongly coupled at that scale, this level of agreement is 
surprisingly good.

\finalize{\begin{sloppypar}}
The\marginpar{summary}
\acro{AdS}/\acro{CFT} models I considered here describe chiral symmetry breaking, 
highly excited mesons, the Higgs branch and heavy-light mesons, 
respectively. They have in common that they are not focused on building 
a perfect \acro{QCD} dual, but instead are used to investigate
particular features of \acro{YM} theory with matter. 
The strategy of keeping a connection to standard \adscft\ with flavours
worked out and the results show either the qualitative behaviour
expected from field theoretic and \acro{SUGRA} considerations or could  
even be matched quantitatively to analytic results in certain limits. 
\finalize{\end{sloppypar}}

\horizrule

\textsf{
As already mentioned this thesis consists of two parts. In the first
part presented so far various aspects of \adscftcorr\ have been
discussed and a number of models extending the \adscftcorr\ to
theories with fundamental quarks have been developed and explored.
The second part is devoted to an analysis of the conformal anomaly in
super-Yang--Mills theories coupled to minimal supergravity in four 
space-time dimensions. This analysis is aimed at providing building
blocks for a future generalisation of the two dimensional
$c$\NB-theorem, see below, to four-dimensional supersymmetric field theories.
}

\horizrule

The\marginpar{trace anomaly} 
conformal anomaly expresses the breaking of conformal invariance
in a classically conformal field theory by quantum effects.
It arises as the trace of the energy-momentum tensor, which---as 
mentioned above---vanishes in a conformally invariant theory, 
and is also called
\emph{trace anomaly}, hence.

An\marginpar{$c$\NB-theorem} 
investigation of the trace anomaly is interesting because
of its potential relation to a four dimensional version of
Zamolodchikov's $c$\NB-theorem \cite{Zamolodchikov:gt}. 
The $c$\NB-theorem is a statement about the irreversibility of
renormalisation group flows connecting two fixed points 
of a quantum field theory in two space-time dimensions.
To be more precise the theorem states the existence of a monotonic
function that at the fixed points, where the $\beta$ functions vanish,
coincides with the trace anomaly coefficient $c$ defined by
\begin{align*}
  \vev{T_m{}^m} &= \frac{c}{24\pi} \cR, 
\end{align*}
where $\cR$ is the scalar curvature. Moreover the coefficient $c$ 
turns up as the central charge of the Virasoro algebra
and in the two point function of the \emt. 

The $c$\NB-theorem is also interesting from a philosophical point of
view, because the $c$-function is interpreted to measure the number of
degrees of freedom along the \acro{RG} flow.  Suppose that one
believes that in the real world this number should be non-increasing
when going to lower energies, a future ``theory of everything'' should
certainly incorporate a function that measures these degrees of
freedom and is monotonic hereby.  While it is not clear that such an
irreversibility theorem should be realised in terms of a
$c$\NB-theorem, the questions remains if there is a class of theories
in four dimensions where an analogous statement to the two dimensional
$c$\NB-theorem can be made. Such\marginpar{4D} a generalisation is
not straight forward since conformal symmetry in four dimensions is
far less powerful because the conformal algebra contains only a finite
number of generators.

In\marginpar{from $a$ to $c$} four dimensions the trace anomaly reads
\begin{align} \label{eq:intro-anomaly}
  \vev{ T_m{}^m } = c\, C^2 - a\, \tilde{\cR}^2 + b\, \cR^2 + f\, \Box \cR, \tag{$\star$}
\end{align}
with $C^2$, $\tilde{\cR^2}$ and $\cR^2$ respectively the
square of the Weyl tensor, the Euler density 
and the square of the Ricci scalar $\cR$.
The first question that arises is which of these
coefficients is to take over the \role\ of the 
two dimensional $c$. 
While $f$ can be removed by adding a local counterterm to the 
quantum effective action, 
$c$ is known to be increasing in some theories
and decreasing in others and $b$ is eliminated by Wess--Zumino consistency
conditions. 
For the remaining coefficient, conventionally denoted ``$a$'', 
there is no known counterexample to $a_\text{UV} > a_\text{IR}$,
though explicit checks can only be performed in certain 
classes of supersymmetric field theories \cite{Anselmi:1997am,Anselmi:1997ys}.
This might be an indication that supersymmetry is a necessary
ingredient for such an $a$\NB-theorem.
The prospect of an $a$\NB-theorem \cite{Cardy:1988cw} has attracted some
interest in the recent past under the name $a$\NB-maximisation
\cite{Intriligator:2003jj}. 

In\marginpar{space-time dependent couplings}
this thesis a different approach inspired by an alternative
proof of the $c$\NB-theorem in two dimensions is chosen \cite{Osborn:1991gm}.
The author of \cite{Osborn:1991gm} couples a quantum field
theory that is conformal to a classical gravity background and investigates
the anomaly arising from that coupling by promoting 
the coupling constants $\la$ to external fields $\la(x)$.

\finalize{\begin{sloppypar}}
This trick yields well-defined operator insertions from 
functional derivations of the generating functional 
with respect to the couplings. A generalisation 
of the Callan--Symanzik equation to Weyl rescalings
is found, which becomes anomalous when Weyl symmetry
is broken upon quantisation. The structure of this
equation is $\Delta_\sigma W = \cA$, where $\Delta_\sigma$
contains a Weyl scaling part and a $\beta$ function part in
analogy to the case of constant couplings and constant scale 
transformations. 
\finalize{\end{sloppypar}}

The\marginpar{anomaly ansatz} shape of the anomaly $\cA$ is
determined by dimensional analysis, yielding an ansatz that is a linear
combination between a set of coefficient functions, which only depend
on the couplings, and a set of basis terms, which depend on the
curvature and derivatives of the couplings. There is only a finite
number of possible basis terms and their coefficient functions can be
perturbatively determined for a particular theory.


Without\marginpar{Wess--Zumino consistency} resorting to a particular
theory, one may nevertheless find constraints between the coefficients
arising from a Wess--Zumino consistency condition
\begin{align*}
  \comm{ \Delta_\sigma }{ \Delta_{\sigma'} } W = 0.
\end{align*}

In two dimensions this consistency condition implies $\beta^i \p_i ( c
+ w_i \beta^i ) = \chi_{ij} \beta^i \beta^j$, where $c$ is the central
charge and $w_i (\la^k)$ and $\chi_{ij}(\la^k)$ are above mentioned
coefficient functions.  $\chi_{ij}$ can be related to the positive
definite Zamolodchikov metric, which is the key ingredient for the
definition of a monotonic $c$-function.



In\marginpar{failure} the four-dimensional case it is such a relation
to a positive definite object that is missing. In particular the
analogous consistency condition for the $a$ coefficient in the four
dimensional trace anomaly \eqref{eq:intro-anomaly} reads
\begin{align*}
  \beta^i \p_i ( a + \tfrac{1}{8} w_i \beta^i ) 
    = \tfrac{1}{8} \chi^g_{ij} \beta^i \beta^j,
\end{align*}
where $\chi^g_{ij}(\la^k)$ is one of the (many) coefficients in the
four-dimensional anomaly ansatz.  There \emph{is} a relation to a
positive definite coefficient $\chi^a$, $\chi^g_{ij} = 2 \chi^a_{ij} +
\text{(other terms)}$, but it is spoiled by the occurrence of extra
terms.

In supersymmetric theories, some of these extra terms are known to
vanish and there might be hope that additional constraints arise from
a local \acro{RG} equation incorporating super-Weyl transformations
that allow the construction of a monotonic $a$-function.  Before
tackling this ambitious task, a first step is to analyse the trace
anomaly in a supersymmetric framework, which is what has been pursued
in the second part of this thesis.

In\marginpar{contents} Chapters \ref{ch:sugra-overview} and
\ref{ch:local-couplings} respectively, I give an introduction to
minimal supergravity in an $\cN=1$ superfield formulation and to the
non-super\-symmetric local renormalisation group technique outlined
above.

In Chapter \ref{ch:susy-trace}, I present superfield versions of the
local \acro{RG} equation, give a complete ansatz for the trace
anomaly, and determine the full set of consistency equations.  I then
discuss the $\cN=4$ case, which gives rise to an interesting puzzle:
In \cite{Osborn:2003vk} by a component approach 
a one-loop result for the trace anomaly of $\cN=4$ \acro{SYM} 
was found to contain a 
conformally covariant operator of fourth order, the Riegert operator
\cite{Riegert:1984kt}, which is reviewed in Section
\ref{sec:riegert-review}. In \cite{Fradkin:1985am} a
supersymmetric version of this operator is given in components, but I
was not able to find a satisfactory superfield version of this
operator. A superfield Riegert operator is known to exist in
new-minimal supergravity \cite{Manvelyan:1995hz}, which however in general is known to be
inconsistent on the quantum level \cite{deWit:1985bn,Shamir:1992ff}.
I discuss the possible origin of that problem, which I
suspect to arise from the impossibility to separate
local $\gr{U}(1)_R$ transformations from super-Weyl transformations
in the minimal supergravity formulation such that a too strong 
symmetry requirement is imposed on the ansatz.\footnote{In new-minimal
supergravity this problem does not arise because $\gr{U}(1)_R$ is
indeed a local symmetry of the theory.}
Nevertheless the extended calculations presented here should provide
a good starting point for further exploration of this fascinating
topic. In the conclusions possible future steps are discussed.




  \Part{Generalizations of AdS/CFT}

  \begin{savequote}[\savequotewidth]
  ``After all, all he did was string together a lot of old, well-known quotations.'' 
  \qauthor{H.\ L.\ Mencken, on Shakespeare}
\end{savequote}

\Chapter{Overview\label{ch:adscft-overview}}

\Section{\acro{QCD}}
The gauge theory of the strong interaction, quantum chromodynamics
(\acro{QCD}), is based on the success of the parton model
\cite{Bjorken:1968dy,Feynman:1973xc}, which describes the
high-energy behaviour of hadrons as bound states of localised but
essentially free particles, to describe the high-energy
hadron spectrum. The other key ingredient
was to realise that an additional \emph{hidden} three-valued quantum
number, \emph{colour}, is needed.

The former means that the theory should be asymptotically free;
i.e.\ the coupling constant becomes small in the ultraviolet regime
(\acro{UV}).  This requirement is only met by Yang--Mills theories,
that means non-Abelian gauge theories.

The latter (hiding the colour) makes plausible a colour dependent
force to form colour singlets only, such that one may assume the
colour symmetry (as opposed to the flavour symmetry) to be gauged. Indeed
lattice calculations demonstrated that \acro{QCD} is \emph{confining},
such that the formation of colour singlets is a consequence of the
dynamics.

The \acro{QCD} Lagrangean\marginpar{\acro{QCD} Lagrangean} describes
an $\gr{SU}(N_c)$ Yang--Mills theory with $N_c=3$ the number of colours
and $N_f=6$ the number of quarks, with a global $\gr{SU}(N_f)_L\times
\gr{SU}(N_f)_R\times \gr{U}(1)_V \times \gr{U}(1)_A$ symmetry that is partly broken by the different mass of the
six quarks, cf.~Table~\ref{tab:quarkmass}.
\begin{table}
\centering
\begin{colortabular}{Cc|Cc|Cc|Cc|Cc}
  \multicolumn{5}{Hc}{\textsc{Quark Masses}}\\
  \cell{c}{Type}     & \cell{c}{$Q$} & \multicolumn{3}{c}{Generations} \\
\bwline
  &&  \textbf{u}       & \textbf{c}            & \textbf{t} \\[-1ex]
  \raisebox{2ex}{up}   & \raisebox{2ex}{$\frac{2}{3}$}  
       & $1.5$ to $4$ MeV & $1.15$ to $1.35$ GeV  & $169$ to $179$ GeV \\ \hline
  && \textbf{d}       & \textbf{s}            & \textbf{b} \\[-1ex]
  \raisebox{2ex}{down} & \raisebox{2ex}{$-\frac{1}{3}$} 
        & $4$ to $8$ MeV   & $80$ to $130$ MeV     & $4.1$ to $4.4$ GeV  \\
\bwline
\end{colortabular}
\caption[Quark Masses (\acro{PDG})]{Quark masses (Particle Data Group \cite{PDBook})
\label{tab:quarkmass}}
\end{table}
It is given by
\begin{align}
  \Lag_\text{QCD} &= - \half \tr F_{mn} F^{mn} + \sum_i^{N_f} \bar{q}_i (i\gamma^m D_m - m_i) q_i  \label{eq:qcdlag}\\  
  F_{mn} &= \p_m A_n - \p_n A_m + i\,g \comm{A_m}{A_n} \notag \\  
  D_m q_i &= (\p_m - i\,g A_m) q_i \notag \\
  A_m &= A_m{}^a T^q \notag \\
  \comm{T^a}{T^b} &= i\,f^{abc} T^c \notag
\end{align}
The
$N_c^2-1=8$ fields $A_m{}^a$ are called \emph{gluons}, the $N_f=6$
quark fields $q_i$ are the Dirac fermions $u,d,s,c,b,t$.  The global
flavour symmetry is explicitly broken by (the inequality of) the
masses $m_i$, though they can be assumed to be realised approximately for
the isospin group $\gr{SU}(2)_f$ or even (including the strange quark)
$\gr{SU}(3)_f$.
The corresponding transformation and algebra as well as 
Noether current and charge read
\begin{align}
  \delta q^i &= i \alpha^a t^a_{ij} q_j, &
  \comm{t^a}{t^b} &= i f^{abc} t^c, \notag \\
  J^a_\mu &= \bar{q}_i \gamma_\mu t^a_{ij} q_j, \\
  Q^a  &= \int d^3x J^a_0, &
  \comm{Q^a}{Q^b} &= i f^{abc} Q^c, \notag
\end{align}
where for $SU(3)_f$ the generators $t^a=\frac{\la^a}{2}$ 
are usually expressed by the eight Gell-Mann matrices $\la^a$.

Furthermore\marginpar{axial transformation}
the Lagrangean is invariant under an overall $\gr{U}(1)_V$
vector symmetry $q \mapsto \e^{i \alpha} q$, often
also referred to by \emph{baryon number} symmetry.
The massless version of \eqref{eq:qcdlag} 
is in addition invariant under the $\gr{U}(1)_A$
\emph{axial transformations}
$q \mapsto \e^{i \beta \gamma_5} q$ giving rise to a
second copy of the flavour symmetry group,
\begin{align}
  \delta q^i &= i \alpha^a t^a_{ij} q_j,  &
  J^{5\,a}_\mu &= \bar{q}_i \gamma_\mu t^a_{ij} q_j, \\
  Q^{5\,a}  &= \int d^3x J^{5\,a}_0, &
  \comm{Q^{5\,a}}{Q^{5\,b}} &= i f^{abc} Q^{5\,c}.  
\end{align}
Together they form the chiral symmetry group
$\gr{SU}(N_f)_L \times \gr{SU}(N_f)_R$, whose 
generators and corresponding algebra are given by
\begin{align}
  Q^a_L &= \half( Q^a - Q^{5\,a} ), &
  Q^a_R &= \half( Q^a + Q^{5\,a} ), \notag \\
  \comm {Q^a_L}{Q^b_L} &= i f^{abc} Q^c_L,&
  \comm {Q^a_R}{Q^b_R} &= i f^{abc} Q^c_R, \\
  \comm {Q^a_L}{Q^b_R} &= 0. \notag
\end{align}
When switching on mass terms this symmetry is not exact
anymore and the associated charges, while still obeying the algebra,
are not conserved; i.e.~become time dependent.

\Section{\texorpdfstring{$\cN=4$ Super-Yang--Mills}{N(susy)=4 Super-Yang-Mills} Theory}
While classically Yang--Mills theories are conformally
invariant, this is no longer true upon quantisation and
the conformal symmetry becomes anomalous.  It turns out that it is
actually quite hard to find a field theory that is conformally invariant
on the quantum level and it comes as a surprise that $\cN=4$
\acro{SYM}, whose formulation was first achieved by compactifying ten
dimensional $\cN=1$ \acro{SYM} on a six dimensional torus, 
actually preserves a larger symmetry
group than its higher dimensional ancestor and has
vanishing $\beta$ functions to all orders in perturbation theory \cite{Sohnius:1981sn}.  

Consequently from the commutators of supercharges and the
generator of special conformal transformation, an additional set of
(so-called conformal) supercharges is generated. From the perspective
of \adscftcorr{} this doubling of supercharges is quite important
since $\cN=4$ has therefore the same number of supercharges as five
dimensional maximally supersymmetric super\emph{gravity}. The full
superconformal algebra is $\gr{SU}(2,2|4)$, where its bosonic
subgroups are $\gr{SU}(2,2) \simeq \gr{SO}(2,4)$, the conformal
group in four dimensions, and $\gr{SU}(4)_\text{R}$, the
R\NB-symmetry group. 

Being\marginpar{multiplets}
maximally supersymmetric, $\cN=4$ \acro{SYM} consists entirely
of one multiplet, the $\mathcal{N}=4$ gauge multiplet. In $\cN=1$
language, this corresponds to one gauge multiplet plus three chiral
multiplets.\footnote{In an attempt to embrace both naming
  conventions used in \acro{SUSY}, \emph{multiplets} are denoted
  chiral, gauge or hyper in conjunction with the number of
  supersymmetries. \emph{Super fields} on the other hand shall always
  mean $\cN=1$ language and will be distinguished by their constraint
  (none, chiral, real, linear) and transformation behaviour of the
  lowest component (scalar, spinor, vector, tensor, density).  
} So the field content is one vector, four chiral fermions and three complex
scalars. 
As the gauge and \acro{SUSY} generators commute, all fields are in
the adjoint representation. Two of the chiral superfields form an
$\cN=2$ hypermultiplet, while the other chiral superfield together with
the $\cN=1$ gauge multiplet forms an $\cN=2$ gauge multiplet.
 
In $\mathcal{N}=1$ superfield language the Lagrangean\marginpar{Lagrangean}
reads 
\begin{equation}
  \Lag 
    = \int d^4\theta \tr \left( \bar{\Phi}^i \e^{2V} \Phi^i \e^{-2V} \right) 
    + \biggl[ \frac{1}{4g^2} \int d^2\theta \, W_\alpha W^\alpha 
                           + \int d^2\theta \, W + \text{c.c.} \biggr],
\end{equation}
where the gauge field strength is given by 
$W_\alpha = - \frac{1}{8} \bar{D}^2 (\e^{-2V} D_\alpha \e^{2V})$
and the superpotential is
\begin{align}
  W = \tr \Phi^3 \comm{\Phi^1 }{\Phi^2 }.
\end{align}

%
%

\Section{Type \acro{IIB} Supergravity\label{sec:iib-sugra}}
There are only two maximally supersymmetric supergravity theories in
ten dimensions, called type \acro{IIA} and type \acro{IIB}. Both are $N=2$
\acro{SUGRA}s and contain (among others) two chiral gravitini, but
\acro{IIA} is non-chiral in the sense that these fermions have
opposite chirality while \acro{IIB} has gravitini of the same
chirality.  The particle content of the latter is given by Table~\ref{tab:IIB}.


\begin{table}
\centering
\begin{colortabular}{Cc|Cc|Cc}
  \multicolumn{3}{Hc}{\textsc{\acro{IIB} \acro{SUGRA} Particle Content}}\\
  \cell{c}{Symbol} & \cell{c}{\#\acro{DOF}}  & \cell{c}{Field} \\
\bwline
  $G_{AB}$               &  $35_B$ & metric --- graviton \\ \hline
  $C + i\vphi$            &  $ 2_B$ & axion --- dilaton \\ \hline
  $B_{AB} + i C_{2\,AB}$  &  $56_B$ & rank 2 antisymmetric\\ \hline
  $C_{4\,ABCD} $          &  $35_B$ & antisymmetric rank 4 \\ \hline
  $\psi _{A \alpha}^{1,2}$ & $112_F$ & two Majorana--Weyl gravitini \\ \hline
  $\lambda_\alpha^{1,2}$  &  $16_F$ & two Majorana--Weyl dilatini\\ \bwline
\end{colortabular}
\caption[\acro{IIB} \acro{SUGRA} Particle Content]{\acro{IIB} \acro{SUGRA} Particle Content \protect\cite{D'Hoker:2002aw}
\label{tab:IIB}}
\end{table}

\acro{IIB} contains a self-dual five-form field $\tilde{F}_5 := F_5 -
\half C_2 \wedge H_3 + \half B\wedge F$, $F_5 := dC_4$, which makes it
hard to write down an action from which all equations of motion may be
derived.\footnote{See
  \protect\cite{Dall'Agata:1998wh,Dall'Agata:1998va} for recent
  attempts to improve this situation.}

Often in the literature \cite{Polchinski:1998rr,D'Hoker:2002aw}, the
following action is used,\footnote{The conventions employed here are:\\
  $A_p=\frac{1}{p!} A_{A_1\dots A_p}$, $(dA_{p+1})_{A_1\dots A_{p+1}}
  = (p+1) \p_{[A_1} A_{A_2\dots A_{p+1}]}$, and\\
  $|F_p|^2=\frac{1}{p!} F_{A_1\dots A_p}F^{A_1\dots A_p}$.  }
augmented by the self-duality condition $\tilde{F}_5 = \hodge
\tilde{F}_5$, which has to be imposed additionally on the equations of
motion and where $\hodge$ denotes the Hodge dual.\marginpar{\acro{IIB}
  action}
\begin{align}
  S_\text{IIB} &= 
     \frac{1}{2\kappa^2} \int d^{10}x \sqrt{G_\text{E}} \bigl\{ 
        R_\text{E} 
        - \frac{ \p_A \bar\tau\, \p^A \tau }{2(\im \tau)^2}
        - \quart \abs{F_1}^2 
        - \half \abs{G_3}^2 
        - \quart\abs{\tilde F_5}^2 \bigl\}\nonumber \\
  &\relphantom{=}
    - \frac{1}{4i\kappa^2} \int C_4 \wedge \bar G_3 \wedge G_3,
\end{align}
where the expressions in order of appearance are the determinant of
the metric, the Ricci scalar $R_{E}$, axion--dilaton field $\tau := C +
i\e^{-\vphi}$ composed of the axion $C$ and the dilaton $\vphi$, field
strength $F_1:=dC$ and $G_3 := \sqrt{\im \tau} (F_3 - i H_3)$ with
$F_3 := dC_2$ and $H_3=dB$.  The complex objects have been introduced
to make manifest an additional rigid $\gr{SL}(2,\mathds{R})$
\marginpar{$\gr{SL}(2,\mathds{R})$} symmetry of type \acro{IIB}
\acro{SUGRA}, which transforms
\begin{align}
   \tau &\mapsto \frac{a\tau+b}{c\tau+d}, &
   \det \begin{pmatrix}a&b\\c&d\end{pmatrix} &= 1, \label{eq:sl-two-R}\\
   G_3 &\mapsto  \frac{c\bar\tau+d}{\abs{c\tau+d}} G_3,
\end{align}
and leaves invariant the other fields.

Many\marginpar{equations of motion} also prefer to follow the
historic approach \cite{Green:1982tk, Howe:1983sr, Schwarz:1983wa,
  Schwarz:1983qr, Green:1987mn} of writing down the equations of
motion only, which restricted to the graviton, axion, dilaton, and
four-form Ramond--Ramond potential read:
\begin{align}\label{eq:IIB-eom}
\begin{split}
  R_{AB} &= \e^{2\vphi} \p_A C\, \p_B C + \p_A \vphi\, \p_B \vphi \\
        &\relphantom= + \tfrac{1}{2 \cdot 4!} \tilde F_{A C_2 \dots C_5} \tilde F_B{}^{C_2\dots C_5},
\end{split} \notag\\
  \nabla_A \nabla^A C &= -2 (\nabla_A C)(\nabla^A \vphi), \\
  \nabla_A \nabla^A \vphi &= \e^{2\vphi} (\nabla_A C)(\nabla^A C),  \notag\\
  \p_{[A_1} (C_4)_{A_2\dots A_5]} &= 
     \veps_{A_1\dots A_5}{}^{A_6\dots A_{10}} 
     \p_{A_6} (C_4)_{A_7\dots A_{10}}, \notag
\end{align}
\finalize{\begin{sloppypar}}
where by convention the total anti-symmetric Levi-Civita symbol 
takes values $\pm \sqrt{-\det G_E}$ for all indices lowered
(and accordingly $\pm \sqrt{-\det G_E}^{-1}$ for all indices
raised).
\finalize{\end{sloppypar}}


\Subsection{\texorpdfstring{$p$-brane}{p-brane} Solutions\label{sec:p-branes}}
There is a particular class of solutions to the supergravity equations
of motion \eqref{eq:IIB-eom} that preserve half of the supersymmetry
and the subgroup $\gr{SO}(1,p)\times\gr{SO}(9-p)$ of the
ten dimensional Lorentz group. Additionally they have a non-trivial
$C_{p+1}$ charge coupled to the supergravity action by
\begin{align}
  S_{p} \sim \int dC_{p+1}.
\end{align}
These solutions are called $p$\NB-branes.  They
\marginpar{$p$\NB-brane ansatz} are determined by the ansatz
\begin{align}
  ds^2 &= H(y)^\ia \eta_{\mu\nu} dx^\mu dx^\nu 
          + H(y)^\ib (dy^2 + y^2 d\Omega_{5}^2)
\end{align}
with $\eta_{\mu\nu}$ the $(p+1)$\NB-dimensional Minkowski metric,
$d\Omega_{8-p}^2$ the line element of the $(8-p)$\NB-dimensional
unit sphere and constants $\ia$, $\ib$ to be determined by the
equations of motion. 
The directions $x$ are referred to as world-volume or longitudinal
coordinates, while $y$ are called transversal.

Since\marginpar{$3$\NB-brane solution}
to this thesis, the most relevant $p$\NB-branes are $3$\NB-branes,
their full solution in terms of bosonic supergravity fields is given,
\begin{align} \label{eq:d3sol}
  ds^2 &= H(y)^{-1/2} \eta_{\mu\nu} dx^\mu dx^\nu 
        + \mrlap{H(y)^{1/2} (dy^2 + y^2 d\Omega_{8-p}^2),} \notag \\
  \Phi &= \Phi_0 = \text{const}, &   C &= \text{const}, \notag \\
  B_{AB} &= C_{2,AB} = 0, \\
  C_4 &= H(y)^{-1} dx^0 \wedge \dots \wedge dx^3,\notag\\
  H(y) &= 1 + \sum_i \frac{L^4}{\abs{\vec{y}-\vec{y}_i}}, &
  L^4 &= 4\pi g_s N \alpha^{\prime 2}, \notag
\end{align}
for a distribution of 3-branes at positions $y_i$.  Close to the
origin of a single brane $\abs{\vec{y}-\vec{y}_i}\ll L^4$, the $1$ in
the warp factor\marginpar{near-horizon geometry} can be neglected such that
the geometry becomes approximately $\AdS_5 \times \mf{S}^5$.

\Section{D\NB-branes}
A D$p$\NB-brane is a $(p+1)$\NB-dimensional hypersurface in the target
space of string theory, where open strings can end
\cite{Polchinski:1995mt,Polchinski:1996fm}.  Their discovery
integrates some features of superstring theory and supergravity that
would have been puzzling without them.  Firstly,\marginpar{boundary
  conditions} the open string admits two kinds of boundary conditions,
\begin{alignat*}{4}
  &\text{Dirichlet}\qquad & X^i(\tau,\sigma) &= \text{const}, \\
  &\text{Neumann}\qquad   & \p_\sigma X^i(\tau,\sigma) &= 0.
\end{alignat*}

However from a \naive\ point of view, Dirichlet boundary conditions
have to be considered unphysical as they break Lorentz invariance
and---worse---make the open strings loose momentum trough their
endpoints. With the discovery of T\NB-duality
\cite{Kikkawa:1984cp,Dai:1989ua, Giveon:1994fu, Alvarez:1993qi} it
became apparent that one could transform from one kind of boundary
condition to the other and it was no longer possible to exclude
Dirichlet boundary conditions a priori.  In the D\NB-brane picture,
momentum conservation can be restored by assuming the D\NB-branes as
dynamical objects can absorb the above mentioned momentum flow.

Secondly, $p$\NB-brane solutions\footnote{$p$-branes are domain wall solutions 
  of \acro{SUGRA}, see Section~\ref{sec:p-branes} for details.
}  of \acro{SUGRA} are interpreted as 
the low energy effective objects corresponding to D$p$-branes.

Thirdly, it was realised early \cite{Paton:1969je}, that it is possible
to attach gauge group factors to the end points of open strings.
These Chan--Paton factors have a natural explanation as encoding which
brane in a stack of coincident branes the string is attached to.

\Subsection{Abelian}
For a single D$p$\NB-brane this factor is a $\gr{U}(1)$ in accordance
with the fact, that the massless modes of open string theory form a
$(p+1)$\NB-dimensional $\gr{U}(1)$ \acro{SYM} with one vector, $9-p$
real scalars, whose \acro{VEV}s describe the position of the brane,
and fermionic superpartners, which shall be ignored in the following.
For constant field strengths $F_{ab}$, $F=\frac{1}{2} F_{ab} \,dX^a
\wedge dX^b$, by resummation it is possible to determine the action to
all orders in $\alpha'$ \cite{Leigh:1989jq} to be the first
(Dirac--Born--Infeld, \acro{DBI})\marginpar{Dirac--Born--Infeld} part of
\begin{align} \label{eq:abelian-dbi}
\begin{split}
  S_\text{D$p$} = 
    &- T_p \int d^{p+1}\xi\, \e^{-\vphi} 
               \sqrt{-\det P[G+B]_{ab} + 2\pi\alpha' F_{ab}} \\
    &\pm T_p \int P\bigl[\tsum C_n \e^B \bigl] \e^{2\pi\alpha' F},
\end{split}
\end{align}
which couples the brane to the massless Neveu--Schwarz (\acro{NS})
sector of closed string theory while the second (Wess--Zumino,
\acro{WZ})\marginpar{Wess--Zumino} part determines the coupling of
the brane to the massless Ramond--Ramond (\acro{RR}) sector. 
The index conventions are depicted in Table~\ref{tab:brane-indices},
while the fields are explained in Section~\ref{sec:iib-sugra}.

\begin{table}
\centering
\begin{colortabular}{*{2}{Bc}}
	\multicolumn{2}{Hc}{\textsc{Index Conventions}}\\ 
	\rowcolor{white}
	longitudinal&transversal \\ \bwline
	\cell{Cc|}{$X^{a,b,\dots}$}&\cell{Cc}{$X^{i,j}$}\\ \hline
	\multicolumn{2}{Cc}{$X^{A,B,\dots}$}\\ \bwline
\end{colortabular}
\caption[Transversal vs.\ Longitudinal Coordinates]{Index conventions for ambient space, world volume
and transversal coordinates
\label{tab:brane-indices}}
\end{table}

The prefactor $T_p$ is given by
\begin{align}
  T_p = \frac{2\pi}{g_s (2\pi \ell_s)^{p+1}}, 
\end{align}
with $g_s$ the string coupling and $\ell_s$ the string length.

Throughout this thesis, for explicit calculations the Kalb--Ramond
field will be assumed to vanish.  As will be commented on below, the
Wess--Zumino term allows coupling to---with respect to the brane's world
volume---lower dimensional \acro{RR} potentials
\emph{if the gauge field has a non-trivial Chern class}.  The only
\acro{RR} potential in the backgrounds discussed here, will be
$C_4$ associated to the five-form flux always present in the \adscftcorr.
In the particular case of a D7\NB-brane, the Wess--Zumino term then
reads
\begin{align}
  S_{D7-WZ} = T_p \int d^8\xi\, P[C_4] \wedge F \wedge F. \label{eq:d7wz}
\end{align}

\Subsection{Non-Abelian}
$N$ parallel D\NB-branes describe a $\gr{U}(1)^N$ gauge theory. When
these branes approach one another, strings stretched between different
branes become light and the gauge symmetry is promoted to $\gr{U}(N)$.
Generalising to the case of $\gr{U}(N)$ is straight forward in the
case of D9\NB-branes,\footnote{ Apart from the additional complication
  of finding the correct series expansion, which is non-trivial due to
  ordering ambiguities.  }  which does not require a generalised
pull-back\marginpar{non-Abelian pull-back} and thus requires merely
an additional trace over gauge indices.  The action of D$p$\NB-branes
of arbitrary world volume dimension $p+1$ can then be determined by
T\NB-duality, which transforms the T\NB-dualized direction from
longitudinal to transversal and vice versa.  The result 
\cite{Myers:1999ps} in string frame is\marginpar{D$p$~action}
\begin{align}
\begin{split}
  S_\text{D$p$} = &- T_p \int d^{p+1} \xi \str \left[ 
    \e^{-\vphi} \sqrt{ \det Q} \sqrt{-\det P[ \tilde E ]_{ab} + 2 \pi \alpha' F_{ab}}
    \right]\\ &\pm T_p \int \str \left[ P[ \e^{i (2\pi\alpha') \iphi \iphi} \tsum C_n \e^B ] \e^{2\pi\alpha'F}\right] \label{eq:nonabelian-dbi} ,
\end{split}
\end{align}
where ``$\str$'' is a trace operation that shall also take care of any
ordering ambiguities in the expansion of the non-linear action. Its
name (``symmetrised trace'') is reminiscent of an ordering
prescription suggested by \cite{Tseytlin:1997cs}, which however is not
valid beyond fifth order. Throughout this thesis, an expansion to
second order will be sufficient and no ordering ambiguities appear at
all.

The following abbreviations have been introduced:
\begin{subequations}
\begin{align}
  \tilde E_{AB} &:= E_{AB} + E_{Ai}(Q^{-1}-\delta)^{ij} E_{j B} \\
  E_{AB} &:= G_{AB}+B_{AB},\\
  Q^i{}_j &:= \delta^i_j + i \gamma \comm{\Phi^i}{\Phi^k} E_{kj}, \label{eq:overview-abbrevQ} \\
  (Q^{-1}-\delta)^{ij} &:= 
     \left[ (Q^{-1})^i{}_k - 
            \delta^i_k \right] E^{kj}, \\
  \gamma &:= 2\pi\alpha',\\
  \iphi\iphi f^{(n)} &:= \frac{1}{2(n-2)!} \comm{\Phi^i}{\Phi^j} 
    f^{(n)}_{ji A_3 \dots A_n} dx^{A_3} \wedge \dots \wedge dx^{A_n},
\end{align}
\end{subequations}
where $f^{(n)}$ is an arbitrary $n$\NB-form field acted upon by
$\iphi$, the interior product with $\Phi^i$. $E^{ij}$ is the inverse of $E_{ij}$ 
(as opposed to the transversal components of $E^{AB}$).

In particular static gauge is chosen,
\begin{align}
   X^a &= \xi^a,&
   X^i &= \gamma \Phi^i(\xi^a),
\end{align}
which means transversal coordinates $X^i$ are in one-to-one correspondence
to the scalar fields $\Phi^i$.
Then the pull-back of an arbitrary ambient space tensor $T_{A_1\dots A_n}$ 
can recursively be defined by
\begin{align}
  P[T_{A_1\dots A_n}]_{a_1\dots a_n} := 
     P[T_{a_1 A_2 \dots A_n}]_{a_2 \dots a_n} 
     + \gamma (\D_{a_1} \Phi^i) P[T_{i A_2 \dots A_n}]_{a_2 \dots a_n}, 
\end{align}
which yields for the combined metric/Kalb--Ramond field
\begin{align}
  P\bigl[\tilde E \bigr]
    &:= \tilde{E}_{ab} + \gamma \tilde{E}_{a i} \D_b \Phi^i 
                   + \gamma \tilde{E}_{i b} \D_a \Phi^i 
      + \gamma^2 \tilde{E}_{ij} \covD_a \Phi^i \D_b \Phi^j .
\end{align}
$\D_a$ denotes the gauge covariant derivative. 

Finally $E_{ab}$ still may contain a functional dependence on
the non-commutative scalars $\Phi$ and is to be understood 
as being defined by a non-Abelian Taylor expansion \cite{Garousi:1998fg}\marginpar{Taylor expansion}
\begin{align}
  E_{ab}(\xi^a) = \exp [ \gamma \Phi^i \partial_{X^i} ] E_{ab}(\xi^a,X^i)\bigr|_{X^i=0}.
\end{align}

Again the Wess--Zumino part shall be given for the eight dimensional case; 
i.e.~a stack of D7\NB-branes,
\begin{equation}
\begin{split}
  S_{WZ} = T_7 \int \str \biggl\{ 
    & P[C_8] 
    + \gamma P[i \gamma \iphi\iphi C_8 + C_6 ]\wedge F\\
    &+ \frac{\gamma^2}{2} P[ (i\gamma \iphi\iphi)^2 C_8
            + i \gamma \iphi\iphi C_6 + C_4] \wedge F \wedge F\\
    &+ \frac{\gamma^3}{3!} P[ (i\gamma \iphi\iphi)^3 C_8 + 
                            (i\gamma \iphi\iphi)^2 C_6 \\
    &\hphantom{+\frac{\gamma^3}{3!} P[} 
     + i\gamma \iphi\iphi C_4 + C_2] \wedge F \wedge F \wedge F
\biggr\},\\
\end{split}
\end{equation}
where $B$ has been assumed to vanish.
For a 3\NB-brane background, there is only a four-form potential and accordingly
the Wess--Zumino part is given by
\begin{align}
  S_{WZ} &= T_7 \int \str \frac{\gamma^2}{2} P[C_4] \wedge F \wedge F + 
                            \frac{i\gamma^4}{3!} P[ \iphi\iphi C_4] \wedge F \wedge F \wedge F. 
  \label{eq:nonabelian-d7-wz}
\end{align}

While \eqref{eq:nonabelian-dbi} encodes the high non-linearity of 
a D\NB-brane action in a compact manner, it is often not suited 
for explicit calculations and needs to be expanded.

\Subsection{Quadratic Action\label{sec:quaddbi}}
As both the non-Abelian scalars and the field strength carry $\gamma$ 
as a prefactor,
it is tempting to think of it as an expansion parameter, keeping track
of the order. However in equation \eqref{eq:overview-abbrevQ} in front of the
commutator there is a factor of $\gamma$ where following
this logic a factor of $\gamma^2$ should be expected.\footnote{Furthermore some 
authors prefer to use factors of $\alpha'$ to obtain D3\NB-transversal 
coordinates with mass dimension 1, thus modifying the manifest $\alpha'$
dependence even though in physical observables such redefinitions cancel 
of course.}

To\marginpar{quadratic order} 
avoid these pitfalls and unambiguously define what is meant by ``quadratic 
order'', a parameter $\veps$ shall be thought to accompany $\gamma$ in each of the
equations of the last Section with the sole exception of \eqref{eq:overview-abbrevQ}, where
an $\veps^2$ is included in front of the commutator. 
Then, the order $\veps^n$ denotes a total of $n$ fields of $\Phi$ or $F_{ab}$ in a term.

Pulling out a factor $E_{ab}(\veps=0)$ (which shall also not depend on 
transverse directions $X^i$ as they come with an $\veps$) from the \acro{DBI} part of 
the D\NB-brane action defines a matrix $M(\gamma)$ according to 
\begin{align}
  S_\text{DBI} = &- T_p \int d^{p+1} \xi \str \left[ 
    \e^{-\vphi} \sqrt{\det Q} \sqrt{-\det E_{ab}(0)} \sqrt{\det M(\veps)} \right],
\end{align}
which has the property $M(0)=\one$ and is given by
\begin{align}
  M(\veps)^a{}_b &= E^{ac}(\veps=0)\, \Bigl( P[ \tilde E(\gamma) ]_{cb} + \veps \gamma F_{cb} \Bigr).
\end{align}
$E^{ac}$ is the inverse of $E_{ac}$.
An expansion in $\veps$ is performed according to 
\begin{align}
\begin{split}
  \sqrt{\det M(\veps)} = 1 
     + \frac{\veps}{2} \tr\left(M'(0)\right) 
     + \frac{\veps^2}{4} \biggl[
          &  \tr\left(M''(0)\right) -\tr\left(M'(0)^2\right) \\
          & +\tfrac{1}{2} \tr^2 \left( M'(0)\right) \biggr]
            + \mathcal{O}(\veps^3),
\end{split}
\end{align}
where
\begin{align}
  M'(0) &= \gamma E^{ac}\Phi^i \partial_{X^i} E_{cb}
           + E^{ac} ( \gamma E_{kb} \D_c \Phi^k 
                     + \gamma E_{ck} \D_b \Phi^k )
           + \gamma E^{ac} F_{cb}, \label{eq:Mprime}\\\displaybreak[0]
\begin{split}
  M''(0) &= \gamma^2 E^{ac} \Phi^i \Phi^j \partial_{X^i}\partial_{X^j} E_{cb} \\
    &\relphantom{=} + 2 \gamma^2 E^{ac} \Phi^i \partial_{X^i} ( E_{kb} \D_c \Phi^k 
                                                  + E_{ck} \D_b \Phi^k ) \\ 
    &\relphantom{=} + E^{ac} [ E_{ci} ( 2i \gamma \comm{\Phi^i}{\Phi^j} - E^{ij} ) E_{jb}
                       + 2 \gamma^2 E_{ij} \D_c\Phi^i \D_b\Phi^j ].
\end{split} \label{eq:Mprimeprime}
\end{align}
All quantities on the right hand sides of \eqref{eq:Mprime} and \eqref{eq:Mprimeprime}
are to be understood as having
$\veps$ set to zero. In particular this means the right hand sides are
evaluated at vanishing transversal coordinates $X^i=0$. 

For\marginpar{quadratic \acro{DBI}}
a diagonal metric and vanishing Kalb--Ramond field, the \acro{DBI} part 
of the action up to quadratic order simplifies dramatically,
\begin{align}
\begin{split}
  S_\text{DBI} &= -T_p \int d^{p+1}\xi \str \e^{-\vphi} \sqrt{-\det G_{ab}}    
     \biggl[ 1 + \text{(lin.)} \\ &\qquad\qquad\qquad
      + \tfrac{\gamma^2}{2} G^{ab} G_{ij} \D_a \Phi^i \D_b \Phi^j  
             + \tfrac{\gamma^2}{4} G^{ac} G^{bd} F_{ab}F_{cd} \\ &\qquad\qquad\qquad
             + \tfrac{\gamma^2}{4} (G^{ab} \p_{X^i}\p_{X^j} G_{ab}) \Phi^i \Phi^j
     \biggr], \label{eq:dbi-quad}
\end{split}
\end{align}
where the following terms vanish unless the transversal coordinates
enter the metric linearly,
\begin{align}
\begin{split}
  \text{(lin.)} &:= \tfrac{\gamma}{2} \tr \mathcal{M} 
             - \tfrac{\gamma^2}{4} \tr \mathcal{M}^2
             + \tfrac{\gamma^2}{8} \tr^2 \mathcal{M}, \\
  \mathcal{M}^a{}_c &:= G^{ab}\Phi^i \p_{X^i} G_{bc}.
\end{split}
\end{align}

\Section{\adscft\ Correspondence\label{sec:adscft}}
The \adscftcorr\ (Anti-de~Sitter/Conformal Field Theory) is the
statement of two seemingly different theories to be equivalent.  These
theories are ten dimensional Type \acro{IIB} string theory on an
$\AdS_5 \times \mf{S}^5$ space-time background and four dimensional
$\cN=4$ extended supersymmetric $SU(N_c)$ Yang--Mills theory. The
latter is a (super)conformal field theory with coupling constant
$g_{YM}^2=g_s$, where $g_s$ is the string coupling.  The string theory has
$N_c$ units of five-form flux through the $\mf{S}^5$, which is related 
to the equal curvature radii $L$ of the $\AdS_5$ and $\mf{S}^5$ by
$L^4 = 4\pi \ell_s^4 g_s N_c$, where $\ell_s = \sqrt{\alpha'}$ is the
string length.
This equivalence is supposed to hold for arbitrary values of $N_c$
and the coupling constants, but since string theory on $\AdS_5\times \mf{S}^5$
is not well-understood, it is usual to take two consecutive limits that make
a supergravity description valid but still leave the duality non-trivial. 
\begin{figure}
\centering
\EPSinclude[width=10cm]{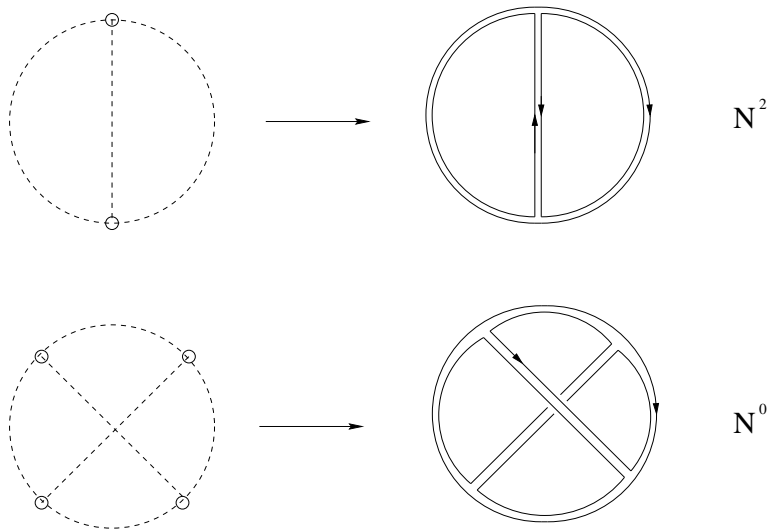}
\caption[Double Line Representation]{\label{fig:intro-thooft}Double
  Line Representation: Non-planar diagrams are suppressed by powers of
  $N_c^2$ \cite{Aharony:1999ti}}
\end{figure}

The\marginpar{\tHooft\ limit}
first limit to take is the \tHooft\ large $N_c$ limit, with $N_c
\to \infty$ while $\la := g_{YM}^2 N_c$ is kept fixed, in which the
field theory reorganises itself in a topological expansion.  This can
be seen by using a double line representation for Feynman diagrams
assigning a line to each gauge index, such that fields in the adjoint
are equipped with two indices, while fields in a vector representation
carry a single line. The diagrams, see
Figure~\ref{fig:intro-thooft}, then correspond to polyhedrons, which
contribute with a power of $N_c$ that is suppressed by the diagram's
genus and the polyhedrons are interpreted as triangulating the string
world sheet, though the exact nature of this triangulation is still
to be understood. Due to $g_s = \la/N_c$ the strict \tHooft\ limit 
corresponds to considering classical string theory on 
$\AdS_5 \times \mf{S}^5$. At the same time the \tHooft\ coupling takes
over the \role\ of the field theoretic coupling constant.

In\marginpar{small curvature}
the second limit $\ell_s \to 0$, the curvature radius is assumed
to be large compared to the string length $\ell_s \ll L$. 
This
corresponds to the low energy limit where supergravity becomes an
effective description. On the field theory side this implies a large
\tHooft\ coupling
\begin{align}
  1 \ll \frac{L^4}{\ell^4} = 4 \pi \la
\end{align}
and a strongly coupled theory therefore, indicating that \adscft\
is a weak-strong duality. This means that one theory in its perturbative
regime is dual to the other theory in the strong coupling regime, which renders
the duality both extremely useful and hard to proof.

While\marginpar{correlation functions}
on the one hand the supergravity version is the weakest form of
the \adscft\ conjecture, it is the most useful version for practical
calculations on the other hand.  The
equivalence of both theories to be expressed by 
\begin{align}
  \vev{ \exp \int d^4x \, \phi^{(0)} \cO }_\text{CFT} &=
     \exp \bigl\{ - S_\text{SUGRA}[ \phi ] \bigr\} 
     \biggr|_{\phi({\p\AdS}) = \phi^{(0)},}
\end{align}
where the field theoretic operator $\cO$ is coupled to the
boundary value $\phi_0$ of an associated supergravity field $\phi$,
which is determined by the supergravity equations of motion
and the boundary condition.

This\marginpar{bulk vs.\ boundary}
implicitly introduces the notion of the conformal field theory
being defined on the boundary of $\AdS_5$, where one may imagine
the $\AdS_5$ space being build up from slices of Minkowski spaces 
parallel to the boundary and fibred over a fifth (``radial'') direction $y$.
The line element reads
\begin{align}
  ds_{\AdS_5\times \mf{S}^5} = \frac{y^2}{L^2} dx_{1,3}^2 
     + \frac{L^2}{y^2} dy^2 + L^2 d\Omega_5^2.
\end{align} 
For the metric to be invariant under rescalings of the coordinates
on the boundary $x$, the radial direction has to transform reciprocal,
which means that $y$ transforms as an energy and is interpreted
as the renormalisation scale of the boundary theory. 
Considering domain wall solutions it is actually possible to 
represent field theoretic renormalisation group flows on the 
supergravity side \cite{Freedman:1999gp,Karch:1999pv},
establishing the fact that the interior
of the \AdS\ space may be interpreted as the infrared (\acro{IR})
and the boundary as the ultraviolet (\acro{UV}) of the field theory.

By\marginpar{operator map}
the standard \adscft\ dictionary supergravity fields, 
$\phi$ being solutions to differential equations of second order, 
encode actually two field theoretic objects, whose conformal
dimension can be read off from the asymptotic behaviour,
\begin{align}
  \phi(y\to\infty) \sim \mathcal{J} y^{\Delta-4} + \vev{\cO} y^{-\Delta},
\end{align}
where the radial direction is interpreted
as the renormalisation scale.
%
%
%
The first, non-normalisable part corresponds to a field
theoretic source and has conformal dimension $4-\Delta$; 
the normalisable part yields the corresponding \acro{VEV}
of mass dimension $\Delta$.
A simple example shall illustrate this. For the bilinear
operator $\bpsi\psi$, the dual supergravity field
has the asymptotic behaviour
\begin{align}
  \phi(y\to\infty) \sim \frac{m}{y} + \frac{c}{y^3},
\end{align}
where $m$ is the mass term of field $\psi$ and $c$ the bilinear
quark condensate $\vev{\bpsi\psi}$.  The difficult part is to find out
which supergravity fields correspond to which field theoretic
operators.  For $\half$-\acro{BPS} states, which correspond to superconformal
chiral primary operators, the situation is simpler because they are
determined by their transformational behaviour under the large global
symmetry group $\gr{SU}(2,2|4)$.  On the field theory side its bosonic
subgroup $\gr{SO}(2,4)\times \gr{SU}(4)\simeq \gr{SO}(2,4)\times
\gr{SO}(6)$ is realised as the conformal and R-symmetry group, 
while it corresponds to the isometry group on the
supergravity side.

From\marginpar{D3\NB-branes}
a string theoretical perspective, the correspondence can be
understood as two different effective descriptions
of a D3\NB-brane stack, namely 
as a Yang--Mills theory from an open string 
perspective and a $p$-brane solution from a closed string perspective.
In the latter case, the $\AdS_5\times \mf{S}^5$ geometry arises from a 
near-horizon limit. 
The picture of \adscft\ being two descriptions of a D3\NB-brane stack
turns out to be particularly useful when adding additional branes
to include fundamental fields into the duality. This shall be the
topic of the next Chapter.



  \begin{savequote}[\savequotewidth]
  We used to think that if we knew one, we knew two, because one and
  one are two. We are finding that we must learn a great deal more
  about ``and''.  
  \qauthor{Sir Arthur Eddington}
\end{savequote}

\Chapter{Spicing with Flavour\label{ch:flavour}}
While the \adscftcorr{} has been a remarkable progress in the
understanding of the \tHooft{} large $N_c$ limit \cite{tHooft:1973jz},
a need to extend the Maldacena conjecture beyond $\cN=4$
super-Yang--Mills (\acro{SYM}) theory was soon felt, see 
\cite{Witten:1998zw} for a most prominent example.
Since $\cN=4$ \acro{SYM} contains only one multiplet, the
gauge field forces its representation on all other fields in the
theory. As a consequence, also the fermions transform under the
adjoint\marginpar{only adjoints} representation, and thus do not
describe quarks.

There have been early attempts to augment the boundary theory with
fundamental fields by including D7\NB-branes in an $\AdS_5 \times
\coset{\mf{S}^5}{\mathds{Z}_2}$ geometry 
\cite{Aharony:1998xz,Fayyazuddin:1998fb}. The orientifold was introduced
to satisfy a tadpole cancellation condition, but the dual $\cN=2$
boundary theory had gauge group $\gr{Sp}(N)$.  In order to obtain
an $\gr{SU}(N_c)$ gauge theory for the description of large $N_c$
cousins of quantum chromodynamics (\acro{QCD}), \cite{Karch:2002sh}
dropped the orientifold from the setup. This was justified by the fact,
that the probe D7\NB-brane wraps a contractible $\mf{S}^3$ cycle on
the $\mf{S}^5$ and does not lead to a tadpole, hence.
In \cite{Karch:2002sh} it was shown that the string mode corresponding to the direction
in which the $\mf{S}^3$ slips from the $\mf{S}^5$ has negative
mass square, but satisfies (saturates) the Breitenlohner--Freedman bound 
and does not introduce an instability.

%
In this Chapter,
the main ideas of \cite{Karch:2002sh} will be reviewed, before calculating
the meson spectrum of a field theory dual to a more general geometry 
in the next Chapter.

\Section{Motivation}
Conventional \adscftcorr{} can be understood as two different limits (see 
the introductory Chapter)
of the same object, namely a stack of $N_c$ coincident D3\NB-branes in
string theory.  The choice on which of those $N_c$ branes an open
string may end, is reflected by the $\gr{SU}(N_c)$ symmetry of the
dual field theory.  The number of ways to attach both ends to the stack is
$N_c^2-N_c$, indicating that the field describing
the open string is in the adjoint representation. When including
another, non-coincident brane in this setup, a string connecting it to
the stack has $N_c$ choices and thus describes a field transforming
under the vector representation of the gauge group.  Another perhaps
less heuristic way to understand this scenario, is to return to the
\tHooft{} expansion. If one takes the intuition about the field
theory's reorganisation into a triangulation of the closed string
world sheet\marginpar{world sheet triangulation} serious, then
apparently, fundamental fields will provide boundaries that lead to a
triangulation of the open string world sheet.  In this sense,
augmenting the \adscftcorr{} by additional branes, which exactly
provide these open strings, extends the correspondence from an
open-closed duality to a full string duality.

While the inclusion of D3 or D5\NB-branes leads to fundamental fields on
the boundary of \AdS\ that are confined to a lower dimensional defect
(so-called ``defect \acro{CFT}s''), the addition of D7\NB-branes
\marginpar{why D7} provides space-time filling fields in the
fundamental representation. Furthermore it breaks supersymmetry by a
factor of two; from $\cN=4$ to $\cN=2$ on the four-dimensional field
theory side by inclusion of an $\cN=2$ fundamental hypermultiplet
given rise to by the light modes of strings with one end on the D3s
and one on the D7s.

\Section{Probe Brane}
In order to maintain the framework of conventional \adscftcorr,
\cite{Karch:2002sh} neglected the gravitational backreaction of the
D7\NB-branes on the geometry, which was justified by requiring the
number $N_f$ of D7\NB-branes to be sufficiently small.  The
contribution of the $N_c$ D3\NB-branes and the $N_f$ D7\NB-branes to the
background fields is of order $g_s$ times their respective number.  So
as long as $N_c \gg N_f$,\marginpar{probe limit}
the geometry is dominated by the D3\NB-branes
and the D7\NB-branes are approximately probe branes.  In the strict
$N_c\to\infty$ limit, which comes with the supergravity description of
\adscft{}, this approximation becomes exact.\footnote{
It should be noted that meanwhile there are supergravity solutions
that include the backreaction of the D7\NB-branes \cite{Burrington:2004id}.
}

This is analogous to the so-called quenched approximation
\marginpar{quenched approximation} in lattice \acro{QCD}, where the
action of the gauge bosons on the matter field is included, while the
action of the matter on the bosons is neglected.

\begin{table}
\centering
\begin{colortabular}{*{10}{Bc}}
        \multicolumn{10}{Hc}{Coordinates}\\
        \rowcolor{white}
        0&1&2&3&4&5&6&7&8&9\\ \bwline
        \multicolumn{4}{Cc}{D3}&&&&&&\\ \hline
        \multicolumn{8}{Cc|}{D7}&&\\ \hline
        \multicolumn{4}{Cc}{$x^{\mu,\nu,\dots}$}&
                \multicolumn{4}{|Cc|}{$y^{m,n,\dots}$ } &
                \multicolumn{2}{Cc}{$z^{i,j,\dots}$}\\ \hline
        &&&&\multicolumn{6}{|Cc}{$r$}\\ \hline
        &&&&\multicolumn{4}{|Cc|}{$y$}&&\\ \hline
	\multicolumn{8}{Cc|}{$X^{a,b,\dots}$}&&\\ \hline
	\multicolumn{10}{Cc}{$X^{A,B,\dots}$}\\ \bwline
\end{colortabular}
\caption[D3/D7-brane Embedding]{
\label{tab:katz-karch-embedding}
D3- and D7\NB-brane embedding in the 
$\AdS_5 \times \mf{S}^5$ geometry. 
The D7\NB-branes (asymptotically) 
wrap an $\AdS_5 \times \mf{S}^3$.
The Table also summarises the index conventions used throughout 
this part of the thesis.}
\end{table}

The metric of $\AdS_5 \times \mf{S}^5$ can be 
written as
\begin{align}
  \begin{aligned}
    ds^2 &= \frac{r^2}{L^2} \eta_{\mu\nu} dx^\mu dx^\nu
          + \frac{L^2}{r^2} (d\sqvec{y} + d\sqvec{z} ) \\
         &= \frac{r^2}{L^2} \eta_{\mu\nu} dx^\mu dx^\nu 
          + \frac{L^2}{r^2} dr^2 + L^2 d\Omega_5^2, 
\end{aligned}\label{eq:ads5-s5}
\end{align}
where the index conventions as well as the embedding of the D7\NB-branes
have been summarised in Table~\ref{tab:katz-karch-embedding}. The
multiplication of vectors is supposed to denote contraction with a
Euclidean metric, that means $d\sqvec{y} = \sum_{4,5,6,7} dy^m dy^m$,
$d\sqvec{z} = \sum_{8,9} dz^i dz^i$.  There are three qualitatively
different types of directions: $x$ denote the world volume coordinates
of the D3s, $y$ the coordinates transversal to the D3s and
longitudinal to the D7s, and $z$ the coordinates transversal to both
kinds of branes.  Since $y$ and $z$ are on the same footing in the
metric, assigning $z$ to the $8,9$\NB-plane is arbitrary, but
manifestly breaks the $\gr{SO}(6) \simeq \gr{SU}(4)_R$ isometry group
\marginpar{isometry group} to $\gr{SO}(4) \times \gr{SO(2)} \simeq
\gr{SU}(2)_L \times \gr{SU}(2)_R \times \gr{U}(1)_R$, where the
orthogonal groups represent rotational invariance in the coordinates
$y$ and $z$, respectively.  In the case of coincident D3 and D7
branes, the hypermultiplet stemming from the strings stretched between
the two stacks is massless, such that there is no classical scale
introduced into the setup and conformal symmetry is maintained in the
strict probe limit. Then the R-symmetry of the field theory is
$\gr{SU}(2)\times \gr{U}(1)_R$.

When\marginpar{embedding} separating the stacks in the 
$z$-plane, the $\gr{SO}(2)_{8,9}\simeq\gr{U}(1)_R$ group is explicitly broken,
though one may use the underlying symmetry to parametrise this breaking as
\begin{align}
  z^8 &= 0, & 
  z^9 &= \tilde{m}_q. \label{eq:constant-embedding}
\end{align}
%
%
%
Since this introduces a scale into the setup, namely a hypermultiplet mass
$m_q=\tilde{m}_q/(2\pi\alpha')$, it is not to be
expected that conformal symmetry, and hence
\AdS\ isometry, can be maintained. The R-symmetry of the field
theory becomes $\gr{SU}(2)_R$ only, which
is in accordance with the geometric symmetry breaking above.

Indeed, the induced metric\marginpar{induced metric} on the D7s reads
\begin{align}
  \begin{aligned}
    ds^2 &= \frac{y^2+\tilde{m}_q^2}{L^2} \eta_{\mu\nu} dx^\mu dx^\nu
          + \frac{L^2 y^2}{y^2+\tilde{m}_q^2} d\sqvec{y} \\
         &= \frac{y^2+\tilde{m}_q^2}{L^2} \eta_{\mu\nu} dx^\mu dx^\nu 
          + \frac{L^2}{y^2+\tilde{m}_q^2} dy^2 + \frac{L^2 y^2}{y^2+\tilde{m}_q^2} d\Omega_3^2, 
          \label{eq:KMMWindD7}
\end{aligned}
\end{align}
which towards the boundary at $\abs{y}\to\infty$, with $y^2 \equiv
\abs{y}^2 := \vec{y}\vec{y}$, approximates $\AdS_5 \times
\mf{S}^3$, reflecting the fact that a quark mass term is 
a relevant deformation that is suppressed in the
ultraviolet. 

This\marginpar{radius = energy} is in accordance with the usual picture of the radial direction 
$r=\sqrt{y^2+\tilde{m}_q^2}$ of the \AdS\ space describing the
energy scale of the field theory, where approaching the interior of
\AdS\ from the boundary corresponds to following a 
renormalisation group flow from the ultraviolet (\acro{UV}) to the infrared (\acro{IR}).

\begin{figure}
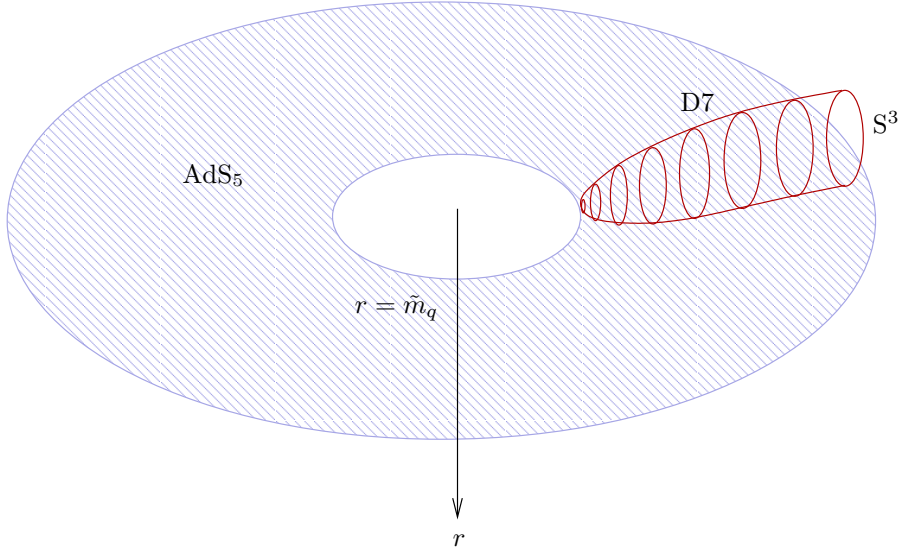

\centering
\PSfraginclude[trim=-3 -3 0 0]{width=12cm}{flow}
\caption[Terminating D7-brane]{The D7\NB-brane wraps an $\mf{S}^3$ on the internal
$\mf{S}^5$ which slips towards a pole and shrinks to zero
size. From the five dimensional point of view, the brane terminates
at a certain radius, but there is no boundary associated to this
ending. (Figure taken from \cite{Babington:2003vm})\label{fig:brane-cigar}}
\end{figure}

When the renormalisation scale is lowered below the quark mass, the
quarks should drop out of the dynamics.
This happens when reaching the radius $r=\tilde{m}_q$ in the ambient space,
which corresponds to the interior of the D7s at $y=0$, where the
D7\NB-branes stop from a five dimensional perspective, although as
depicted in Figure~\ref{fig:brane-cigar} there is no boundary
associated to this ending.

When\marginpar{conformal limit} 
$\tilde{m}_q=0$, the $\gr{U}(1)_R$ and $\gr{SO}(2,4)_{\AdS}$ 
symmetry are restored and the D7s fill the whole of the ambient
$\AdS_5$,
which suggests that conformal symmetry is restored. 
However, this is only true in the strict probe limit, as otherwise
contributions to the beta function of order $N_f/N_c$ occur
\cite{Karch:2002sh,Kruczenski:2003be}.

\Section{Analytic Spectrum}
\Subsection{Fluctuations of the Scalars}
The spectrum of the undeformed D3/D7 system described above admits 
analytic treatment at quadratic order \cite{Kruczenski:2003be} and therefore sets the baseline for
the numerical determination of meson spectra in the more complicated
setups of the following Chapters.

From equations \eqref{eq:d3sol}, \eqref{eq:abelian-dbi} and \eqref{eq:d7wz}
the D7\NB-brane action in a background of D3\NB-branes reads
\begin{align}
\begin{split}
  S_{D7} &= - T_7 \int d^8\xi \sqrt{-\det( P[G]_{ab} + (2\pi\alpha')F_{ab} ) }\\
        &\relphantom{=} + \frac{2\pi\alpha'}{2} T_7 \int P[C_4]\wedge F\wedge F,
\end{split} \\
  C_4 &= \frac{r^4}{L^4} dx^0 \wedge \dots \wedge dx^3,  \label{eq:c4}
\end{align}
where $P$ is the pullback to the world-volume of the D7\NB-branes and
$r^2 = y^2+z^2$.

For\marginpar{fluctuations about the embedding}
fluctuations of the scalars, the Wess--Zumino term contributes only
at fourth order (with $(\text{scalar})^2\cdot F^2$). From the action
and for an embedding according to 
\begin{align}
  z^8 &= 0 + (2\pi\alpha') \delta z^8(\xi), & 
  z^9 &= \tilde{m}_q + (2\pi\alpha') \delta z^9(\xi). \label{eq:kmmwscalarfluct}
\end{align}
the expansion of the action to quadratic order \eqref{eq:dbi-quad} yields
\begin{align}
  \Lag &= \sqrt{- \det g_{ab}} ( 1 + \half (2\pi\alpha')^2 g_{ij} g^{ab} \p_a z^i \p_b z^j ), \label{eq:KMMWquadlag}
\end{align}
where the fact that metric admits a diagonal form has been used.  For
the induced D7 metric \eqref{eq:KMMWindD7}, the Lagrangean
\eqref{eq:KMMWquadlag} reads at quadratic order
\begin{align}
  2 (2\pi\alpha')^{-2} \Lag 
      &= y^3 \sqrt{\det(\hat{g})} [ 
            \eta^{\mu\nu} (\p_\mu \delta z^8)(\p_\nu \delta z^8) \\ &
            + \pfrac{L^2}{y^2 + \tilde{m}_q^2}^2 (\p_y \delta z^8 )^2  
            + \hat{g}^{\ia\ib} (\p_\ia z^8)(\p_\ib z^9) + (z^8 \vs z^9) ], \notag
\end{align}
with\marginpar{equation of motion} $\hat{g}_{\ia\ib}$ the metric on
the three sphere and the equation of motion
\begin{align}
  \frac{L^4}{(y^2+\tilde{m}_q^2)^2} \p^\mu \p_\mu \delta z^i + y^{-3} \p_y (y^3 \p_y \delta z^i) + y^{-2} \hat{\nabla}^\ia \hat{\nabla}_\ia \delta z^i &= 0, &
  i &= 8,9, \label{eq:KMMW-sc-eom}
\end{align}
where $\hat{\nabla}_\ia$ is the covariant derivative on the
unit $\mf{S}^3$.  An ansatz\marginpar{radial equation} for separation of variables $\delta
z^i(x^\mu,y,\mf{S}^3)=\zeta^i(y) \e^{ik\cdot x} \mathcal{Y}^\spinl(\mf{S}^3)$,
with $\hat{\nabla}^\ia \hat{\nabla}_\ia \mathcal{Y}^\spinl =
-\spinl(\spinl+2)\mathcal{Y}^\spinl$, $\ell \in \mathds{N}_0$ yields 
\begin{gather}
  \biggl[ \p_{\tilde{y}}^2 + \frac{3}{\tilde{y}}\p_{\tilde{y}} +
  \frac{\tilde{M}_s^2}{(1+\tilde{y}^2)^2} - \frac{\spinl (\spinl +2
    )}{\tilde{y}^2} \biggr] \zeta^i(\tilde{y}) = 0, \label{eq:KMMWradialeom} \\
  \tilde{y} = \frac{y}{\tilde{m}_q}, \qquad \tilde{M}_s^2 = - \frac{k^2 L^4}{\tilde{m}_q^2}, 
\end{gather}
where a rescaling has removed all explicit scale dependencies.
Requiring regularity at the origin, the radial equation
\eqref{eq:KMMWradialeom} can be solved uniquely in terms of a
hypergeometric function,
\begin{align}
\begin{split}
  \zeta^i(y) &= \frac{y^\spinl}{(y^2+\tilde{m}_q^2)^{n+\spinl+1}} 
  \hyperFtwoone \bigl(-(n+\spinl+1),-n; \spinl+2; -y^2 / \tilde{m}_q^2 \bigr),\\ \label{eq:KMMWsol}
  M_s^2 &= -k^2 = \frac{4\tilde{m}_q^2}{L^4} (n+\spinl+1)(n + \spinl+2), 
\end{split}
\end{align}
with the discretisation condition $n \in \mathds{N}_0$ from
normalisability.  Note that the spectrum becomes degenerate in the
conformal $\tilde{m}_q\to 0$ limit.  The conformal dimension of the
boundary operator dual to the solution above, can be read off from its
scaling behaviour with respect to the radial coordinate. In
\cite{Kruczenski:2003be} the \acro{UV} behaviour\marginpar{\acro{UV}
  behaviour} is determined from \eqref{eq:KMMWsol}, but one may
instead simply discuss the radial equation \eqref{eq:KMMWradialeom},
which for large $\tilde{y}$ becomes approximately
\begin{align}
  \biggl[ \p_{\tilde{y}}^2 + \frac{3}{\tilde{y}}\p_{\tilde{y}} - \frac{\spinl (\spinl +2
    )}{\tilde{y}^2} \biggr] \zeta^i(\tilde{y}) = 0. 
\end{align}
Its solutions are of the form $\zeta^i(\tilde{y})= A \tilde{y}^\spinl + B
\tilde{y}^{-\spinl-2}$, which contradicts the \naive\ \adscft\ expectation of
$\tilde{y}^{\Delta-4} + \tilde{y}^{-\Delta}$ as can be seen from taking the sum of the
exponents.  This is due to the appearance of a determinant factor
$\sqrt{-\det g_{ab}}\sim \tilde{y}^3$, which imposes
a non-canonical normalisation\marginpar{non-canonical normalisation}
on the kinetic term. So the generic
behaviour should be $\tilde{y}^{p+\Delta-4} + \tilde{y}^{p-\Delta}$ and subtracting
the exponent of the non-normalisable solution, which corresponds to a
field theory source, from that of the normalisable one, which
corresponds to a vacuum expectation value, it can be seen that
\begin{align}
\begin{split}
  & - (\spinl + 2) - \spinl = (p-\Delta) - (p+\Delta-4) = -2 \Delta + 4 \\
  &\implies \Delta = \spinl + 3.
\end{split}
\end{align}

\Subsection{Fluctuations of the Gauge Fields\label{sec:flavour-gauge-fluct}}
The equations of motion for the gauge fields read
\begin{align}
  \p_a ( \sqrt{-\det g_{cd}} F^{ab}) - \frac{4\rho(\rho^2+\tilde{m}_q^2)}{L^4} \veps^{b \ib\ic} \p_\ib A_\ic = 0,
\end{align}
with $\veps^{\ia\ib\ic}$ taking values $\pm1$, and $0$ when the
free index $b$ is none of the angular $\mf{S}^3$ directions. 

Expanding the equation of motion yields
\begin{align}
\begin{split}
  \biggl[ (g_{xx})^{-1} \p_\mu \p^\mu 
     &+ y^{-3} \p_y (y^3 (g_{yy})^{-1} \p_y) 
     + \tilde{\nabla}_\ia \tilde{\nabla}^\ia \biggr] A_\nu\\
  &- \p_\nu \biggl[ (g_{xx})^{-1} \p_\mu A^\mu 
             + y^{-3} \p_y (y^3 (g_{yy})^{-1} A_y)
             + \tilde{\nabla}^\ia A_\ia \biggr] = 0 ,
\end{split}\\ \displaybreak[1] 
  \biggl[ (g_{xx})^{-1} \p_\mu \p^\mu 
     &+ \tilde{\nabla}_\ia \tilde{\nabla}^\ia \biggr] A_y
      - \p_y \biggl[ (g_{xx})^{-1} \p_\mu A^\mu 
             + \tilde{\nabla}^\ia A_\ia \biggr] = 0, \label{eq:ay} \\
\begin{split}
  \biggl[ (g_{xx})^{-1} \p_\mu \p^\mu 
     &+ y^{-3} \p_y (y^3 (g_{yy})^{-1} \p_y) 
     + \tilde{\nabla}_\ia \tilde{\nabla}^\ia \biggr] A_\id \\
  &- \p_\id \biggl[ (g_{xx})^{-1} \p_\mu A^\mu 
             + y^{-3} \p_y (y^3 (g_{yy})^{-1} A_y)
             + \tilde{\nabla}^\ia A_\ia \biggr]\\
  & - C'_4 \, \tilde{g}_{\id\ia} \veps^{\ia\ib\ic} \p_\ib A_\ic = 0 , 
\end{split}\label{eq:aS3}
\end{align}
each of which has to be satisfied for a particular ansatz.  For the
components $(A_\mu, A_y, A_\ia)$, the first two should transform under
$\gr{SO}(4)_{4567}$ as scalars, while the last should transform as a
vector and accordingly be built up from vector\marginpar{spherical
  harmonics} spherical harmonics. The simplest choice is
$\tilde{\nabla}^\ia \mathcal{Y}^\spinl$, which transforms in the
$(\frac{\spinl}{2},\frac{\spinl}{2})$. The other two possibilities are
$\mathcal{Y}^{\spinl,\pm}_\ia$, which transform in the
$(\frac{\spinl\pm1}{2},\frac{\spinl\mp1}{2})$ and obey
\begin{align}
  \tilde{\nabla}^2 \mathcal{Y}^{\spinl,\pm}_\ia 
       - 2 \delta_\ia^\ib \mathcal{Y}^{\spinl,\pm}_\ib 
      &= - (\spinl+1)^2 \mathcal{Y}^{\spinl,\pm}_\ia, \\
  \veps_{\ia\ib\ic} \tilde{\nabla}_\ib \mathcal{Y}^{\spinl,\pm}_\ic 
      &= \pm (\spinl \pm 1) \mathcal{Y}^{\spinl,\pm}_\ia, \\
 \tilde{\nabla}^\ia \mathcal{Y}^{\spinl,\pm}_\ia &= 0.
\end{align}
The modes containing $\mathcal{Y}^{\spinl,\pm}$ should not mix with the 
others since they are in a different representations. The following types
of solutions can be obtained:
\begin{subequations}
\begin{align}
  &\text{Type I}\pm& A_\ia &=\phi_I^\pm(y) \e^{ikx}\mathcal{Y}^{\spinl,\pm}_\alpha, &
    A_\mu &= A_y = 0, \label{eq:kmmw-type-I} \\
  &\text{Type II} & 
    A_\mu &= \xi_\mu \phi_{II}(y) \e^{ikx} \mathcal{Y}^\spinl,&
    A_y  &= A_\ia = 0, \qquad k_\mu \xi^\mu = 0, \label{eq:kmmw-type-II} \\
  &\text{Type III} & 
    A_y &=  \phi_{III}(y) \e^{ikx} \mathcal{Y}^\spinl, &
    A_\ia &= \tilde{\phi}_{III}(y) \e^{ikx} \tilde{\nabla}_\ia \mathcal{Y}^\spinl.
    \label{eq:kmmw-type-III}
\end{align}
\end{subequations} 
Type II and III come from recognising that in the gauge $\p_\mu A^\mu = 0$,
$A_\mu$ does not appear in \eqref{eq:ay} and \eqref{eq:aS3}, and can therefore be
treated independently. Kruczenski et al.\ argue that modes not satisfying the
gauge condition are either irregular or have a polarisation parallel to the
wave vector $k$; i.e.~can be brought to the gauge $\p_\mu A^\mu = 0$.

The simplest radial equation arises from the ansatz II,
\begin{align}
  \biggl[ \p_{\tilde{y}}^2 + \frac{3}{\tilde{y}}\p_{\tilde{y}}  +
  \frac{\tilde{M}_{II}^2}{(1+\tilde{y}^2)^2} - \frac{\spinl (\spinl +2
    )}{\tilde{y}^2} \biggr] A_a = 0. \label{eq:KMMW-vec-eom}
\end{align}
Up to the polarisation vector, this is the same equation as 
\eqref{eq:KMMW-sc-eom} and therefore produces a degeneracy of 
the\marginpar{mass spectrum} mass spectrum,
\begin{align}\label{eq:kmmw-typeII-mass}
 \tilde{M}_{II}^2 = \tilde{M}_s^2 = 4 (n+\spinl+1)(n+\spinl+2), \qquad n,\spinl \ge 0, 
\end{align}
with the same conformal dimension $\Delta = \spinl + 3$.

For type III and I$\pm$, an analogous calculation yields the mass formulae and conformal 
dimensions of the corresponding \acro{UV} operators,
\begin{align}
  \tilde{M}_{I+}^2  &= 4 (n+\spinl+2)(n+\spinl+3),& 
             \Delta &= \spinl + 5& 
            \spinl &\ge 1,\\
  \tilde{M}_{I-}^2  &= 4 (n+\spinl)(n+\spinl+1),  & 
             \Delta &= \spinl + 1& 
            \spinl &\ge 1,\\
  \tilde{M}_{III}^2 &= 4 (n+\spinl+1)(n+\spinl+2),&
             \Delta &= \spinl + 3& 
            \spinl &\ge 1,
\end{align} 
with $n\ge0$ in all cases.

The\marginpar{matching of representations}
full mesonic mass spectrum is given in Table~\ref{tab:KMMWspectrum},
were the Dirac fermions needed to fill the states into massive $\cN=2$
supermultiplets have been added.
Since the $\gr{SU}(2)_L$ group commutes with the supercharges, all
states in the same supermultiplet should be in the same representation
with respect to the left quantum number. Indeed redefining $\spinl$ in 
such a manner that the $\gr{SU}(2)_L$ representations are the same 
also makes the mass coincide.
This argument cannot be applied to the right quantum number, for the 
supercharges are not singlets under the $R$-symmetry. (Although 
the spectrum is symmetric under swapping the \role{}s of the 
left and right group, which corresponds to considering an anti-D7\NB-brane, that
is the opposite sign in front of the Wess--Zumino term.)
\begin{table}
\centering
\begin{colortabular}{ClCc|>{$}Cc<{$}|>{$}Cc<{$}|>{$}Cc<{$}|>{$}Cc<{$}}
  \multicolumn{6}{Hc}{\textsc{\acro{IIB} \acro{SUGRA} Particle Content}}\\
  \multicolumn{2}{c}{Type} & 
  \cell{c}{$\gr{SU}(2)_R$} & \cell{c}{$\gr{U}(1)_R$} &
  \cell{c}{}&
  \cell{c}{$\Delta-\spinl$} 
  \\ \bwline
  1 scalar  &(I-) & \frac{\spinl+2}{2} & 2 & \spinl\ge0 & 2 \\ \hline
  2 scalars &(s)  & \frac{\spinl}{2}   & 0 & \spinl\ge0 & 3 \\ \hline
  1 vector  &(II) & \frac{\spinl}{2}   & 0 & \spinl\ge0 & 3 \\ \hline
  1 scalar  &(III)& \frac{\spinl}{2}   & 0 & \spinl\ge1 & 3 \\ \hline
  1 scalar  &(I+) & \frac{\spinl-2}{2} & 0 & \spinl\ge2 & 4 \\ \hline
  1 Dirac   &(F1) & \frac{\spinl+1}{2} & 1 & \spinl\ge0 & \tfrac{5}{2}  \\ \hline
  1 Dirac   &(F2) & \frac{\spinl-1}{2} & 1 & \spinl\ge1 & \tfrac{9}{2}  \\ \bwline
\end{colortabular}
\caption[Mesonic Spectrum in $\AdS_5\times\mf{S}^5$]{Mesonic Spectrum
  in $\AdS_5\times\mf{S}^5$. The Dirac fermions are deduced from
  Supersymmetry. $\Delta$ is the conformal
  dimension of the corresponding \acro{UV} operator and the representations
  have been shifted to have the same $\gr{SU}(2)_L$ spin $\frac{\spinl}{2}$ 
  and therefore the same mass $\tilde{M}^2 = 4 (n + \spinl +1)(n+\spinl+2)$, $n \ge 0$. 
  \label{tab:KMMWspectrum}}
\end{table}
\Section{Operator Map}
As has been seen, the fluctuation modes of the D7\NB-brane organise
themselves in $\cN=2$ multiplets, which are made of a chiral primary
field and descendants. The mode with highest $\gr{SU}(2)_R$ quantum number
is the scalar of type (I-). The choice of the corresponding primary operator
\marginpar{lowest primary}
is restricted by the requirement of containing exact two hypermultiplet fields
in the fundamental representation, being in the same representation 
$(\frac{\spinl}{2}, \frac{\spinl+2}{2})_0$ and having conformal dimension
$\Delta=\spinl+2$. For $\spinl=0$ this merely admits the unique combination
\begin{align}
  \cO^I &= \psi^\ia \sigma^I_{\ia\db} \bpsi^\db,
\end{align}
with the Pauli matrices $\sigma^I$. The higher chiral primary
\marginpar{Kaluza--Klein primaries} in the Kaluza--Klein tower, can be
obtained by including the adjoint operators obtained a the subset
$Y^{4,5,6,7}$ of the six adjoint scalars of the $\cN=4$ multiplet by
traceless symmetrisation,\footnote{The four scalars belong to the
  $\cN=2$ hypermultiplet.}
\begin{align}
  \chi_\spinl &= Y^{(i_1,}\dots Y^{i_{\spinl})}.
\end{align}
The operators $\chi_\spinl$ transforms under $\gr{SU}(2)_L \times
\gr{SU}(2)_R \times \gr{U}(1)_R$ as $(\frac{\spinl}{2}, \frac{\spinl}{2})_0$,
which in the combination 
\begin{align}
  \cO_\spinl^I &= \psi \chi_\spinl \sigma^I \bpsi,
\end{align}
gives a $(\frac{\spinl}{2}, \frac{\spinl+2}{2})_0$ of conformal
dimension $\Delta = \spinl+2$.  The other operators can be obtained
from acting with supercharges on those chiral primaries.



  \begin{savequote}[\savequotewidth]
  Where a calculator on the \acro{ENIAC} is equipped with 18,000 vacuum tubes
  and weighs 30 tons, computers in the future may have only 1,000
  vacuum tubes and perhaps weigh 1.5 tons.  
  \qauthor{unknown, ``Popular Mechanics'', March 1949}
\end{savequote}

\setlength{\chapterwidth}{0.8\textwidth}
\Chapter{First Deformation: Geometry\label{ch:dilatondriven}}
\defaultchapterwidth

\Section{Chiral Symmetry Breaking}
While much progress has been made in the sector of \adscftcorr, it has
proved difficult to find a realistic holographic dual of \acro{QCD}.
There are many reasons, which range from practical---working with
ten-dimensional supergravity equations---to principle: The ultraviolet
(\acro{UV}) regime is weakly coupled, which corresponds to strong
coupling (large curvature) on the \AdS\ side and hence the requirement
of quantising string theory on that background. Furthermore models
discussed so far contain only one scale and cannot provide a
separation of supersymmetry (\acro{SUSY}) breaking and confinement
scale $\Lambda_{QCD}$.

Despite those obstacles \adscftcorr\ has been remarkably successful in 
capturing many aspects of \acro{QCD}. In this Chapter,
such an aspect will be the important feature of\marginpar{chiral symmetry}
chiral symmetry breaking,
which shall be described holographically. 
Since supersymmetry prohibits chiral symmetry breaking 
as a non-vanishing chiral \acro{VEV} violates D\NB-flatness,
the
background geometry has to be deformed in such a way that \acro{SUSY}
is broken. At the same time it is desirable not to loose contact
to the well tested framework of \adscft. It is therefore crucial
to look at a geometry that in the ultraviolet approaches $\AdS_5\times \mf{S}^5$.

Here\marginpar{dilaton deformed backgrounds}
this will be achieved by preserving in the whole space time
an $\gr{SO}(1,3) \times \gr{SO}(6)$ isometry. There
are two \acro{IIB} supergravity
backgrounds in the literature \cite{Constable:1999ch,Gubser:1999pk,Kehagias:1999tr},
which satisfy this condition. The implications of the background
by Constable--Myers \cite{Constable:1999ch} have been studied
in \cite{Babington:2003vm}. Here the focus shall be on the
background found by 
Kehagias--Sfetsos and (independently) Gubser.

In analogy to the undeformed case of the previous Chapter, a D7\NB-brane
embedding parallel to the D3s will be considered and its scalar 
and vector fluctuations 
be studied. By diagonalising the fields, the discussion of multiple
D7\NB-branes reduces to several identical copies of the single brane case
and has therefore no impact on the mass spectrum. 
There is however the important difference that a D7\NB-brane
\emph{stack} admits non-trivial gauge configurations such that 
the Wess--Zumino term $C_4\wedge F\wedge F$ \emph{can} contribute.
The effect of non-trivial $F \wedge F$ will be studied
in the next Chapter, the Wess--Zumino term will be assumed to vanish 
for now and an Abelian Dirac--Born--Infeld action (\acro{DBI}) can safely 
be considered therefore.

\begin{figure}
\centering
  \EPSinclude[trim=-10 -10 -10 -10, height=9cm]{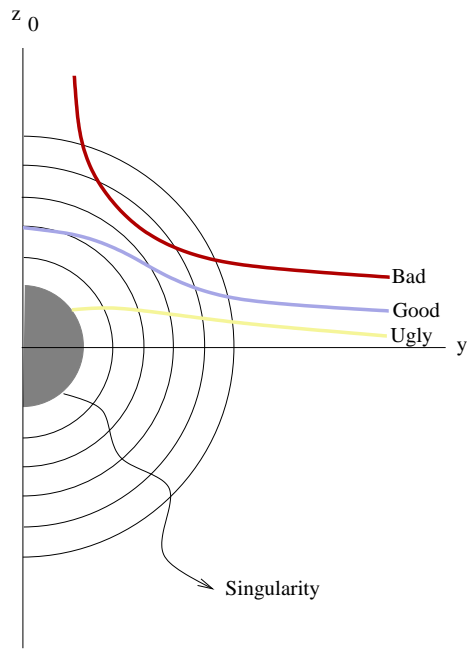}
  \caption[Regularity Conditions]{\label{fig:goodbadugly} Possible
    solutions for the D7 embeddings. The half circles correspond to
    constant energy scale $\mu$. The ``bad'' solution cannot have an
    interpretation as a field theoretic flow. The ``ugly'' solution
    hits the singularity (filled circle at the centre) and can thus
    not be relied on. (Plot taken from \cite{Babington:2003vm})}
\end{figure}

As\marginpar{chiral condensate vs.\ \acro{SUSY}}
has been explained in Section~\ref{sec:adscft}, the quark mass $m_q$ and
chiral quark condensate $c$ form the source/\acro{VEV} pair that is
described by the \acro{UV} values of scalar fields on the brane.
(Which in the string picture describe the transversal position of
the brane.)  In the supersymmetric scenario, the only solutions that
have a field theoretic interpretation require $c=0$ for all $m_q$. In
particular, this implies that there is no chiral quark condensate in
the limit $m_q\to0$ and no dynamical chiral symmetry breaking, hence.
Basically the problem is that in terms of geometry a chiral
condensate corresponds to a brane bending outward and behaving
irregular towards the interior of \AdS.
%
%
%
Since the radial direction of
the \AdS\ space corresponds to the energy scale in the field theory,
such a bending means that the field theory flows to the \acro{IR}
\emph{and comes back} as is shown (``Bad'') in Figure~\ref{fig:goodbadugly}.  
Clearly this is an unphysical behaviour.
The effect of the deformed background is that the D7\NB-brane
experiences attraction from the singularity and bends inward
compensating the effect of the boundary value $c$. 
This compensating is highly sensitive to the exact value of
the chiral condensate as a function of the quark mass, which 
completely fixes the functional dependence.

\begin{figure}
\centering%
\ifBeautifulForm
\else
\EPSinclude[width=12cm]{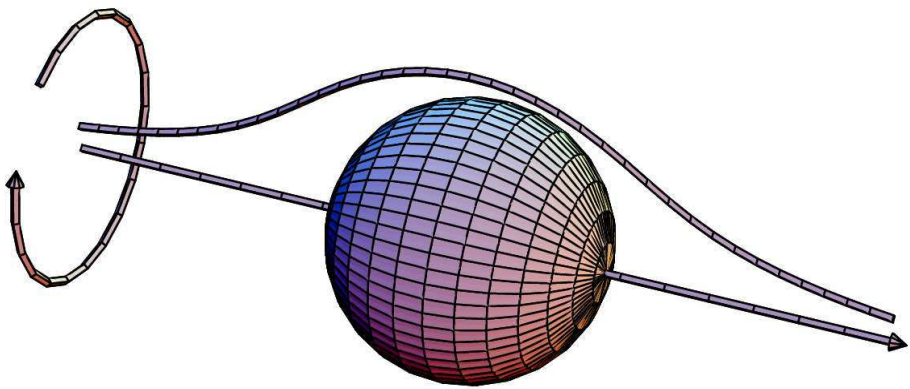}\vspace{1ex}
\fi

\caption[Breaking of the Axial Symmetry]{
  Spontaneous breaking of the $\gr{U}(1)_A$ symmetry, which 
  rotates (circle) the $8,9$\NB-plane, by the zero quark mass solution.  
  Zero quark mass means that the asymptotic separation between
  the brane embedding and the $y$-axis (large axis) vanishes.
  Non-vanishing quark mass means explicit and thereby not spontaneous
  symmetry breaking.\label{fig:gubser3d}
}
\end{figure}

In\marginpar{isometry} the previous Chapter, it was explained how adding D7\NB-branes to the
\adscftcorr\ breaks the $\gr{SO}(6)_{4\dots9} \simeq \gr{SU}(4)_R$
isometry of the six D3 transversal coordinates to an
$\gr{SO}(4)_{4567} \times \gr{SO}(2)_{89}$ isometry, which
corresponds to $\gr{SU}(2)_L \times \gr{SU}(2)_R \times \gr{U}(1)_R$,
with $\gr{SU}(2)_R \times \gr{U}(1)_R$ the R-symmetry group of the
$\cN=2$ superconformal Yang--Mills theory.\footnote{To be precise, the
  $\gr{SO}(2)_{89}$ corresponds to the $\gr{U}(1)_A$ axial symmetry,
  while the $\gr{U}(1)_R$ R\NB-symmetry is $\diag[ \gr{SO}(2)_{45}
  \times \gr{SO}(2)_{67} \times \gr{SO}(2)_{89}]$. Breaking of
  $\gr{SO}(2)_{89}$ implies breaking of the axial and R-symmetry
  simultaneously.}  Giving a mass to the $\cN=2$ hypermultiplet
corresponds to separating the two brane stacks and breaking the
conformal symmetry.  This has two effects: On the field theory side,
the breaking of conformal symmetry reduces the R-symmetry to
$\gr{SU}(2)_R$, on the supergravity side it breaks the rotational
invariance in the $8,9$\NB-plane associated to the $\gr{U}(1)_R$.  Now this
breaking acquires an additional interpretation in the limit $m_q\to0$,
where this $\gr{U}(1)$ is present in the \acro{UV}, but is broken
\emph{dynamically} by the branes bending away from the symmetry axis,
cf.~Figure~\ref{fig:gubser3d}: The symmetry spontaneously broken
by the chiral condensate is the $\gr{U}(1)_A$ axial symmetry.

Since\marginpar{Goldstone mode} determining the chiral symmetry
breaking behaviour is equivalent to finding correct D7\NB-brane
embeddings, one may go one step further and also find
fluctuations about these embeddings, which corresponds to meson
excitations in the correct field theoretic vacuum.  For vanishing
quark mass, the bilinear quark condensate breaks the axial symmetry
spontaneously and the associated meson becomes massless providing a
holographic version of the Goldstone theorem.

It should be noted that the explicit breaking of the $\gr{U}(1)_A$ by
an instantonic anomaly, which in \acro{QCD} is responsible for the
$\eta'$ to be heavy, is suppressed in the large $N_c$ limit. In that
sense the holographic $\eta'$ is more similar to a Pion even though it
is not related to the breaking of the chiral $\gr{SU}(N_f)_L \times
\gr{SU}(N_f)_R$ to its diagonal subgroup. Therefore in particular for
comparison with experimental data the Pion mass is a more appropriate
choice.

This Chapter~\marginpar{contents} is organised as follows. First, 
the Dirac--Born--Infeld (\acro{DBI}) action and 
the equations of motion describing the D7\NB-brane
embedding and fluctuations about the vacuum solution
will be derived. Then the background by Gubser, Kehagias--Sfetsos (\acro{GKS}) will be
shortly reviewed and transformed into a convenient 
coordinate system. The undeformed supersymmetric scenario
will be compared with a numerical evaluation of the 
chiral symmetry breaking and meson spectrum in the dilaton deformed 
background. Additionally, the behaviour of 
strongly radially excited mesons will be discussed.

\Section{\acro{DBI} to Quadratic Order}
Consider the diagonal background metric
\begin{align}
  ds^2 &= g_{xx}(y,z) dx^2_{1,3} + g_{yy}(y,z) (dy^2 + y^2 d\Omega_3^2) + \hat g_{zz}(y,z) (dz^2 + z^2 d\theta^2),  \label{eq:dilatonmetric}
\end{align}
which may be written as
\begin{align}
  g_{(10)} &= \diag ( g_{xx} \one_{1,3},\, g_{yy},\, g_{yy}\, y^2 \, \tilde g_{\alpha\beta},\, 
                      \hat g_{zz},\, \hat g_{\theta\theta} ), 
\end{align}
where $\tilde g_{\ia\ib}$ is the metric on the unit three sphere, and it holds 
\begin{align}
  \hat g_{\theta\theta} = z^2 \hat g_{zz}.
\end{align}
In the case $g_{yy}=\hat{g}_{zz}$, the radial direction of the warped
\AdS\ space can be expressed as $r^2=y^2 + z^2 =
\sqvec{y} + \sqvec{z}$ with $y, \Omega_3 \mapsto y^5,\dots, y^7$ and $z,
\theta \mapsto z^8 = z \sin \theta, z^9 = z \cos \theta$ a
transformation from respectively spherical or polar to Cartesian
coordinates.

Choosing static gauge,\marginpar{static gauge}
\begin{align}
  x^{0,\dots,3} &= \xi^{0,\dots,3},&
  y^{4,\dots,7} &= \xi^{4,\dots,7},&
  z^{8,\dots,9} &= \phi^{8,9} (\xi^{0,\dots,7}),
\end{align}
the \acro{DBI} action in Einstein frame 
for a D7\NB-brane in this background is given by 
\begin{align}
  S &= \int d^4x\, dy\, d\Omega_3 \sqrt{-g}\e^\varphi 
       \biggl[ \begin{aligned}[t] 1 + \hat g_{zz} g^{ab} (\p_a \Phi) (\p_b \bar\Phi)&\\ \mathrel{+} \half \e^{-\vphi} F_{ab} F^{ab}&\; \smash{\biggr]^{\frac{1}{2}}} \end{aligned} 
\end{align}
where expansion to second order in the
scalar fields $\Phi, \bar\Phi = \phi^9 \pm i \phi^8$ 
and field strength has been performed.
The remaining determinant is 
\begin{align}
  \sqrt{-g} &= y^3 \sqrt{\tilde g}\, (g_{xx}g_{yy})^2.
\end{align}

\Section{Quadratic Fluctuations}
Expanding an action
\begin{align}
  S = \int d^8\xi\, \Lag(\phi^i,\p_a \phi^i)
\end{align}
into small fluctuations $\delta\phi^i$ around a solution $\phi_0$
of the Euler--Lagrange-equations yields
\begin{align}
  \phi^i &= \phi_0^i + \veps \delta\phi^i, \\
  S &= \int d^8\xi\, \Lag_0 + \frac{1}{2} \veps^2 
      \biggl[ \frac{\p^2 \Lag }{\p(\p_a \phi^i ) \p (\p_b \phi^j)} \biggr]_{\veps=0} 
      (\p_a \delta \phi^i ) (\p_b \delta \phi^j ) .
\end{align}
Note that the above
statement is merely the Legendre criterion for
an extremal solution of a variational principle, which is a minimum
if the parenthesised expression above is positive definite.

In accordance with the previous Chapter, where dependence on $x$ was
associated to massive excitations and dependence on the spherical
coordinates $\Omega_3$ gave rise to Kaluza--Klein states, the embedding
of the D7 that forms a ground state should only depend on the radial
direction $y$. For fluctuations about a \marginpar{vacuum
  solution}\emph{vacuum solution} $\phi_0=\phi_0(y)$, $F_0^{ab} \equiv
0$, the quadratic expansion in scalar and vector fluctuations yields
\begin{align}
  S = \int d^4x&\, dy\, d\Omega_3 \sqrt{-g}\e^{\vphi}  
       \sqrt{1 + |\phi'_0(y)|^2 } \biggl[ \nonumber\\
       1&+ \frac{1}{2} \pfrac{g^{ab} \hat g_{ij}}{1 + |\phi'(y)|^2} 
             (\p_a \delta \phi^i)(\p_b \delta \phi^j) \\
       &- \frac{1}{2} \pfrac{g^{yy} \hat g_{ij} (\p_y \phi_0^i) (\p_y \delta \phi^j)}
                           {1 + |\phi'(y)|^2} ^2 
       + \frac{1}{4}\, \frac{F_{ab}F^{ab}}{1 + |\phi'_0(y)|^2} \, \biggr], \nonumber\\
\intertext{with}
\begin{split}
   \sqrt{-g} &= y^3 (g_{xx} g_{yy})^2 \sqrt{\tilde g}, \\
  |\phi'_0(y)|^2 &:= \hat g_{ij}\, g^{ab} (\p_a \phi_0^i) (\p_b \phi_0^j),
\end{split}
\end{align}
and $F_{ab}F^{ab}$ expressed solely in terms of fluctuations $\delta A_m$ about the trivial
background $A_m\equiv0$.

For\marginpar{Cartesian vs.\ polar}
numerics, expressing the scalar fluctuations in terms of Cartesian
fields $z^8,z^9$ has some advantages.\footnote{In particular, the
  excitation number $n$ of the meson tower \eqref{eq:KMMWsol}
  corresponds to the number of zeros of the solution to the radial
  equation \eqref{eq:KMMWradialeom}, which provides a good check
  whether a meson solution was accidentally skipped.}  From the field
theoretic point of view, expressing the fluctuations in polar
coordinates $z \e^{i\theta} = z^9+iz^8$ is more natural, because the
fluctuations of the pseudo-Goldstone mode correspond then exactly to
rotations of the $\gr{U}(1)_A$.  Since both approaches yield the same
results due to the infinitesimal nature of the fluctuations, the polar
coordinate formulation will be chosen here.

For $\Phi = z \e^{i\theta}$, $z = z_0(y) + \delta\sigma(\vec{x},\vec{y})$
and $\theta = 0 + \delta\pi(\vec{x},\vec{y})$, expansion of the \acro{DBI} action to
quadratic order in the fluctuations yields
\begin{align}
\begin{split}
  S = \int d^4x&\, dy\, d\Omega_3 \sqrt{\tilde g}\, \e^{\vphi} y^3 (g_{xx} g_{yy})^2  
       \sqrt{1 + z_0'(y)^2 } \biggl[ \\
       1&+ \frac{1}{2} \frac{g^{ab} \hat g_{\theta\theta}  (\p_a \delta\pi)(\p_b \delta\pi) 
                             +g^{ab} \hat g_{zz}  (\p_a \delta\sigma)(\p_b \delta\sigma) }{1 + z_0'(y)^2} \\
        &- \frac{1}{2} \pfrac{ g^{yy} \hat g_{zz} (\p_y z_0) (\p_y \delta\sigma)  }
                           {1 + z_0'(y)^2}^2 
       + \frac{1}{4}\, \frac{F_{ab}F^{ab}}{1 + |z_0'(y)|^2} \, \biggr], 
\end{split} \label{eq:abel-dbi-quad-fluct}
\end{align}
where $(z_0')^2=\hat{g}_{zz} g^{yy} (\p_y z_0)^2$.

\Section{Equations of Motion}

\Subsection{Vacuum}
From\marginpar{embedding}
\eqref{eq:abel-dbi-quad-fluct} by setting $\delta\sigma = \delta\pi = 0$, 
the equation describing the D7 embedding in terms of $z_0(y)$ is obtained,
\begin{align} \label{eq:vaceom}
\begin{split}
  \frac{d}{dy} \biggl[ \frac{ g^{yy} g_{zz} \cF(y,z_0) }{\sqrt{1 + z_0'(y)^2 } }\, z_0'(y) \biggr]
    &= g^{yy} g_{zz} \sqrt{1 + z_0'(y)^2 } \frac{\p}{\p z_0} \cF(y, z_0),\\
  \cF &= \e^{\vphi} y^3 (g_{xx} g_{yy})^2  .
\end{split}
\end{align}

\Subsection{Pseudoscalar Mesons}
\finalize{\begin{sloppypar}}
The\marginpar{Goldstone mode} pseudoscalar mesons correspond to
fluctuations along the $\gr{U}(1)_A$ and---as shall be seen
below---become massless for vanishing quark mass. They are thus
(pseudo-) Goldstone bosons, which become true Goldstones for
$m_q\to0$. Their equations of motion are
\finalize{\end{sloppypar}}
\begin{align}
   \p_a \biggl[ \frac{\sqrt{|\tilde g|} \cF(y,z_0)}{\sqrt{ 1 + z_0'(y)^2}} \hat g_{\theta\theta} g^{ab} \p_b \delta\pi \biggr] = 0,
\end{align}
which for the ansatz $\delta\pi = \delta\pi(y) \e^{ik\cdot x} \mathcal{Y}^\spinl(\mf{S}^3)$ and $M^2_\pi = - k^2$ read
\begin{align}\label{eq:radial-pseudoscalar-eom}
   \frac{\sqrt{1 + z_0'(y)^2} }{\cF} \p_y \biggl[ \cF \frac{g^{yy}\hat g_{\theta\theta} }{ \sqrt{1+ z_0'(y)^2} } \p_ y \delta\pi \biggr] & \\
     + \biggl[M_\pi^2 \hat g_{\theta\theta} g^{xx} -  \spinl(\spinl+2) \frac{\hat g_{\theta\theta} g^{yy}}{y^2} \biggr]& \delta\pi = 0, \nonumber
\end{align}
with the same shorthand $\cF$ as in \eqref{eq:vaceom}.

\Subsection{Scalar Mesons}
These\marginpar{Higgs mode} correspond to fluctuations in the radial direction transverse to the $\gr{U}(1)_A$. 
The equations of motion for the scalar mesons are
\begin{align}
   \p_a \biggl[ \frac{\sqrt{|\tilde g|} \cF(y,z_0)}{\sqrt{ 1 + z_0'(y)^2}} \hat g_{zz} g^{ab} \p_b \delta\sigma \biggr] = 
     \p_y  \biggl[ \frac{\sqrt{|\tilde g|} \cF(y,z_0)}{\sqrt{ 1 + z_0'(y)^2}^3} (\hat g_{zz} g^{yy})^2 (\p_y z_0)^2 \p_y \delta\sigma \biggr],
\end{align}
which for the ansatz $\delta\sigma = \delta\sigma(y) \e^{ikx} \mathcal{Y}^\spinl(\mf{S}^3)$ and $M^2_\sigma = - k^2$ become
\begin{align}\label{eq:radial-scalar-eom}
   \frac{ \sqrt{1 + z_0'(y)^2}}{\cF} \p_y \biggl[ \cF  \frac{\hat g_{zz} g^{yy}}{\sqrt{ 1+ z_0'(y)^2} }\biggl(1-\frac{\hat g_{zz} g^{yy} z_0'(y)^2}{ 1 + z_0'(y)^2}\biggr) \p_ y \delta\sigma \biggr] &\\
     + \biggl[ \hat g_{zz} g^{xx} M_{\sigma}^2 - \spinl(\spinl+2) \frac{\hat g_{zz} g^{yy}}{y^2} \biggr] & \delta\sigma = 0. \nonumber
\end{align}
Again it holds $\cF(y,z_0) = \e^{\vphi} y^3 (g_{xx} g_{yy})^2$.

\Subsection{Vector Mesons}
In accordance with Section~\ref{sec:flavour-gauge-fluct}, vector mesons can be obtained
from the D7\NB-brane gauge fields whose equations of motion are
\begin{align}
  \p_a \biggl[ \frac{ \sqrt{\tilde{g}}\, y^3 (g_{xx} g_{yy})^2 }{\sqrt{1 + z_0'(y)^2}}  F^{ab} \biggr] = 0
\end{align}
for solutions with no components on the $\mf{S}^3$, $\delta A_\alpha=0$. 
The  ansatz $\delta A_\nu=\xi_\nu\, \delta\rho(y) \e^{ik\cdot x} \mathcal{Y}^\spinl(\mf{S}^3)$,
where the polarisation vector $\xi_\nu$ satisfies $k_\mu \xi_\mu=0$, yields
\begin{align} \label{eq:radial-vector-eom}
  \frac{\sqrt{1 + z_0'(y)^2}}{y^3 (g_{xx} g_{yy})^2}
  \p_y \biggl[ 
     \frac{y^3 g_{xx} g_{yy}}{\sqrt{1 + z_0'(y)^2}}
   \p_y \delta\rho \biggr]& \\
  + \biggl[(g^{xx})^2 M_\rho^2  - \frac{\spinl(\spinl+2)}{y^2} \biggr] & \delta\rho = 0. \nonumber
\end{align}

\Section{Backgrounds}

\Subsection{\texorpdfstring{$\AdS_5\times \mf{S}^5$}{AdS(5) \texttimes\ S(5)}}
In this Section, it is demonstrated that the holographic description of 
the undeformed, supersymmetric case \cite{Kruczenski:2003be}
shows no chiral symmetry breaking. 
To describe the field theoretic vacuum, the embedding should neither depend
on $x$, which gives rise to a massive excitation, nor on the coordinates
of the internal $\mf{S}^3$, which gives rise to Kaluza--Klein states.
Using the $\gr{SO}(2)_{89}$ symmetry, one may choose the coordinate
system such that the embedding is simply $z^9 = z_0(y)$. 

Then the linearised equation of motion \eqref{eq:KMMWradialeom} is given by 
\begin{align}
  \biggl[ \p_{\tilde{y}}^2 + \frac{3}{\tilde{y}}\p_{\tilde{y}} \biggr] z_0(y) = 0
\end{align}
with $\tilde{M}=\spinl=0$. The full (as opposed to only asymptotic) solutions\footnote{Keep in mind
  that these are only solutions expanded to quadratic order. For the
  Abelian case one can do better, expand the determinant to all orders
  and keep the square root unexpanded. However, the outcome does not
  change.} are of the form
\begin{align}
 z_0(y) &= m + c\,y^{-2},
\end{align}
with the conformal dimension of the dual operator $\bpsi\psi$ given by
$\Delta=\spinl+3=3$.  For $c=0$, this is the constant embedding chosen
by Kruczenski et al.\ \cite{Kruczenski:2003be} and presented in the
previous Chapter.  For $c\neq0$, the solution diverges when
approaching the centre $y\to0$ of the D7\NB-branes.  This by itself is
still a valid D7\NB-brane embedding in the supergravity sense. However,
it does not have an interpretation\marginpar{\acro{IR} regularity} as
a field theoretic renormalisation group flow, because the D7\NB-brane
embedding cannot be expressed as a (one-valued) function of the radial
variable $r^2 = y^2 + z_0(y)^2$, which corresponds to the energy
scale. This is also depicted as the ``bad'' solution in
Figure~\ref{fig:goodbadugly}.


\Subsection{\acro{GKS} Geometry}
A particular solution to
the type \acro{IIB} supergravity equations of motion \eqref{eq:dilatonmetric}
that preserves $\gr{SO}(1,3)\times\gr{SO}(6)$ isometry was
found by Gubser \cite{Gubser:1999pk} (and independently 
by Kehagias--Sfetsos \cite{Kehagias:1999tr}), who chose an appropriate
warped diagonal ansatz for the metric, a Freund--Rubin ansatz
for the five-form flux and took only the dilaton as a 
non-constant supergravity field with a radial dependence.

The solution presented in \cite{Gubser:1999pk,Kehagias:1999tr} takes the form\footnote{In the original publication $B^2/24$ is used instead of $B^2$ to parametrise the deformation.}
\begin{align}
  ds_{10}^2 &= \e^{2\sigma} dx_{1,3}^2 + \frac{L^2 d\sigma^2}{1+ B^2 \e^{-8\sigma}} + L^2 d\Omega_5^2, \\
  \vphi - \vphi_0 
    &= \sqrt{\frac{3}{2}} \arcoth \sqrt{1+B^{-2} \e^{8\sigma}}, \label{eqn:dilaton-sigma}
\end{align}
where due to  $\arcoth x = \frac{1}{2} \ln \frac{x+1}{x-1}$ the dilaton $\vphi$ 
may be written as
\begin{align}
  \vphi - \vphi_0 
     &= \sqrt{\frac{3}{8}} \ln \frac{ \sqrt{1+B^{-2} \e^{8\sigma}}+1 }{ \sqrt{1+B^{-2} \e^{8\sigma}} - 1 }.
\end{align}
These coordinates are such that
\begin{center}
\begin{tabular}{llll}
\acro{IR} & $\sigma \to -\infty$ & singularity in the far interior, \\
\acro{UV} & $\sigma \to +\infty$ & boundary,
\end{tabular}
\end{center}
where there is a naked singularity in the infrared.

For\marginpar{$\gr{SO}(6)$ manifest coordinates} calculating the
meson spectrum in a background, it is more convenient to work in a
coordinate system that brings the metric exactly to the
$\gr{SO}(1,3)\times\gr{SO}(6)$ manifest form \eqref{eq:dilatonmetric}.
This can be achieved by the coordinate transformation
\begin{align}
 \e^{2\sigma} &= \sqrt{\frac{B}{2r_0^4}} \, r^2 \sqrt{ 1 - \frac{r_0^8}{ r^8 }}, \label{eq:gubser-trafo}
\end{align}
which yields
\begin{align}\label{eq:gubser-geometry}
  \begin{split}
  ds_{10}^2 &= g_{xx}(r) dx_{1,3}^2 + g_{yy}(r) (dr^2 + r^2 d\Omega_5^2), \\
  g_{xx}(r) &= \frac{r^2}{L^2} \sqrt{1-\frac{r_0^8}{r^8}},\\
  g_{yy}(r) &= g_{zz} = \frac{L^2}{r^2}, \\
  \vphi - \vphi_0 &= \sqrt{\frac{3}{2}} \ln \frac{r^4+r_0^4}{r^4-r_0^4}.
  \end{split}
\end{align}
Note that additionally $x$ has been rescaled such that $g_{xx}$ reproduces the
canonical normalisation of the asymptotic \AdS\ that is approached for
$r\to\infty$ and $r_0$ is the minimum value of $r$ where the infrared
singularity resides.  

For computations it is convenient to rescale the
coordinates by $r_0$ such that effectively $r_0\mapsto1$; i.e.\ all
equations become independent of $r_0$.  In this frame the quark mass is
measured in units of $r_0 T$, with $T$ the string tension, and the
meson mass in units $L^{-2} r_0$. As will be shown below, for large quark
masses the supersymmetric results of the undeformed $\AdS_5\times
\mf{S}^5$ are reproduced, such that $M \sim m_q$. Due to
\begin{align} \label{eq:gubserunits}
  \frac{M L^2}{r_0} = \text{const.}\cdot \frac{(2\pi\alpha')m_q}{r_0}
\end{align}
the supersymmetric limit $r_0\to0$ allows direct identification of the
numerical constant with that of equation \eqref{eq:KMMWsol}.  The
situation is more complicated for the similar background of
Constable--Myers, see Section \ref{sec:CM}, where by rescaling the
deformation parameter cannot be entirely removed from the equations of
motion, such that it also enters the numerical constant.  Moreover in
that background the units depend on the deformation parameter in such
a way that it does not cancel in a relation similar to
\eqref{eq:gubserunits}.

\Section{Chiral Symmetry Breaking in \acro{GKS}\label{sec:gubser-chiral-sym}}
For \eqref{eq:gubser-geometry}, the equation of motion \eqref{eq:vaceom} 
for the vacuum solution $z = z_0(y)$ is given by
\begin{gather}
  \frac{d}{dy} \biggl[\frac{y^3 f}{\sqrt{1 + z_0'(y)^2 }} z_0'(y) \biggr]
    = y^3 \sqrt{1 + z_0'(y)^2 } \frac{\p}{\p z_0} f , \label{eq:gubser-vac-eom} \\[2ex]
  f = \frac{ (r^4+1)^{(1+\Delta/2)} (r^4-1)^{(1-\Delta/2)} }{r^8}, \qquad r^2 = y^2+ z_0(y)^2, \qquad
  \Delta = \sqrt{6}.\nonumber
\end{gather}
The constant $\Delta$ has been defined for convenient comparison to a background
by Constable--Myers, cf.~Chapter~\ref{ch:heavylight}, 
and should not be mixed up with the conformal dimension.

\begin{figure}
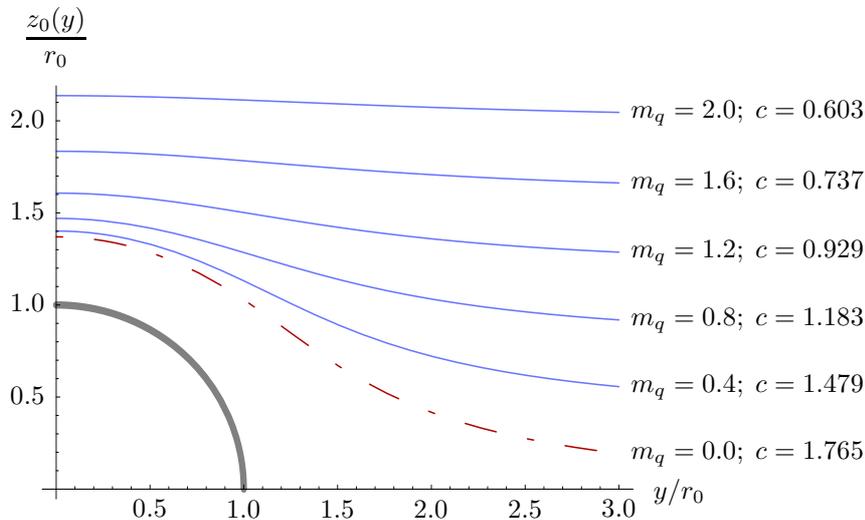

\centering
\PSfraginclude[trim=-5 -20 -120 -25]{width=12cm,height=75mm}{gubser-vac}%
\caption[D7\NB-brane Embeddings]{
  The Figure shows regular D7 embeddings
  with different quark mass. The embedding coordinate 
  $z_0(y)$ is a radial coordinate in the $8,9$\NB-plane.
  While all solutions break the rotational $\gr{U}(1)_A$ symmetry in 
  that plane, the zero quark mass solution (dashed) does so \emph{spontaneously}.
  \label{fig:gubser-vac}
}
\end{figure}

\begin{figure}
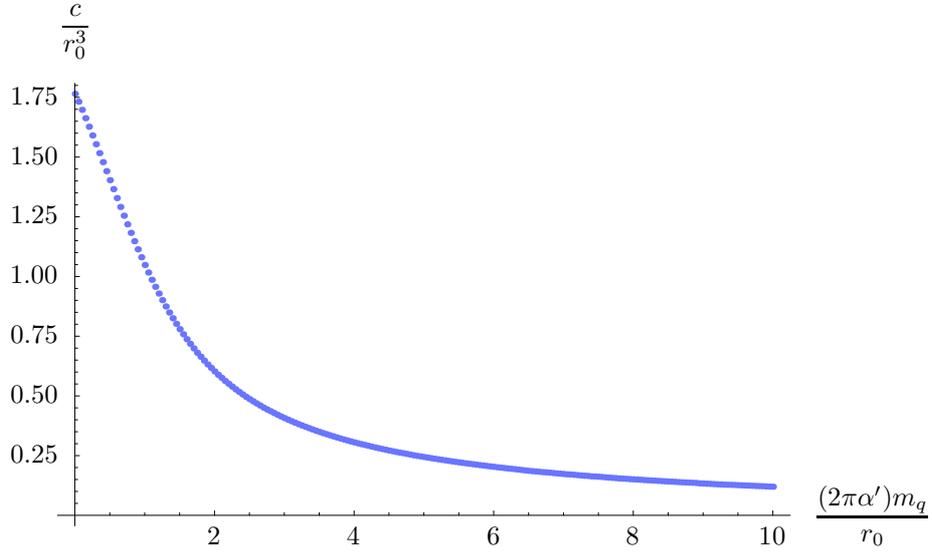
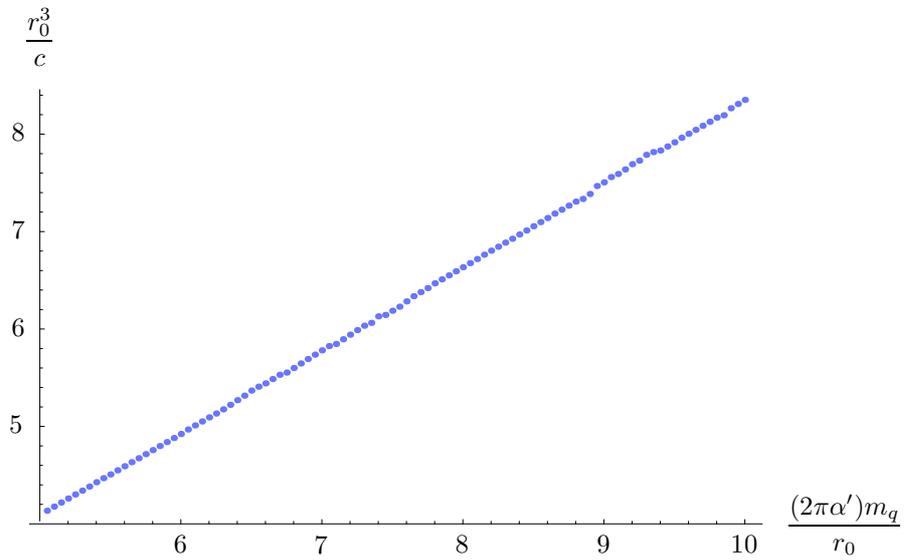

\centering
\subfigure[]{%
  \PSfraginclude[trim=-5 -10 -40 -20]{width=12cm}{gubser-m-vs-c}%
}\\
\subfigure[]{%
  \PSfraginclude[trim=-5 -10 -40 -40]{width=12cm}{gubser-m-vs-c-linear}%
}
\caption[Chiral Condensate in Dilaton Deformed Background]{
  The first plot shows the chiral condensate $\langle\bpsi\psi\rangle$ 
  as a function of the quark mass $m_q$ as determined by regularity
  requirements for the D7 embedding. For large quark mass $m_q$ the 
  chiral condensate behaves like $c\sim \frac{1}{m_q}$ in accordance 
  with predictions from effective field theory. 
  \label{fig:gubser-m-vs-c}}
\end{figure}

Since the background \eqref{eq:gubser-geometry} approaches
$\AdS_5\times\mf{S}^5$ towards the boundary, it does not come as a
surprise that the \acro{UV} behaviour of $ z_0(y)$ is given by
$m_q+c y^{-2}$ with $m_q$ the quark mass and $c$ the bilinear quark
condensate as in the supersymmetric case. In the infrared there are
still two solutions of qualitatively different behaviour: One is
divergent and cannot correspond to field theoretic vacuum therefore,
the other approaches a constant. However, the infrared dynamics is
modified such that the pair in the \acro{UV} mixes while going to the
\acro{IR}.  Whereas in the supersymmetric case the \acro{UV} solution
with $c=0$ corresponded one-to-one to the regular behaviour in the
\acro{IR}, now for each value of $m_q$ there is only one value of $c$
such that the combined solution mixes into a regular one in the
\acro{IR}.  Such regular solution have been determined numerically and
are plotted in Figure~\ref{fig:gubser-vac}.  Each of the solutions is
determined by a pair of quark mass and quark condensate.  These pairs
also determine the quark condensate as a function of the quark mass as
is shown in Figure~\ref{fig:gubser-m-vs-c}.

The\marginpar{qualitative behaviour} possible outcomes for
arbitrary combinations of $m_q$ and $c$ are depicted in Figure
\ref{fig:goodbadugly}: The solution can hit the singularity (denoted
``ugly'', since the supergravity approximation fails when coming to
close to the singularity), the solution may diverge (denoted ``bad'',
because it cannot correspond to a field theoretic flow), or the
solution may reach a constant value for the embedding coordinate 
$z_0(y)$ at $y=0$, denoted ``good''. In terms of the ambient space radial
coordinate $r^2 = y^2 + z_0^2$ the D7\NB-brane ``ends'' at $r= z_0(0)$
by the $\mf{S}^3$ slipping from $\mf{S}^5$ of the background and
shrinking to zero size at a pole of the $\mf{S}^5$,
cf.~Figure~\ref{fig:brane-cigar}.  There\marginpar{stability} is a
tachyon associated to this slipping mode, but its mass obeys
(saturates) the Brei\-ten\-loh\-ner--Freed\-man bound \cite{Breitenlohner:1982jf} and does not
lead to an instability hence.

One might however worry about
regular solutions reaching the singularity. For the
discussion of whether this may happen, it is advantageous to shift the
point of view to the infrared.  

\begin{figure}
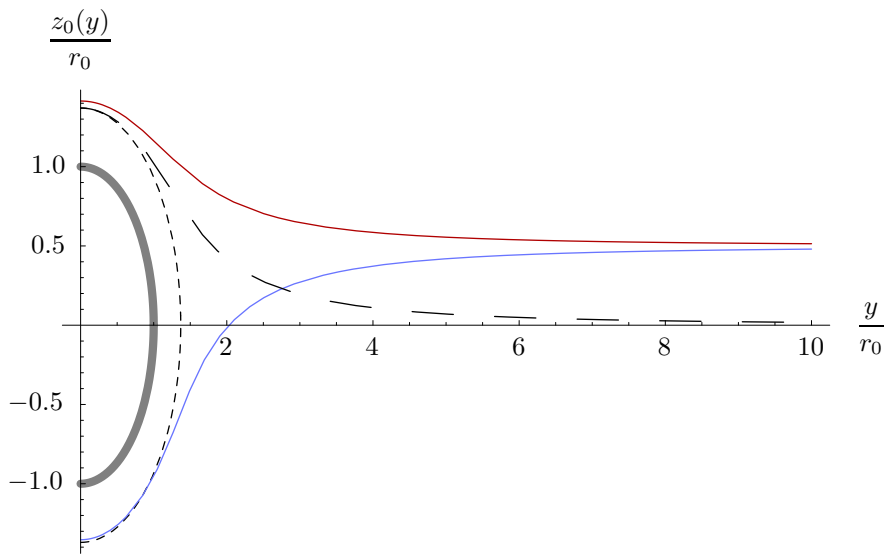

\centering
  \PSfraginclude[trim=-10 -10 -20 -25]{width=12cm}{gubser-negmass}
\caption[Singularity Shielding]{
  Two solutions of the same quark mass and the zero quark mass solution
  (dashed) are depicted. The zero mass solution exactly 
  avoids the region between the inner circle, which is the 
  singularity, and dashed outer ``shielding'' circle. 
  Of the two massive solutions, only the one with the larger 
  action enters the shielded region, cf.~fig.~\ref{fig:gubser-actions}.
  \label{fig:gubser-shielding}}
\end{figure}

Starting\marginpar{singularity shielding} at a \emph{finite} value in
the \acro{IR}, there has to be a \emph{unique} flow to the \acro{UV},
which fixes the correct combination of $m_q$ and $c$, since one also
needs the \acro{IR}-\emph{divergent} solution to create arbitrary
combinations of $m_q$ and $c$.  As has been explained above, $z_0(0)$
sets the scale were the quarks drop out of the dynamics. So one
generically may expect that a large quark mass corresponds to a large
value of $ z_0(0)$.  Starting at distances closer to the singularity
generates solutions with smaller quark mass till one reaches a
limiting solution at $ z_0(0)\approx 1.38$ that corresponds to
vanishing quarks mass.  Going even closer to the singularity gives
rise to a spurious negative quark mass.  Due
to the $\gr{SO}(2)_{89}$ present, these solutions are in fact positive
mass solutions with negative\marginpar{negative condensate} quark condensate, as can be seen by
rotating around the $y$-axis, see Figure~\ref{fig:gubser3d}.  This
assigns two potentially valid solutions to each positive quark
mass.\footnote{ The situation is to some extent analogous to asking
  which is the shortest route connecting two points on a sphere. The
  answer is a grand circle, which however also provides the longest
  straight route.  } However solutions that do not come closer to the
singularity than the zero quark mass solution have a smaller potential
energy $V=-\Lag$,
cf.~Figure~\ref{fig:gubser-actions}, and are therefore physical.
This realises some sort of screening mechanism preventing
solutions from entering the region between the zero-quark mass
solution and the singularity, cf.~Figure~\ref{fig:gubser-shielding}.
The physical solutions outside have a positive quark condensate.

\begin{figure}
\centering
  \PSfraginclude[trim=-10 -10 -40 -20]{width=12cm}{gubser-actions}
  \caption[D7\NB-brane Action and Physical Solution]{Potential Energy
    of D7\NB-brane embeddings as function of the quark mass:  Since
    the energy itself is infinite what is actually plotted is the
    finite difference to the action of the zero quark mass solution
    defined as follows,\vspace{-1em}\\\noindent
    \begin{minipage}{\linewidth}
    \begin{align*}
      E(m_q) - E(0) = - \Delta S = - \lim_{Y\to\infty} \int\limits_0^Y
      \Lag(m_q) - \Lag(m_q=0) dy.
    \end{align*}
    \end{minipage}
    The physical solutions have smaller energy and are farther from
    the singularity than the zero-mass solution.
    \label{fig:gubser-actions}}
\end{figure}

Having\marginpar{vanishing quark mass} established the conditions that
determine the chiral condensate as a function of the quark mass---the
result is plotted in Figure~\ref{fig:gubser-vac}---the case $m_q=0$
will be discussed in more detail now.  $ z_0(y)\equiv0$ is obviously a
solution of the equations of motion, which does however reach the
singularity.  To obtain a solution exhibiting chiral symmetry breaking
behaviour, it is necessary to either start at a suitable value in the
\acro{IR}, thus forcing the solution to behave as desired, or to start
with an infinitesimal deviation from $ z_0\equiv m_q=0$ in the
\acro{UV}. This situation is analogous to calculations of the
magnetisation in solid state physics, where spontaneous symmetry
breaking is initiated by an arbitrarily small but non-vanishing
external $B$ field.

The conclusion is that indeed spontaneous chiral symmetry breaking is
observed in this geometry and one may wonder about the appearance of
an associated Goldstone mode.

\begin{figure}
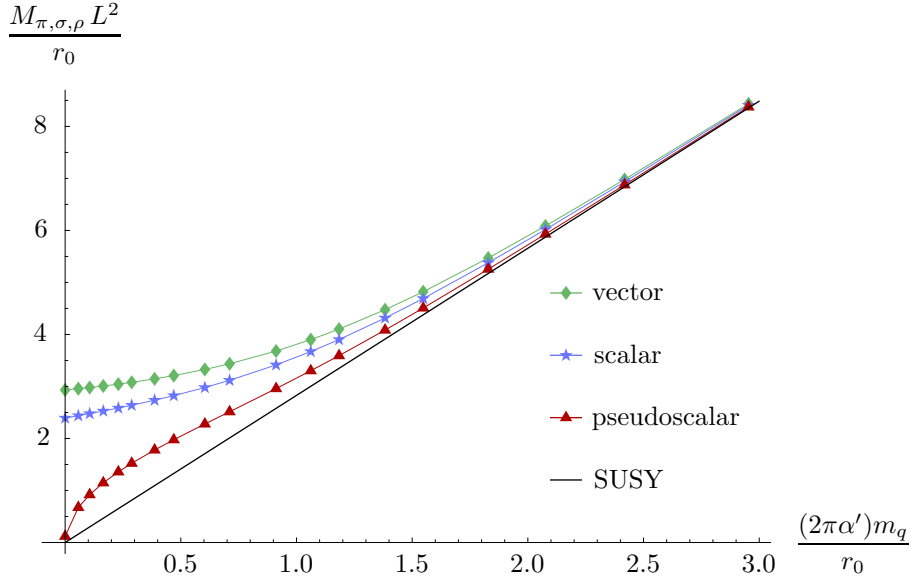
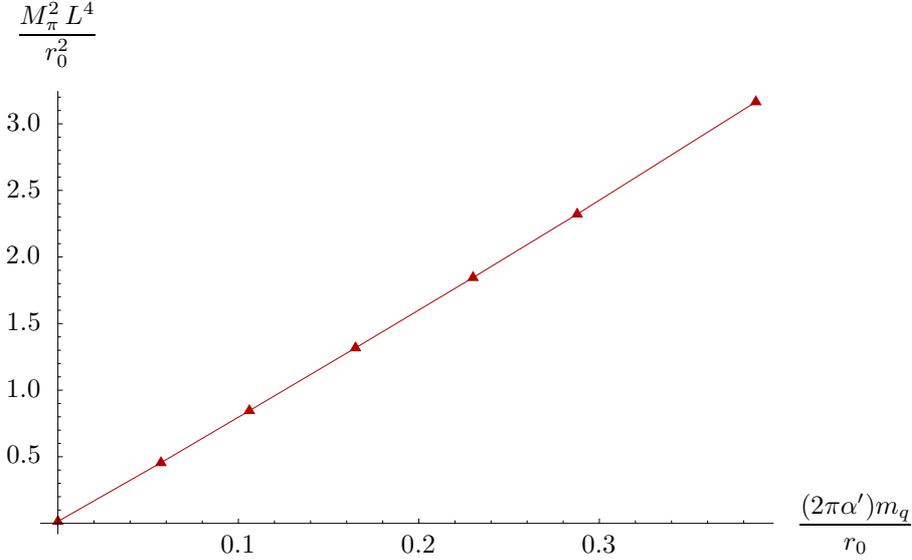

\centering
\subfigure[Lightest mesons]{
  \PSfraginclude[trim=-5 -20 -25 -25]{width=12cm}{gubser-lowmesons}
}
\subfigure[\label{fig:gubser-gmor}Pseudoscalar meson and \acro{GOR} relation: $M_\pi \sim \sqrt{\strut m_q}$]{
  \PSfraginclude[trim=-5 -20 -40 -25]{width=12cm}{gubser-gmor}
}
\caption[Lightest Scalar, Pseudoscalar and Vector Meson]{
  Plot (a) shows the lightest vector, scalar and pseudoscalar meson (in order of
  decreasing mass). While the scalar and vector meson retain a mass, 
  the pseudoscalar meson becomes massless and therefore a true Goldstone boson 
  in the limit $m_q\to0$. Furthermore its mass exhibits a square root behaviour
  as predicted from effective field theory, plot~(b).
  For large quark masses, supersymmetry is restored and the analytic
  \acro{SUSY} result $M(n=0,\spinl=0)=\frac{2m_q}{L^2}\sqrt{2}$ is 
  reproduced (black straight line).
  \label{fig:gubser-lowmeson}}
\end{figure}

\Section{Mesons\label{sec:gubser-mesons}}
The meson spectrum\marginpar{spectrum} is determined by finding regular and normalisable
solutions to the equations of motion arising from fluctuations about
the brane embedding.  These equations, given in
\eqref{eq:radial-pseudoscalar-eom}, \eqref{eq:radial-scalar-eom} and
\eqref{eq:radial-vector-eom}, can be solved in analogy to the case of
the embedding equation \eqref{eq:vaceom}, which has been discussed in
the previous Section. The solutions of the meson equations have a
boundary behaviour of generic type $c_1+c_2/y^2$.
 In contrast to the
embedding solutions, where regularity fixed $c_2$ as a function of $c_1$,
the fluctuations should always be normalisable, such that the
solutions behave as $y^{-2}$ towards the boundary. The remaining overall
factor $c_2$ is undetermined because the equations of motion are linear.
The requirement of
regularity in the infrared can then only be satisfied by a discrete set of
values for the meson mass $M$, which determines the spectrum.  The
result for the lowest lying meson modes is depicted in
Figure~\ref{fig:gubser-lowmeson}.

With\marginpar{Goldstone} these results it is possible to return to
the question of a holographic realisation of Goldstone's theorem.  For
the following discussion, it is important to keep in mind that the
supergravity approximation in \adscftcorr\ implies being in the
$N_c\to\infty$ limit, where the $\gr{U}(1)_A$ axial symmetry is
non-anomalous in the field theory.  A look at the large $N_c$ limit of
\acro{QCD}, where the $\eta'$ becomes massless and thus a true
Goldstone boson, inspires to look for the corresponding
(pseudo\NB-)\linebreak[0]Goldstone meson in this geometric setup.  

A massless embedding with \acro{UV} behaviour $ z_0(y) \sim c\,y^{-2}$
restores the $\gr{U}(1)_A \simeq \gr{SO}(2)_{89}$ symmetry in the
\acro{UV} and therefore shows spontaneous symmetry breaking.  In
particular that means that the embedding solution $ z_0
\e^{i\theta_0}$, has an undefined angle $\theta_0$ at the boundary,
which acquires an arbitrary value along the flow, picked out spontaneously by the
dynamics. Clearly any fluctuation in the $\theta$ angle simply
corresponds to a rotation into an---because of the presence of the
$\gr{U}(1)$---equivalent but different value of $\theta_0$. Since these
values are all equivalent, the fluctuation in the $\theta$ direction
should be a flat direction in the potential and correspond to a
massless meson.  

When the $\gr{U}(1)_A$ symmetry is explicitly broken
in the \acro{UV} by the quark mass ($ z_0 \sim m_q + c(m_q)\,y^{-2}$),
fluctuations in the angular direction do not rotate into an equivalent
embedding and are therefore expected to become massive.
Figure~\ref{fig:gubser-lowmeson} shows that this holographic version
of the Goldstone theorem is indeed realised. 
Furthermore beyond a certain quark mass, supersymmetry\marginpar{\acro{SUSY}} is restored
and the meson masses become degenerate.
For small quark mass, Figure~\ref{fig:gubser-gmor}, 
accordance with a prediction from effective field theory is found,
the Gell-Mann--Oakes--Renner (\acro{GOR}) relation \cite{Gell-Mann:1968rz}
\begin{align}
  M_\pi^2 &= \frac{m_q \vev{\bpsi\psi}}{N_f f_\pi^2}.
\end{align}

\Section{Highly Excited Mesons\label{sec:highexcite}}
In this Section inspired by a similar analysis in \cite{Schreiber:2004ie}, 
the highly excited meson spectrum in the present background 
shall be investigated. 
In \adscft\ this corresponds to considering mesons with large
radial excitation number $n$. According to \cite{Glozman:2004gk}
the semi-classical approximation becoming valid in this limit
gives rise to a restoration of chiral symmetry, because its
breaking resulted from quantum effects at one-loop order 
which are suppressed for $S\gg\hbar$.

\begin{figure}
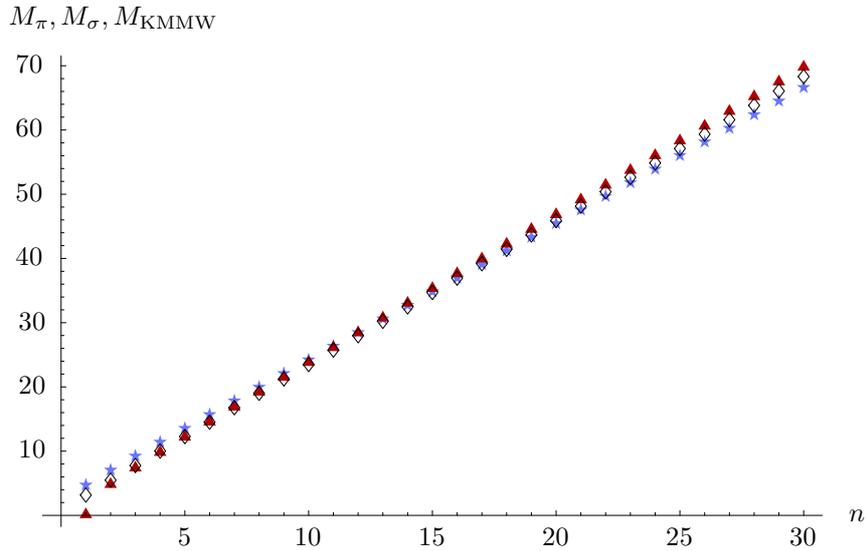
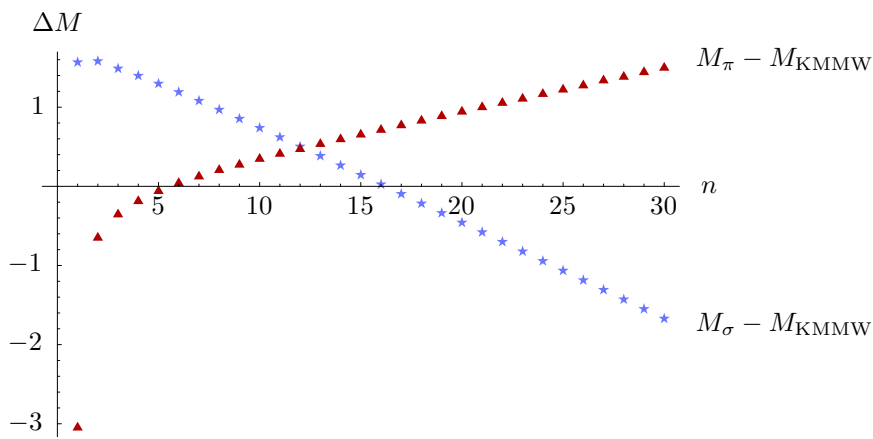

\centering
\subfigure[Highly Excited Mesons]{
  \PSfraginclude[trim=-5 -10 -20 -20]{width=12cm}{gubser-highexcite1}
}
\subfigure[Difference to \acro{SUSY} Case]{
  \PSfraginclude[trim=-5 -10 -90 -20]{width=12cm}{gubser-highexcite2}
}
\caption[Highly Excited Mesons (Overview)]{These plots show highly
  radially excited mesons for the $m_q=0$ embedding (with $r_0=L=1$ for numerics). 
  For the 
  analytically solvable \acro{SUSY} case, this 
  corresponds to $n\gg1$ and therefore 
  $M_\text{KMMW}=2\sqrt{(n+\spinl+1)(n+\spinl+2)}\sim 2n$. 
  While the proportionality to $n$ is preserved in the deformed case,
  the overall slope of the \acro{SUSY} case is different 
  and has been adjusted by multiplying $M_\text{KMMW}$ by $1.15$ for 
  comparison. The difference to this rescaled mass as depicted in 
  plot (b) suggests that the mass of the scalar and pseudoscalar
  mesons \emph{is not degenerate} in the limit of large excitations.
  \label{fig:gubser-highmeson}}
\end{figure}

\cite{Schreiber:2004ie} found the rather generic behaviour 
\begin{align}
  M_n &\sim n,&
  n &\gg 1, \label{eq:gubser-highexcite-holo}
\end{align}
for holographic duals of \acro{QCD}-like theories. This
is not in accordance with field theoretic expectations \cite{Shifman:2005zn}, 
which can be derived from simple \marginpar{scaling arguments}scaling arguments: 
The length of the flux tube spanned between two ultra-relativistic quarks of energy
$E = p = M_n /2$ is
\begin{align}
  L \sim \frac{M_n}{\Lambda_{QCD}^2},
\end{align}
such that from the quasi-classical quantisation condition
\begin{align}
  \int p \,dx \sim p\, L \sim \frac{M_n^2}{\Lambda_{QCD}} \sim n,
\end{align}
one reads off
\begin{align}
  M_n \sim \sqrt{n}.
\end{align}
This is in contradiction to the results \eqref{eq:gubser-highexcite-holo}
and also the numerical results one obtains for the \acro{GKS} background shown
in Figure~\ref{fig:gubser-highmeson}. However this behaviour
was to be expected since it is also found in the analytic 
spectrum of Kruczenski et\ al. 

A\marginpar{chiral symmetry restoration}
to some extent related question is whether the \emph{difference}
$\delta M_n$ of the scalar and pseudoscalar meson mass shows the right
field theoretic behaviour, which has been predicted to be $|\delta M_n|
\lesssim n^{-3/2} \Lambda_{QCD}$ with alternating sign for $\delta
M_n$ \cite{Shifman:2005zn}.

While the analytic supersymmetric case fulfils this requirement
trivially $\delta M_n = 0$, interestingly this seems not to be the case 
for the \acro{GKS} background
as can be seen in
Figure~\ref{fig:gubser-highmeson}. Actually $\delta M_n$ even could not be
shown to vanish at all in the limit $n\to\infty$ implying that neither chiral
symmetry nor supersymmetry is restored.  

\begin{figure}
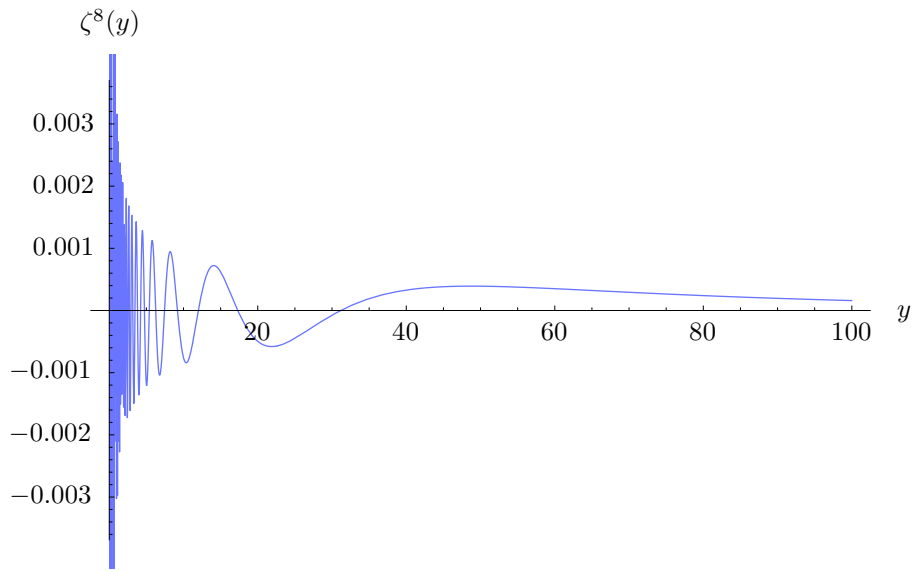
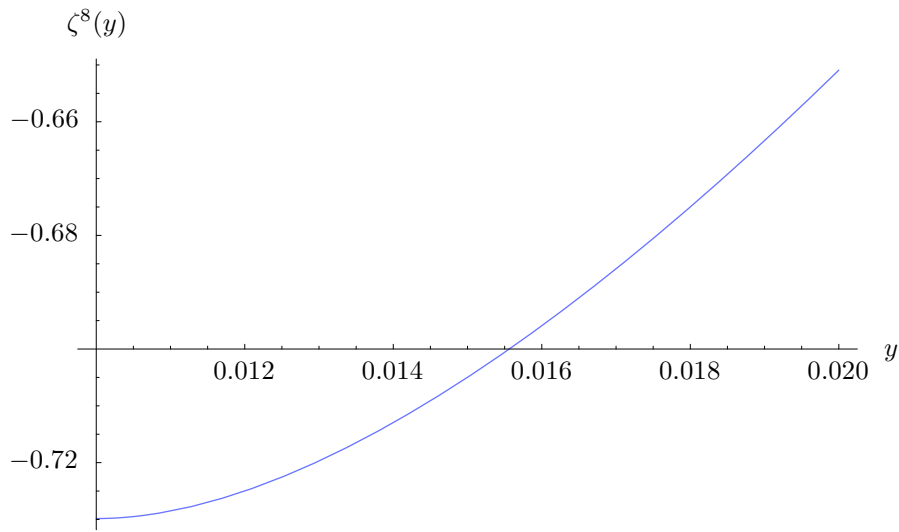

\centering
\subfigure[Strong \acro{IR} dependence]{
  \PSfraginclude[trim=-15 -10 -10 -20]{width=12cm}{gubser-zeta5-far}
}
\subfigure[\acro{IR} Regularity]{
  \PSfraginclude[trim=-15 -10 -10 -10]{width=12cm}{gubser-zeta5-zoom}
}
\caption[Close-up of a Highly Radially Excited Meson]{Pseudoscalar meson solution with excitation
  number $n=49$; i.e.~the solution plotted in (a) has 49 zeros.
  Most of them concentrate in the far \acro{IR}, but the 
  solution is still smooth close to the centre (b).  
  Increasing the excitation number scans the \acro{IR} in more
  detail, where scalar and pseudoscalar meson mass are different. 
  Therefore it is not expected to find mass degeneracy when 
  increasing $n$ further. (Note that Cartesian fluctuations as
  opposed to polar fluctuations in $z$ and $\theta$ have been plotted.
  The mass spectrum is independent of this choice.) \label{fig:gubser-zeta}}
\end{figure}

Having a closer look at the
behaviour of such highly excited mesons,
cf.~Figure~\ref{fig:gubser-zeta}, one notices that the effect of large
radial excitation is that the interior of the holographic space
corresponding to the field theory's infrared is probed more densely.
This suggests that for highly excited mesons in such a holographic
description infrared effects might indeed not be sufficiently
suppressed.  On the other hand it seems surprising that mass
degeneracy is not restored contrary to the case of large quark mass,
where the mesons end up in the supersymmetric regime and do become
degenerate as has been demonstrated in
Figure~\ref{fig:gubser-lowmeson}. 

Currently the method for calculating the meson spectrum inherently 
requires expansion to quadratic order in fluctuations. It would
certainly be interesting to extend this procedure to include higher
order contributions and reexamine the question of whether 
at least restoration of mass degeneracy can be achieved in the
limit of highly radial excitation.

\begin{savequote}[\savequotewidth]
  When I'm working on a problem, I never think about beauty. I think
  only how to solve the problem. But when I have finished, if the
  solution is not beautiful, I know it is wrong.  
  \qauthor{R.\ Buckminster Fuller} 
\end{savequote}

\Chapter{Second Deformation: Gauge~Fields\label{ch:higgsbranch}}

\Section{Introduction}

In this Chapter the meson spectrum of the Higgs branch of the 
particular $\cN=2$
super-Yang--Mills (\acro{SYM}) theory \eqref{eq:n2-lag} that can be
described by a D3/D7-brane system \cite{Karch:2002sh} in the framework
of \adscftcorr\ \cite{Maldacena:1998re,Witten:1998qj,Gubser:1998bc}
will be determined.  The analogous calculations for the Coulomb branch
can be performed analytically \cite{Kruczenski:2003be}, see
Chapter~\ref{ch:flavour}, and can be made contact to in the cases of
zero and infinite Higgs vacuum expectation value (\acro{VEV}).

The work presented here is intrinsically a generalisation of the D3/D7
system of Chapter~\ref{ch:flavour} to the case of \emph{more than one} D7
brane, which corresponds to having multiple quark flavours.  In
particular, an additional effect that goes beyond simply having
multiple copies of the Abelian case is considered.  On the
supergravity side it arises from the Wess--Zumino term in the D7\NB-brane
action, allowing four-dimensional instanton configurations to be
classical solutions of the D7\NB-brane gauge fields. On the field theory
side this corresponds to switching on a vacuum expectation value
(\acro{VEV}) for the fundamental hypermultiplet. The field theory is
therefore on the Higgs branch.

In the following Sections, the dual field theory will be presented and
the exact notion of ``Higgs branch'' (which actually is a mixed
Coulomb--Higgs branch) will be clarified. A short review of the
\acro{BPST} instanton solution is given.

The equations of motions are derived that determine the vector meson
spectrum, which is calculated numerically and discussed analytically
in the limits of small and large Higgs \acro{VEV}.  Finally the
operator dictionary is explained and the fluctuations corresponding to
scalar mesons are shown to fall into the same supermultiplets.

\Section{Conventions}
The main difference between this Chapter and the preceding ones 
is the use of a non-Abelian D7\NB-brane action to extend the 
analysis of the \acro{SUSY} D3/D7 system to the sector of
two flavours ($N_f=2$).  Therefore, the introduction of non-Abelian 
gauge covariant derivatives 
\begin{align*}
  \covD_a &= \partial_a + g A_a, \\
  F_{ab} &= \partial_a A_b - \partial_b A_a + g \comm{A_a}{A_b}, \\
\end{align*}
can no longer be avoided and in addition to the index conventions of
Table~\ref{tab:indices}, a few notations have to be established.
%
\begin{table}
\centering
\begin{colortabular}{*{10}{Bc}}
        \multicolumn{10}{Hc}{Coordinates}\\
        \rowcolor{white}
        0&1&2&3&4&5&6&7&8&9\\ \bwline
        \multicolumn{4}{Cc}{D3}&&&&&&\\ \hline
        \multicolumn{8}{Cc|}{D7}&&\\ \hline
        \multicolumn{4}{Cc}{$x^{\mu,\nu,\dots}$}&
                \multicolumn{4}{|Cc|}{$y^{m,n,\dots}$ } &
                \multicolumn{2}{Cc}{$z^{i,j,\dots}$}\\ \hline
	&&&&\multicolumn{3}{|Cc|}{$y^{M,N,\dots}$}&&&\\ \hline
        &&&&\multicolumn{6}{|Cc}{$r$}\\ \hline
        &&&&\multicolumn{4}{|Cc|}{$y$}&&\\ \hline
	\multicolumn{8}{Cc|}{$X^{a,b,\dots}$}&&\\ \hline
	\multicolumn{10}{Cc}{$X^{A,B,\dots}$}\\ \bwline
\end{colortabular}
%
\caption{\label{tab:indices}Index Conventions}
\end{table}

The indices $M,N,\dots$ will also be used as $\gr{SU}(2)$ generator indices, 
with the convention $\veps_{456}=1$ and the Hermitean Pauli matrices
\begin{align*}
  (T_4,T_5,T_6) &= \left(	\begin{pmatrix} 0& 1\\ 1&0 \end{pmatrix},
  			\begin{pmatrix} 0&-i\\ i&0 \end{pmatrix},
			\begin{pmatrix} 1& 0\\-1&0 \end{pmatrix} \right),\displaybreak[0]\\
  T_M T_N &= i\veps_{MNK} T_K, \qquad
  \tr {T_M T_N} = 2\delta_{MN}, 
\end{align*}
which allows to introduce the (anti-Hermitean) quaternion
 basis\marginpar{quaternion basis}
\begin{align}
\sigma_{4,5,6,7} = ( i T_{4,5,6}, \one ).
\end{align}

The reader shall be reminded that in this basis $\gr{SO}(4)_{4 5 6 7}$
transformations of $y^m$ can be also written as
\begin{align}
  y^m \sigma_m \mapsto y^m U_L \sigma_m U_R,
\end{align}
with $U_L$ and $U_R$ elements of $\gr{SU}(2)_L$ and  $\gr{SU}(2)_R$
respectively. Since the vector $(0,0,0,y^7)$ is invariant under
transformations $U_L=(U_R)^{-1}$, rotations in the first 
three coordinates $\gr{SO}(3)_{456}$ can be
identified with the diagonal subgroup $\diag [\gr{SU}(2)_L \times 
\gr{SU}(2)_R]$.

\Section{Dual Field Theory}
On the string theory side, the setup discussed here is that
of a stack of D3\NB-branes and a parallel stack of D7s.
In the decoupling limit, this amounts to considering
type \acro{IIB} supergravity (\acro{SUGRA}) on 
$\AdS_5\times \mf{S}^5$ with $N_f$ probe D7\NB-branes,
which is dual to an $\cN=2$ gauge theory obtained
from coupling $N_f$ $\cN=2$ hypermultiplets in the fundamental
representation to $\cN=4$ $\gr{SU}(N_c)$ \acro{SYM} \cite{Karch:2002sh}.

In $\cN=1$ language the
Lagrangean of the dual field theory is
\begin{align}\label{eq:n2-lag}
\begin{split}
  \Lag 
    &= \int d^4\theta \tr \left( \bar{\Phi}^i \e^{2V} \Phi^i \e^{-2V} 
        \mathrel{+} Q_i^\dagger \e^{2V} Q^i 
        + \tilde Q_i \e^{2V} \tilde Q^{i\dagger} \right) \\&\quad
    + \biggl[ \frac{1}{4g^2} \int d^2\theta \, W_\alpha W^\alpha 
                           + \int d^2\theta \,W + \text{c.c.} \biggr]
%
\end{split}
\end{align}
where the chiral fields $\Phi_{1,2,3}$ and the gauge field $V$ build up the 
$\cN=4$ adjoint hypermultiplet, which in turn can be split into an $\cN=2$
adjoint hypermultiplet composed of $\Phi_{1,2}$ and an $\cN=2$ adjoint 
gauge multiplet of $V$ and $\Phi_3$. $Q^i$ and $\tilde Q_i$ are the $N_f$ chiral fields
that build up the $\cN=2$ fundamental hypermultiplet, and the superpotential is
\begin{align}
W= \tr (\epsilon_{IJK}\Phi_I\Phi_J\Phi_K) + \tilde Q_i (m_q+\Phi_3) Q^i.  \label{eq:d3-d7-superpot}
\end{align}

At\marginpar{stability} finite $N_c$ this theory is not
asymptotically free, and the corresponding string background suffers
from an uncancelled tadpole.  However, in the strict probe limit
$N_f/N_c \to 0$, the contributions to the \tHooft\ couplings $\beta$
function, which scale like $N_f/N_c$, are suppressed. Furthermore
the
dual AdS string background has no tadpole problem because the
probe D7\NB-branes wrap a contractible $\mf{S}^3$. Although contractible,
the background is stable, since the tachyon associated with
shrinking the $\mf{S}^3$ satisfies (saturates) the
Brei\-ten\-loh\-ner--Freed\-man bound \cite{Breitenlohner:1982jf}. Moreover
the $\AdS_5 \times \mf{S}^3$ embedding has been shown to be
supersymmetric \cite{Skenderis:2002vf}.

\Subsection{Higgs Branch}
In terms of $\cN=2$ multiplets, the theory consists of an adjoint gauge and
hypermultiplet, which form the $\cN=4$ hypermultiplet of
$\cN=4$ $\gr{SU}(N_c)$ \acro{SYM}, and $N_f$ fundamental
hypermultiplets.  When the scalars of the latter acquire a \acro{VEV}, the
theory is on the Higgs branch.

While the scalars $\phi_{1,2}$ of the adjoint hypermultiplet independently
may also
have a \acro{VEV}, \acro{VEV}s of the $\cN=2$ gauge multiplet's scalar $\phi_3$
prohibit a \acro{VEV} for the fundamental hypermultiplets. Refining the
discussion for the components gives rise to the mixed Coulomb--Higgs
branch. The superpotential in $\cN=1$ language is\footnote{There are
  three adjoint $\cN=1$ chiral fields $\Phi_{1,2,3}$ with lowest
  components $\phi_{1,2,3}$ and one real field $V$, which forms an
  $\cN=2$ gauge multiplet with $\Phi_3$. The $N_f$ chiral fields $Q^i$
  and $\tilde{Q}_i$ make up the $\cN=2$ fundamental hypermultiplet and
  have lowest components $q^i$ and $\tilde{q}_i$.}
\begin{align}
  W= \tr (\epsilon_{IJK}\Phi_I\Phi_J\Phi_K) 
     + \tilde Q_i (m_q+\Phi_3) Q^i \tag{\ref{eq:d3-d7-superpot}},
\end{align}
with index $i$ enumerating the $N_f=2$ hypermultiplets.

Assume that a small number $k$ of the components of $\phi_3$ obtain a
\acro{VEV},
\begin{align}\label{eq:higgs-coulomb}
  (\phi_3)_{N_c\times N_c} = \begin{pmatrix} 0& & & & & \cr &\ddots & & & &  \cr
        & &0 & & & \cr & & &-v & &  \cr & & & &\ddots & \cr
        & & & & & -v
\end{pmatrix},
\end{align}
which is dual to separating out $k$ D3\NB-branes from the stack, and
moreover that these \acro{VEV}s are exactly such that some of the
components of $m+\langle\phi_3\rangle$ vanish, $v=m$, which is dual to
the singled out D3\NB-branes coinciding with the D7\NB-branes.  Then
F-flatness conditions $\tilde{q}_i (\phi_3+m) = (\phi_3+m)q^i = 0$
permit the corresponding $2k$ components of the fundamental
hypermultiplet to also acquire a non-vanishing \acro{VEV}
\begin{align}
  (q^i)_{N_c \times 1} &= \begin{pmatrix}
          0 \cr \vdots \cr 0 \cr \alpha_1^i \cr
               \vdots \cr \alpha_k^i 
       \end{pmatrix},&
  (\tilde q_i)_{1 \times N_c} &= \begin{pmatrix} 
                  0 & \cdots & 0 & \beta_{1i} & \cdots &  \beta_{ki} \end{pmatrix}.
\end{align} 
These \acro{VEV}s, which are further constrained by
additional F- and D-flatness conditions, are the string theory dual
of the D3\NB-branes that coincide with the D7\NB-branes \emph{to be
dissolved}\marginpar{dissolved branes} \cite{Douglas:1995bn} in the D7\NB-branes and form
instantons in the gauge fields of the D7s. This process is caused
by the Wess--Zumino coupling $S_{WZ} \sim \int P[C_{(4)}] \wedge F
\wedge F$.
Note that the backreaction of the dissolved D3\NB-branes can only be neglected
when their number $k$ is small in comparison to $N_c$. Specifically in 
this Chapter $k=1$ will be assumed. 

Taking into account the breaking of $\gr{SU}(2)_R \times \gr{SU}(2)_f$ 
to its diagonal subgroup, which is mediated by the instanton configuration
on the supergravity side, the structure of the \acro{VEV}s is as follows
\begin{align}\label{eq:higgs-coulomb-1}
  (\phi_3)_{N_c\times N_c} &= \begin{pmatrix} 0& & & & & \cr &\ddots & & & &  \cr
        & &0 & & & \cr 
        & & & & & -m
       \end{pmatrix},&
  (\mathfrak{q}^i_\ia) &= \begin{pmatrix} 0 \cr \vdots \cr 0 \cr \veps_{i\ia} \Lambda \end{pmatrix},
\end{align}
with 
$\ia=1,2$ the $\gr{SU}(2)_R$ index 
and $\mathfrak{q}_1 = q$, $\mathfrak{q}_2 = \tilde{q}$.


\Section{Supergravity}
\Subsection{Instantons}
In Yang--Mills (\acro{YM}) theories, instantons arise as finite action
solutions from the semi-classical approximation to path integrals,
which requires to find all solutions that minimise the Euclidian
action.  These solutions, (anti-)self dual gauge field configuration
of arbitrary topological charge $k$, can be found from a set of
algebraic equations, the so-called \acro{ADHM} constraints due to
Atiyah, Drinfeld, Hitchin and Manin.  These equations are non-linear
and cannot be solved in general because of their complex structure,
though there has been recent progress in \adscft\ inspired large $N_c$
considerations \cite{Dorey:1999pd}. In particular the four dimensional
\acro{ADHM} constraints arise from D and F-flatness conditions of
D$(p+4)$\NB-branes probed by D$p$\NB-branes
\cite{Witten:1995gx,Douglas:1996uz}.

Although the \acro{ADHM} formalism works for all non-exceptional groups, the
focus here will be on $\gr{SU}(N)$ theories in Euclidian space.
Consider the following action,
\begin{align}
  S = - \frac{1}{2} \int d^4x \tr F_{mn}^2 + i \theta k, \label{higgs-inst-action}
\end{align}
with the topological charge and field strength
\begin{align}
  k &:= -\frac{g^2}{16 \pi^2} \int d^4y \tr F_{mn}\hodge F_{mn},\qquad k \in \mathds{Z},\displaybreak[0]\\
  F_{mn} &:= \p_m A_n - \p_n A_m + g \comm{A_m}{A_n},\\
  \hodge F_{mn} &:= \half \veps_{mnkl} F_{kl}
\end{align}
and anti-Hermitean gauge field $A_m$ such that the covariant derivative reads
$\covD_m = \p_m + g A_m$.

Instantons with negative topological charge, also known as
anti-instan\-tons, will not be considered here. The action is minimised
by self dual solutions
\begin{align}
\begin{split}
  \hodge F_{mn} &= \pm F_{mn},\\
  \implies S &= - 2\pi i k\tau \quad k>0,
\end{split}
\end{align}
with the complex coupling $\tau = \frac{4\pi i}{g^2} + \frac{\theta}{2\pi}$.

The self-dual $\gr{SU}(2)$ instanton solution, 
also known as the Belavin--Polyakov--Shvarts--Tyupkin (\acro{BPST}) instanton \cite{Belavin:1975fg}, is 
given by
\begin{align}
	\Ainst_n &= g^{-1} \frac{2 (y_m-Y_m) \sigma_{mn}}{(y-Y)^2+\Lambda^2}, &
	F^\text{inst}_{mn} &= g^{-1} \frac{4 \rho^2 \sigma_{mn} }{((y-Y)^2+\Lambda^2)^2},
\end{align}
with the instanton moduli $\Lambda$ (size) and $Y^m$ (position). 
The Lorentz generators are given by
\begin{align}  
	\sigma_{mn}	&= \frac{1}{4} (\sigma_m\bar\sigma_n-\sigma_n\bar\sigma_m),&
	\bar\sigma_{mn} &= \frac{1}{4} (\bar\sigma_m\sigma_n-\bar\sigma_n\sigma_m),
\intertext{%
and it holds
}
	\sigma_{mn}	&= \frac{1}{2} \veps_{mnkl}\sigma_{kl},&
	\bar\sigma_{mn} &= -\frac{1}{2} \veps_{mnkl}\bar\sigma_{kl}.
\end{align}
The above identification of gauge indices with vector indices
expresses the instanton breaking the $\gr{SU}(2)_L \times \gr{SU}(2)$
to its diagonal subgroup, with $\gr{SU}(2)_L$ from the double covering
group of the Euclidian Lorentz group $\gr{SO}(4)$ and $\gr{SU}(2)$ the
gauge group.

The \acro{BPST} instanton falls off slowly for large distances, which
creates convergence problems of various integrals. A well known
solution in the instanton literature is the use of a singular gauge
transformation
\begin{align}
  U(y) &:= \frac{\sigma_m (y-Y)^m}{\abs{y-Y}}, \label{eq:higgs-singular-gauge-trafo}
\end{align}
which transforms the non-singular instanton solution to a singular one,
\begin{align}
  A_n = g^{-1} \frac{2 \Lambda^2 (y-Y)_m \bsigma_{mn} }{ (y-Y)^2 [(y-Y)^2 + \Lambda^2]}, \label{eq:singinst}
\end{align}
that has better large distance behaviour. This particular gauge
transformation also associates $\gr{SU}(2)_R$ with the gauge group,
such that \eqref{eq:singinst} breaks the $\gr{SU}(2)_L \times
\gr{SU}(2)_R \times \gr{SU}(2)$ to $\gr{SU}(2)_L \times \diag
[\gr{SU}(2)_R \times \gr{SU}(2)]$.  Note that also in the instanton
literature a known consequence of
\eqref{eq:higgs-singular-gauge-trafo} is \emph{the modification of
  boundary terms}.  Therefore consequences for the \adscft\ dictionary
are also to be expected.


\Subsection{D7-brane Action}
As a reminder the $AdS_5 \times \mf{S}^5$ background as given in
\eqref{eq:ads5-s5}, \eqref{eq:c4} is
\begin{align}
    ds^2 &= H^{-1/2}(r) \,\eta_{\mu\nu}dx^\mu dx^\nu + %
           H^{1/2}(r)\, ( d\vec{y}^{\,2} + d\vec{z}^{\,2} ), 
\end{align}
with
\begin{align}
    H(r) &= \frac{L^4}{r^4},       & r^2 &= \vec{y}^{\,2} + \vec{z}^{\,2}, \\
    L^4 &= 4\pi g_s N_c (\alpha')^2,\qquad  & \vec{y}^{\,2} &= \sum_{m=4}^{7} y^m y^m, \\
    C^{(4)}_{0123} &= H^{-1} ,     & \vec{z}^{\,2} &= (z^8)^2+(z^9)^2, \\[2ex]
    \e^{\vphi}&=\e^{\vphi_\infty}=g_s.
\end{align}
The constant embedding 
\begin{align}
  z^8 &= 0, &
  z^9 &= \tmq 
\end{align}
defines the distance $\tmq=(2\pi\alpha')m_q$ between the D3 and D7\NB-branes
and therefore determines the mass $m_q$ of the fundamental hypermultiplet. 

Moreover it yields the induced metric \eqref{eq:KMMWindD7}
\begin{align}\label{D7geom}
\begin{split}
  ds^2_{\text{D7}} &=
     H^{-1/2}(r)\, \eta_{\mu\nu}dx^\mu dx^\nu +
     H^{1/2}(r)\, d\vec{y}^{\,2},  \\[1ex]
  r^2 &= y^2 +  (2\pi\alpha')^2 m_q^2, \qquad y^2 \equiv y^m y^m
\end{split}
\end{align}
on the D7\NB-brane.

At quadratic order, the non-Abelian \acro{DBI} action \eqref{eq:dbi-quad}
and the Wess--Zumino term \eqref{eq:nonabelian-d7-wz}
are respectively
\begin{align}
\begin{split}
  S_\text{DBI} &= -\mu_7 \int d^{p+1}\xi \str \e^{-\vphi} \sqrt{-\det G_{ab}}    
     \biggl[ \tfrac{\la^2}{2} \D_a \Phi_i \D^a \Phi^i  
             + \tfrac{\la^2}{4}  F_{ab}F^{ab}
     \biggr]\\
     &= - \tfrac{T_7 \gamma^2}{4} \int d^4x\, d^4y\; \tr \Bigl[ -
         \begin{aligned}[t]
            & 2 H(r) \covD_\mu \Phi \covD_\mu \bar\Phi   
              + 2 \covD_m \Phi \covD_m \bar \Phi + \\
            & H(r) F_{\mu\nu} F_{\mu\nu} + 2 F_{m\nu}F_{m\nu} +\\
            & H^{-1}(r) F_{mn} F_{mn} \Bigr],
         \end{aligned}
\end{split}\displaybreak[1]\\
\begin{split}
  S_{WZ} &= T_7 \int \str \frac{\gamma^2}{2} P[C^{(4)}] \wedge F \wedge F\\
        &=  T_7 \frac{\gamma^2}{4} \int \tr
               H^{-1}(r) F_{mn}
            \tfrac{1}{2}F_{rs} \;
            dx^0 \wedge \dots dx^3 \wedge 
               \underbrace{dy^m \wedge dy^n \wedge dy^r \wedge dy^s}
                         _{=\veps_{mnrs} \; dy^4 \wedge dy^5 \wedge dy^6 \wedge dy^7} \\
         &= T_7 \frac{\gamma^2}{4} \int d^4x\, d^4y\; H^{-1}(r) \tr F_{mn} \; \hodge F_{mn},
\end{split}
\end{align}
where $\Phi,\bPhi=\Phi^9 \pm i \Phi^8$, $\gamma=2\pi\alpha'$
 and the Hodge dual is $\hodge
F_{mn}:=\half \veps_{mnrs} F_{rs}$, with the epsilon symbol $\veps_{4567}=1$.
All indices have been lowered and are now contracted with a Minkowski
metric $\eta_{ab}=(\eta_{\mu\nu},\delta_{mn})$. 
This will be true for all subsequent expressions in this Chapter,
providing a convenient framework for the discussion of solutions that
are self-dual with respect to the flat metric $\delta_{mn}$. 

These\marginpar{\acro{DBI}/\acro{WZ} conspiracy}
solutions arise because there is a (known,
cf.~\cite{Witten:1995gx,Douglas:1996uz}) correspondence between
instantons and the Higgs branch.  The discussion in this thesis will be
confined to quadratic order,\footnote{The explicit expanded form of
  the non-Abelian \acro{DBI} action is only known to finite order,
  cf.~\cite{Oprisa:2005wu} for terms at sixth order.  The existence of
  instanton solutions puts constraints on unknown higher order terms
  \cite{Guralnik:2004ve,Guralnik:2005jg}.} where the \acro{DBI} and
Wess--Zumino term due to $F_{mn} ( F_{mn} - \hodge F_{mn} ) =
2F^-_{mn}F^-_{mn} $ complement one another to yield
\begin{align}
\begin{split}
  S &= - \tfrac{T_7 \gamma^2}{4} \int d^4x\, d^4y\; \tr \Bigl[ -
         \begin{aligned}[t]
            & 2 H(r) \covD_\mu \Phi \covD_\mu \bar\Phi   
              + 2 \covD_m \Phi \covD_m \bar \Phi + \\
            & H(r) F_{\mu\nu} F_{\mu\nu} + 2 F_{m\nu}F_{m\nu} +\\
            & 2 H^{-1}(r) F^-_{mn} F^-_{mn} \Bigr]. 
         \end{aligned}
\end{split} \label{eq:higgs-quad-action}
\end{align}
This action is extremised by the configuration
\begin{gather}
\begin{split}
  &\boxed{ F^-_{mn} = 0, } \label{eq:selfdual-bg} \\
  \Phi=\tmq,&\qquad 
  F_{\mu\nu}=F_{mn}= 0, 
\end{split}
\end{gather}
which is manifestly self-dual with respect to the D3-transversal 
flat metric $\delta_{mn}$.
The particular background configuration that will be investigated here,
\begin{align}
  A_m   &= \frac{2\Lambda^2\bsigma_{mn}y_n}{y^2(y^2+\Lambda^2)},&
  A_\mu &= 0, &  
  \Phi_0 &= \tilde{m}_q, 
\end{align} 
takes the singular gauge instanton \eqref{eq:singinst} 
as an ansatz for \eqref{eq:selfdual-bg} that brings the
correct boundary behaviour for the \adscft\ dictionary
as will be seen below.
 

\Section{Meson Spectrum}
 
Now the meson spectrum for fluctuations about the above
background shall be calculated. Obviously there should
be massless mesons corresponding to changes of 
the instanton moduli, size ($\Lambda$) and position 
(not explicit in the above ansatz, since the instanton 
is simply located at $y_m=0$).  These will be ignored and
concentration will be instead on the more interesting
fluctuations of the gauge fields and scalars. 
The simplest modes are vector fluctuations of type~II,
cf.~eq.~\eqref{eq:kmmw-type-II}, and scalar fluctuations, 
both in the same supermultiplet and in the
the lowest representation of  $\gr{SU}(2)_L \times \diag [\gr{SU}(2)_R \times \gr{SU}(2)_f]$.
In particular this means that the fluctuations will
be assumed to be independent of angular variables
in the D3\NB-transversal/D7\NB-longitudinal coordinates;
i.e.~in the language of the analytically solvable scenario of Chapter 
\ref{ch:flavour}: $\spinl=0$.

\Subsection{Vector Fluctuations}
In accordance with the coordinate splitting $X_a = x_\mu,y_m$
performed in the action~\eqref{eq:higgs-quad-action},
fluctuations of the form $\Afluct := A - \Ainst$ will be considered.
The simplest ansatz for gauge fluctuation, which at the same time
is most interesting due to describing vector mesons, is given by
``Type~II'' fluctuation \eqref{eq:kmmw-type-II} in the language of Kruczenski et~al., 
see Chapter~\ref{ch:flavour}.
This particular ansatz is non-trivial in the D3-longitudinal
components only, such that the simplest non-Abelian choice is
a singlet under $\gr{SU}(2)_L$ and a triplet under $\diag[\gr{SU}(2)_R\times \gr{SU}(2)_f]$.
An obvious ansatz is given by
\begin{gather} \label{eq:higgs-vec-ansatz}
  \Afluct_\mu{}^{(a)} = \xi_\mu(k) f(y) \e^{ik_\mu x_\mu} T^a,
  \qquad y^2 \equiv y^m y^m,
\end{gather}
and 
\begin{align}
  A_\mu &= \Afluct_\mu,&
  A_m   &= \Ainst_m.
\end{align}
The Euler--Lagrange equations
\begin{align}
  \partial_\mu \frac{\partial\Lag}{\partial\partial_\mu A_\nu^M} 
    + \partial_m \frac{\partial\Lag}{\partial\partial_m A_\nu^M}
    - \frac{\partial\Lag}{\partial A_\nu^M} &= 0, \label{eqn:eomgreek} \\
  \partial_\mu \frac{\partial\Lag}{\partial\partial_\mu A_n^M} 
    + \partial_m \frac{\partial\Lag}{\partial\partial_m A_n^M}
    - \frac{\partial\Lag}{\partial A_n^M} &= 0 \label{eqn:eomlatin}
\end{align}
for the action \eqref{eq:higgs-quad-action} are
\begin{align}
  \covD_\mu \left(H F_{\mu\nu} \right) + \covD_m F_{m\nu} &= 0, \label{eqn:eomgreekexpanded}\\
  \covD_\mu F_{\mu n} + 2\, \covD_m \left[ H^{-1} F^-_{mn} \right] \label{eqn:eomlatinexpanded}&=0.
\end{align}
To linear order the former becomes $\p_\mu\Afluct_\mu=0$, 
which is solved by $k_\mu \xi_\mu = 0$, while the latter
reads 
\begin{align}
\begin{split}
H \partial_\mu \partial_\mu \Afluct_\nu  
  + \partial_m \partial_m \Afluct_\nu + g\; \partial_m \comm{ \Ainst_m }{ \Afluct_{\nu} }\qquad& \\
  + g \comm{ \Ainst_m }{ \partial_m \Afluct_\nu }
  + g^2 \comm{ \Ainst_m }{ \comm{ \Ainst_m }{ \Afluct_{\nu} } }
  &= 0,
\end{split} \label{eqn:flucteom}
\end{align}
which for the ansatz \eqref{eq:higgs-vec-ansatz} yields
\begin{align}
  0 = \biggl[ \frac{M^2 L^4}{(y^2 + (2\pi\alpha')^2 m_q^2)^2}
                - \frac{8 \Lambda^4}{y^2(y^2+\Lambda^2)^2}
                + \frac{1}{y^3}\partial_y (y^3 \partial_y ) \biggr] f(y),
                \label{eqn:ansatzeom}
\end{align}
where $M^2 = -k_\mu k_\mu$ in accordance with having chosen a
Minkowski metric with mostly plus convention for contraction
of flat indices. 

For numerics it is convenient to join the two parameters 
quark mass and instanton size by rescaling according to
\begin{align}
        \tilde y &\equiv \frac{y}{2\pi\alpha' m_q}, &
        \tilde\Lambda &\equiv \frac{\Lambda}{2\pi\alpha' m_q}, &
        {\tilde M}^2 &\equiv \frac{M^2 L^4}{(2\pi\alpha' m_q)^2},
\end{align}
such that equation \eqref{eqn:ansatzeom} becomes
\begin{align}
  0 = \biggl[ \frac{{\tilde M}^2}{({\tilde y}^2 + 1)^2}
                - \frac{8 {\tilde\Lambda}^4}{{\tilde y}^2({\tilde y}^2
                +{\tilde\Lambda}^2)^2}
                + \frac{1}{{\tilde y}^3}\partial_{\tilde y}
                ({\tilde y}^3 \partial_{\tilde y} ) \biggr] f({\tilde y}).
                \label{eqn:eomrescaled}
\end{align}
\begin{figure}
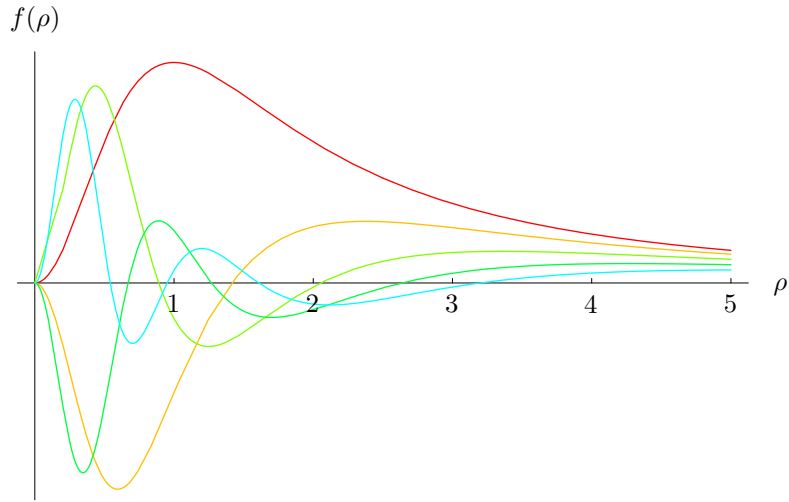
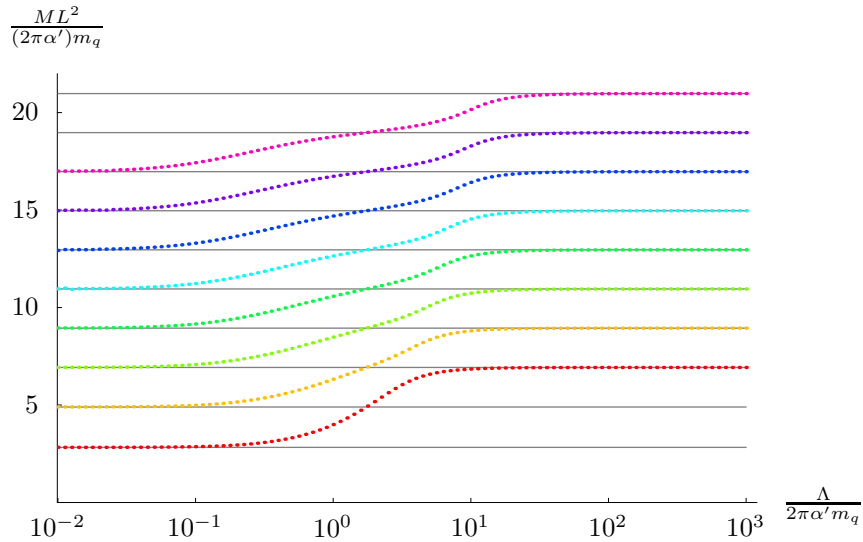

\centering
  \subfigure[Regular solutions of \eqref{eqn:eomrescaled} in %
             arbitrary units for $\Lambda=2\pi\alpha' m$]{
    \PSfraginclude{width=11cm}{fluct-sols}
  }\\
  \subfigure[Numerically determined meson masses.]{\label{fig:higgs-mesonmass-B}
    \PSfraginclude{width=12.5cm}{meson-ladder}
  } 
\caption[Meson Masses on the Higgs Branch]{\label{fig:higgs-mesonmass}
  Each dotted line represents a regular solution of the equation of
  motion, corresponding to a vector multiplet of mesons. 
  Plot (a) shows the five regular solutions of \eqref{eqn:eomrescaled} 
  corresponding to the lightest meson masses in (b). The units on 
  axis of ordinate in (a) are arbitrary because \eqref{eqn:eomrescaled} 
  is a linear equation. The vertical axis in (b)
  is $\sqrt{\lambda}M/m_q$ where $M$ is the meson mass, $\lambda$
  the \tHooft\ coupling and $m_q$ the quark mass. The horizontal axis
  is $v/m_q$, where $v= \Lambda/2\pi \alpha' $ is the Higgs \acro{VEV}.
  In the limits of zero and infinite instanton size (Higgs VEV), 
  the spectrum (grey horizontal lines)
  obtained analytically in Chapter~\ref{ch:flavour} is recovered.
}
\end{figure}

At\marginpar{operator map} 
large  $\tilde y$ (\ref{eqn:eomrescaled}) has two linear
independent solutions whose asymptotics are given by
$\tilde y^{-w}$ with $w=0,2$.  The normalisable
solutions corresponding to vector meson states behave as
$\tilde y^{-2}$ asymptotically. From  standard \adscftcorr, 
one expects $w=\Delta$ and
$w=4-\Delta$, where $\Delta$ is the \acro{UV} conformal dimension
of the lowest dimension operator with the quantum numbers of the
vector meson.  However, the kinetic term does not have a standard
normalisation; i.e.~the radial component of the Laplace operator
appearing in the equation above is not (only)
$\partial_{\tilde y}^2$, and consequently an extra factor of
$\tilde y^\alpha$, for some $\alpha$, appears in the expected
behaviour; so the exponents actually read $w=\alpha+\Delta, 
\alpha+4-\Delta$.
From the difference it is read off that $\Delta=3$. The dimensions
and quantum numbers are those of the $\gr{SU}(2)_f$ flavour current,
\begin{gather} \label{flavour}
  {\cal J}_\mu^b
    = - \bar \psi^{\pm i} \gamma_\mu \sigma^b{}_{ij} \psi_{\mp}{}^{j}
    + \bar q^{\alpha i } \stackrel{\leftrightarrow}{D}_\mu
           \sigma^b{}_{ij}\, q_\alpha{}^j \, ,
\end{gather}
with $\alpha$ the $\gr{SU}(2)_R$ index and $i,j$ the flavour indices.
This current has $\gr{SU}(2)_R\times \gr{SU}(2)_L \times \gr{U}(1)$  quantum
numbers $(0,0)_0$.

The asymptotic behaviour of the supergravity solution is
\begin{align}
\Afluct^{\mu}_{b (a) } =
\xi^{\mu}(k)\e^{ik\cdot x}f(\tilde y)\delta_{ab} \sim
\tilde y^{-2} \bra{a,\xi,k} \mathcal{J}^{\mu}_b(x) \ket{0},
\end{align}
where $\mathcal{J}^{\mu}$ is the $\gr{SU}(2)_f$ flavour current and
$\ket{a,\xi,k}$ is a vector meson with polarisation $\xi$, momentum
$k$, and flavour triplet label $a$. Note that the index $b$ in
$\Afluct^{\mu}_{b (a)}$ is a Lie algebra index, whereas the index
$(a)$ labels the flavour triplet of solutions.

The meson spectrum\marginpar{meson spectrum} is numerically
determined by a shooting technique using interval bisection to find
the values $\tilde M^2$ that admit solutions to
\eqref{eqn:eomrescaled} that are regular ($c_2=0$ for \acro{IR}
behaviour $c_1\tilde{y}^2 + c_2 \tilde{y}^{-4}$) and normalisable
($c_1=0$ for \acro{UV} behaviour $c_1 + c_2 \tilde{y}^{-2}$). The
result for the lowest lying modes is shown in
Figure~\ref{fig:higgs-mesonmass}.

In passing\marginpar{why singular gauge} it is noted that the second term in
\eqref{eqn:eomrescaled}, which comes from the $g^2$ term in the equation
of motion \eqref{eqn:flucteom}, is roughly the instanton squared and
up to numerical constants would have been $y^2/(y^2+\Lambda^2)^2$ for
the instanton in non-singular gauge. This contribution would have
changed the \acro{UV} behaviour of $f(y)$ and therefore prohibited
to make contact to the \acro{SUSY} case in the limit of zero instanton
size, where \eqref{eqn:flucteom} can be solved analytically. 

In\marginpar{asymptotics} 
the limit of infinite instanton size, one might expect the same
spectrum since the field strength 
vanishes locally. This corresponds to 
infinite Higgs \acro{VEV}
 in the field theory, which reduces the gauge group from 
$\gr{SU}(N_c)$ to $\gr{SU}(N_c-1)$.  This difference is negligible
in the large $N_c$ limit and one might expect to return to the origin
of moduli space.  However there is a non-trivial shift of the
spectrum, which makes the flow from zero to infinite Higgs \acro{VEV}
not quite a closed loop as can be
seen in Figure~\ref{fig:higgs-mesonmass-B}.  

Since at both ends the analytic spectrum in reproduced, it should
be possible to capture this behaviour in the equation of 
motion \eqref{eqn:flucteom}. 
Indeed a simultaneous treatment of both cases can be achieved by
rewriting \eqref{eqn:flucteom} in the suggestive form
%
\begin{align}
  0 = \biggl[ \frac{{\tilde M}^2}{({\tilde y}^2 + 1)^2}
                - \frac{\spinl (\spinl +2 )}{{\tilde y}^2}
                + \frac{1}{{\tilde y}^3}\partial_{\tilde y}
                ({\tilde y}^3 \partial_{\tilde y} ) \biggr]
  f({\tilde y}), \label{eq:higgs-asymp}
\end{align} 
with $\spinl = 0,2$ for zero or infinite $\tilde \Lambda$ respectively.  

This is the same equation \eqref{eq:KMMWradialeom} that was
found for fluctuations about the trivial background $A^a=0$,
but $\spinl$ was given rise to by excitations on the 
internal manifold. The ansatz was
\begin{align}
{\Afluct}^{\mu} = \xi^{\mu}(k)\e^{ik_{\mu}x^{\mu}}f(y) {\cal Y}^\spinl(\mf{S}^3),
\end{align}
with ${\cal Y}^\spinl(\mf{S}^3)$ the scalar spherical harmonics on $\mf{S}^3$
transforming under $(\frac{\spinl}{2},\frac{\spinl}{2})$ representations 
of $\gr{SU}(2)_L \times \gr{SU}(2)_R$.  \cite{Kruczenski:2003be} found
that \eqref{eq:higgs-asymp} can be solved analytically in terms
of a hypergeometric function \eqref{eq:KMMWsol} parametrised by $n$ and $\spinl$, which 
by regularity and normalisability become quantised and yield the
discrete spectrum
\begin{align} 
  {\tilde M}^2 & = 4(n+\spinl+1)(n+\spinl+2), \qquad n,\spinl \ge 0.
  \tag{\ref{eq:kmmw-typeII-mass}}
\end{align}

For intermediate values of the instanton size, a flow connecting
the analytically known spectra is expected and could be confirmed numerically,
see Figure~\ref{fig:higgs-mesonmass-B}.
\begin{figure}
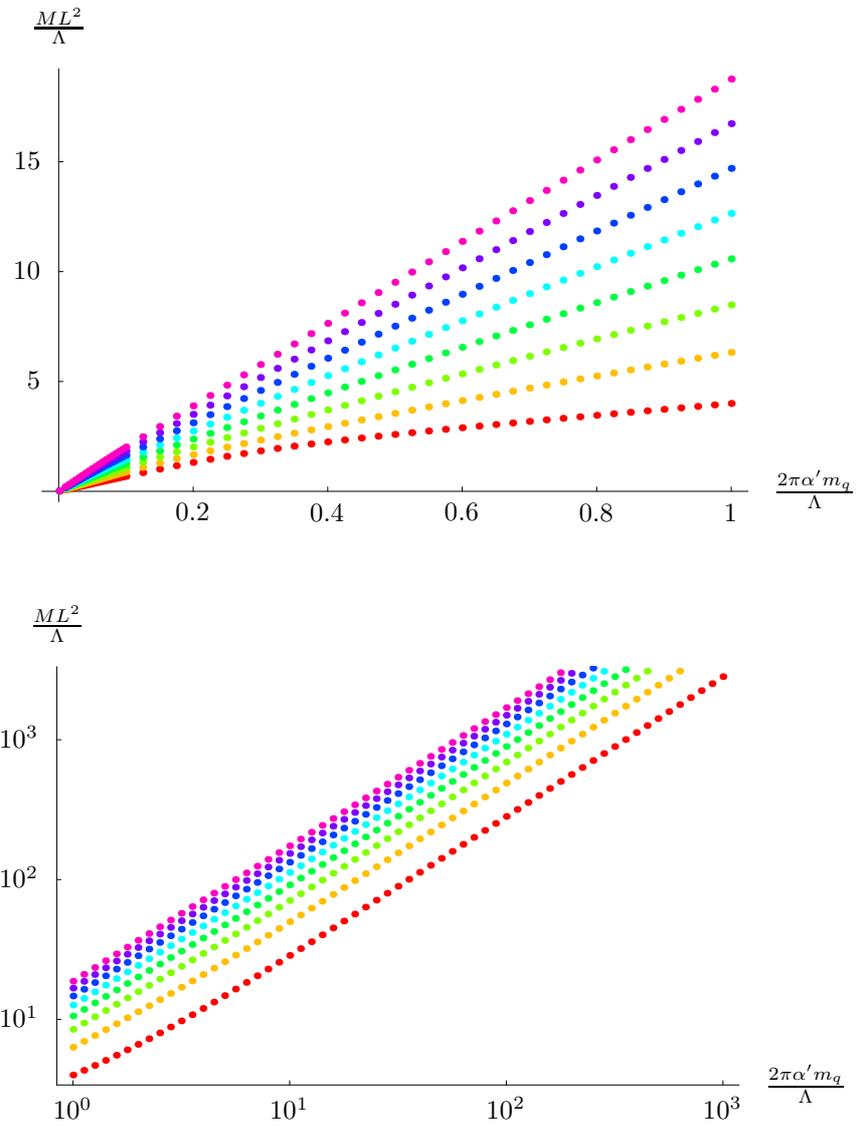

\centering
  \subfigure{  
    \PSfraginclude{width=12cm}{meson-garb}
  }
  \subfigure{
    \PSfraginclude[trim=20 0 0 0]{width=12cm}{meson-garb-log}
  }
\caption[Meson Masses on the Higgs Branch 2]{\label{fig:higgs-mesonmass2}
        Numerical results for the meson mass spectrum as function of the quark
        mass. Both for $m_q/\Lambda \rightarrow 0$ and for 
        $m_q/\Lambda\rightarrow \infty$,
        the curves become linear, however with different slope. The
        asymptotic slopes correspond to the constant values approached 
        in Figure
        \protect\ref{fig:higgs-mesonmass-B}.}
\end{figure}

It\marginpar{singular gauge revisited} remains to comment on how it
is possible to continuously transform a spherical harmonic in the
$(0,0)$ of the unbroken $\gr{SU}(2)_L\times\gr{SU}(2)_R$ into one that
transforms under the $(1,1)$, while $\gr{SU}(2)_L$ is unbroken along
the flow.\footnote{$\gr{SU}(2)_R \times \gr{SU}(2)_f$ is broken to
  $\diag [\gr{SU}(2)_R \times \gr{SU}(2)_f]$ except at zero and
  infinite Higgs \acro{VEV}.}  The solution to this puzzle is that the
instanton in singular gauge does not vanish in the limit of large
instanton size, while in non-singular gauge it does. So\marginpar{closing the loop} the spectrum
at large instanton size is related to the one at zero instanton size
exactly by the singular gauge transformation
\eqref{eq:higgs-singular-gauge-trafo},
which 
reads
\begin{align}
U = \frac{y^m \sigma^m}{ \abs{y} }. 
\end{align}
This gauge transformation is \emph{large}. While in the instanton literature
it is merely employed as a computational trick to improve convergence
of numerical calculations for large distance from the instanton core, 
in this setup it has physically observable consequences because
the large distance behaviour is related to the conformal dimension of
boundary operators. It also does not leave the global charges under
$\gr{SU}(2)_L \times \gr{SU}(2)_R \times \gr{SU}(2)_f$ invariant: Acting
on the ansatz \eqref{eq:higgs-vec-ansatz}, the singular gauge transformation 
\eqref{eq:higgs-singular-gauge-trafo} yields
\begin{align}\label{eq:ltwo}
 {\Afluct}_\mu{}^{(a)} &= \xi_\mu(k) f(y) \e^{ik_\mu x_\mu} \bigl[
    \frac{y^m y^n}{y^2} \sigma^m T^a\bar\sigma^n \bigr]. 
\end{align}
The parenthesised expression should be the $\spinl=2$ spherical
harmonic.  Due to $\sigma^m T^a\bar\sigma^n$ being traceless, there
is indeed no singlet contribution. Moreover a spherical harmonic
should be independent of $\abs{y}$ as is true for $\frac{y^m
  y^n}{y^2}$.  With $\hat{g}^{ij}$ the metric on the three
sphere
it holds
\begin{align}
 \p_m\p_m \mathcal{Y}^\spinl= y^{-2}
\hat{\nabla}_i \hat{g}^{ij} \hat{\nabla}_j \mathcal{Y}^\spinl =
-\spinl(\spinl+2)\,y^{-2}\, \mathcal{Y}^\spinl,
\end{align}
which is also satisfied by \eqref{eq:ltwo}.

\Subsection{Scalar Fluctuations}
The mesons arrange themselves in massive $\cN=2$ multiplets, some
of which are obtained by different, scalar ans\"atze for the gauge 
fluctuations \eqref{eq:higgs-vec-ansatz}.
In addition, there arise mesons from fluctuations of the scalars in 
\eqref{eq:higgs-quad-action}.
For these the equation of motion reads
\begin{align}
  H \p_\mu \p_\mu \Phi &+ \covD_m \covD_m \Phi = 0,
\intertext{where}
\begin{split}
  \covD_m \covD_m \Phi &= \partial_m\partial_m \phi +  \comm{\Ainst_m}{\partial_m \Phi} \\
     &\quad + \partial_m \comm{\Ainst_m}{\Phi} 
            +  \comm{\Ainst_m}{\comm{\Ainst_m}{\Phi} },
\end{split}
\end{align}
which coincides with the equation of motion for the gauge field
\eqref{eqn:flucteom} except for the vector index present. Therefore
the same ansatz up to a polarisation vector
\begin{equation} \label{scalaransatz}
  \Phi^{(a)} = f(\tilde y)\e^{i k_\mu x_\mu}  T^a
\end{equation}
yields exactly the same radial differential equation
\eqref{eqn:eomrescaled} and the same mass spectrum, Figure
\ref{fig:higgs-mesonmass}.

The scalar fluctuations (\ref{scalaransatz}) are dual to  the
descendant $QQ(q_i \bar q^i)$ of the scalar bilinear $q_i \bar
q^i$, which  has conformal dimension $\Delta=3$. At $\Lambda =0$
the scalar bilinear is in the $(0,0)$ representation of the
unbroken $\gr{SU}(2)_L\times \gr{SU}(2)_R$ symmetry.



  \begin{savequote}[\savequotewidth]
If little else, the brane is an educational toy.
\qauthor{Tom Robbins (up to a small typo)}
\end{savequote}

\Chapter{Heavy-Light Mesons\label{ch:heavylight}}

This Chapter is similar in spirit to the D3-D7 systems discussed
so far, though different in implementation. The reason is that 
while fundamental fields are still assumed to arise from D7
branes in a---possibly deformed---\AdS\ space, 
the requirement to describe quarks of vastly different mass, 
as needed for heavy-light mesons, makes those quarks
arising from a stack of \emph{coincident} D7\NB-branes being no
longer a good approximation. 
In this regard, heavy-light mesons are intrinsically \emph{stringy}
and cannot be captured by the \acro{DBI} techniques discussed 
in the previous Chapters.  Unfortunately as full quantised 
string theory on \AdS\ is not well understood, the question arises
of how to transfer such features into a supergravity framework.

Here idealised heavy-light mesons will be considered, composed of 
a massless and a very massive quark,
such that in an appropriate background, the light quark
may exhibit dynamical chiral symmetry breaking, while the heavy quark
does not. For now, let us stick with the \AdS\ case.
Clearly the geometric picture is that of two parallel
(probe) D7\NB-branes in a background determined by a stack of D3\NB-branes.
The different quark masses correspond to the two different 
separations of the D7\NB-branes from the D3 stack. Strings describing heavy-light mesons
now differ from light-light and heavy-heavy ones, whose ends
are attached to the respective same brane, by being stretched
between the two different D7\NB-branes. 
In the limit where the heavy quark is much heavier than the light quark,
henceforth called \emph{large separation limit}, the string becomes
very long and admits a classical description. 

To\marginpar{effective point-particle action}
 obtain a description both simple and 
similar to the examples studied so far, the ansatz of a rigid non-oscillating
string is chosen that moves in the \AdS\ radial direction along the D7\NB-branes,
with the essential assumption that oscillations and longitudinal movement
are suppressed in the large separation limit.\footnote{On the field theory
side at large separation; i.e.~large quark mass $m_H$, effects distinguishing
vector from scalar mesons are suppressed by $m_H^{-1}$.  Indeed the formalism
described here is not capable of capturing such a difference and meson
masses are thus manifestly degenerate.} 
Integration of the Polyakov action along the string 
can then be performed, yielding effectively
a centre-of-mass movement weighted by a factor from 
averaging over the geometry between the two D7s.
To obtain a field equation, \naive\ quantisation is performed, which
results in a modified Klein--Gordon equation. (In a Minkowski space,
this procedure yields the conventional, unmodified Klein--Gordon equations.)
After the \AdS\ case, the discussion will be moved on to 
the dilaton deformed
background by \acro{GKS} introduced in Chapter~\ref{ch:dilatondriven}
and a similar background by Constable--Myers.
Both exhibit chiral symmetry breaking. 
While these are known to be far from perfect \acro{QCD} gravity duals,
experience shows that even simple holographic models can reproduce
measured mass values with an accuracy of 10--20\%.
Assuming the two respective quark flavours associated to the D7\NB-branes
being up and bottom, the mass of the rho ($u\bar{u}$) and upsilon ($b\bar{b}$) 
meson can be used to fix all scales in the theory and yield a numerical
prediction for the B meson mass, which indeed is less than 20\% from the
experimental value.

\Section{Heavy-Light Mesons in \texorpdfstring{$\AdS_5\times \mf{S}^5$}{AdS(5) \texttimes\ S(5)}}

\begin{figure}[t]
\centering
\EPSinclude[width=8cm]{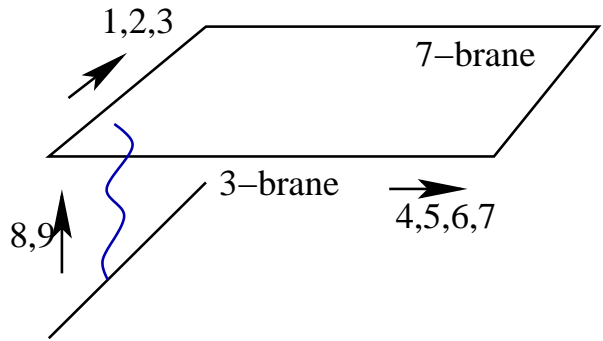}
\caption[D3/D7 Geometry]{\label{fig:hl-d3d7}The geometry of the D3-D7 
system under consideration \cite{Erdmenger:2006bg}.}
\end{figure}

\begin{figure}[t]
\centering
  \EPSinclude[height=4.5cm]{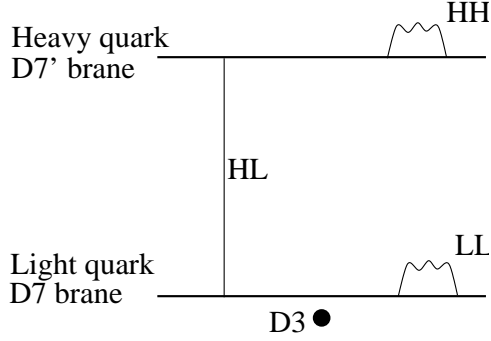}
\caption[Heavy-Light Brane Configuration]{The brane configuration 
  including both a heavy and a light   quark.  The 77 and $7'7'$ 
  strings are holographic to light-light and  heavy-heavy mesons 
  respectively. Heavy-light mesons are
  described by strings between the two D7\NB-branes.
  \label{fig:hl-stretch}}
\end{figure}

As shown in Chapter~\ref{ch:flavour}, quarks can be introduced into
the \adscftcorr\ by augmenting the D3 stack with a stack of probe
D7\NB-branes \cite{Karch:2002sh}.  The backreaction of the $N_f$
D7\NB-branes on the $\AdS_5 \times \mf{S}^5$ geometry
\eqref{eq:ads5-s5} formed by the $N_c$ D3\NB-branes may be neglected
as long as $N_f\ll N_c$; i.e.\ $N_f$ is kept fixed in the \tHooft\ limit.
\begin{align} \label{eq:hl-ads5-s5}
  ds^2    &= \frac{r^2}{L^2} \eta_{\mu\nu} dx^\mu dx^\nu 
          + \frac{L^2}{r^2} dr^2 + L^2 d\Omega_5^2, 
\end{align}
This corresponds
to the quenched approximation of lattice gauge theory on the field theory side.
The D7\NB-branes wrap an $\AdS_5 \times \mf{S}^3$ geometry when coincident with the
D3s. When separated the corresponding $\cN=2$ hypermultiplet acquires a mass 
and the D7\NB-branes wrap a geometry 
\begin{align}
    ds^2 
         &= \frac{y^2+\tilde{m}_q^2}{L^2} \eta_{\mu\nu} dx^\mu dx^\nu 
          + \frac{L^2}{y^2+\tilde{m}_q^2} dy^2 + \frac{L^2 y^2}{y^2+\tilde{m}_q^2} d\Omega_3^2, 
\end{align}
which is only asymptotically $\AdS_5 \times \mf{S}^3$ and does not
fill the complete $\AdS_5$ background, but instead terminates from the
five-dimensional point of view and drops from the \acro{IR} dynamics.
This configuration is shown in Figure~\ref{fig:hl-d3d7}.
The\marginpar{spectrum}
meson spectrum can be determined analytically \cite{Kruczenski:2003be}
and the  degenerate mass of the scalar and pseudoscalar meson is given by
\begin{align}
  M_s^2 &= \frac{4\tilde{m}_q^2}{L^4} (n+\spinl+1)(n + \spinl+2).
\end{align}
\finalize{\begin{sloppypar}}
These\marginpar{two flavours}
 mesons are build up from quarks carrying all the same mass; i.e.~they
form ``light-light'' or ``heavy-heavy'' mesons depending on the distance
$\tmq = (2\pi\alpha')m_q$ between the D7\NB-branes and the D3 stack.
When considering two D7\NB-branes with \emph{different} distances 
$\tilde{m}_L$ and $\tilde{m}_H$ to the
D3 stack, there are accordingly two towers of mesons $M_H$ and $M_L$
whose lightest representatives have a mass ratio of $\frac{m_L}{m_H}$
and which come from strings having attached both ends to the same brane.
The configuration is shown in Figure~\ref{fig:hl-stretch}.
Strings stretched between the two branes should then form a set
of mesons composed of a heavy and a light quark.
\finalize{\end{sloppypar}}

In the limit $m_H \gg m_L$ the string becomes very long and will be
assumed to be in the semi-classical limit, where quantum effects
to the unexcited string can be neglected. The string described
here will therefore approximate above mesons, which by construction
will be degenerate. 

The\marginpar{Polyakov} gauge-fixed Polyakov action will be taken as a starting point
\begin{align} 
  S_P = - \frac{T}{2} \int d\sigma \, d \tau \, G_{\mu \nu} 
            ( - \dot{X}^\mu \dot{X}^\nu + X'{} ^{\mu} X'{} ^{\nu}), \label{eq:polyakov}
\end{align}
such that the constraints
\begin{align} 
  G_{\mu \nu} \dot{X}^\mu X'{}^\nu &=0, &
  G_{\mu \nu}(\dot{X}^\mu \dot{X}^\nu + X'{}^\mu X'{}^\nu)&=0,
\end{align}
have to be taken into account.

The two D7\NB-branes are assumed to be separated from the D3 stack
in the same direction $\theta=0$; i.e.~the string connecting
them will obey $\sigma=z$, where $\sigma$ is the spatial world sheet 
coordinate and
$z \e^{i\theta}=z^9 + i z^8$.
While the string will be allowed to move along the world volume
of the D7s, it shall be stiff such that integration over $\sigma$
can be performed to generate an effective point particle action.
With the embedding\marginpar{embedding}
\begin{align}
X^A &= (x^\mu(\tau), y^m(\tau), z^8=0, z^9=\sigma),
\end{align}
which implies $\dot{X}^a X'_a=0$ automatically,  
and the $\AdS_5\times \mf{S}^5$ geometry \eqref{eq:hl-ads5-s5}, the
Polyakov action reads
\begin{align} \label{eq:hl-ads-action} 
   S_P &= - \frac{T}{2} \int d \tau \int\limits_{\tml}^{\tmh} d \sigma
     \left[ - \frac{y^2 + \sigma^2}{L^2} \dot x^\alpha \dot x_\alpha 
            - \frac{L^2}{(y^2 + \sigma^2)} \dot y^i \dot y_i 
            + \frac{L^2}{(y^2 + \sigma^2)} \right],
\end{align}
where $y\equiv\abs{y}\equiv\sqrt{\sum_{i=4,5,6,7} (y^i)^2}$. 
Integrating over $\sigma$ yields
\begin{align} 
   S_P &= - \frac{T}{2} \int d \tau \left[ 
               - f(y) \dot{x}^2 - g(y) \dot{y}^2 + g(y) 
          \right], \label{eq:hl-pointaction}
\end{align}
with (choosing $\tml=0$)
\begin{align} 
  f(y) &= \frac{1}{L^2} \bigl( y^2 \tmh +\frac{1}{3} \tmh^3\bigr) , &
  g(y) &= \frac{L^2}{y} \arctan \frac{\tmh}{y} . 
\end{align}
The remaining constraint equation $G_{\mu \nu}(\dot{X}^\mu \dot{X}^\nu + X'{}^\mu X'{}^\nu)=0$
is
\begin{align}
  \frac{y^2 + \sigma^2}{L^2} \dot x^\alpha \dot x_\alpha 
            + \frac{L^2}{(y^2 + \sigma^2)} \dot y^i \dot y_i 
            + \frac{L^2}{(y^2 + \sigma^2)}  = 0,
\end{align}
which gives
\begin{align} \label{eq:hl-legendre}
\frac{1}{f(y)} p_{x}^2 &+\frac{1}{g(y)} p_{y}^2 + T^2 g(y) =0, \\ 
  p_x^\ia &:= \frac{\p \Lag}{\p \dot{x}_\ia}, \nonumber \\
  p_y^i &:= \frac{\p \Lag}{\p \dot{y}_i} \nonumber
\end{align}
when integrating over $\sigma$. 
The same calculation for Minkowski space gives $f(y)=g(y)=\tmh$, such
that one obtains $E^2=m^2+p^2$.  For \AdS\ space the mass $m$ depends
on the position of the string $y$ via the factors $f(y)$ and $g(y)$, 
which average over the geometry between the two D7\NB-branes.

\begin{figure}
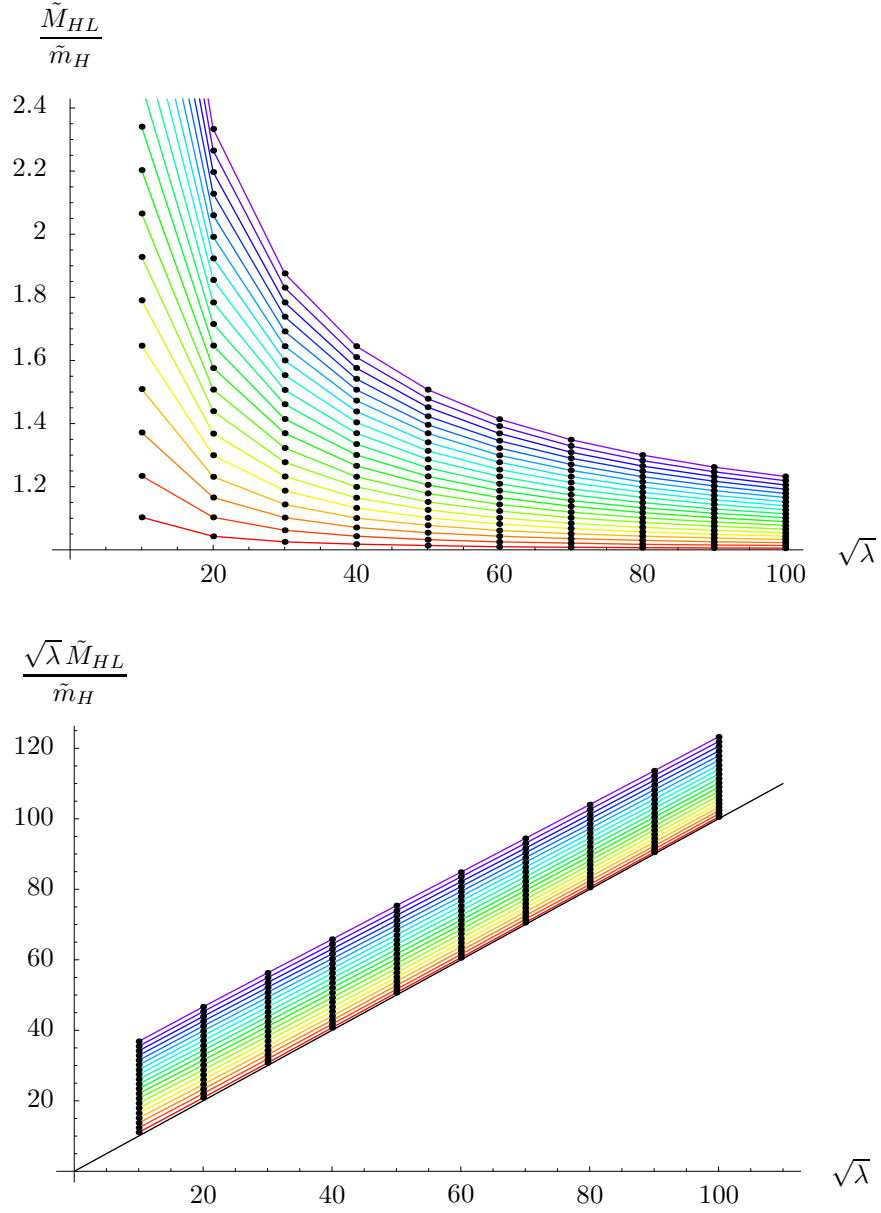

\centering
\PSfraginclude[trim=65 35 60 25]{ width=12cm}{hl-ads-converge}\\
\PSfraginclude[trim=25 20 55 15]{ width=12cm}{hl-ads-linear}
\caption[Heavy-Light Meson Spectrum in \AdS]{
  The mass ratio of the heavy-light meson and the heavy quark
  mass (the light quark is taken to be massless) 
  as a function of the \tHooft\ coupling for the \AdS\ background. 
  In the large $\lambda$ limit, $M_{HL} L^2/(2\pi\alpha'\,m_H)$ 
  behaves as $1 + \text{const.}/\sqrt{\lambda} + \mathcal{O}(\lambda^{-1})$.
  The black line in the second plot corresponds to $M_{HL} L^2=(2\pi\alpha')m_H$.
  \label{fig:hl-ads-large}}
\end{figure}
\begin{figure}
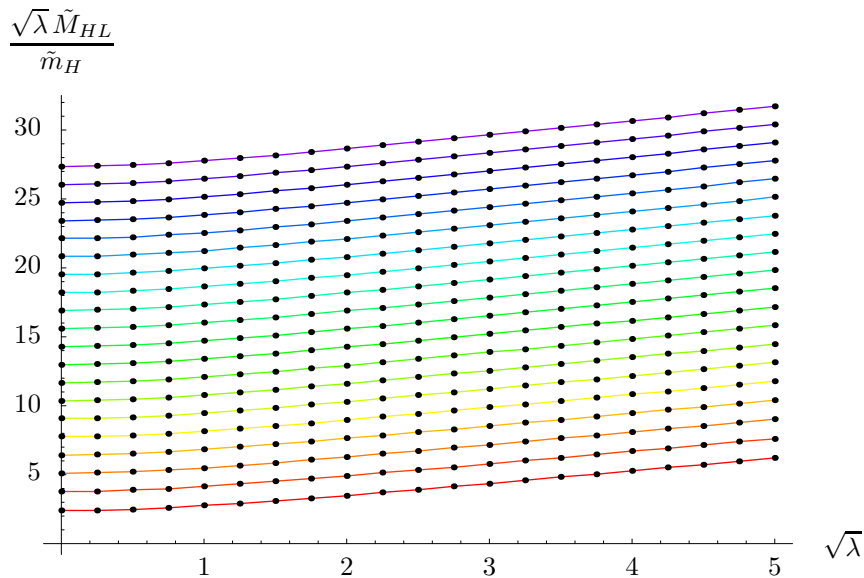

\centering
\PSfraginclude[trim=25 20 55 15]{ width=12cm}{hl-ads-small}
\caption[H/L Meson Spectrum in \AdS\ (small \tHooft\ coupling)]{
The heavy-light meson spectrum in \AdS\ for small \tHooft\ coupling
with vanishing mass for the light quark. The mass ratio behaves
as $\text{const.}/\sqrt{\lambda}+\mathcal{O}(\lambda)$. 
Note however that the supergravity approximation is not reliable in this
regime.\label{fig:hl-ads-small}
}
\end{figure}

For\marginpar{equation of motion}
the quantisation prescription $p \mapsto -i\p$, the following
modified Klein--Gordon equation is obtained
\begin{align}
   \left[\Box_x^2 
       + \frac{f(y)}{ g(y)} \nabla_y^2 
       - T^2 g(y) f(y) \right] \phi(\vec{x},\vec{y}) = 0. \label{eq:hl-ads-eom}
\end{align}
The usual procedure for this kind of equations is to find the correct
background solution, which by assumption only depends on the radial
direction $y$ and find fluctuations about this solution.  By a
separation ansatz these fluctuations can be seen to be a plain wave in
the $x$ direction and spherical harmonics in the angular coordinates
$\Omega_3(y^{4,5,6,7})$. The remaining equation for the radial
coordinate $y$ often has to be solved numerically.  

In the \acro{UV}
limit $y\to\infty$, \eqref{eq:hl-ads-eom} is dominated by the Laplace
operator in the $y$ directions due to $\frac{f}{g} \sim y^4$ and $f\,g
\to 1$, such that
\begin{align}
  \nabla_y^2 \phi = 0.
\end{align}

When $\phi$ only depends on $y$, the solution has the form 
required to couple to the \acro{VEV} and source of a 
heavy-light quark bilinear $\bpsi_H \psi_L$.
\begin{align} 
  \phi(y\to\infty) = \tilde{m}_{HL} + \frac{c_{HL}}{y^2}+\dots 
\end{align}
However\marginpar{trivial vacuum}
there are no heavy-light mass mixing term and no 
heavy-light bilinear condensate in \acro{QCD}, so $\phi(y)\equiv0$
is chosen.

Assuming a singlet under $\gr{SU}(2)_L \times \gr{SU}(2)_R$, 
the ansatz for linearised fluctuations about above vacuum
solution reads 
\begin{align}
   \phi &= 0 + h(y) \e^{ik \cdot x}, &
   M_{HL}^2 &= -k^2,
\end{align} 
where $h(y)$ shall be regular in the \acro{IR} and normalisable
$h(y\to\infty)\sim y^{-2}$.  Only for a discrete set of values for
$M_{HL}$ this requirement can be satisfied.  For numerics it is
convenient to employ rescaled coordinates
$y=\tilde{m}_H\,\tilde{y}$, such that equation \eqref{eq:hl-ads-eom}
reads
\begin{align} \label{eq:hl-ads-num}
  \left[ \frac{\pi}{\la} \frac{\tilde{y}^3 +
      \frac{\tilde{y}}{3}}{\arctan \frac{1}{\tilde{y}}}
    \nabla_{\tilde{y}}^2 + \left( \tilde{y} + \frac{1}{3\tilde{y}}
    \right) \arctan \frac{1}{\tilde{y}} +
    \frac{M^2 L^4}{\tilde{m}_H^2} \right] h(\tilde{y}) &= 0 .
\end{align}
The \tHooft\ coupling $\la$ arises from $R^4/(2\pi\alpha')=g_sN_c/\pi$.
The mass ratios yielding regular normalisable solutions to \eqref{eq:hl-ads-num}
have been plotted in Figures \ref{fig:hl-ads-large} and \ref{fig:hl-ads-small}.
It can be read off
\begin{align}
  \frac{M_H}{m_H} &= \frac{2\pi\alpha'}{L^2} \Bigl[ 1 + \frac{\text{const.}}{\sqrt{\lambda}} + \mathcal{O}(\lambda^{-1}) \Bigr].
\end{align}
In the large $\lambda$ limit, $\tilde{M}_{HL} = \tilde{m}_H$ is approached
in agreement with the \naive\ expectation of the meson mass being equal to
the string length times its tension. For comparison in Figure \ref{fig:hl-ads-small} the 
mass ratio is plotted for small values of the \tHooft\ coupling, where 
supergravity is not a reliable approximation anymore.

\Section{Dilaton Flow Geometries}
The $\cN=2$ \acro{SYM} considered so far provides a basis for studying
meson spectra since it gives analytic expressions for solutions and
masses consisting of identical quarks.  However it does not capture a
number of phenomenologically relevant features like chiral symmetry
breaking since chiral symmetry breaking requires \acro{SUSY} breaking.
The setup discussed now improves at least in that regard by providing
a simple geometry that describes a non-supersymmetric dual of a large
$N_c$ \acro{QCD}-like theory and thus exhibits dynamical chiral
symmetry breaking.

The first background discussed is the dilaton deformed background by
Gubser, Kehagias--Sfetsos, which has been described in Chapter~\ref{ch:dilatondriven}. It
is demonstrated that the semi-analytic prediction of the \AdS\ case is
reproduced in the large heavy-quark limit.  Then the same procedure is
applied to the similar geometry of Constable and Myers, but it turns
out that in this setup the heavy-light meson spectrum does not
approach the \AdS\ spectrum in a similar manner.

\Subsection{\acro{GKS} Geometry}
Let me remind the reader that the \acro{GKS} geometry is given by,
cf.~\eqref{eq:gubser-geometry},
\begin{align}\label{eq:hl-gubser-geometry-reviewed}
  \begin{split}
    ds_{10}^2 &= g_{xx}(r) dx_{1,3}^2 + g_{yy}(r) (d\sqvec{y} + d\sqvec{z}), \\
    g_{xx}(r) &= \frac{r^2}{L^2} \sqrt{1-r^{-8}},\\
    g_{yy}(r) &= g_{zz} = \frac{L^2}{r^2}, \\
    \e^\vphi &= \e^{\vphi_0} \pfrac{r^4+1}{r^4-1}^{\sqrt{\frac{3}{2}}},\\
    r^2 &= \sqvec{y} + \sqvec{z},
  \end{split}
\end{align}
where Einstein frame has been used and the coordinates have been rescaled
such that infra-red singularity resides at $r=1$. 
The coordinates $y^{4,5,6,7}$ and $z^{8,9}$
are on equal footing and can be interchanged by $\gr{SO}(6)$ transformations
until probe D7\NB-branes, which  break the $\gr{SO}(6)$
to $\gr{SO}(4)\times \gr{SO}(2)$, are  introduced to obtain quarks.
The D7\NB-branes are embedded according to 
$z=\abs{z^9+iz^8}= z_0(y)$, which yields the following equation of motion
\begin{gather} \label{eq:hl-gubser-vaceom-reviewed}
  \frac{d}{dy} \biggl[\frac{y^3 f}{\sqrt{1 + z_0'(y)^2 }} z_0'(y) \biggr]
    = y^3 \sqrt{1 + z_0'(y)^2 } \frac{\p}{\p z_0} f , \\[2ex]
  f = \frac{ (r^4+1)^{(1+\Delta/2)} (r^4-1)^{(1-\Delta/2)} }{r^8}, \qquad r^2 = y^2+ z_0(y)^2, \qquad
  \Delta = \sqrt{6}.\nonumber
\end{gather}
\begin{figure}
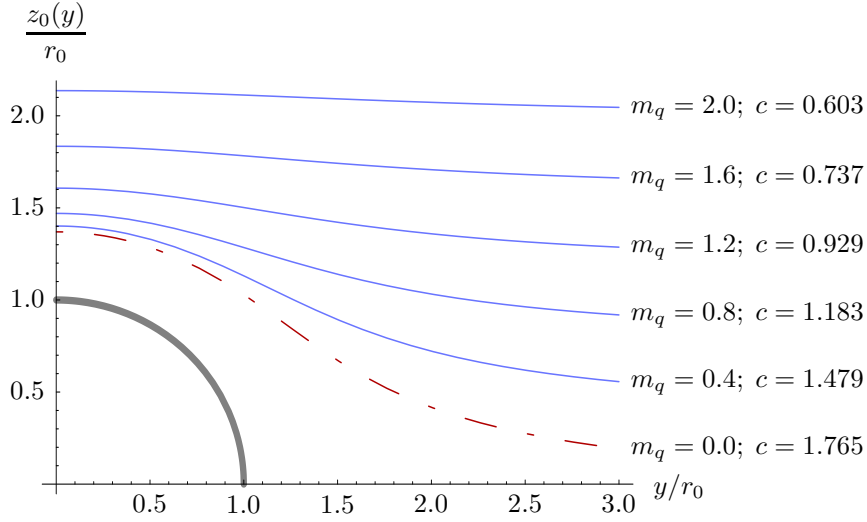

\centering
\PSfraginclude[trim=-5 -20 -120 -25]{width=12cm,height=75mm}{gubser-vac}%
\caption[Vacuum Embeddings]{
  Chiral symmetry breaking embeddings in the \acro{GKS} geometry.
  \label{fig:hl-gubser-vac}
}
\end{figure}
At large $y$, solutions to \eqref{eq:hl-gubser-vaceom-reviewed}
take the form 
\begin{align}
 z_0 = \frac{\tilde m_q}{r_0} + \frac{c}{r_0^3 y^2} + \dots \, , \label{eq:hl-cm-asymp}
\end{align}
which by standard \adscft\ duality corresponds to a source of
conformal dimension $1$ and a \acro{VEV} of conformal dimension $3$ in
the field theory.  The former corresponds to the quark mass
$m_q=\tilde m_q/(2\pi\alpha')$ and describes the asymptotic separation
$\tilde m_q$ of the D3 and D7\NB-branes, the latter is the bilinear quark
condensate $c\sim\vev{\bpsi\psi}$. The factor of $r_0$, which gives the 
position of the singularity, arises from the coordinate rescaling used
to remove $r_0$ from the metric and equations of motion. 

Requiring\marginpar{regular embeddings} regularity in the
\acro{IR} by $\p_y z_0(0)=0$ fixes the quark condensate as a function
of the quark mass, see Section~\ref{sec:gubser-chiral-sym}. Some
regular solutions to \eqref{eq:hl-gubser-vaceom-reviewed} are plotted in
Figure~\ref{fig:hl-gubser-vac}, which provide the D7 embeddings
that are used as the boundary conditions for the heavy-light string in the
following.

The Polyakov action \eqref{eq:polyakov}, which due to being in string
frame requires additional factors of $\e^{\vphi/2}$, reads for this background
\begin{align} 
   S_P &= - \frac{T}{2} \int d \tau \int\limits_{z_0(m_L)}^{z_0(m_H)} dz_0
     \left[ - \e^{\vphi/2} g_{xx} \dot x^\alpha \dot x_\alpha 
            - \e^{\vphi/2} g_{yy} \dot y^i \dot y_i 
            + \e^{\vphi/2} g_{yy} \right] ,
\end{align}
with the metric factors and dilaton from \eqref{eq:hl-gubser-geometry-reviewed}.

One obtains again an equation of motion of the form 
\begin{align}
   \left[\Box_x^2 
       + \frac{f(y)}{ g(y)} \nabla_y^2 
       - T^2 g(y) f(y) \right] \phi(\vec{x},\vec{y}) = 0, \label{eq:hl-gubser-eom}
\end{align}
where the coefficients $f(y)$ and $g(y)$ this time are given by
\begin{align} 
  f(y)
      &= \int\limits_{ z_0(m_L)}^{ z_0(m_H)} d z_0\,\e^{\vphi/2} g_{xx},&
  g(y)
      &=  \int\limits_{ z_0(m_L)}^{ z_0(m_H)} d z_0\, \e^{\vphi/2} g_{yy}.
 \label{eq:hl-gubser-fg}
\end{align}

\begin{figure}
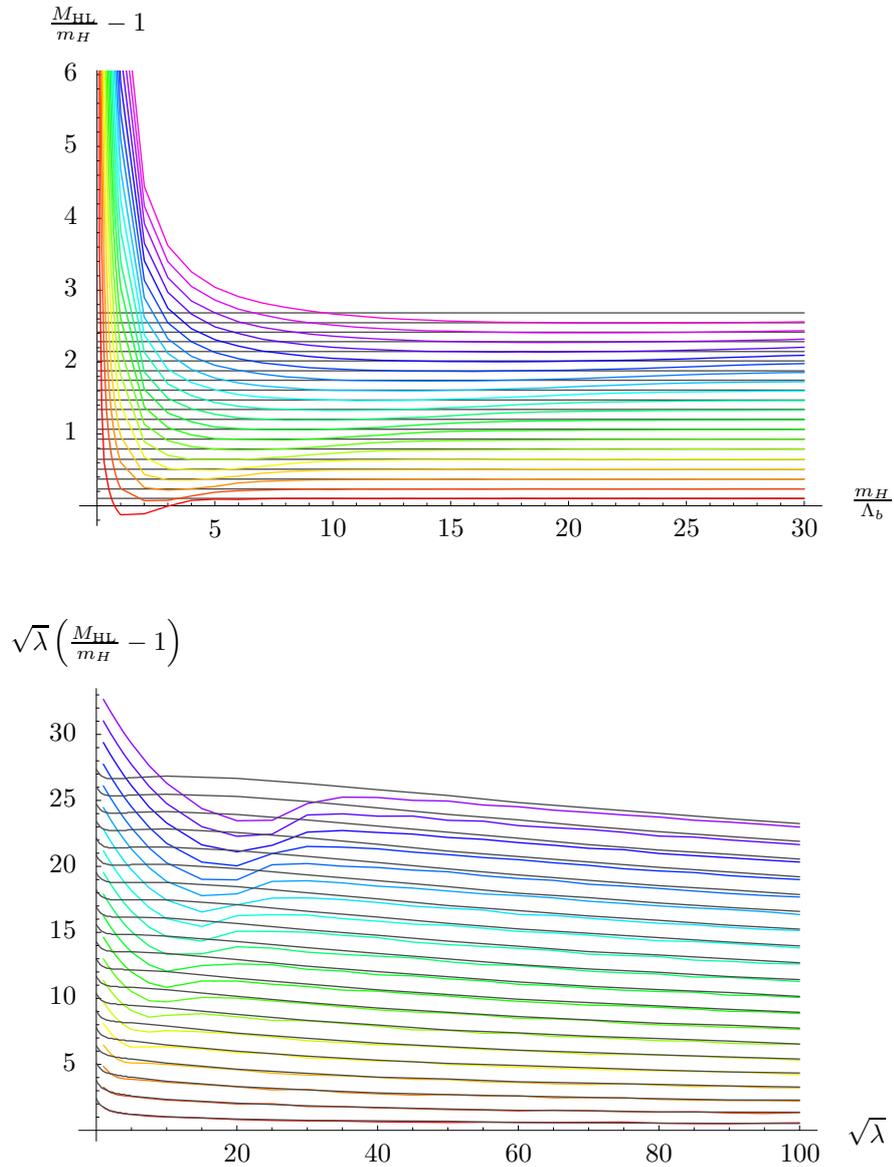

\centering
\PSfraginclude[trim=10 20 55 10]{ width=12cm}{hl-gubser-xmas-compare-lambda100-all}\\
\PSfraginclude[trim=22 20 55 10]{ width=12cm}{hl-ads-gubser-comparison}
\caption[Binding Energy in \acro{GKS} Background]{
   The binding energy of the heavy-light meson masses as a function of the
   heavy quark mass for $\lambda=100$ (first plot) and as a function
   of the \tHooft\ coupling for $m_H=11.50\,\Lambda$ (second plot).
   The respective \AdS\ values are shown as gray lines in the background
   and are approached in the limit of large values of the heavy quark mass, 
   while for small values effects of the chiral symmetry breaking are seen.
   \label{fig:hl-gubser-mesons}}
\end{figure}

The integration limits in \eqref{eq:hl-gubser-fg}; i.e.~the positions
of the D7\NB-branes, are given by the solutions to
\eqref{eq:hl-gubser-vaceom-reviewed}, which are only known
numerically, such that $f(y)$ and $g(y)$ also require numerics.  

For an ansatz describing a field theoretic vacuum $\phi\equiv\phi_0(y)$, 
equation \eqref{eq:hl-gubser-eom} has the same \acro{UV}
behaviour as the \AdS\ case, $\phi_0(y\to\infty) \sim \tilde{m}_{HL} + c_{HL}\,y^{-2}$,
where $\tilde{m}_{HL}$ corresponds to heavy-light mass mixing term and $c_{HL}$ 
to a heavy-light quark condensate. Because both are absent in \acro{QCD}, 
fluctuations about the trivial vacuum $\phi_0(y)\equiv0$ are considered.
Plot~\ref{fig:hl-gubser-mesons}\marginpar{fluctuation ansatz} shows the mass spectrum of
normalisable, regular solutions
\begin{align}
  \delta\phi = \phi(y) \e^{ik\cdot x}
\end{align}
as it can be obtained from
\begin{align}
   \left[ \frac{M_{HL}^2}{\Lambda^2}
       + \frac{\pi}{\la} \frac{\hat{f}(y)}{\hat{g}(y)} \nabla_y^2 
       - g(y) f(y) \right] \phi(\vec{x},\vec{y}) = 0
   \label{eq:hl-gubser-eom-num}  
\end{align}
with $\Lambda=r_0/(2\pi\alpha')$ the \acro{QCD} scale.  $\hat{f}$ and
$\hat{g}$ can be obtained from \eqref{eq:hl-gubser-fg} by setting
$L=1$.  The light quark mass $m_L$ has been set to zero to describe a
quark experiencing dynamical chiral symmetry breaking, while the large
quark mass $m_H$ is varied.

The spectrum obtained is very similar to that of the \AdS\ geometry.
To make the deviations caused by the deformation more visible,
the binding energy has been plotted. 
In Figure \ref{fig:hl-gubser-mesons} it is shown for
$\lambda=100$  as a function of the quark mass. It is also
shown as a function of the \tHooft\ coupling with the 
(for now arbitrary value of the) heavy quark mass 
$m_H=11.50\,\Lambda$. The binding energy\marginpar{\acro{SUSY}
  restoration} approaches its \AdS\ values for $m_H\to\infty$, but
highly excited mesons converge more slowly.  Both features can be
understood from the spectrum of light-light/heavy-heavy mesons in
Chapter~\ref{ch:dilatondriven}. The higher the quark mass, the higher
is the energy scale, where the brane ``ends'' and decouples from the
spectrum. At high energies supersymmetry is restored and the
light-light mesons become degenerate.  While the effect is the same
for the heavy-light mesons, that argument is not quite true anymore
since the light quark has been set to be massless all the time---at
least one end of the string stays close to \acro{IR} region. However
the centre of mass of the heavy-light string moves farther away from
the interior of the space when the heavy quark mass grows.  The
effective averaging of the geometry in \eqref{eq:hl-gubser-fg} takes
into account more and more of the geometry far from the centre, which
is nearly \AdS.

At the same
time highly excited mesons probe the \acro{IR} more densely as has
been seen in Section~\ref{sec:highexcite}, so they require the string
to be stretched much more to allow neglecting the vicinity of the singularity.

\Subsection{\texorpdfstring{Constable--Myers'}{Constable-Myers'} Background\label{sec:CM}}
The particular geometry considered here is a dilaton deformed 
\AdS\ geometry introduced in \cite{Constable:1999ch}, which 
has been employed by \cite{Babington:2003vm,Babington:2003up}
to describe chiral symmetry breaking in \adscft. Like the background
of the previous Section it is a warped $\AdS_5\times \mf{S}^5$ geometry
with a running dilaton that preserves $\gr{SO}(1,3)\times\gr{SO}(6)$
isometry. 

The background is given by
\begin{align}
  ds^2 &= H^{-1/2} X^{\delta/4} dx_{1,3}^2 + \mrlap{
          H^{1/2} X^{(2-\delta)/4} Y 
              (d\sqvec{y} + d\sqvec{z}),} \nonumber\\
  H &= X^\delta - 1, &
  X &= \frac{r^4 + b^4}{r^4 - b^4}, &
  Y &= \frac{r^4 - b^4}{r^4}, \nonumber\\
  \e^{2\vphi} &= \e^{2\vphi_0} X^{\Delta}, &
  C_{(4)} &= \mrlap{H^{-1} dx_0 \wedge \dots \wedge dx_3,} \nonumber\\
  \delta &= \frac{L^4}{2 b^4}, &
  \Delta^2 &= 10 - \delta^2, 
\end{align}
with $r^2 = \sqvec{y} + \sqvec{z}$.
$R$ and $b$ are free parameters and will be set to $1$ for the numerics,
since that allows to make contact with \cite{Babington:2003vm}, where 
the same choice has been made.
The authors of \cite{Babington:2003vm} embedded the D7\NB-branes according to 
$z=\abs{z^9+iz^8}= z_0(y)$ and obtained the following equation of motion 
\begin{align} \label{eq:hl-cm-eommc}
  \frac{ d}{d y} \left[ 
    \frac{\e^{\vphi} \mathcal{G}(y, z_0)}{\sqrt{ 1 + (\p_y z_0)^2}} (\p_y z_0)\right] 
  = \sqrt{ 1 + (\p_y z_0)^2} 
  \frac{\p}{\p z_0} \left[ \e^{\vphi} { \mathcal G}(y, z_0)
  \right],
\end{align} 
where 
\begin{align}
  \mathcal{G}(y, z_0) 
    =  y^3 \frac{((y^2 + z_0^2)^2 + 1)^{1+\Delta/2} ( (y^2 + z_0^2)^2 - 1)^{1-\Delta/2} }{(y^2 + z_0^2)^4}.
\end{align} 
This is the same equation as \eqref{eq:hl-gubser-vaceom-reviewed}
albeit with a free parameter $\Delta$, which in the \acro{GKS} geometry has
the fixed value $\sqrt{6}$.  The asymptotic
behaviour and their field theoretic interpretation are the same as
for the \acro{GKS} background and have been reviewed in the previous Section.
Note however that only the
particular combination $\e^\vphi \sqrt{-g}$ appearing in the equation
for the vacuum embedding \eqref{eq:hl-cm-eommc} coincides in both backgrounds.
On the level of meson spectra, the results for light-light mesons are 
similar but not identical to those in \acro{GKS}.

\begin{figure}
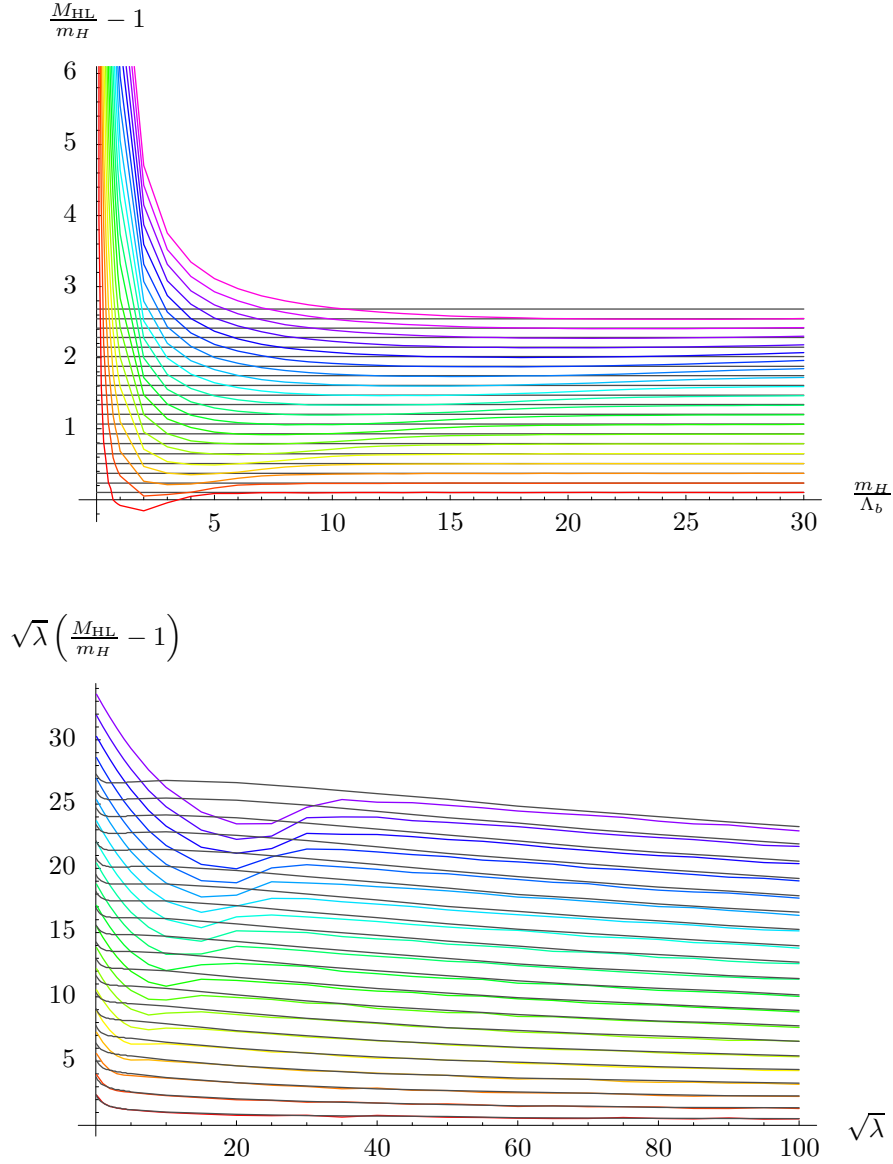

\centering
\PSfraginclude[trim=10 20 55 10]{ width=12cm}{hl-cm-xmas-compare-lambda100-all}\\
\PSfraginclude[trim=22 20 55 10]{ width=12cm}{hl-comparison}
\caption[Binding Energy in Constable--Myers' Background]{
   The binding energy of the heavy-light meson masses as a function of the
   heavy quark mass for $\lambda=100$ (first plot) and as a function
   of the \tHooft\ coupling for $m_H=12.63/\Lambda_b$ (second plot).
   The respective \AdS\ values are shown as gray lines in the background
   and are approached in the limit of large values of the heavy quark mass, 
   while for small values effects of the chiral symmetry breaking are seen.
   \label{fig:hl-cm-mesons}}
\end{figure}

Expanding the \acro{DBI} action \eqref{eq:abelian-dbi} to quadratic
order in fluctuations \eqref{eq:abel-dbi-quad-fluct} yields
\eqref{eq:radial-vector-eom} for a vector meson ansatz, that is an
ansatz of the form $A_\mu = \xi_\mu \delta\rho(y) \e^{ik\cdot x}$,
$M_\rho^2=-k^2$ for the D7 gauge field. The vector meson radial
equation \eqref{eq:radial-vector-eom} reads for the Constable--Myers background
\begin{align}\label{eq:hl-cm-vector-eom-2}
   \partial_y \bigl(K_1(y)\partial_y \delta\rho(y) \bigr)+M_\rho^2 K_2(y) \delta\rho(y)=0,
\end{align}
with
\begin{align}
  K_1 &= X^{1/2} y^3 (1+z_0'^2)^{-1/2}, & 
  K_2 &= H X^{1-\delta/2} Y^2 y^3 (1+ z_0'^2)^{-1/2}
\end{align}
and
\begin{align} \label{eq:hl-cm-XY}
  X &= \frac{(y^2+z_0^2)^2 + 1}{(y^2+z_0^2)^2 - 1}, &
  Y &= \frac{(y^2+z_0^2)^2 - 1}{(y^2+z_0^2)^2}. 
\end{align}
The Polyakov action
\begin{align} 
  S_P = - \frac{T}{2} \int d \tau \left[- f(y) \dot{x}^2 - g(y) \dot{y}^2 + g(y) \right]
\end{align}
preserves its \AdS\ form but the coefficients are now
\begin{align} 
  f(y)
       &= \int\limits_{z_0(m_L)}^{z_0(m_H)} dz_0\,
         (X^{1/2}-1)^{-1/2} X^{\Delta+\frac{1}{8}} , \\
  g(y)
       &=  \int\limits_{z_0(m_L)}^{z_0(m_H)} dz_0\,
          Y (X^{1/2}-1)^{1/2}
          X^{\Delta+\frac{3}{8}} ,
\end{align}
with $X$, $Y$ defined in \eqref{eq:hl-cm-XY} and the integration limits are
given by the solutions to \eqref{eq:hl-cm-eommc}.

Scalar fluctuations of the form $\phi = 0 + \delta\phi(y) \e^{ik\cdot x}$ yield 
\begin{align} 
  \left[ \frac{M_{HL}^2}{\Lambda_b^2} + \frac{(2\pi\alpha')^2}{b^4} 
   \frac{f(y)}{ g(y) } \nabla_y^2 - g(y) f(y) \right] \phi = 0, \label{eq:hl-cm-num}
\end{align}
with $\Lambda_b = b/(2\pi\alpha')$ the \acro{QCD} scale and
$(2\pi\alpha')^2/b^4 = 2\pi\delta/\lambda$.  
For boundary conditions $\p_y
\delta\phi(0)=0$ and $\delta\phi(y\to\infty)\sim c y^{-2}$ 
equation \eqref{eq:hl-cm-num}
determines
the meson spectrum. Since it is very similar to the \AdS\ spectrum, 
the binding energy, which demonstrates the deviations more clearly,
has been plotted in  Figure~\ref{fig:hl-cm-mesons}
for massless light quark.

\Section{Bottom Phenomenology\label{sec:bmesons}}

There has been a number of attempts to apply holographic methods to
phenomenological models \cite{Erlich:2005qh,DaRold:2005zs}, even for
the Constable--Myers background of the previous Section
\cite{Evans:2006dj}, successfully reproducing light quark meson data
with an accuracy better than 20\%.
That shall be motivation enough to
compare the heavy-light spectra calculated here with the bottom quark
sector of \acro{QCD}; i.e.~the massless quark in the setup above
will be assumed to play the \role\ of an up quark, while the
heavy quark, which  will lie in the \AdS-like region, 
will be interpreted as a bottom quark. 

In\marginpar{shortcomings}
that regime supersymmetry will be restored and the 
field theory will be strongly coupled even though \acro{QCD} dynamics
should be perturbative at this energy scale. 
These are respective consequences of the background being too simple 
(though a background exhibiting separation of scales is not known yet)
and an intrinsic feature of the \acro{SUGRA} version of \adscft\ that
can only be overcome by a full string treatment, which is currently
out of reach.

\begin{figure}
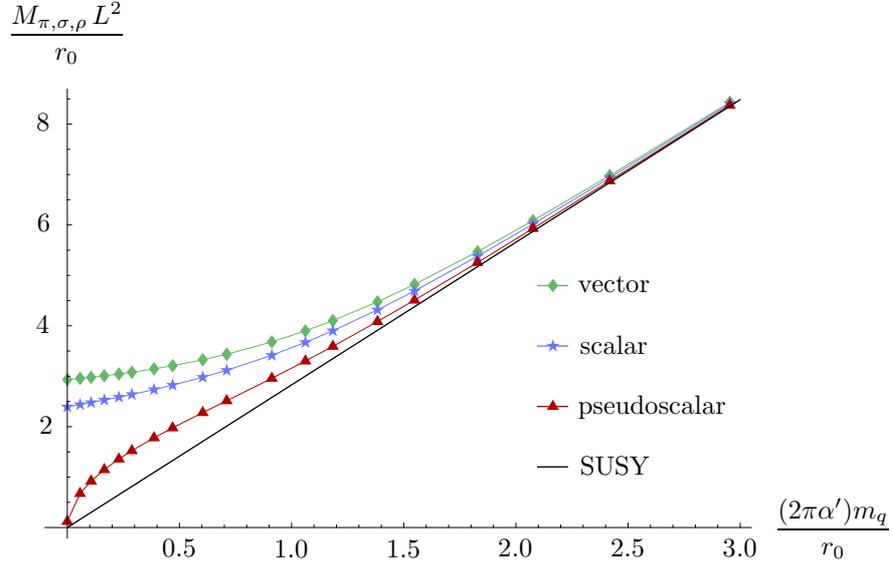

\centering
  \PSfraginclude[trim=-10 -10 -30 -30]{width=12cm}{gubser-lowmesons}
\caption[Mesons in \acro{GKS} (reviewed)]{
  Lightest pseudoscalar, scalar and vector mesons in the dilaton
  deformed geometry (\acro{GKS}). The vector mode for the massless quark is
  interpreted as a Rho meson, while for the heavy quark mass it yields
  the the Upsilon.  See also Section~\ref{sec:gubser-mesons}.
  \label{fig:hl-gubser-review}}
\end{figure}
\begin{figure}
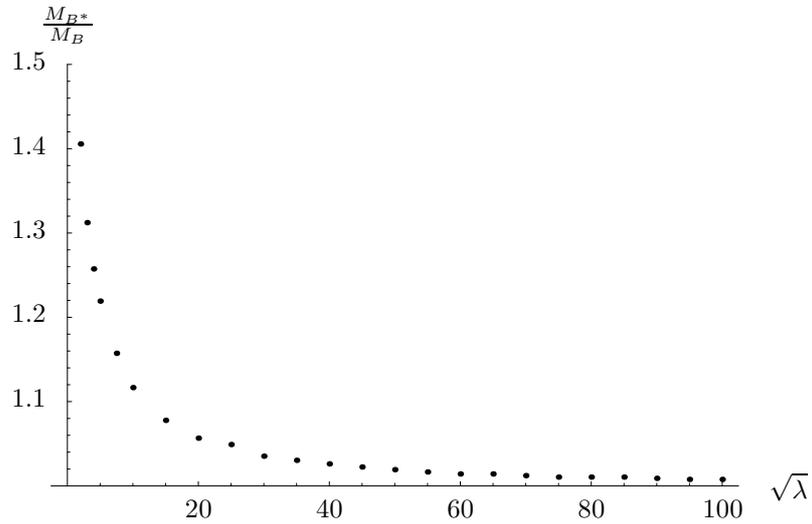

\centering
  \PSfraginclude[trim=55 35 55 35]{width=12cm}{hl-gubser-b-bstar-ratio}
\caption[B/B\textsuperscript{*} Mass Ratio]{
  Ratio of the mass of the lowest and first excited heavy-light meson mode for
  the \acro{GKS} and Constable-Myers background. (They really do look exactly the same,
  since the different units expressing the different dependence on the respective 
  deformation parameter cancel in the ratio.) 
  For large \tHooft\ parameter the ratio approaches 1, while the physical 
  B/B\textsuperscript{*} ratio
  (which is 1.01) is reached at $\lambda\approx82$.\label{fig:hl-gubser-bbstar}}
\end{figure}

The scales of the theory will be fixed by identifying the mass of the
lowest vector meson mode with the Rho and Upsilon mesons,
which are chosen as input data since they are less
sensitive to the light quark mass than the pseudoscalar modes roughly
corresponding to the Pion, cf.~Figure~\ref{fig:hl-gubser-review}
and~Section~\ref{sec:gubser-mesons} for details.

From Figure~\ref{fig:hl-gubser-review} the $\rho$ mass for
the \acro{GKS} background is read off to be 
$M_\rho L^2/r_0 = 2.93$.  Preserving the physical ratio
\begin{align}
  M_\Upsilon / M_\rho = 9.4\,\text{GeV} / 770\,\text{MeV,}
\end{align}
the $\Upsilon$ mass has to be $M_\Upsilon L^2/r_0=35.8$ and the heavy quark mass
can be read off to be $m_b=12.7\, \Lambda$.

The \tHooft\ parameter can be determined from the physical ratio of the
mass of the Rho and the B meson by searching for the value of $\lambda$
for which the numerical value of the lowest heavy-light excitation
satisfies
\begin{align}
  \pfrac{M_B}{M_\rho}^\text{phys} 
     = \pfrac{M_{HL}(\la)}{\Lambda}^\text{num}  \pfrac{r_0}{M_\rho L^2}^\text{num} 
       \sqrt{\frac{\lambda}{\pi}}.
\end{align}
Unfortunately this yields a value of the \tHooft\ coupling of $\lambda=2.31$.
As can be seen in Figure \ref{fig:hl-gubser-bbstar} the mass ratio of the predicted  
B and B\textsuperscript{*} meson reaches its physical value of approximately
1.01 only for very large $\lambda$. Identifying $M_{HL}$ with the
physical quark mass $M_B=5279$~MeV, one obtains a \acro{QCD} scale 
of 225~MeV\@.

With respect to the B mass ratio, the situation is slightly better for the 
background by Constable and Myers,
where the same procedure yields a prediction of $\lambda=5.22$. 
While it is not clear if this value is sufficient for the large $\lambda$
approximation inherent in the employed formulation of the \adscftcorr,
it gives a prediction for $M_{B^*} = 6403$~MeV, which is 20\% 
larger than the measured value of $5325$~MeV\@. Again a much
larger value of the \tHooft\ coupling would be required to achieve a better
agreement. For the \acro{QCD} scale on obtains $\Lambda_b=340$~MeV\@, 
which is a little too high. With $m_H = 12.63\,\Lambda_b$ the physical
b quark mass is predicted to be $4294$~MeV.

The overall agreement with experiment is far from perfect. However this
does not come as a surprise since the b quark mass ($m_b \approx 12\,\Lambda$ 
in both backgrounds)
is far in the supersymmetric
regime: Restoration of supersymmetry takes place approximately at $m_q \approx 1.5\,\Lambda$
as can be seen in Figure \ref{fig:hl-gubser-review}.
In other words a string connecting a brane describing a light quark and this
``b quark'' has about 80\% of its length in the supersymmetric region, 
which is a good approximation of pure \AdS. 
The only way to improve this situation would be to use a (yet unknown)
background that allows to separate the \acro{SUSY} breaking scale from
the \acro{QCD} scale.




  \Part{Space-time Dependent Couplings}

  \begin{savequote}[\savequotewidth]
  Supersymmetry is the greatest invention since the wheel.
  \qauthor{A.~Oop, ``A supersymmetric version of the leg'', Gondwansaland predraw, to be discovered \cite{Gates:1983nr}}
\end{savequote}

\Chapter{Supergravity Overview\label{ch:sugra-overview}}
The second part of this thesis is devoted to the discussion of 
the conformal anomaly in supersymmetric field theories, in particular
supersymmetric Yang--Mills theories. 

The approach chosen is an extension to superfields of the space-time
dependent coupling techniques Osborn \cite{Osborn:1991gm} applied to
non-super\-sym\-me\-tric theories coupled to a gravity background in order
to give an alternative proof of Zamolodchikov's $c$\NB-theorem,
cf.~Chapter~\ref{ch:local-couplings}.  Consequently a coupling to
supergravity will have to be considered and its superfield formulation
shall be reviewed in this Chapter.

In Chapter~\ref{ch:susy-trace} a discussion of the supersymmetric
conformal anomaly will be given.


%

\Section{Conventions}
To establish notations, a few basic ingredients for
supersymmetry are reviewed in the shortest possible manner.
Throughout this part, a dotted/undotted Weyl spinor notation 
is being used. 

The simplest double covering representation of the
Lorentz group can be constructed as follows.  An arbitrary vector
$v^\ada$ transforms under a Lorentz transformation $\Lambda^a{}_b \in
\gr{SO}(1,3)$ according to
\begin{align}
  x^a \mapsto x'^a = \Lambda^a{}_b x^b. \label{eq:lorentz}
\end{align}
The\marginpar{double covering}\index{double covering group} 
double covering group $\gr{SL}(2,\mathds{C})$ 
transforms the same vector according to
\begin{align}
  \sigma_a{}^\ada x^a \mapsto 
     (U^\ia{}_\ib \sigma_a{}^{\ib\db} U^\dagger{}^\da{}_\db) x^a 
     \equiv \sigma_a{}^\ada x'^a,
\end{align}
with $U$ the element of the double covering group
chosen such that $x'^a$ coincides with the definition \eqref{eq:lorentz}. The
matrices $\sigma^a:=(\one,\vec{\sigma})$ are 
the Pauli matrices augmented by the unity matrix. 
As an aside, the ``1 to 2'' relation of the
two representations can be easily seen from the fact that for any
$U$ being a solution to $(U^\ia{}_\ib \sigma_a^{\ib\db}
U^\dagger{}^\da{}_\db) = \sigma_b{}^\ada \Lambda^b{}_a  $, $-U$ is also a solution.
The group $\gr{SL}(2,\mathds{C})$ leaves invariant the antisymmetric 
tensors\marginpar{symplectic metric} $\veps_{\ia\ib}$ and $\veps_{\da\db}$, defined by
\begin{align}
  \veps_{12} &= \veps_{\dot1\dot2} = -1, &
  \veps^{12} &= \veps^{\dot1\dot2} = 1,
\end{align}
where the epsilon symbols with raised indices constitute the
respective inverse matrices by $\veps^{\ia\ib}\veps_{\ib\ic}=\delta^\ia_\ic$.
Since for any element $U$ of $\gr{SL}(2,\mathds{C})$ it holds
the relation $\veps^{\ia\ib}=\veps^{\ic\id} U_\ic{}^\ia U_\id{}^\ib$,
the combination $\veps^{\ia\ib} \psi_\ia \psi_\ib$ is invariant under
$\psi_\ia \mapsto U_\ia{}^\ib \psi_\ib$ and therefore a Lorentz scalar.
In other words,
the epsilon matrices can be used to obtain contragradiently transforming
representations according to
\begin{align}
  \psi^\ia &=  \veps^{\ia\ib} \psi_\ib, &
  \psi_\ia &= \veps_{\ia\ib} \psi^\ib, \\
  \bpsi^\da &= \veps^{\da\db} \bpsi_\db, &
  \bpsi_\da &= \veps_{\da\db} \bpsi^\db.
\end{align}
For the sake of brevity, an indexless\marginpar{indexless notation}
notation is often employed for
contracted adjacent objects, where different conventions are being used
for dotted and undotted indices,
\begin{align}
  \psi \chi &:= \psi^\ia \chi_\ia, &
  \bpsi \bchi &= \bpsi_\da \bchi^\da . 
\end{align}
This particular choice has the advantage that $\overline{\psi\chi} =
\bpsi \bchi$.

It is common to introduce
\begin{align}
  x^\ada := \tsigma_a{}^\ada x^a,
\end{align}
with $\tsigma_a^{\ada}=\veps^{\ia\ib}\veps^{\da\db}(\sigma_a)_{\ib\id}$,
and convert back and forth between the two representations using the
relations
\begin{align}
  (\sigma_a)_{\ia\dc} (\tsigma_b)^{\ib\dc} + (\sigma_b)_{\ia\dc} (\tsigma_a)^{\ib\dc} 
    = - 2 \eta_{ab} \delta_\ia{}^\ib, \\
  (\tsigma_a)^{\ic\da} (\sigma_b)_{\ic\db} + (\tsigma_b)^{\ic\da} (\sigma_a)_{\ic\db} 
    = - 2 \eta_{ab} \delta_\db{}^\da, 
\end{align}
which imply
\begin{align}
  x^a &= -\half (\sigma^a)_\ada x^\ada, &
  x^a x_a &= -\half x^\ada x_\ada.
\end{align}

A\marginpar{Gra\ss{}mann parity}\index{Gra\ss{}mann parity} superspace is
defined to be a space with coordinates $x^\ada$ of even Gra\ss{}\-mann
parity and $\theta^\ia$, $\btheta^\da=(\theta^\ia)^\dagger$ of odd
Gra\ss{}\-mann parity; i.e.\ anticommuting. The Gra\ss{}\-mann parity of a
quantity $q$ is symbolised by $\#q$ and capital Latin letters are used
to denote collective indices; e.g.\ the supercoordinates are
labelled\footnote{This convention implies that components of a
  tensorial object $t_{A_1\dots A_n}$ have a varying number of
  indices. Commas will be used to separate index pairs $\ada,\ib\dc$
  whenever this disambiguation is necessary.}
$z^A=(x^\ada,\theta^\ia,\btheta_\da)$ and transform under the
$(\half,\half)$, $(\half,0)$ and $(0,\half)$ representations,
respectively.\footnote{The latter are (complex) Weyl spinors as
  opposed to Dirac spinors, which are composed of two Weyl spinors. }
Arbitrary irreducible representations $(\frac{m}{2},\frac{n}{2})$
\index{irreducible representation} are given by symmetric tensors
\begin{align}
\psi^{\ia_1,\dots,\ia_m,\db_1,\dots,\db_n} \equiv
\psi^{\{\ia_1,\dots,\ia_m\},\{\db_1,\dots,\db_n\}}, 
\end{align}
where the weight
is chosen such that \panti sym\-me\-tri\-sa\-tion\marginpar{\panti
  symmetrisation}\index{symmetrisation}\index{anti-symmetrisation} is
idempotent,
\begin{align}
  \psi_{\{\ia_1,\dots,\ia_N\}} 
     &= \frac{1}{N!} \sum \psi_{\pi(\ia_1),\dots,\pi(\ia_N)}, \\
  \psi_{ [\ia_1,\dots,\ia_N]}
     &= \frac{1}{N!}  \sum \sign (\pi) \psi_{\pi(\ia_1),\dots,\pi(\ia_N)},
\end{align}
and \panti symmetrisation is performed over only those indices enclosed
in braces that are not additionally enclosed in a pair of vertical 
bars $|\;|$. 
From the spin-statistics theorem follows that any \emph{physical} field 
$\psi^{\ia_1,\dots,\ia_m,\db_1,\dots,\db_n}$ has Gra\ss{}mann parity $m+n \pmod 2$.

Partial superderivatives $\p_A=(\p_\ada,\p_\ia,\bp^\da)$ are defined by
\begin{align}
  \gcomm{\p_A}{z^B} = (\p_A z^B ) := \delta_A{}^B
\end{align}
where the ($\mathds{Z}_2$-)\emph{graded commutator}\marginpar{graded commutator}\index{graded!commutator} is
defined by
\begin{align}
  \gcomm{A}{B} := AB - (-1)^{\#A\,\#B} BA
\end{align}
and obeys the graded Leibniz rule and Jacobi\marginpar{Leibniz, Jacobi} identity\index{graded!Leibniz rule!graded}\index{graded!Jacobi identity}
\begin{gather}
  \gcomm{A}{B\,C} = \gcomm{A}{B}C + (-1)^{\#A\,\#B} B \gcomm{A}{C},\\
   (-1)^{\#A\,\#C} \gcomm{A}{\gcomm{B}{C}} 
      + (\text{cyclic $A\mapsto B\mapsto C$})
      = 0. \label{eq:def:jacobi}
\end{gather}
The partial derivatives in a flat superspace satisfy 
\begin{align}
  \gcomm{\p_A}{\p_B} &= 0.
\end{align}

A\marginpar{components}\index{superfield!components} superfield $f(x,\theta,\btheta)$ on $\mathds{R}^{4|4}$ can be defined by 
a Taylor expansion in the non-commuting coordinates according to
\begin{align}
\begin{split}
  f(z^A) &= 
     A(x) + \theta^\ia \psi_\ia(x) + \btheta_\da \bpsi^\da(x)\\&\relphantom{=}
     + \theta^2 F(x) + \btheta^2 \bar F(x) + \theta \sigma^a \btheta V_a(x) \\&\relphantom{=}
     + \btheta^2 \theta^\ia \la_\ia(x) + \theta^2 \btheta_\da \bla^\da(x) 
     + \theta^2 \btheta^2 G(x),
\end{split}
\end{align}
where the respective coefficients are called \emph{components}.
Mass dimension and Gra\ss{}mann parity of the superfield are by definition 
given by the respective property of the lowest component $A$.
This definition of a superfield can be extended to include tensorial fields
by simply promoting the components to tensors.

In a similar manner a superfield can be defined on $\mathds{C}^{4|2}$, 
which is build up from four complex ($y^a$) 
and two anticommuting ($\theta^\ia$) coordinates. For the remaining part of this
introduction, these two superspaces
will be referred to as the real ($\mathds{R}^{4|4}$) and complex ($\mathds{C}^{4|2}$) superspace respectively. 
The real superspace is a subspace of the complex superspace, embedded 
according to
\begin{align}
  y^a=x^a+i\theta\sigma^a\btheta.
\end{align}
By\marginpar{chiral superfields}\index{superfield!chiral} this relation holomorphic superfields 
can be defined on the real superspace (where they are known as \emph{chiral superfields}) according to
\begin{align}
\begin{split}
  \Phi(x,\theta,\btheta) 
    &\hphantom{:}= \Phi(x+i\theta\sigma^a\btheta,\theta) 
    = \e^{iH} \Phi(x,\theta) \\
  H &:= \theta\sigma^a\btheta \p_a,
\end{split}
\end{align}
where $H$ has been defined with future generalisations in mind.
(The current choice of $H$ has the unique property of making 
super-\Poincare\ transformations on both spaces coincide, thus 
providing the only \Poincare\ invariant embedding of $\mathds{R}^{4|4}$
into $\mathds{C}^{4|2}$.)

The property $\bp \Phi(y)=0$ can be rewritten as\marginpar{flat
  covariant derivative}\index{covariant derivative!flat superspace}
\begin{subequations}\label{eq:flatcov}
\begin{align}
  \bar D_\da \Phi(x,\theta,\btheta) &= 0, &
  \bar D_\da := \e^{iH} (-\bp_\da) \e^{-iH} &= -\bp_\da - i \theta^\ia \p_\ada. 
\end{align}
Analogously, for an antichiral field it holds 
\begin{align}
  D_\ia \Phi(x,\theta,\btheta) &= 0, &
  D_\ia := \e^{-iH} (\p_\ia) \e^{iH} &= \p_\ia + i \theta^\ia \p_\ada.
\end{align}
\end{subequations}
The set of derivatives $D_A = (\p_a, D_\ia, \bar D^\da)$ has the
property of commuting with the supersymmetry generators and mapping a
tensor superfield into a tensor superfield with respect to the Lorentz
group. Hence, they are called (flat) \emph{covariant} derivatives.
The observant reader has noticed the unusual sign in front of
$\bp_\da$ in definition \eqref{eq:flatcov}, which is related to
convenient complex conjugation properties as will be explained below.
While partial derivatives obey trivial (anti-)commutation rules, this
is no longer true for covariant derivatives ($\acomm{D_\ia}{\bar
  D_\da}=-2i\p_\ada$), and consequently special attention has to be
paid to the reordering upon complex conjugation, in particular
Hermitean and complex conjugation no longer coincide.

The\marginpar{conjugation} Hermitean conjugate $\cO^\dagger$ and transpose $\cO^\transpose$ 
of an operator $\cO$ are respectively defined by
\begin{align}
  \int \overline{\cO^\dagger \chi} \psi &:= \int \bar\chi \cO \psi, \\
  \int ( \cO^\transpose \chi ) \psi &:= (-1)^{\#\chi\,\#\cO} \int \chi \cO \psi, 
\end{align}
which additionally allows to define the complex conjugate by
\begin{align}
\cO^* := (\cO^\dagger)^\transpose.
\end{align}
In particular, these definitions imply the following reorderings
\begin{alignat}{2}
  &(\cO_1\dots \cO_N)^\dagger 
    &&= \cO_N^\dagger\dots \cO_1^\dagger , \\
  &(\cO_1\dots \cO_N)^\transpose 
    &&= (-1)^{\#\cO_1\,\#\cO_2} \cO_N^\transpose\dots  \cO_1^\transpose ,\\
  &(\cO_1\dots \cO_N)^*
    &&= (-1)^{\#\cO_1\,\#\cO_2} \cO_1^*\dots  \cO_N^* .
\end{alignat}
%
\begin{table}[t]
\centering
\begin{colortabular}{>{$}Cc<{$}|>{$}Cc<{$}|>{$}Cc<{$}|>{$}Cc<{$}}
   \multicolumn{4}{Hc}{\textsc{Conjugations}}\\
     \multicolumn{1}{c}{$\cO$}& 
     \multicolumn{1}{c}{$\cO^\dagger$} & 
     \multicolumn{1}{c}{$\cO^*$} & 
     \multicolumn{1}{c}{$\cO^\transpose$} \\ \bwline
     \cO_1\cdots \cO_n &
     \cO_n^\dagger \cdots \cO_1^\dagger &
     \pi_{\#F} \, \cO_1^* \cdots \cO_n^* &
     \pi_{\#F} \, \cO_n^\transpose \cdots \cO_1^\transpose \\ 
   \hline
     \psi^\ia & \bar\psi^\da & \bar\psi^\da & \psi^\ia \\
   \hline
     \psi^{\ia_1\dots\ia_m \db_1\dots \db_n} & 
     \bar\psi^{\db_n\dots \db_1 \ia_m\dots\ia_1} & 
     \pi_n\pi_m \bar\psi^{\db_n\dots \db_1 \ia_m\dots\ia_1}& 
     \pi_n\pi_m \psi^{\ia_m\dots\ia_1 \db_n\dots \db_1}\\
   \hline
     \p_a & - \p_a & \p_a & -\p_a\\ 
   \hline
     \p_\ia & \bp_\da & - \bp_\da & -\p_\ia\\ 
   \hline
     D_a & - D_a & D_a & -D_a\\ 
   \hline
     D_\ia & -\bar D_\da & \bar D_\da & -D_\ia \\ \bwline
\end{colortabular}
\caption[Definition of Conjugations]{\label{tab:conjugation}
  Definition of the Hermitean and complex conjugate as well as 
  transposition (from left to right).
  The symbol\\\noindent 
  \begin{minipage}{\textwidth}
  $$\pi_m := (-1)^{\floor{\frac{m}{2}}} = (-1)^{\half m(m-1)}$$\vskip0ex
  \end{minipage}
  denotes the sign change induced by reversing the order of 
  $m$ anticommuting objects while $\#F$ is the number of fermionic terms 
  in the corresponding expression.}
\end{table}
%
From 
\begin{align}
  \acomm{(\bp_\da)^\dagger}{(\bar z^\db)^\dagger}
    = \acomm{\bp_\da}{\bar z^\db}^\dagger
    = (\delta_\da{}^\db)^\dagger 
    &= \delta_\ia{}^\ib 
    = \acomm{\p_\ia}{z^\ib},\\
  -\comm{(\p_a)^\dagger}{(z^a)^\dagger}
    = \comm{\p_a}{z^b}^\dagger
    = (\delta_a{}^b)^\dagger 
    &= \delta_a{}^b 
    = \comm{\p_a}{z^b}
\end{align}
one may deduce
\begin{align}
  (\p_a)^\dagger &= - \p_a , \\
  (\p_\ia)^\dagger &= \bp_\da,
\end{align}
while the transpose $\p_A^T = -\p_A$ is determined by partial
integration. So complex conjugation of a spinor partial derivative involves
an additional minus sign compared to other fermionic objects. As complex
conjugation is an operation which will be employed quite frequently when
working directly with the supergravity algebra, the definition of 
covariant spinor derivatives \eqref{eq:flatcov} involves an additional minus
sign for compensation. The conjugation rules are summarised in Table 
\ref{tab:conjugation}. As one can see, for the case of (anti-)commuting 
objects---``numbers''---Hermitean conjugation and complex conjugation are
the same. 

In the supergravity literature,\marginpar{conventional traps} the use of different
notations and conventions is quite common. In particular it crucially
depends on the task to be performed, which conventions are the most
suitable.  This thesis follows closely the conventions of
\cite{Buchbinder:1998qv}, which contain the potential trap that for an
antisymmetric tensor
\begin{align}
  \psi_{\ia\ib} \sim \veps_{\ia\ib}
\end{align}
the corresponding contragradient tensor reads 
\begin{align}
  \psi^{\ia\ib} = \veps^{\ia\ic} \veps^{\ib\id} \psi_{\ic\id} 
                \sim \veps^{\ia\ic} \veps^{\ib\id} \veps_{\ic\id} = - \veps^{\ia\ib}
\end{align}
as a consequence of the conventions used for raising and lowering
operators.

The other major source of this compilation \cite{Gates:1983nr} uses an
\emph{imaginary} symplectic metric, which introduce a relative minus
sign for complex conjugation of contragradient indices. Additionally,
there appears a minus sign in the \emph{complex} conjugation of
spinorial covariant superderivatives $D_\ia = (\bar D_\da)^\dagger = -
(\bar D_\da)^*$. Furthermore, quadratic quantities
$D^2$ contain a factor of one half, which materialises upon partial integration.

\Section{Superspace Supergravity}
In analogy to the non-supersymmetric case, a pseudo-Riemannian
supermanifold is defined by an atlas of maps from open sets of points
on the supermanifold to coordinates in flat superspace.  When there is
curvature, in general more than one map is required to cover the whole
manifold and the maps are distorted in the sense, that a non-Minkowski
metric is needed to capture this distortion in terms of those
superspace coordinates, which shall be called \emph{world} or
\emph{curved}\marginpar{world vs.\ tangent} coordinates coordinates
$z^M=(z^m,\theta^\mu,\btheta_\dm)$.  To each point of the
supermanifold one may attach a \emph{tangent} superspace (also
referred to as \emph{flat}), whose coordinates are called
$z^A=(z^a,\theta^\alpha,\btheta_\da)$.  The distinction of flat vs.\
curved will also be made in referring to the indices only as indicated
in Table~\ref{tab:sugra-indices}. 


\begin{table}
\centering
\newcommand\myempty{}
\ifx\bwline\myempty
\begin{colortabular}{BlBcBc} 
   \multicolumn{3}{Hc}{\textsc{SUGRA Index Conventions}}\\
        \rowcolor{white}
        &$c$-coordinates ($x$)&$a$-coordinates ($\theta$)\\
        &\cell{Cc|}{$m,n,\dots$}&
        \cell{Cc}{$\mu,\nu,\dots$}\\\hline
        \smash{\raisebox{2ex}{world}}&
        \multicolumn{2}{Cc}{$M,N,\dots$}
        \\ \arrayrulecolor{white}\hline\arrayrulecolor{bgcolor}
        &\cell{Cc|}{$a,b,\dots$}&
        \cell{Cc}{$\alpha,\beta,\dots$}\\\hline
        \smash{\raisebox{2ex}{tangent}} &\multicolumn{2}{Cc}{$A,B,\dots$}
\end{colortabular}
\else
{
   \renewcommand{\arraystretch}{1.5}%
\begin{tabular}{|l|cc|} 
  \multicolumn{3}{c}{\textsc{SUGRA Index Conventions}}\\
        \cell{c}{}&\cell{c}{$c$-coordinates ($x$)}&\cell{c}{$a$-coordinates ($\theta$)}\\ \hline
        &\cell{c|}{$m,n,\dots$}&
        \cell{c|}{$\mu,\nu,\dots$}\\
        \smash{\raisebox{2ex}{world}}&
        \multicolumn{2}{c|}{$M,N,\dots$}
        \\ \hline
        &\cell{c|}{$a,b,\dots$}&
        \cell{c|}{$\alpha,\beta,\dots$}\\
        \smash{\raisebox{2ex}{tangent}} &\multicolumn{2}{c|}{$A,B,\dots$} \\ \hline
\end{tabular}
}
\fi
\caption{\label{tab:sugra-indices}
  Superfield Supergravity Index Conventions}
\end{table}

Superspace supergravity requires\marginpar{doubled Lorentz} 
a tangent space formulation, where
superspace general coordinate transformations, realised as gauged 
curved superspace translations, are augmented by an additional set of 
superlocal Lorentz transformations acting on the tangent space only.
The reason is that without this doubling spinors
can only be realised non-linearly, which is inconvenient
\cite[p.~235]{Gates:1983nr}.

A first order differential operator, expressed as 
\begin{align}
  K = K^M \p_M + \half K^{ab} M_{ab}
    = K^M \p_M + K^{\ia\ib} M_{\ia\ib} +  K^{\da\db} \bM_{\da\db},
\end{align}
therefore allows to define covariant transformation under combined
supercoordinate and super-Lorentz transformations according to 
\begin{align}
  X \mapsto \e^K X \e^{-K}. \label{eq:def:cov}
\end{align}

The $\al{sl}(2,\mathds{C})$ versions\marginpar{Lorenz generators}\index{Lorentz generators!\al{sl}{2,\mathds{C}}}
$M_{\ia\ib} = \half (\sigma^{ab})_{\ia\ib}M_{ab}$ and
$\bM_{\da\db} = \half (\tsigma^{ab})_{\da\db}M_{ab}$ 
of the Lorentz generator $M_{ab}$ act on the
corresponding indices (i.e.~only on indices
of the same kind) according to 
\begin{align}
  M_{\ib\ic} \psi_{\ia_1 \dots \ia_n} 
    &= \half \sum_i ( 
         \veps_{\ia_i \ib} 
            \psi_{\ic \ia_1 \dots \not\ia_i \dots \ia_n} +
         \veps_{\ia_i \ic} 
            \psi_{\ib \ia_1 \dots \not\ia_i \dots \ia_n} ), \\
  \bM_{\db\dc} \psi_{\da_1 \dots \da_n} 
    &= \half \sum_i ( 
         \veps_{\da_i \db} 
            \psi_{\dc \da_1 \dots \not\da_i \dots \da_n} +
         \veps_{\da_i \dc} 
            \psi_{\db \da_1 \dots \not\da_i \dots \da_n} ). 
\end{align}
In particular, it holds
\begin{align*}
  M_\bc \psi_\ia &= \half (\veps_\ab \psi_\ic + \veps_\ac \psi_\ic), \\
  M_\bc \psi^\ia &= \half (\delta^\ia_\ib \psi_\ic + \delta^\ia_\ic \psi_\ib), \\
  M_\ab \psi^\ib &= \tfrac{3}{2} \psi_\ib.
\end{align*}

In analogy\marginpar{curved covariant derivatives}\index{covariant derivative!curved superspace}
to ordinary gravity (with torsion) one may define a derivative
\begin{align}
  \D_A &= E_A + \Omega_A
\end{align}
that transforms covariantly under \eqref{eq:def:cov} by adding a 
vierbein field $E_A:=E_A{}^M \p_M$ and a superconnection 
\begin{align}
\Omega_A:= \half \Omega_A{}^{BC}M_{BC} 
         = \Omega_A{}^{\ib\ic} M_{\ib\ic} + \Omega_A{}^{\db\dc} \bM_{\db\dc}.
\end{align}
%
The vierbein obeys the algebra\marginpar{anholonomy}\index{anholonomy coefficients}
\begin{align}
  \gcomm{E_A}{E_B} &= C_{AB}{}^C E_C, \label{eq:anholo} \\
  C_{AB}{}^C &= (E_A E_B{}^M - (-1)^{\#A\,\#B} E_B E_A{}^M) E_M^C,
\end{align}
where $C_{AB}{}^C$ are the supersymmetric generalisation of anholonomy
coefficients.  The non-degenerate supermatrix $E_A{}^M$ can be used to
convert between world and tangent indices according to
\begin{align}
  V_A = E_A{}^M V_M,
\end{align}
and the bosonic submatrix $E_a{}^m$ is the well known vierbein
field of gravity obeying
\begin{align}
  \eta_{ab} &= g_{mn}E_a{}^m E_b{}^n.
\end{align}

The covariant derivatives form an algebra\marginpar{curvature, torsion}
\begin{align}
  \gcomm{\D_A}{\D_B} &\hphantom{:}= T_{AB} + R_{AB}, \label{eq:def:sugra:algebra} \\
  T_{AB} &:= T_{AB}{}^C \p_C, \\
  R_{AB} &:= \half R_{AB}{}^{bc} M_{bc} 
          = R_{AB}{}^{\ib\ic} M_{\ib\ic} + R_{AB}{}^{\db\dc} \bM_{\db\dc}, 
\end{align}
\finalize{\begin{sloppypar}}
with $T_{AB}= -(-1)^{\#A\,\#B} T_{BA}$ the supertorsion and
$R_{AB}= -(-1)^{\#A\,\#B} R_{BA}$ the supercurvature, which may be
completely expressed in terms of the supertorsion as a consequence of
the Bianchi identities. The latter are just the 
Jacobi identities \eqref{eq:def:jacobi}  for the algebra 
\eqref{eq:def:sugra:algebra}.  
\finalize{\end{sloppypar}}

\Section{Non-minimal Supergravity}
The algebra above is a highly reducible representation of
supergravity. To extract the physical degrees of freedom a number of
constraints has to be imposed. One distinguishes between conventional
constraints\marginpar{conventional
  constraints}\index{constraints!non-minimal \acro{SUGRA}}
\begin{subequations}
\begin{align}
  &
  \begin{aligned}
  T_{\ia\db}{}^\ic &= T_{\ia\db}{}^{\dc} = R_{\ia\ib}{}^{cd} = 0,\;\\
  T_{\ia\db}{}^c &= -2i \sigma^c_{\ia\db} 
  \end{aligned}
  \biggr\} \iff \D_\ada = -2i \acomm{\D_\ia}{\bD_\da}, \\
  &\,T_{\ia\ib}{}^\ic 
    = T_{\da\db}{}^{\dc} 
    = T_{\ia,\ib\{\db,}{}^\ib{}_{\ic\}} 
    = T_{\ia,\{\ib}{}^{\db,}{}_{\ic\}\db} = 0,
\end{align}
which are equivalent to redefinitions of the algebra's constituents,
and representation preserving constraints\marginpar{representation preserving constraints}
\begin{align}
  T_{\ia\ib}{}^c =  T_{\da\db}{}^c 
    = T_{\ia\ib}{}^{\dc} = T_{\da\db}{}^{\ic} = 0,
\end{align}
\end{subequations}
which imply the existence of (anti-)chiral superfields by ensuring
the Wess--Zumino consistency condition
\begin{align}
  \bD_\da \chi &= 0 \implies \acomm{\bD_\da}{\bD_\db} \chi = 0.
\end{align}
While the Bianchi identities\marginpar{solving of Bianchi identities} are trivially fulfilled by the
unconstrained derivatives, this is no longer true, when introducing
constraints whose consequences for the remaining torsion fields have
to be evaluated.  Since this procedure is straight-forward, it will
not be reproduced here due to the length of the calculation and the
fact, that it may be found in the literature
\cite{Gates:1983nr,Buchbinder:1998qv,Bagger:1990qh,Wess:1992cp} 
under the name of ``solving the Bianchi identities''.

After solving the Bianchi identities, 
all torsions and curvatures can be expressed in terms of a few
basic fields,
\begin{align}
  T_\ia &:= (-1)^{\#B} T_{\ia B}{}^B, \\
  G_\ada &:= i T^{\ib,}{}_{\ib\da,\ia} + i T^{\db,}{}_{\ia\db,\da}, \\
  R &:= \tfrac{1}{12} R^{\da\db}{}_{\da\db}, \\
  W_{\ia\ib\ic} &:= \half T_{\{ia}{}^{\db,}{}_{\ib|\db|,\ic\}}, 
\end{align}
where $R$ and $\bR$ are chiral and antichiral superfields, 
$G_{\ada}$ is real, and $T_\ia$, $W_{\ia\ib\ic}$ are 
complex superfields, all of which are subject to a set of 
Bianchi identities and obey the so-called 
``non-minimal supergravity algebra''.

\Subsection{Algebra and Bianchi identities}
The non-minimal\marginpar{defining relations} supergravity algebra is defined by the following 
three (anti-)commutators,
\begin{align}
  \acomm{\D_\ia}{\bD_\da} &= -2i \D_\ada, \\
  \acomm{\D_\ia}{\D_\ib} &= -4 \bR M_{\ia\ib}, \\
\begin{split}
  \comm{\bD_\da}{ \D_{\ib\db} } 
    &= \veps_{\da\db} \biggl[ \half \bT^\dc \D_{\ib\dc} 
         - i ( R + \tfrac{1}{8} \bD_\dc \bT^\dc - \tfrac{1}{16} \bT^2 ) \D_\ib\\
    &\hphantom{= \veps_{\da\db} \biggl[ }
         - i \bpsi_\ib{}^\dc \bD_\dc 
         + i (\bD^\dd - \half \bT^\dd ) \bpsi_\ib{}^\dc \bM_{\dd\dc} \\
    &\hphantom{= \veps_{\da\db} \biggl[ }
        + 2i X^\ic M_{\ib\ic} - 2i W_\ib{}^{\ic\id} M_{\ic\id} \biggr]
         - i (\D_\ib R) \bM_{\da\db}.
\end{split}
\end{align}
The missing relations\marginpar{implications} can be determined from the Bianchi identities (see below) 
and complex conjugation. 
\begin{align}
  \acomm{\bD_\da}{\bD_\db} &= 4 R \bM_{\da\db}, \\
\begin{split}
  \comm{\D_\ia}{ \D_{\ib\db} } 
    &= \veps_{\ia\ib} \biggl[ \half T^\ic \D_{\ic\db} 
         + i ( \bR + \tfrac{1}{8} \D^\ic T_\ic - \tfrac{1}{16} T^2 ) \bD_\db\\
    &\hphantom{= \veps_{\ia\ib} \biggl[ }
         + i \psi^\ic{}_\db \D_\ic 
         + i (\D^\id - \half T^\id ) \psi_\db{}^\ic M_{\id\ic} \\
    &\hphantom{= \veps_{\ia\ib} \biggl[ }
         - 2i \bar X^\dc \bM_{\db\dc} + 2i \bW_\db{}^{\dc\dd} M_{\dc\dd} \biggr]
         + i (\bD_\db \bR) M_{\ia\ib},
\end{split}\\
  \comm{ \D_\ada }{ \D_{\ib\db} } 
     &= \ihalf \acomm{\comm{\D_\ia}{\D_{\ib\db}} }{ \bD_\da }  + 
        \ihalf \acomm{\comm{\bD_\da}{\D_{\ib\db}} }{ \D_\ia },  
\end{align}
with the abbreviations 
\begin{align}
  \psi_\ada &= G_\ada - \tfrac{1}{8} \D_\ia \bT_\da - \tfrac{1}{8} \bD_\da T_\ia, \\
\begin{split}
  X_\ia &= 
     \tfrac{1}{12} \biggl[ (\bD_\dc - \half \bT_\dc) (\bD^\dc - \half \bT^\dc) 
                            - 4 R \biggr] T_\ia \\&\relphantom{=}
     + \tfrac{1}{12} \biggl[ \begin{aligned}[t] 2 \psi_\ada & +
                             (\bD_\da - \half \bT_\da) (\D_\ia - \half T_\ia) \\&
                             + \half (\D_\ia - T_\ia) (\bD_\da - \half \bT_\da) \biggr]
                             \bT^\da.
                            \end{aligned}
\end{split}
\end{align}

The Bianchi identities\marginpar{Bianchi identities} expressed in terms of the supertorsion components read
\begin{align}
\begin{split}
  \bD_\da R = 0,  \qquad 
  G_a &= \bG_a, \qquad
  W_{\abc} = W_{\{\abc\}}, \\
  \D_\ia T_\ib + \D_\ib T_\ia &= 0,  \\
   (\bD^\da - \half \bT^\da ) \psi_\ada &= \D_\ia R, \qquad
  (\bD_\da - \half \bT_\da ) W_\abc = 0, \\
  (\D^\ic - T^\ic) W_\abc 
     &= \ihalf (\D_\ia{}^\da - \ihalf (\D_{\ia} \bT^\da)) \psi_{\ib\da} + ( \ia \vs \ib ).
\end{split}
\end{align}

\Subsection{Partial Integration}
From the supergravity algebra \eqref{eq:def:sugra:algebra} it can be 
shown that 
\begin{align}
(-1)^{\#A} E^{-1} \D_A V^A 
  - (-1)^{\#B} V^A T_{AB}{}^B 
  = (E^{-1} V^A) \lvec{E}_A,
\end{align}
which implies
\begin{align}
  \int d^8z\, E^{-1} (\D_\ada - (-1)^{\#B} T_{aB}{}^B) V^\ada &= 0, \displaybreak[0]\\
  \int d^8z\, E^{-1} (\D_\ia + T_\ia) V^\ia &= 0,\\
  \int d^8z\, E^{-1} (\bD_\da + T_\da) V^\da &= 0.
\end{align}
$E^{-1} := \sdet^{-1} E_A{}^M $ is the real superspace analogue of $\sqrt{-g_{mn}}$.

Clearly it is a natural alternative to consider the combination $\D_\ia + T_\ia$ as
the basic covariant derivative. Then $T_\ia$ takes over the \role\ of a $\gr{U}(1)_R$
connection, an approach chosen in \cite{Gates:1983nr}.

\Subsection{Superdeterminant}
In the last Section the superdeterminant has been introduced and its definition
shall follow here, belatedly. In analogy to the non-supersymmetric case
the superdeterminant is the exponential of the logarithm's supertrace
\begin{align}
  \sdet A^M{}_N &:= \exp \str \ln A^M{}_N,
\end{align}
where the\marginpar{supertrace} supertrace is
\begin{align}
  \str A^M{}_N := (-1)^{\#M} A^M{}_M,
\end{align}
which is cyclic and invariant under a suitably defined
supertransposition 
\begin{align}
(A^{\text{sT}})_M{}^N &:= (-1)^{\#N+\#M\#N} A^N{}_M.
\end{align}

For practical calculations, the following theorem is much more
important
\begin{align}
  z^{\prime M} &= \e^{-K} z^M, & K &= K^M {\strut{\p}}_M, \\
  \sdet \frac{\p z^{\prime M}}{\p z^N} &= (1\cdot \e^{\lvec{K}}), &
  \lvec{K} &= K^M {\strut\lvec{\p}}_M.
\end{align}
The right\marginpar{right operator} partial derivative ${\strut\lvec{\p}}_M$ in $\lvec{K}$
acts on the components $K^M$ and everything to the left of $\lvec{K}$,
such that
\begin{align}
  \lvec{K} &= (-1)^{\#M} {\strut\lvec{\p}}_M K^M + (-1)^{\#M} (\p_M K^M).
\end{align}
Additionally the following rule holds
\begin{align}
  (1\cdot \e^{\lvec{K}})(\e^K \Phi) = (\Phi\cdot \e^{\lvec{K}}). \label{eq:leftexp}
\end{align}
Proofs for any of these statements can be found
in the literature, in particular \cite{Buchbinder:1998qv}.

\Subsection{Super-Weyl Transformations\label{sec:non-min-weyl}}
While the algebra of the previous Sections is by construction invariant
under general supercoordinate and superlocal Lorentz transformations, 
it is in addition invariant under transformations of the vierbein
of the form
\begin{align}
  E_\ia  &\mapsto L E_\ia, \\
  \bE_\da  &\mapsto \bL \bE_\da, \\
  E_\ada &\mapsto L\bL E_\ada, \\
  E & \mapsto (L\bL)^2 E,
\end{align}
which are easily seen to  represent Weyl transformation of the bosonic vierbein 
component, when restricting $L$ to (the real part of) its lowest component.
The unconstrained complex superfield 
$L=\exp (\half \Delta + \ihalf \kappa)$ parametrises mixed superlocal scale transformations (by~$\Delta$) and superlocal chiral transformations (by~$\kappa$).
The latter can also be understood as local $\gr{U}(1)_R$ transformations.

The elements of the non-minimal supergravity algebra transform under
this symmetry as
\begin{align}
  \D_\ia &\mapsto L \D_\ia - 2(\D^\ib L) M_{\ia\ib}, \\
  \bD_\da &\mapsto \bL \bD_\da - 2 (\bD^\db \bL) \bM_{\da\db}, \\
  T_\ia &\mapsto L T_\ia + \D'_\ia \ln (L^4 \bL^2), \\
  R &\mapsto -\quart( \bD^2 - 4R) \bL^2. 
\end{align}

\Subsection{Prepotentials}
As a consequence of Frobenius' theorem, the vierbein field,
which is subject to the constraint \eqref{eq:anholo} can 
be decomposed into unconstrained superfields $F$, $W$ and $N_\ia{}^\mu$,
called \emph{prepotentials},
\begin{align}
  E_\ia &= F N_\ia{}^\mu \e^W \p_\mu \e^{-W}, \qquad \det N_\ia{}^\mu = 1, \\
  \bE_\da &= - \bF \bN_\da{}^{\dot{\mu}} \e^\bW \bp_{\dot{\mu}} \e^{-\bW}.
\end{align}
Because the ``superscale'' field $F$ 
has been introduced to allow the choice
$\det N_\ia{}^\mu = 1$, it is also the only prepotential that transforms under
super-Weyl transformations: $F\mapsto LF$.
Under coordinate transformations induced by $K = K^M \p_M = \bar{K}$, all prepotentials
transform covariantly,
\begin{align}
   F' &= (\e^K F), &
   (N_\ia{}^\mu)' &= (\e^K N_\ia{}^\mu ), &
   W' &= (\e^K W),
\end{align}
while only $N_\ia{}^\mu$ transforms under superlocal transformations
\begin{align}
   (N_\ia{}^\mu)' &= (\e^{\half K^{ab} M_{ab}}) N_\ia{}^\mu.
\end{align}

While all supergravity superfields can be expressed in terms of prepotentials,
only the two simple expressions
\begin{align}
  T_\ia &= E_\ia \ln [ E F^2 (1\cdot \e^{\lvec{W}}) ], \label{eq:prepoT} \\
  R &= -\quart \hat{\bE}_{\dot{\mu}} \hat{\bE}^{\dot{\mu}} \bF^2 \label{eq:prepoR}
\end{align}
shall be given here with the \emph{semi-covariant vierbein}\marginpar{semi-covariant vierbein} $\hat{E}$ defined
by
\begin{align}
  \hat{E}_\ia &:= F^{-1} E_\ia, & 
     \hat{E}_\ia &=: N_\ia{}^\mu \hat{E}_\mu \nonumber\\
  \hat{\bE}_\da &:= \bF^{-1} \bE_\da, \label{eq:semi-cov-vierbein}\\
  \hat{E}_\ada &:= \ihalf \acomm{\hat{E}_\ia}{\hat{\bE}_\da} .\nonumber
\end{align}
There is an additional prepotential $\vphi$, the chiral compensator,
that can be chosen to take over the  \role\ of $F$, see Section~\\ref{sec:superweyltrafos}.

\Section{Minimal Supergravity}
From the non-minimal supergravity algebra, a formulation
containing less auxiliary fields may be obtained by 
setting $T_\ia = 0$.
This has a number of consequences: The algebra simplifies 
considerably, $W_{\ia\ib\ic}$ becomes a chiral field and
super-Weyl transformations can be formulated using a chiral
parameter field.

\Subsection{Algebra and Bianchi Identities}
\finalize{\begin{sloppypar}}
The minimal supergravity algebra is determined by the 
three \panti commutators $\acomm{\D_\ia}{\bD_\da}$,
$\acomm{\D_\ia}{\D_\ib}$, $\comm{\D_\ia}{\bD_{\ib\db}}$,
which are listed below with some of their straight-forward
implications 
\finalize{\end{sloppypar}}
\begin{subequations}\label{eq:minsugra}
\begin{align}
  \bigl\{ \D_\ia, \bD_\da \bigr\} &= -2i \D_\ada, \\ 
  \bigl\{ \D_\ia, \D_\ib \bigr\} &= -4 \bR M_{\ia\ib}, \label{eqn:ac-cov-D}\\
  \bigl\{ \bD_\da, \bD_\db \bigr\} &= 4 R \bM_{\da\db},\label{eqn:ac-cov-bD}\\
  \D_\ia \D_\ib &= \half \veps_\ab \D^2 -2 \bR M_\ab, \\
  \bD_\da \bD_\db &= - \half \veps_{\da\db} \bD^2 + 2 R \bM_{\da\db}, \\
  \D_\ia \D^2 &= 4\bR \D^\ib(\veps_\ab + M_\ab ), \\
   \D^2\D_\ia &= -2 \bR \D^\ib(\veps_\ab + M_\ab ), \displaybreak[0]\\
  \bigl[ \D^2, \bD_\da \bigr] &= 
     -4 ( G_\ada + i\D_\ada)\D^\ia + 4 \bR \bD_\da  \\*
     &\relphantom{=} -4 (\D^\ic G^\id{}_\da )M_{\ic\id} + 8 \bW_\da{}^{\dc\dd}\bM_{\dc\dd},\nonumber\\
  \bigl[ \bD^2, \D_\ia \bigr] &= 2i \bigl[\bD^\da, \D_\ada \bigr] + 4i \D_\ada\bD^\da \\*
     &= 
     -4 ( G_\ada - i\D_\ada)\bD^\da + 4 R \D_\ia  
     -4 (\bD^\dc G_\ia{}^\dd{} )\bM_{\dc\dd} + 8 W_\ia{}^{\ic\id}M_{\ic\id},\nonumber\\
  \bigl[ \bD_\da, \D_{\ib\db} \bigr] 
      &= -i \veps_{\da\db} (R\D_\ib + G_\ib{}^\dc \bD_\dc)\\*
      &\relphantom{=}-i (\D_\ib R) \bM_{\da\db} 
        + i \veps_{\da\db} (\bD^\dc G_\ib{}^\dd) \bM_{\dc\dd} 
        - 2i \veps_{\da\db}W_\ib{}^{\ic\id}M_{\ic\id}, \nonumber \\
  \bigl[ \bD^\db, \D_{\ib\db} \bigr] 
      &= -2i (R\D_\ib + G_\ib{}^\dc \bD_\dc)  
        + 2i (\bD^\dc G_\ib{}^\dd) \bM_{\dc\dd} 
        - 4i W_\ib{}^{\ic\id}M_{\ic\id}, \\
  \bigl[ \D_\ia, \D_{\ib\db} \bigr] 
       &=i \veps_\ab (\bR \bD_\db + G^\ic{}_\db \D_\ic)\\*&\relphantom{=}
         +i(\bD_\db \bR) M_\ab - i \veps_\ab (\D^\ic G^\id{}_\db) M_\cd 
         +2i \veps_\ab \bW_\db{}^{\dc\dd} \bM_{\dc\dd}, \nonumber \\
  \bigl[ \D^\ib, \D_{\ib\db} \bigr] 
       &=2i (\bR \bD_\db + G^\ic{}_\db \D_\ic)
         - 2i (\D^\ic G^\id{}_\db) M_\cd 
          +4i \bW_\db{}^{\dc\dd} \bM_{\dc\dd}, \displaybreak[0] \\
  \bigl[ \D^2, \bD^2 \bigr] 
      &= \bigl[\D^2,\bD_\da\bigr] \bD^\da - \bD^\da\bigl[\D^2,\bD_\da\bigr] \\
      &=  8i G_\ada \D^\ada 
         - 4i \D_\ada \bigl[\D^\ia,\bD^\da\bigr]\nonumber\\*&\relphantom{=}
         - 4(\D^\ia R)\D_\ia + 4 (\bD_\da \bR)\bD^\da \nonumber\\*&\relphantom{=}
         - 8 R \D^2 + 8 \bR \bD^2 \nonumber\\*&\relphantom{=}
         - 8(\D^\ic G^\id{}_\da)\bD^\da M_{\ic\id} 
         + 8(\bD^\dc G^{\ia\dd}) \D_\ia \bM_{\dc\dd} \\*&\relphantom{=}
         - 16 W^{\ia\ic\id} \D_\ia M_{\ic\id} 
         + 16 \bW_\da{}^{\dc\dd} \bD^\da \bM_{\dc\dd} \nonumber\\*&\relphantom{=}
         -8 (\D^\ib W_\ib{}^{\ic\id})M_{\ic\id} 
         +8 (\bD_\da \bW^{\da\dc\dd})\bM_{\dc\dd}. \nonumber
\end{align}
\end{subequations}
%
In minimal \acro{SUGRA} $R$ and $W_{\abc}$ are chiral fields, $G_\ada$ is real.
\begin{subequations}
\begin{align}
  G_a &= \bG_a, \\
  \bD_\da R &=0, \\ 
  \bD_\da W_\abc &=0, &  W_\abc &= W_{(\abc)}.
\end{align}
The remaining identities also simplify dramatically,
\begin{align}
  \bD^\da G_\ada &= \D_\ia R,\\     
  \D^\ia G_\ada &=  \bD_\da \bR,\\ 
  \D^\ic W_\abc &= \ihalf \D_\ia{}^\da G_{\ib\da} 
                    + \ihalf \D_\ib{}^\da G_\ada.
\end{align}
\end{subequations}
Some trivial consequences of the above identities are
\begin{align}
  \bD_\da G^\ada &= -\D^\ia R, \\  
  \D_\ia G^\ada &= -\bD^\da \bR, \\
  \D^\ada G_\ada &= \ihalf (\D^2R - \bD^2 \bR),\\
\begin{split}
  (\D^2\la)(\bD^2\bla) 
     &= 4 \,G_\ada (\D^\ia \la)(\bD^\da \bla) 
       + 8 (\D_\ada \la)(\D^\ada \bla) \\&\relphantom{=}
       + \text{(total derivative)}. 
\end{split}\label{eq:candidate-relation}
\end{align}

\Subsection{Chiral Projector and d'Alembertian}
As a consequence of \eqref{eqn:ac-cov-bD} as long as $R\neq0$, 
$\bD^2 U$ is no longer chiral ($U$ being an arbitrary superfield). 
But for tensor superfields carrying no \emph{dotted} indices the
following operator gives a covariantly chiral superfield. 
\begin{align}
  \bD_\da (\bD^2 - 4 R) U_{\ia_1\dots\ia_n} 
     &= 0  \qquad \forall\text{ undotted tensor superfield }U
\end{align}
Evidently the flat space limit, $R\to0$ restores the usual
chirality property of $\bD^2 U$.

Since chiral scalar superfields will play an important \role\ 
in this thesis, the commutators \eqref{eq:minsugra} acting
on chiral scalar fields are worked out explicitly in 
appendix~\ref{ch:chiralalgebra}.
The combination $\bD^2 - 4 R$ is also known as the 
\index{chiral projector}\emph{chiral projector} .

From the chiral projector a generalisation of the 
d'Alembert operator\marginpar{\panti chiral d'Alembertian}\index{d'Alembertian!\panti chiral}
to the space of \panti chiral superfields can be given.
The \panti chiral d'Alembertian $\Box_+$ ($\Box_-$) is defined 
by
\begin{align}
  \Box_+ &:= (\D^a+iG^a) \D_a +\quart (R\D^\ia + (\D^\ia R)) \D_\ia, \\
  \Box_- &:= (\D^a-iG^a) \D_a +\quart (\bR\bD_\da + (\bD_\da \bR)) \bD^\da,
\end{align}
and maps to \panti chiral fields as long as it acts on \panti chiral fields.
In this case $\Box_+$ ($\Box_-$) may be rewritten in the following manner,
\begin{align}
  \Box_+ \la &= \tfrac{1}{16} (\bD^2 - 4 R) \D^2 \la, \\
  \Box_- \bla &= \tfrac{1}{16} (\D^2 - 4 \bR) \bD^2 \bla, 
\end{align}
which makes manifest the (anti-)chirality property.

Also note that $\bD^2\D^2 \la= 16(\Box_+ + \quart R \D^2) \la$.

\Subsection{Super-Weyl Transformations\label{sec:superweyltrafos}}
The condition $T_\ia = 0$ is only invariant under a subset
of the mixed super-Weyl/local $\gr{U}(1)_R$ transformations
discussed in Section~\ref{sec:non-min-weyl}.
To ensure that $0$ maps to $0$ under those transformations, from
\begin{align}
   0 = T_\ia \mapsto L T_\ia + L \D_\ia \ln (L^4\bL^2) = 0,
\end{align}
the condition $\D_\ia \ln (L^4\bL^2) = 0$ is read off. 
Consequently the parameter $L$ is restricted to be of the form
\begin{align}
\begin{split}
  L &= \exp(\half \sigma - \bsigma), 
  \qquad \bD_\da \sigma = \D_\ia \bsigma = 0,\\
  \bL &= \exp(\half \bsigma - \sigma).\\
\end{split}
\end{align}

The minimal supergravity fields transform according to
\begin{align}
  \D_\ia' &= L\D_\ia - 2(\D^\ib L)M_\ab ,\\
  R' &=  -\quart \chiproj \bL^2 ,\\
  G'_\ada &= L\bL G_\ada + \half \bL \D_\ia \bD_\da L - \half L \bD_\da \D_\ia \bL \\
  W'_\abc &= L^2\bL W_\abc,
\end{align}
or in terms of $\sigma$ and $\bsigma$,
\begin{align}
  \D_\ia' &= \e^{\half\sigma-\bsigma} ( \D_\ia - (\D^\ib \sigma) M_\ab) ,\\
  R' &= -\quart \e^{-2\sigma} [\chiproj \e^{\bsigma}], \\
  G'_\ada &= \e^{-(\sigma+\bsigma)/2} \bigl[ G_\ada 
             +\half (\D_\ia \sigma)(\bD_\da \bsigma) 
             + i\bigl( \D_\ada (\bsigma - \sigma )\bigr) \bigr] , \label{eqn:weyl-var-G} \\
  W'_\abc &= \e^{-3\sigma/2} W_\abc .
\end{align}

Formulating\marginpar{chiral compensator}\index{chiral compensator}
$\bT_\da=0$ in terms of prepotentials \eqref{eq:prepoT} yields
the important equation
\begin{align}
  \bE_\da \vphi &= 0, &
  \vphi^{3} &:= E^{-1} \bF^{-2} (1\cdot \e^{\lvec{\bW}})^{-1},\label{eq:def-chi-comp}
\end{align}
where the exponent of ``$3$'' is for convenience as is seen in the 
next equation. Since for any scalar 
$\bD_\da \equiv \bE_\da$, the field $\vphi$ is chiral and transforms
under generalised super-Weyl transformations into
\begin{align}
   \vphi^3 \mapsto [(L\bL)^{-2} E^{-1}] [\bL^{-2} \bF^{-2}] (1\cdot \e^{\lvec{\bW}})^{-1} 
      = L^{-2} \bL^{-4} \vphi^3 = (\e^{\sigma} \vphi)^3. 
\end{align}
This makes $\vphi$ the compensating field for super-Weyl transformations. Accordingly
it is called \emph{chiral compensator}.


\Subsection{Chiral Representation and Integration Rule}
Performing the picture changing operation
\begin{align}
  \tilde{V} &= \e^{-\bW} V, \\
  \tilde{\D}_A &= \e^{-\bW} \D_A \e^{\bW} = \tE_A{}^M \p_M + \half \tilde{\Omega}_A^{bc}M_{bc},
\end{align}
and 
additionally going to the gauge $N_\ia{}^\mu = \delta_\ia{}^\mu$
introduces the so-called \emph{chiral representation}. 
The important feature of the chiral representation is that 
the spinorial vielbein $\tilde{\bE}_\da = -\bF \bp_\da$ takes
a most simple form, while $\tE_\ia$ and complex conjugation 
are more complicated than in the \emph{vector representation} used so far.
The determinant of the vierbein becomes
\begin{align}
   \tE^{-1} = (E^{-1}\e^{-\lvec{W}}),
\end{align}
such that
\begin{align}
  \int d^8z \, \tE^{-1} \tilde{\Lag}
  = \int d^8z \, (E^{-1}\e^{-\lvec{W}}) \e^{-W} \Lag
  \stackrel{\eqref{eq:leftexp}}{=} \int d^8z \, E^{-1} \Lag. 
  \label{eq:realspace-chiralrep}
\end{align}

In chiral representation, equations \eqref{eq:def-chi-comp} and \eqref{eq:prepoR} read
\begin{align}
  \tvphi^3 \bF^2 &=  \tE^{-1}, \\
  \tR &= \quart \bp_{\dot{\mu}} \bp^{\dot{\mu}} \bF^2,
\end{align} 
which combined yield
\begin{align}
  \tvphi^3 \tR &= \quart \bp_{\dot{\mu}} \bp^{\dot{\mu}} \tE^{-1}, \\
  \implies \tvphi^3 \tilde{\Lag}_c 
     &= \quart \bp_{\dot{\mu}} \bp^{\dot{\mu}} \biggl(\frac{\tE^{-1}}{\tR} \tilde{\Lag}_c\biggr).
\end{align} 
This gives the important \emph{chiral integration rule}
\begin{align}
  \int d^6z \,\tvphi^3 \tilde{\Lag}_c &= \int d^8z \,\frac{\tE^{-1}}{\tR} \tilde{\Lag}_c \stackrel{\eqref{eq:realspace-chiralrep}}{=}  \int d^8z \frac{E^{-1}}{R} \Lag , \label{eq:chi-int-rule}
\end{align}
due to $d^2\btheta = \quart  \bp_{\dot{\mu}} \bp^{\dot{\mu}}$.

\Section{Component Expansion}
\Subsection{Superfields and First Order Operators}
In supergravity as opposed to flat supersymmetry, the (non-linearised)
components of a superfield are given in terms of covariant derivatives
$\D_\ia$ and $\bD_\da$ and are in one-to-one correspondence to the
coefficients in the usual $\theta, \bar\theta$ expansion of a
superfield.
\begin{align}
  f \proj &&
  \D_\ia f \proj && 
  \bD_\da f\proj \nonumber\\
  -\quart \D^2 f \proj && 
  -\quart \bD^2 f \proj &&
  \half \bigl[ \D_\ia, \bD_\da\bigr] f \proj \label{eq:comp} \\
  -\quart \D_\ia \bD^2 f \proj && 
  -\quart \bD_\ia \D^2 f \proj &&
  -\tfrac{1}{32} \bigl\{\D^2,\bD^2 \bigr\} f \proj  \nonumber
\end{align}
Here, the notation 
\begin{align}
  f \proj &:= f(x,\theta=0,\btheta=0)
\end{align}
has been introduced.

For arbitrary super\emph{fields} $f_1$ and $f_2$, it holds
\begin{align}
  (f_1 f_2)\proj = f_1\proj f_2\proj,
\end{align}
which obviously can no longer be true when $f_1$ is an \emph{operator} containing 
derivatives on anticommuting coordinates.

The space projection of a general first order differential operator
\begin{align}
  \cO = \cO^M(z) \partial_M + \cO^{ab}(z) M_{ab}
\end{align}
is defined to be
\begin{align}
  \cO\proj = \cO^M \proj \partial_M + \cO^{ab}\proj M_{ab}.
\end{align}

Acting with such an operator on an arbitrary superfield (with Lorentz
indices of $f$ suppressed), one immediately sees that
\begin{align}
  (\cO f)\proj &= (\cO^M \partial_M f)\proj + (\cO^{ab}M_{ab}f)\proj \nonumber \\
               &= \cO^M \proj \partial_M f\proj + \cO^{ab} \proj M_{ab}f\proj\\
               &= (\cO \proj f)\proj \nonumber
\end{align}
is different from
\begin{align}
   \cO\proj f\proj &= \cO^m \proj \partial_m f \proj 
                     + \cO^{ab}\proj M_{ab}f \proj .
\end{align}

Using pure superspace methods, it is possible (though tedious)
to show, that in Wess--Zumino gauge the vector derivative 
has the following expansion,
\begin{align}
  \D_\ada \proj &= \nabla_\ada \proj + \half \Psi_{\ada,}{}^\ib \D_\ib \proj 
    + \half \bPsi_{\ada,\db} \bD^\db \proj,
\end{align}
with $\Psi$ the gaugino field strength. As a simple example, the expansion
of $\D_\ada f$ is given,
\begin{align}
\begin{split}
  (\D_\ada f)\proj &= (\D_\ada\proj f)\proj \\
    &= \nabla_\ada (f\proj) 
    + \half \Psi_{\ada,}{}^\ib \bigl((\D_\ib f) \proj \bigr)
    + \half \bar\Psi_{\ada,\ib} \bigl((\bD^\ib f) \proj\bigr). \label{eq:vecD-expand}
\end{split}
\end{align}
More complicated combination of the derivatives $\D_\ia$, $\bD_\da$ 
and $\D_\ada$ acting on a field require rearrangement such that
the leftmost derivative is of vector type. Then the above rule (with
$f$ containing the remaining derivatives) can be used to recursively 
reduce the superspace derivatives to space-time covariant derivatives $\nabla_\ada$
until only expressions containing component combinations \eqref{eq:comp} 
of the spinorial derivatives are left over.
Due to the three-folding caused by each application of \eqref{eq:vecD-expand}, 
let alone the required rearrangement of vector derivatives to the left,
even terms with a relatively small number of derivatives may grow dramatically.
The situation is (slightly) better when one is not interested in terms containing
the gaugino field strength. Therefore, the operator $\bose$ shall denote
space-time projection while simultaneously neglecting all gravitational 
fermionic and auxiliary fields. 

\Subsection{Supergravity Fields}
The derivation of the component expansion in Wess--Zumino gauge is 
rather involved and only the final expression shall be reproduced 
here. The real part of the prepotential $W$ can be gauged away, but
requiring instead the condition
\begin{align}
  \exp (\bW^n \p_n) x^m = x^m + i \cH^m(x,\theta,\btheta) \qquad
  \cH^m = \bar{\cH}^m 
\end{align}
defines the gravitational Wess--Zumino gauge, also called gravitational
superfield gauge.
In this gauge, the gravitational degrees of freedom are encoded in
the gravitational
superfield $\cH^m$ and the chiral compensator $\hat{\vphi}(x,\theta)$.
\begin{align}
  \cH^m &= 
    \theta \sigma^a \btheta e_a{}^m 
    + i \btheta^2 \theta^\ia \psi^m{}_\ia
    - i \theta^2 \btheta_\da \bPsi^{m\da}
    + \theta^2 \btheta^2 A ^m \nonumber \\
  \hat{\vphi}^3 &= e^{-1} (1-2i\theta\sigma_a\bPsi^a + \theta^2 B) \qquad 
  \hat{\vphi} = \e^{-\bW} \vphi \\
  \hat{\bar{\vphi}}^3 &= e^{-1} (1-2i\btheta\tsigma_a\Psi^a + \btheta^2 \bB) \nonumber
\end{align}

In Wess--Zumino gauge, the spinorial semi-covariant vierbein fields
\eqref{eq:semi-cov-vierbein} coincide with the partial derivatives and
can therefore be used to extract the components of the above
gravitational superfields just as in flat supersymmetry.

The spinorial semi-covariant vierbein fields 
$\hat{E}_\ia$, $\hat{\bE}_\da$ 
were defined by just pulling out a factor of $F$ from the covariant spinorial
derivatives $\D_\ia$, $\bD_\da$. 
In addition without proof, for the prepotential $F$ it holds
\begin{align}
  F \proj &= 1, &
  \hat{E}_\ia F &= -\ihalf \bPsi^{\ia\db}{}^\db,
\end{align}
such that 
\begin{align}
  \D_\ia \cO \proj 
    &= \hat E_\ia \cO \proj, \nonumber\\
  -\quart \D^2 \cO \proj 
    &= - \quart \hat E^2 \cO \proj 
       + \ihalf  \bPsi^\ia{}_{\db,}{}^\db
         \D_\ia \cO \proj .
\end{align}
This allows to write down the chiral compensator's components 
in terms of covariant derivatives
\begin{subequations}
\begin{align}
  \vphi^3 \proj &= e^{-1}, \\
  \D_\ia \vphi^3 \proj &= -2i e^{-1} (\sigma^a \bar \Psi_a)_\ia, \\
  -\quart \D^2 \vphi^3 \proj &= e^{-1} ( B - \bPsi \tsigma \sigma \bPsi ),
\end{align}
\end{subequations}
where $\bPsi \tsigma \sigma \bPsi = - \bPsi^\ia{}_{\db,}{}^{\db} \bPsi_{\ia\dc,}{}^{\dc}$.
In other words
\begin{subequations}
\begin{align}
  \vphi \proj &= e^{-1/3} \\
  \D_\ia \vphi \proj &= -\frac{2}{3} i e^{-1/3} (\sigma^a \bPsi_a)_\ia \\
  -\quart \D^2 \vphi \proj &= \frac{1}{3} e^{-1/3} ( B - \frac{1}{3} \bPsi \tsigma \sigma \bPsi )
\end{align}
\end{subequations}

For the chiral supertorsion component:
\begin{subequations}
\begin{align}
    \bR \proj &= \frac{1}{3} \mathbf{B}, \qquad
    \mathbf{B} = 
       B + \half \bar\Psi^a\tilde\sigma_a\sigma_b\bar\Psi^b 
         + \half \bar\Psi^a \bar\Psi_a, \\
   \bD_\da \bR\proj &= \frac{4}{3} \bPsi_{\da\db,}{}^\db 
       + \frac{i}{3} \mathbf{B} \Psi^\ib{}_{\da,\ib}, \\
\begin{split}
   \bD^2 \bR \proj &= \frac{2}{3} (\cR + \frac{i}{2} \veps^{abcd}\cR_{abcd} ) 
       + \frac{8}{9} \bar{\mathbf{B}}\mathbf{B} \\&\relphantom{=}
       - \frac{2}{9} \mathbf{B} (\Psi^a \sigma_a \tsigma_b \Psi^b + \Psi^a \Psi_a )
           \\&\relphantom{=}
       + i \bD_\da \bR \proj (\tsigma_b \Psi^b)^\da 
       + \frac{2i}{3} \Psi^{\ia\da,\ib} \D_{\{\ia} G_{\ib\},\da} \proj,
\end{split} \label{eqn:scalarcurv}
\end{align} 
\end{subequations}
where $\cR$ denotes the Ricci scalar, tensor or Riemann tensor, respectively.

For the real supertorsion component:
\begin{subequations}
\begin{align}
  G_a \proj &= \frac{4}{3} \mathbf{A}_a,  \\
\begin{split}
  \mathbf{A}_a &= A^a + \frac{1}{8} \veps^{abcd}\mathcal{C}_{bcd} 
       - \frac{1}{4} ( \Psi_a \sigma_b \bPsi^b + \Psi^b \sigma_b \bPsi_a) \\&\relphantom{=}
       - \frac{1}{2} \Psi^b \sigma_a \bPsi_b 
       + \frac{i}{8} \veps^{abcd} \Psi_b \sigma_c \bPsi_d ,
\end{split}\\
  \bD_{\{\da}G^\ib{}_{\db\}} \proj
    &= -2 \Psi_{\da\db,}{}^\ib 
       + \frac{i}{3} \bar{\mathbf{B}} \bPsi^\ib{}_{\{\da,\db\}} ,
       \\
  \bD_{\{\da} \D^{\{\ic}G^{\id\}}{}_{\db\}} \proj 
      &= 2 E^{\ic\id}{}_{\da\db} 
         + 2 i \Psi^{\{\ic}{}_{\{\da,}{}^{\id\}} \bD_{\db\}} \bR \proj 
         - i \Psi_{\ia\{\da,}{}^\ia \D^{\{\ic} G^{\id\}}{}_{\db\}} \proj \nonumber \\&\relphantom{=}
         + 2i \bPsi_{\ia\{\da,\db\}} W^{\ia\ic\id} \proj 
         + \frac{2}{3} \mathbf{B} (\tsigma^{ab})_{\da\db} \Psi_a{}^\ic \Psi_b{}^\id ,
\end{align} 
\end{subequations}
with
\begin{subequations}
\begin{align}
  E_{ab} &:= \quart \bigl( 
     2 \tcR_{ab} + \tfrac{i}{2} (
          \veps_{acde}\cR^{cde}{}_b + 
          \veps_{bcde}\cR^{cde}{}_a 
          -\half \eta_{ab} \veps^{cdef} \cR_{cdef} ) \bigr), \\
  \tcR_{ab} &:= \half (\cR_{ab} + \cR_{ba}) - \quart \eta_{ab} \cR 
             = \cG_{\{ab\}} + \quart \eta_{ab} \cR.
\end{align}
\end{subequations}

\Subsection{Full Superspace Integrals}
Using the chiral integration rule \eqref{eq:chi-int-rule},
any real superspace integral can be reduced to a chiral 
one.
\begin{align}
\begin{split}
  S &= \int d^8z \,E^{-1} \Lag \\
    &= \int d^8z \, \frac{E^{-1}}{R} \underbrace{\bigl(- \quart\bigr) (\bD^2 - 4R) \Lag}_{=: \Lag_c }
\end{split}
\end{align}

Then\marginpar{density formula}\index{density formula} 
the following manipulations, which crucially depend
on the semi-covariant vierbein coinciding \eqref{eq:semi-cov-vierbein} 
with the partial derivatives in Wess--Zumino gauge, lead to the
\emph{density formula} 
\begin{align}
\begin{split}
  S &= \int d^6z \, \hat\vphi^3 \hat\Lag_c 
     = \quart \int d^4x \, \p^\ia \p_\ia ( \hat\vphi^3 \hat\Lag_c ) 
     = -\quart \int d^4x \, \hat E^2 (\vphi^3 \Lag_c) \proj \\
    &= -\quart \int d^4x \, \vphi^3 \proj \hat E^2 \Lag_c \proj 
         + 2 \D^\ia \vphi^3 \proj \D_\ia \Lag_c \proj 
         + \hat E^2 \vphi^3 \proj \Lag_c \proj \\
    &= \int d^4x \, \vphi^3 \proj (-\quart \D^2 \Lag_c ) \proj 
         - \quart \D^\ia \vphi^3 \proj \D_\ia \Lag_c \proj 
         + B \Lag_c \proj.
\end{split}
\end{align}
where $B= \mathbf{B}- \half \bar\Psi^a\tilde\sigma_a\sigma_b\bar\Psi^b - \half \bar\Psi^a \bar\Psi_a = -\quart \D^2 \vphi^3 + e^{-1} \bPsi \tsigma \sigma \bPsi$.


 
  \begin{savequote}[\savequotewidth]
  I adore simple pleasures. They are the last refuge of the complex.
  \qauthor{Oscar Wilde, ``The Picture of Dorian Gray'' }
\end{savequote}

\Chapter{Space-Time Dependent Couplings\label{ch:local-couplings}}

This Chapter is meant to give a short introduction
into the space-time dependent couplings technique and its
application to a proof of Zamolodchikov's $c$\NB-theorem in
two dimensions. Additionally the four dimensional trace
anomaly and some of the problems encountered
when trying to extend the theorem to four dimensions are discussed.

\Section{Weyl Transformations}

\Subsection{Conformal Killing Equation}
A Weyl transformation is a rescaling of the metric by
a space-time dependent factor 
\begin{equation}
  g_{mn} \mapsto \e^{-2\sigma} g_{mn}.
\end{equation}
Upon restriction to flat space these transformations generate the
conformal group, which locally preserves angles.

Using
\begin{align}
\begin{split}
  \delta g_{mn} &= -2\sigma g_{mn}, \\
  \delta x^m &= \xi^m, \label{eq:infinitesimal-gc}\\
  \delta dx^m &= (\p_n \xi^m) dx^n,
\end{split}
\end{align}
the requirement of invariance of the line element
\begin{align}
  \delta(ds^2) \seteq 0 = [-2\sigma g_{mn} + \p_m \xi_n + \p_n \xi_m ] dx^m dx^n 
\end{align}
amounts to the \emph{conformal Killing vector}\marginpar{conformal Killing vector} equation
\begin{equation}
\begin{split}
  \partial_m \xi_n + \partial_n \xi_m &= \tfrac{2}{d} \partial_k \xi^k g_{mn}, \\
  \sigma &= \tfrac{1}{d} \partial_k \xi^k,
\end{split}
\label{EqnConformalKilling}
\end{equation}
where $d$ is the dimension of space time.

Under \eqref{eq:infinitesimal-gc}, the action transforms as follows,
\begin{align}
\begin{split}
  \delta S 
    &= \int d^dx \, \frac{\delta S}{\delta g_{mn}} \delta g_{mn} \\
    &= \int d^dx \, \bigl[ -\half T^{mn}] [ -2 \sigma g_{mn}], 
\end{split}
\end{align}
which demonstrates that for conformal invariance the trace of
the energy-momentum tensor has to vanish.

As an aside, in two dimensions after Wick rotation the conformal Killing vector equation
becomes the Cauchy--Riemann\marginpar{Cauchy--Riemann} system, such that 
conformal transformations are given by 
holomorphic or antiholomorphic functions.
Decomposing these functions by a Laurent expansion demonstrates that the two dimensional
conformal group has infinitely many generators, which form the Witt/Virasoro algebra.

The four dimensional case is generic and will be discussed below.
\Subsection{Conformal Algebra in \texorpdfstring{$d>2$}{d>2}}
In $d>2$ dimensions in Minkowski space, infinitesimal conformal transformations are given
by
\begin{gather}
  \xi^a(x) = a^a + \omega^{ab}x_b + \lambda x^a + ( x^2 b^a - 2 x^a  x_b b^b ) \label{EqnInfinitesimalConformal} \\
\intertext{with the corresponding generators} 
  \delta_C = i a^a P_a + i \omega^{ab}M_{ab} + i \lambda D + i b^a K_a,
\end{gather}
which form the conformal algebra
\begin{equation}
\begin{aligned} {}
[M_{ab}, P_c]
  &= -2 i P_{[a}\eta_{b]c}, &  
[M_{ab}, K_c] 
  &= -2 i K_{[a}\eta_{b]c}, \\
[D,P_a]
  &= -i P_a, &
[D,K_a]
  &= i K_a, \\
[D,M_{ab}] &=0, &
[P_a,K_b] &= 2i(M_{ab}-\eta_{ab}D), \\
[M_{ab}, M_{cd}]
  &= \mrlap{2i\left(\eta_{a[c}M_{d]b}-\eta_{b[c}M_{d]a}\right).}
\end{aligned} \label{EqnConformalAlgebra}
\end{equation}
\begin{table}[t]
\centering
\begin{colortabular}{Cl|>{$}Cl<{$}|>{$}Cc<{$}}
   \multicolumn{3}{Hc}{\textsc{Conformal Transformations}}\\
%
     \multicolumn{1}{c}{Name}& 
     \multicolumn{1}{c}{Group Element} & 
     \multicolumn{1}{c}{Generator} \\ \bwline
  translations       & x^a \mapsto x^a + a^a & P_a  \\ \hline
  (Lorentz\footnotemark) 
  rotations          & x^a \mapsto \Lambda^a{}_b x^b & M_{ab} \\ \hline
  dilation           & x^a \mapsto \lambda x^a & D \\ \hline
  \acro{SCT}\footnotemark   & x^a \mapsto \frac{x^a + b^a x^2}{\Omega_\text{SCR}(x)} & K_a
\\ \bwline
\end{colortabular}
\caption[Conformal Transformations]{\label{tab:conf-trafo}
  Finite Conformal Transformations}
\end{table}%
\addtocounter{footnote}{-1}%
\footnotetext{$\Lambda^c{}_a \eta_{cd} \Lambda^d{}_b = \eta_{ab}$}%
\stepcounter{footnote}%
\footnotetext{Special Conformal Transformation}%
%
This can be identified with the algebra $\mathfrak{so}(d,2)$ 
by defining a suitable $(d+2)\times(d+2)$ matrix
\begin{gather}
  M_{\hat{m}\hat{n}} := 
    \begin{pmatrix}
       M_{mn}                   & \tfrac{1}{2}(K_m - P_m) & \tfrac{1}{2}(K_m + P_m) \\
       -\tfrac{1}{2}(K_m - P_m) & 0                       & -D \\
       -\tfrac{1}{2}(K_m + P_m) & D                       & 0 
    \end{pmatrix}  
\end{gather}
and choosing $\eta_{\hat{m}\hat{n}} = \diag(\eta_{mn},1,-1)$ as metric.
As an aside, the $d$\NB-dimen\-sion\-al conformal algebra is identical
to the $(d+1)$\NB-dimen\-sion\-al $\al{ads}$ algebra
\begin{equation}
  \al{cf}_d \equiv \al{ads}_{d+1} \equiv \al{so}(2,d). \label{EqnCFTequalsADSalgebra}
\end{equation}
The finite transformations\marginpar{finite transformations}
corresponding to the infinitesimal solutions
\eqref{EqnInfinitesimalConformal} are shown in
Figure~\ref{tab:conf-trafo}, where
$\Omega_\text{SCT}(x):=1-\vec{b}\cdot\vec{x}+b^2 \sqvec{x}$ is the scale
factor $\Omega$ of the metric for special conformal transformations,
and $\vec{a}\cdot\vec{b}$ has been used as a short-hand for $\eta_{mn} a^m b^n$.

\Subsection{Weyl Transformations of the Riemann Tensor}
Since superspace supergravity is described using a tangent space
formulation, which has the additional advantage of a metric
$\delta[\eta_{ab}] = 0$ invariant under Weyl transformations, the
transformational behaviour of the Riemann $\cR_{abcd}$ and Weyl
$C_{abcd}$ tensor, Ricci tensor $\cR_{ab}$ and scalar $\cR$, and
covariant derivative $\nabla$ under $\delta[g_{mn}] = -2\sigma g_{mn}$
shall be given in terms of tangent space objects.
\begin{subequations}\label{eq:riemannweyl}
\begin{align}
  \delta[e_a{}^m] &= \sigma e_a{}^m, \\
  \delta[\sqrt{-\det g}] &= \delta[ \det e^{-1} ] = -\sigma d \sqrt{-\det g} = -\sigma d \det e^{-1}, \\
  \delta[\cR^{ab}{}_{cd}] &= \delta{\strut}^{[a}_{[c} \nabla{\strut}^{b]} \nabla{\strut}_{d]}\sigma + 2 \sigma\cR^{ab}{}_{cd}, \\
  \delta[\cR_{abcd}] &= \eta_{[c[a} \nabla_{b]} \nabla_{d]}\sigma + 2\sigma \cR_{abcd}, \\
  \delta[\cR_{ab}] &= \eta_{ab} \nabla^2\sigma + 2\nabla_a\nabla_b \sigma + 2\sigma \cR_{ab}, \\
  \delta[\cR] &= 6\nabla^2\sigma + 2\sigma \cR, \displaybreak[0]\\
\begin{split}
  \delta[\cG_{ab}] &= \delta[\cR_{ab}] - \half \eta_{ab} \delta[\cR] \\
         &= -2 \eta_{ab} \nabla^2 \sigma + 2 \nabla_a\nabla_b \sigma + 2\sigma \cG_{ab},
\end{split} \\
  \delta[C_{abcd}]&= 2\sigma C_{abcd},\displaybreak[0]\\
  \delta[\nabla_a] &= \sigma \nabla_a - (\nabla^b\sigma)M_{ab}, \qquad
  M_{ab} V^c = \delta^c_a V_b - \delta^c_b V_a, \\ 
  \delta[\nabla_a\la] &=\sigma \nabla_a \la, \\
  \delta[\nabla^2\la] &= 2\sigma (\nabla^2\la) + (2-d) (\nabla^a\sigma)(\nabla_a\la), 
\end{align}
\end{subequations}
where $d$ is the space-time dimension, which from now on will be
assumed to be equal to four.

\Subsection{Weyl Covariant Differential Operators\label{sec:riegert-review}}
By definition a field $\psi$ 
is denoted 
\emph{conformally covariant} if it transforms under Weyl transformations
into $\e^{w\sigma} \psi$, that is homogeneously with Weyl weight $w$. 
In particular, it is interesting to have invariant expressions of the
form 
\begin{align}
  \int d^4x \,e^{-1} \chi^* \Delta_{4-2w} \psi,
\end{align}
with $\Delta_{4-2w}$ a differential operator of order $4-2w$ and
$\psi$, $\chi$ are assumed to be Lorentz scalars.

The unique local, Weyl covariant differential operator acting on such
fields $\psi$ and $\chi$ of Weyl weight $1$ is given by
\begin{align}
  \Delta_2 &= \nabla^2 - \tfrac{1}{6} \cR,
\end{align}
which can be easily verified using relations \eqref{eq:riemannweyl}.
It is however entertaining to derive this expression in a slightly different 
manner.
 
General relativity is not invariant under Weyl transformations as can 
be seen from the Einstein--Hilbert action transforming according to
\begin{align}
  \int d^4x\, e^{-1} \cR &\mapsto 
    \int d^4x\, e^{-1} [ 
          \e^{-2\sigma} \cR + 6 (\nabla^a \e^{-\sigma}) (\nabla_a \e^{-\sigma}) 
    ].
\end{align}
Since Weyl transformations form an Abelian group, a parametrisation
may be chosen where two consecutive 
transformations with parameters $\sigma_1$ and
$\sigma_2$ correspond to a single Weyl transformation with parameter $\sigma_1+\sigma_2$.
(Evidently $e_a{}^m \mapsto \e^\sigma e_a{}^m$ is such a parametrisation.)
Replacing the parameter of the first transformation by a field $\phi=e^{-\sigma_1}$ 
of Weyl weight $1$ yields an invariant expression as can be seen from
\begin{align}
  \e^{\sigma_1} e_a{}^m = \phi^{-1} e_a{}^m 
    \mapsto (\e^{-\sigma_2} \phi^{-1}) (\e^{\sigma_2} e_a{}^m) = \phi^{-1} e_a{}^m.
\end{align}
Therefore, the following action is Weyl invariant
\begin{align}
    \int d^4x\, e^{-1} [ \phi^2 \cR + 6 (\nabla^a \phi) (\nabla_a \phi) ]
   &= 6 \int d^4x\, e^{-1} \phi [ \nabla^2 - \tfrac{1}{6} \cR ] \phi 
\end{align}
and the operator $\Delta_2$ has been rederived.  

In \marginpar{compensator} addition the important notion of a
\emph{compensating field}, here $\phi$, has been introduced.
Compensating fields allow incorporating a symmetry into the
formulation of a theory that originally was not part of it.
An analogue procedure is needed to embed \Poincare\ supergravity into
the Weyl invariant supergravity algebra by use of a so-called 
\emph{chiral compensator}.

Unfortunately, the elegant method above does not lend itself to generalisations
and clearly cannot be used to construct a conformally covariant operator for a field of vanishing Weyl weight. 
However a dimensional analysis can be
used to write down a basis for such an operator and determine the 
prefactors from Weyl variation.
The\marginpar{Riegert operator}
following operator due to Riegert \cite{Riegert:1984kt} is the unique conformally covariant
differential operator of fourth order, which because of its importance
for this work will be given in several equivalent forms, 
\begin{align}
\begin{split}
  \Delta_4 &:= 
     \nabla^4 + 2 \cG_{ab} \nabla^a \nabla^b 
     + \tfrac{1}{3} \nabla^a \cR \nabla_a \label{eq:riegert}\\*
  &\hphantom{:}= 
     \nabla^4 + 2 \cG_{ab} \nabla^a \nabla^b 
     + \tfrac{1}{3} (\nabla^a \cR) \nabla_a + \tfrac{1}{3} \cR \nabla^2 \\*
  &\hphantom{:}= 
     \nabla^4 + 2 \cR_{ab} \nabla^a \nabla^b 
     + \tfrac{1}{3} (\nabla^a \cR) \nabla_a - \tfrac{2}{3} \cR \nabla^2 \\*
  &\hphantom{:}= 
     \nabla^4 + 2 \nabla^a \cR_{ab} \nabla^b 
     - \tfrac{2}{3} (\nabla^a \cR) \nabla_a - \tfrac{2}{3} \cR \nabla^2,\\
\end{split}
\end{align}
or partially integrated,
\begin{align}
\begin{split}
  \la'\Delta_4\la &= 
     (\nabla^2\la)(\nabla^2\la')
      - 2 \cG_{ab} (\nabla^a\la)(\nabla^b\la') \\&\hphantom{:=}
      - \tfrac{1}{3} \cR (\nabla^a\la)(\nabla_a\la')
      + \text{(total deriv.)} .
\end{split}
\end{align}

\Section{Zamolodchikov's \texorpdfstring{$c$-Theorem}{c-Theorem} in Two Dimensions}
In a classical theory scale invariance is expected at the ultraviolet
limit where particle masses may be neglected and at the infrared limit
where massive particles decouple from the theory. In this sense the
transition from \acro{UV} to \acro{IR} is irreversible in a classical
theory. For simple theories scale invariance (which implies one 
additional symmetry generator) may be enough to establish 
conformal symmetry (which in two dimensions implies an infinite set
of symmetry generators and is thus a much larger symmetry).
At the quantum level, conformal invariance is often broken. Still
there are many known examples of two dimensional theories which
flow from one conformal fixed point in the \acro{UV} to another
one in the \acro{IR}. In four dimensions the existence of conformal
fixed points is much more difficult to establish. 

The breaking of conformal invariance at the quantum level
is induced by the introduction of a regulator during renormalisation,
which creates a scale $\mu$ that leads to 
%
non-vanishing \emph{anomaly terms} in the trace of the \emt.

Renormalisation group (\acro{RG}) theory describes the change of the
effective Hamiltonian of a theory during the change of scale.
The\marginpar{RG equation}
breaking of scale invariance is described by the \acro{RG} equation
\begin{gather}
  \mu \frac{d}{d\mu}W = \mu \frac{\partial}{\partial\mu}W 
      + \beta^i \frac{\partial}{\partial\lambda^i} W = 0, \\
  \beta^i := \mu \frac{\partial\lambda^i}{\partial\mu}, \\
  W = W(\lambda^i, \mu), 
\end{gather}
where $W$ is the generating functional of the connected Green's functions,
which due to being a formal series expansion of physical observables
is expected to be \acro{RG} invariant, 
that is constant with respect to the scale $\mu$.

\begin{figure}
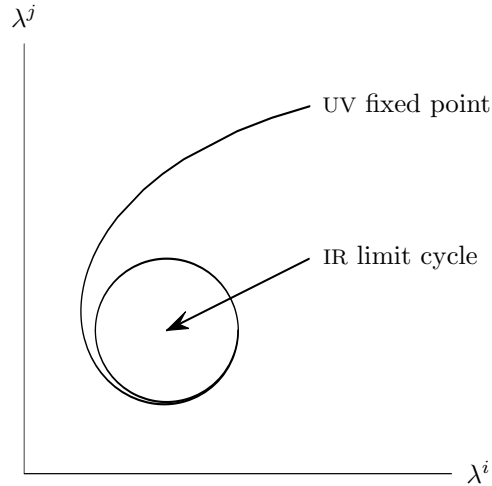

\centering
\PSfraginclude[trim=5 -5 -50 -20]{width=70mm,height=70mm}{limitcycle}
\caption{\label{fig:limit-cycle}Limit Cycle in the Space of Couplings}
\end{figure}
From a mathematical point of view, there is no reason a theory should
not exhibit a complex flow behaviour. In particular the
\acro{RG} flow could approach a limit cycle, see Figure
\ref{fig:limit-cycle}, possibly making the theory increase and
decrease its number of degrees of freedom periodically while going to
lower and lower energies.  Since this is certainly an unphysical
behaviour, a natural question is under which conditions such a
behaviour cannot be displayed by a quantum field theory.


A partial answer to this question was given by Zamolodchikov's fundamental 
theorem~\cite{Zamolodchikov:gt} in two dimensions, which states
the irreversibility of \acro{RG} flows connecting two fixed points 
in two dimensions.

\begin{theorem}[Zamolodochikov 1986]
\pagebreak[0] 
  ``There exists a function $c(g)$ of the coupling constant $g$ in a
  2D renormalisable field theory which decreases monotonically under
  the influence of a renormalisation group transformation.  This
  function has constant values only at fixed points, where $c$ is the
  same as the central charge of a Virasoro algebra of the
  corresponding conformal field theory.''
\end{theorem}

Therefore, it holds
\begin{align}
  c_{UV} \ge c_{IR},
\end{align}
where $c$ is the respective value of central charge at the infrared 
and ultraviolet.

\Section{Conformal Anomaly in Four Dimensions}
Due to its elegance and simplicity, the two-dimensional $c$\NB-theorem
was hoped to soon be generalised to four dimensions, 
but an accepted proof is outstanding for 20 years.

The first obstacle that arises is the question of which quantity
is to take over the \role\ of the two dimensional central charge $c$,
which in two dimensions turns up as the central charge of the 
conformal algebra, as the coefficient of the two point function
of the \emt, and as the anomalous contribution to the trace of
the \emt.

In\marginpar{trace anomaly}
the four dimensional trace anomaly, the following constants appear
\begin{align}
  \vev{ T_m{}^m } = c\, C^2 + a\, \tilde{\cR}^2 + b\, \cR^2 + f \Box \cR, 
\end{align}
where $\cR$ is the scalar curvature (Ricci scalar),
$C^2$ is the square of the Weyl tensor, and $\tilde{\cR}^2$ is the
Euler density,
\begin{align}
  C^2 &:= C_{abcd} C^{abcd} 
       = \cR^{abcd} \cR_{abcd} - 2 \cR^{ab} \cR_{ab} + \tfrac{1}{3} \cR^2, \\
  \tilde{R}^2 &:= \cR^{abcd} \cR_{abcd} - 4 \cR^{ab} \cR_{ab} + \cR^2  .
\end{align}
There are known counter examples for a ``$c$''\NB-theorem in four 
space-time dimensions but that still leaves open the possibility 
of an $a$\NB-theorem \cite{Cardy:1988cw}, which holds in all examples that
permit explicit checking. Since these are supersymmetric theories,
it may well be that supersymmetry is a necessary ingredient for 
the irreversibility of \acro{RG} flows. (As an aside in all known
examples of holographic renormalisation group flows that permit
determination of the anomaly coefficients on both ends of the flow
it holds $c=a$. On the supergravity side monotonicity of the flow
is related to energy conditions as they have to be employed in
causality considerations in Einstein gravity \cite{Freedman:1999gp}.)
Often by an abuse of language the $a$\NB-theorem is also called
$c$\NB-theorem, even though the prefactor of Euler density is
conventionally denoted ``$a$''.

\Section{Local \acro{RG} Equation and the  \texorpdfstring{$c$-Theorem}{c-Theorem}\label{sec:locrg-ctheorem}}
The analysis of this Section will be confined 
to idealised renormalisable field theories that are classically
conformally invariant and involve a set of 
coupling constants $\la^i$ corresponding to local scalar
operators $\cO^i$. 
Due to conformal invariance the coupling constants should have
mass dimension zero such that the operator's mass dimension should be
equal to the space-time dimension. 

When the theory is not conformally invariant on the quantum level
the trace of the \emt\ is non-vanishing and can be expressed in 
terms of some operator basis formed by $\cO^i$
\begin{align}
    \vev{ T_m{}^m } = \beta^i  \vev{ [\cO_i] }, \label{eq:locrg-conv-trace}
\end{align}
where $[\cO_i]$ denotes a (by some renormalisation scheme) well-defined
operator insertion
 and $\beta^i$ are the beta functions associated
to the corresponding couplings $\la^i$. 

When Weyl symmetry is preserved during quantisation, the beta
functions and therefore the trace of the \emt\ vanish.

Promoting\marginpar{operator insertions}
the coupling constants $\la^i$ to fields as
well as the metric,
\begin{align}
   \la^i &\mapsto \la^i(x), \\
   \eta_{mn} &\mapsto g_{mn}(x),
\end{align}
allows to give well-defined expressions for the operators $\cO_i$ 
(the bracket indicating that the operator is well-defined
will be silently dropped, henceforth)
and the \emt,
\begin{align}
  \cO_i(x) &:= \frac{\delta}{\delta\la^i(x)} W, &
  T^{mn}(x) &:= 2\frac{\delta}{\delta g_{mn}(x)} W.
\end{align}
This requires the theory
to be defined for a general curved background metric $g_{mn}$. 
In addition to the counterterms present in the \acro{QFT} on flat space 
with constant couplings, which give rise to the usual running
of couplings, generically there should be now also counterterms $\cA$
depending on the curvature and on $\p_m \la^i$, which vanish in
the limit of constant couplings and metric.
In particular \eqref{eq:locrg-conv-trace} acquires additional 
contributions according to
\begin{align}\label{eq:locrg-trace-anomaly}
  \vev{T_m{}^m} = \beta^i \vev{\cO_i} + \nabla_m \vev{\mathcal{J}^m} + \cA,
\end{align}
with $\mathcal{J}^m$ a local current. In general the trace above is
not a local expression, which is why it was important to introduce
space-time dependent couplings to give a meaning to any products
of finite operators by functional derivatives with respect to 
couplings or the metric. 
The essential assumption is that the \emph{anomaly} $\cA$ stays a 
local expression to all orders, or in other words that the non-local
contribution to the vacuum expectation value of the trace 
is contained in $\vev{ \cO_i }$.

The statement \eqref{eq:locrg-trace-anomaly} can be recast in the
form
\begin{align} \label{eq:locrg-local-rge}
  \Delta^W_\sigma W &= \Delta^\beta_\sigma W 
             - \int d^Dx \sqrt{g} \cA(\sigma, \cR_{abcd}, \p_m \la^i), 
\end{align}
where $W = \ln \int [d\phi] \exp (- S/\hbar) $ is the generating functional
of the connected Green's functions, $\sigma$ is the parameter of 
Weyl transformation generated by $\Delta^W_\sigma$ and
\begin{align}
  \Delta^W_\sigma &:= 2 \int dV \, g^{mn} \frac{\delta}{\delta g^{mn}}, &
  dV &= d^Dx \sqrt{g}, \\
  \Delta^\beta_\sigma &:= \int dV\, \sigma \beta^i \frac{\delta}{\delta\la^i},
\end{align}
with $D$ the number of space-time dimensions.

Equation \eqref{eq:locrg-local-rge} is in effect a local version
of the (anomalous) Callan--Symanzik equation 
\begin{align}
  \biggl[ \mu \frac{\p}{\p\mu} + \beta^i \frac{\p}{\p\la^i} \biggr] W = \cA.
\end{align}
The shape of $\cA(\sigma, \cR_{abcd}, \p_m \la^i)$ is restricted by
power counting and the requirement to vanish in the 
flat space/constant coupling limit, such that in this limit
the local \acro{RG} equation \eqref{eq:locrg-local-rge} reduces 
to the homogeneous Callan--Symanzik equation when imposing the
condition
\begin{align}
  \biggl[ \mu \frac{\p}{\p\mu} 
     + 2 g^{mn} \frac{\delta}{\delta g^{mn}} \biggr] W &= 0,
\end{align}
which is a consequence of \naive\ dimensional analysis.

As a simple example a possible parametrisation of the
ambiguous\footnote{In this formulation the anomaly is of course only
  determined up to partial integrations. Furthermore it is only defined
  up to adding local counterterms to the vacuum energy functional
  $W$.} anomaly in two dimensions is
\begin{align}\label{eq:locrg-2d-anomaly}
  (\Delta^W_\sigma - \Delta^\beta_\sigma )W 
    = \int dV \, \biggl[ \sigma \bigl( \half c \cR + \half \chi_{ij} \p_m \la^i \p^m \la^i \bigr) 
      + (\p_m \sigma) w_i \p^m \la^i \biggr],
\end{align}
with $c$, $\chi_{ij}$ and $w_i$ arbitrary function of the couplings,
which may be determined in a perturbative expansion with the assumption
that the above shape is preserved to all orders, and partial derivatives
 $\p_i := \p_{\la^i}$.

A\marginpar{Wess--Zumino consistency}
further constraint on the anomaly with far less trivial consequences
arises from Weyl transformations being Abelian, which implies
\begin{align}
\comm{\Delta^W_\sigma - \Delta^\beta_\sigma}
     {\Delta^W_{\sigma'} - \Delta^\beta_{\sigma'}} &= 0.
\end{align}
This Wess--Zumino consistency condition renders the 
determination of the trace anomaly an algebraic (cohomological) problem. 

In the case of two dimensions \eqref{eq:locrg-2d-anomaly}
the consistency condition yields
\begin{align}
  \comm{\Delta^W_\sigma - \Delta^\beta_\sigma}
     {\Delta^W_{\sigma'} - \Delta^\beta_{\sigma'}} 
     = \int dV (\sigma' \p_m \sigma - \sigma \p_m \sigma') V^m, \\
  V_m = (\p_m \la^i) ( \p_i ( c + w_j \beta^j) -
          \chi_{ij} \beta^j + ( \p_i w_j - \p_j w_i ) \beta^j )
\end{align}
and therefore the following \emph{coefficient consistency condition} holds
\begin{align} \label{eq:locrg-coeff-consist}
  \beta^i \p_i ( c + w_j \beta^j) = \chi_{ij} \beta^i \beta^j. 
\end{align}
The arbitrariness of $W$ with respect to local functionals
of the fields 
\begin{align}\label{eq:locrg-redef}
  \delta W &= \int dV ( \half b \cR - \half c_{ij} \p_m \la^i \p^m \la^j)
\end{align}
implies for the coefficients
\begin{align} 
  \delta c &= \beta^i \p_i b, &
  \delta \chi_{ij} = \lieD_\beta \chi_{ij} 
    &= \beta^k \p_k c_{ij} + 2 \beta_i \beta^k c_{kj}, \\
  \delta w_i &= - \p_i b + c_{ij} \beta^j, &
  \delta (c + w_j \beta^j) &= c_{ij} \beta^i \beta^j.
\end{align}  
The\marginpar{Zamolodchikov metric}
Zamolodchikov metric $G_{ij}$,
\begin{align}
  G_{ij}(t) &= \frac{1}{8} (x^2)^2 \vev{ \cO_i(x) \cO_j(0)},& 
  t &= \half \ln \mu^2 x^2,
\end{align}
is positive by unitarity (or reflection positivity in Euclidean space).
It can be shown that $G_{ij} = \chi_{ij} + \lieD_\beta c_{ij}$.

Then the function
\begin{align}
  C&:= 3 (c + w_i \beta^i + c_{ij} \beta^i \beta^j ) 
\end{align}
is monotonic by \eqref{eq:locrg-coeff-consist} and positive
definiteness of $G_{ij}$,
\begin{align}
  C' = -\beta^i \p_i C = -3 G_{ij} \beta^i \beta^j < 0.
\end{align}
This is Zamolodchikov's famous $c$\NB-theorem.

Of course there is more to be said about renormalisation scheme
dependence, for details see \cite{Osborn:1991gm}. Here it shall
suffice to mention that equation \eqref{eq:locrg-coeff-consist} is
invariant under \eqref{eq:locrg-redef}.

\Subsection{\texorpdfstring{$a$-Theorem}{a-Theorem}\label{sec:locrg-atheorem}}
The same calculation can be repeated in four space-time dimensions,
giving rise to a \emph{system} of coefficient consistency equations 
\emph{much} more involved than the two dimensional example.
The complete set of anomaly terms and consistency equations
shall not be reproduced here, the interested reader is referred 
to \cite{Osborn:1991gm} instead.

Omitting a number of less interesting terms, a sketch of the 
four dimensional trace anomaly 
is given by
\begin{align}
  [\Delta^W_\sigma - \Delta^\beta_\sigma] W 
     &= \int dV\, \sigma \bigl[ 
          \begin{aligned}[t] &a \,\tilde{\cR}^2 + c \,C^2 + b\, \cR^2 \\ &
          + \half \chi^g_{ij} \,\cG^{mn} \p_m \la^i \p^n \la^j
          + \half \chi^a_{ij} \nabla^2 \la^i \nabla^2 \la^j \\ &
          + \half \chi^b_{ijk} \,\p_m \la^i \,\p^m \la^j \nabla^2 \la^k 
          + \dots \bigr]
          \end{aligned} \\ &\quad
       + \int dV\, \p^m \sigma \bigl[ 
              S_{ij}\, \p_m \la^i \nabla^2 \la^j + \dots
            \bigr], \notag
\end{align}
with $\tilde{\cR^2}$, $C^2$, $\cR^2$, $\cG_{mn}$ the 
Euler density, square of the Weyl tensor and Ricci scalar and the
Einstein tensor, respectively.

The coefficient consistency equation analogue to \eqref{eq:locrg-coeff-consist} 
reads 
\begin{align}
  \beta^i \p_i ( a + \tfrac{1}{8} w_j \beta^j ) 
    &= \tfrac{1}{8} \chi^g_{ij} \beta^i \beta^j .
\end{align}
By virtue of a further consistency equation, 
\begin{align}
  \chi^g_{ij} + 2 \chi^a_{ij} 
     + 2 \p_i \beta^k \chi^a_{kj} + \beta^k \chi^b_{kij} 
     = \lieD_\beta S_{ij}, \label{eq:locrg-chi-a-g}
\end{align}
where $-\chi^a_{ij}$ can be shown to be positive definite, there
might be hope to find a four-dimensional ``a\NB-theorem'', when
getting under control the other coefficients $\chi^b_{kij}$
and $S_{ij}$. 
In the bosonic sector discussed by Osborn, this seems not feasible.
However there might be additional constraints in supersymmetric
theories. This is the topic of the next Chapter.







  \begin{savequote}[\savequotewidth]
  The most exciting phrase to hear in science, the one that heralds
  new discoveries, is not ``Eureka!'' but ``That's funny \dots'' 
  \qauthor{Isaac Asimov}
\end{savequote}

\Chapter{Supersymmetric Trace~Anomaly\label{ch:susy-trace}}

This Chapter generalises the local renormalisation group equation
reviewed in the previous Chapter to a minimal supergravity framework.
A basis for the trace anomaly is found and the consequences of
the Wess--Zumino consistency conditions for super-Weyl transformations
are evaluated.

\Section{\acro{SUSY} Local \acro{RG} Equation\label{sec:susytrace-localrg}}

The (integrated) local Callan--Symanzik (\acro{CS}) equation of the previous Chapter
reads
\begin{align} \label{eq:susytrace-localrge}
\begin{split}
  [\int d^4x \sqrt{-g}\, \sigma(x)&\, 2 g^{mn} \frac{\delta}{\delta g_{mn}}
  + \int d^4x \sqrt{-g}\, \sigma(x) \, \beta^i \frac{\delta}{\delta \la^i(x)} ] W \\
     &=  \int d^4x \sqrt{-g} A(\sigma,\la^i). 
\end{split}
\end{align}
Generically\marginpar{chiral coupling}
the action for a supersymmetric Yang--Mills theory reads
\begin{align}
  S &= \frac{1}{8\pi} \la \int d^6z \tr W^\ia W_\ia + \text{c.c.}, \\
  W^\ia& =-\frac{1}{8} \bar{D}^2 ( \e^{-2V} D_\ia \e^{2V} ),
\end{align}
with $\la$ the coupling constant, which may be complex,
\begin{align}
\la = \frac{4\pi}{g^2} - \frac{i\theta}{2\pi}.
\end{align}
Because
the action is chiral it is natural to promote the complex couplings 
to chiral fields as well.

Coupling to \emph{minimal} supergravity, which is both the simplest
and best explored choice, implies that the Weyl parameter $\sigma(x)$
becomes a chiral field too. Furthermore the supersymmetric
generalisation of the trace of the \emt\ (``supertrace'') is also
chiral and defined by
\begin{align}
  \cT = \vphi \frac{\delta S}{\delta\vphi} .
\end{align}
The supertrace is related to the supercurrent by
\begin{align}
  \bD^\da \cT_\ada = - \frac{2}{3} \D_\ia \cT,
\end{align}
where the supercurrent is defined by
\begin{align}
  \cT_\ada = \frac{\delta S}{\delta \mathbf{H}_\ada},
\end{align}
with $\mathbf{H}_\ada$ corresponding to the gravitational
superfield.\footnote{To be precise, it is the quantum superfield
  associated to the gravitational superfield $H_\ada$ in
  quantum-background splitting.  In Wess--Zumino gauge the lowest
  component of the gravitational superfield $H_\ada$ contains the
  vierbein.}

Accordingly a \acro{SUSY} version of \eqref{eq:susytrace-localrge}
should be given by \cite{Babington:2005vu}
\begin{align}
\biggl[   
  \int d^6z\,\sigma\, \vphi \frac{\delta}{\delta\vphi} -
     \int d^6z\,\sigma\, \beta^i \frac{\delta}{\delta\la^i} 
  + \text{c.c.} 
\biggr] W = A + \text{c.c.},
\end{align}
%
%
where $A$ denotes the anomaly which consists entirely of
terms that contain supergravity fields or depend on a derivative 
of $\la$ or $\bla$,
\begin{align}
  A = \int d^6z \phi^3 \sigma \cA.
\end{align}
Using the differential operators
\begin{align}
  \Delta^W_{\sigma,\bsigma} &:= \Delta^W + \bar\Delta^W, \\
  \Delta^\beta_{\sigma,\bsigma} &:= \Delta^\beta  + \bar\Delta^\beta, \\
  \Delta^W &:= \int d^6z\,\sigma\, \phi \frac{\delta}{\delta\phi},\\
  \Delta^\beta &:= \int d^6z \, \sigma \beta \frac{\delta}{\delta\la},
\end{align}
the \acro{SUSY} local \acro{RG} equation can be recast into
the form
\begin{align}
  (\Delta^W - \Delta^\beta) W &= A + \bar A.
\end{align}
It is convenient to additionally split this local \acro{CS}\marginpar{local CS equation}
equation into a chiral and anti-chiral equation,
\begin{align}
  (\Delta^W - \Delta^\beta ) W &= A, \\
  (\bar\Delta^W - \bar\Delta^\beta ) W &= \bar A, 
\end{align}
which gives rise to the following two Wess--Zumino\marginpar{Wess--Zumino consistency}
consistency conditions,
\begin{align}
  \comm{ \Delta^W_{\sigma} - \Delta^\beta_{\sigma}} 
       { \Delta^W_{\sigma'} - \Delta^\beta_{\sigma'} } W = 0, \\
  \comm{ \Delta^W_{\bsigma} - \Delta^\beta_{\bsigma}}
       { \Delta^W_{\sigma} - \Delta^\beta_{\sigma} } W = 0.
\end{align}
It remains to find a suitable expression for the anomaly $A$.

\Section{Basis for the Trace Anomaly}
In this Section a basis of dimension two operators is constructed
that consists strictly of supergravity superfields (supertorsions) and 
covariant chiral derivatives and furthermore
contains no fields with negative powers.\footnote{Due to the peculiarities 
of curved superspace there is actually a seemingly non-local term namely
$R^{-1}W_{\ia\ib\ic}W^{\ia\ib\ic}$, which is Weyl covariant by itself 
and could be trivially included in the discussion. The expression is related
to the Pontryagin invariant.}

By assumption (see Section~\ref{sec:susytrace-localrg})
the Weyl parameter $\sigma$ and the couplings $\la^i$ are 
chiral scalar fields.

The strategy for finding a basis of dimension two operators is as
follows.
\begin{enumerate}
\item Use the freedom to partially integrate to remove any 
      derivatives on the Weyl parameter $\sigma$.
      The anomaly then has the shape
  \begin{align}
  \Delta^W \Gamma = \int d^8z\, E^{-1} \sigma 
     \cB(\la,\bla) \cdot 
     \cA,
  \end{align}
   with $\cA = \cA(R,\bR,G_\ada,W_\abc,\bW_{\da\db\dc},\D,\bD, \D\la,\bD\bla)$.

\begin{table}
\centering
\begin{colortabular}{Cc*{3}{|Cc}}
\multicolumn{4}{Hc}{Supergravity Fields}\\
\cell{c}{quantity}  & \cell{c}{dimension} & \cell{c}{undotted} & \cell{c}{dotted} \\
\bwline
$R$      & 1   & 0 & 0 \\ \hline
$\bR$    & 1   & 0 & 0 \\ \hline
$\D$     & 1/2 & 1 & 0 \\ \hline
$\bD$    & 1/2 & 0 & 1 \\ \hline
\rlap{$\D_\ada$}\hphantom{$\D$}& 1   & 1 & 1 \\ \hline
$ G$     & 1   & 1 & 1 \\ \hline
$W$      & 3/2 & 3 & 0 \\ \hline
$\bW$    & 3/2 & 0 & 3 \\ \bwline
\end{colortabular}
\caption[Dimensional Analysis for Supergravity Fields]{
  \label{tab:sugrafields}Dimensional Analysis for Supergravity Fields: 
The total dimension of any basis term has to be two, the number of
respective dotted and undotted indices even.
}
\end{table}

\item Expand in derivatives on couplings. Since 
  the overall scaling dimension is supposed to be two, there are at
  most four derivatives and consequently at most four couplings in $\cA$.

  Furthermore since all basis terms for $\cA$ should be scalars, the
  total number of indices should be even (dotted and undotted indices
  respectively). The properties relevant to these simple counting
  arguments are summarised in Table~\ref{tab:sugrafields}.

  The following combinations (bars not yet included) 
  have a chance to yield the right dimension and index structure:
  \begin{gather*}
    2\times R, \quad 2\times G, \quad (1\times R, 2\times \D), \quad
    (1\times G, 2\times \D), \quad 4\times \D.
  \end{gather*}
\end{enumerate}

Taking into account the algebra and Bianchi identities, several
derivatives acting on the same coupling $\la$ 
can be brought to a standard
order. I chose
\begin{gather}
   \D_\ia \la, \quad 
   \D^2 \la, \quad
   \D_\ada \la, \quad
   \D_\ada \D_\ib \la, \quad 
   \D_\ada \D^2 \la, \quad
   \D_\ada \D^\ada \la,
\end{gather}
and accordingly for $\bla$.

In total there arise 38 terms, such that the basis ansatz
for the anomaly reads
\begin{align}
  \cB & \cdot \cA \notag\\
\begin{split}
   &=  b^{(A)} G_\ada G^\ada 
     + b^{(B)} R\bR
     + b^{(C)} R^2 
     + b^{(\bar C)} \bR^2 \\&
     + b^{(D)} (\D^2R) 
     + b^{(\bar D)} (\bD^2\bR) \\&
     + b^{(E)}_i R \D^2 \la^i
     + b^{(\bar E)}_\bi \bR \bD^2 \bla^\bi  \\&
     + b^{(F)}_\bi R \bD^2 \bla^\bi
     + b^{(\bar F)}_i \bR \D^2 \la^i
     + b^{(G)}_i (\D^\ia R)(\D_\ia \la^i) 
     + b^{(\bar G)}_\bi (\bD_\da \bR)(\bD^\da \bla^\bi) \\&
     + b^{(H)}_i G^\ada  \D_\ada \la^i
     + b^{(\bar H)}_\bi G^\ada  \D_\ada \bla^\bi 
     + b^{(I)}_i \D^\ada \D_\ada \la^i  
     + b^{(\bar I)}_\bi \D^\ada \D_\ada \bla^\bi 
\end{split}\notag \displaybreak[0] \\
&
     + b^{(J)}_{ij} R (\D^\ia \la^i)(\D_\ia \la^j) 
     + b^{(\bar J)}_{\bi\bj} \bR (\bD_\da \bla^\bi)(\bD^\da \bla^\bj) \notag \\&
     + b^{(K)}_{ij} \bR (\D^\ia \la^i)(\D_\ia \la^j) 
     + b^{(\bar K)}_{\bi\bj} R (\bD_\da \bla^\bi)(\bD^\da \bla^\bj) \notag\\
\begin{split}&
     + b^{(L)}_{i\bj} G^\ada (\D_\ia \la^i)(\bD_\da \bla^\bj)
     + b^{(M)}_{i\bj} (\D^\ada \la^i) (\D_\ada\bla^\bj) \\&
     + b^{(N)}_{ij} (\D^\ada \la^i) (\D_\ada\la^j)
     + b^{(\bar N)}_{\bi\bj} (\D^\ada \bla^\bj)(\D_\ada\bla^\bj)\\&
     + b^{(O)}_{i\bj} (\D^\ia\la^i)  (\D_\ada \bD^\da \bla^\bj) 
     + b^{(\bar O)}_{\bi j} (\bD^\da\bla^\bi)(\D_\ada \D^\ia \la^j) \\&
     + b^{(P)}_{i\bj} (\D^2\la^i)(\bD^2\bla^\bj) \\&
     + b^{(Q)}_{ij}   (\D^2\la^i)(\D^2\la^j) 
     + b^{(\bar Q)}_{\bi\bj} (\bD^2\bla^\bi)(\bD^2\bla^\bj)
\end{split}\label{eqn:defbase} \displaybreak[0]\\
\begin{split}&
     + b^{(R)}_{ijk}   (\D^\ia \la^i)(\D_\ia\la^j)(\D^2 \la^k)
     + b^{(\bar R)}_{\bi\bj\bk} (\bD_\da \bla^\bi)(\bD^\da\bla^\bj)(\bD^2 \bla^\bk)\\&
     + b^{(S)}_{ij\bk}    (\D^\ia \la^i)(\D_\ia\la^j)(\bD^2 \bla^\bk) 
     + b^{(\bar S)}_{\bi\bj k} (\bD_\da \bla^\bi)(\bD^\da\bla^\bj)(\D^2 \la^k)\\&
     + b^{(T)}_{ij\bk}     (\D_\ada \la^i) (\D^\ia\la^j)(\bD^\da \bla^\bk)
     + b^{(\bar T)}_{\bi\bj k}     (\D_\ada \bla^\bi)(\D^\ia\la^k)(\bD^\da \bla^\bj) \\&
     + b^{(U)}_{ij\bk\bl} (\D^\ia\la^i)(\D_\ia\la^j)(\bD_\db\bla^\bk)(\bD^\db\bla^\bl) \\&
     + b^{(V)}_{ijkl} (\D^\ia\la^i)(\D_\ia\la^j)(\D^\ib\la^k)(\D_\ib\la^l)\\&
     + b^{(\bar V)}_{\bi\bj\bk\bl} (\bD_\da\bla^\bi)(\bD^\da\bla^\bj)(\bD_\db\bla^\bk)(\bD^\db\bla^\bl).
\end{split}\notag
\end{align}
where $b^{(A\dots\bar V)}$ are potentially functions of $\la$ and
$\bla$.\footnote{Note that $b^{(T)}$ and $b^{(\bar T)}$ are the only
  coefficients which potentially can be asymmetric in two indices of
  the same type. As we will see later, the variations \emph{are}
  symmetric, so consistency conditions can only give results for the
  respective symmetric part.}  However, this choice is not minimal as
it still allows for partial integration with respect to $\bD_\da$
because the chiral field $\sigma$ ignores these.  Single derivatives
on $\bla$ cannot be removed by partial integration in general, since a
derivative acting on the coefficient $b$ reproduces the same term
again.

More precisely, due to \marginpar{minimal basis}
\begin{align}
  \int d^8z \, b_\bj (\bD^\da \bla^\bj) \bar X_\da &
     = \int d^8z \, \biggl[  \tilde b_\bj + (\p_\bj \tilde b_\bi) \bla^\bi \biggr] (\bD^\da \bla^\bj) \bar X_\da \notag \\*&
     = - \int d^8z\, \tilde b_\bj \bla^\bj (\bD^\da \bar X_\da),\notag\\
  b_\bj &= \p_\bj (\tilde b_\bi \bla^\bi) \label{coeff:deq}
\end{align}
a basis term with a single derivative on $\bla$ can only be removed
from the tentative basis if a $\tilde b$ obeying \eqref{coeff:deq}
exists; i.e.\ the integrability conditions $\p_\bi b_\bj = \p_\bj
b_\bi$ are fulfilled. This is certainly not true in general, but for
only one coupling or if the theory is invariant under arbitrary
exchange of the coupling constants $\bla^\bi \leftrightarrow
\bla^\bj$, the basis reduces further.

Apart from this complication, removable terms are those which either
have an \emph{outer} $\bD$ derivative (as opposed to one being hidden
behind a $\D^\ia$) or can be brought to that form by using the Bianchi
identities and the supergravity algebra.

The above ``basis'' not being a minimal set of
operators is not really a problem (except for creating a bit of
extra work in the followings), since it will be possible to
consistently set to zero the prefactors to such superfluous
terms belatedly.

\Section{\texorpdfstring{Wess--Zumino}{Wess-Zumino} Consistency Conditions}
It is now time to evaluate the Wess--Zumino consistency conditions
\begin{align}
  \comm{ \Delta^W_{\sigma} - \Delta^\beta_{\sigma}} 
       { \Delta^W_{\sigma'} - \Delta^\beta_{\sigma'} } W = 0, \label{eq:susytrace-deltadelta}\\
  \comm{ \Delta^W_{\bsigma} - \Delta^\beta_{\bsigma}}
       { \Delta^W_{\sigma} - \Delta^\beta_{\sigma} } W = 0.\label{eq:susytrace-deltabardelta}
\end{align}
As shall be seen, all necessary expressions can be 
determined from 
\begin{align} \label{eq:susytrace-basic-delta-square}
( \Delta^W_{\sigma}  - \Delta^\beta_{\sigma} ) ( \Delta^W_{\sigma'} - \Delta^\beta_{\sigma'} ) W,
\end{align}
which requires to calculate the Weyl variation of all 
basis terms as well as to determine the expressions
\begin{gather}
  \Delta^W_{\sigma} (\Delta^W_{\sigma'} - \Delta^\beta_{\sigma'} ) W, 
\qquad
 \Delta^\beta_{\sigma} (\Delta^W_{\sigma'} - \Delta^\beta_{\sigma'} ) W.
\end{gather}
Since the calculation is straight-forward but tedious, the
results have been banned to appendices~\ref{app:weyl-variation}, 
\ref{app:wzcc} and \ref{app:coeffconsist}.

The general structure of \eqref{eq:susytrace-basic-delta-square}
is  
\begin{align}
  ( \Delta^W_{\sigma}  - \Delta^\beta_{\sigma} ) 
  &( \Delta^W_{\sigma'} - \Delta^\beta_{\sigma'} ) W \nonumber\\*&
     = \int d^8z\, E^{-1} \sigma' \bigl\{ 
          \sigma \cF_{0} + (\D^\ia\sigma) \cF_\ia + 
          (\D^2 \sigma) \cF_{2} + (\D^\ada \sigma) \cF_\ada 
        \nonumber \\*&\hphantom{=\int d^8z\, E^{-1} \sigma' \bigl\{ } 
     + (\D_\ada\D^\ia \sigma) \bcF^\da_{3} + (\D_\ada\D^\ada\sigma) \cF_{4} 
       \bigr\}, \label{eq:defF}
\end{align}
where the coefficients $\cF$ can be determined from the intermediate
results in appendix \ref{app:weyl-variation} and are listed in 
appendix \ref{app:wzcc}.

The naming scheme for the anomaly terms has been chosen such that the
calculation of the Weyl consistency conditions only requires
\begin{align}
  \Delta_\sigma \Delta_{\sigma'} W
\end{align}
to be computed by variation. The reader may convince himself that the
other three operator combinations can be determined from the 
following simple set of rules.
\begin{align}
  \Delta_{\sigma'} \Delta_\sigma W       
     &=  ( \Delta_\sigma \Delta_{\sigma'} W )^{\sigma \leftrightarrow \sigma'}; \\[1.5ex]
  \Delta_{\sigma} \bar\Delta_{\bsigma} W 
     &=  ( \Delta_\sigma \Delta_{\sigma'} W )^\tri, \\[1.5ex]
  \begin{split}
      (b^{(x)})^\tri &:= \bar b^{(\bar x)}, \\
      (\sigma')^\tri &:= \bsigma, \\
      (\sigma)^\tri  &:= \sigma, \\
      (\dots)^\tri   &:= (\dots);      
  \end{split}\label{eq:deftri} \\[1.5ex]
  \bar\Delta_{\bsigma} \Delta_{\sigma'} W &= \overline{(\Delta_\sigma \Delta_{\sigma'} W )^\tri}, \label{eq:bardeltadelta} 
\end{align}
where $(\dots)$ denotes anything that is not covered by explicit prior rules.
Note that for the few real terms, it holds $b^{(\bar x)} = b^{(x)}$.

%
So the $\comm{\Delta}{\Delta}$ Wess--Zumino consistency condition \eqref{eq:susytrace-deltadelta}
is 
\begin{multline}
  [ \Delta^W_{\sigma}  - \Delta^\beta_{\sigma},
   \Delta^W_{\sigma'} - \Delta^\beta_{\sigma'} ] W \\
     = \int d^8z\, E^{-1}  (\sigma' \D^\ia \sigma - \sigma \D^\ia \sigma') 
           \bigl\{ 
	      \cF_\ia 
	      - \D_\ia (\cF_2 -\iquart \bD_\da \bcF_3^\da ) \qquad \\
	      + \ihalf \bD^\da ( \cF_\ada - \D_\ada \cF_4 )  
	      + i G_\ada \bcF_3^\da 
	   \bigr\},
\end{multline}
while the $\comm{\Delta}{\bar\Delta}$ Wess--Zumino consistency condition 
\eqref{eq:susytrace-deltabardelta} yields 
\begin{align}
\begin{split}
  \comm{ \Delta^W_{\bsigma}  - \Delta^\beta_{\bsigma} }
       { \Delta^W_{\sigma} - \Delta^\beta_{\sigma} } W& \\
     = \int d^8z\, E^{-1} \biggl[  \sigma \bsigma \,(\text{b}) &+ \sigma(\D^\ada\bsigma)\, (\text{c}) + (\D_\ada \sigma)(\D^\ada\bsigma)\, (\text{d}) \biggr] ,
\end{split}
\end{align}
with $(\text{b})$, $(\text{c})$ and $(\text{d})$ the respective left hand sides of
\begin{center}
\fbox{\fbox{
\begin{minipage}{0.9\textwidth}
\begin{subequations}
\begin{align}
\begin{split}
              &\cF_\ia 
	      - \D_\ia (\cF_2 -\iquart \bD_\da \bcF_3^\da )  \\&\qquad
	      + \ihalf \bD^\da ( \cF_\ada - \D_\ada \cF_4 )
	      + i G_\ada \bcF_3^\da = 0,
\end{split} \label{eq:ss-wz}\\[2ex]
          \bigl\{ &
	       \cF_{0}
	       - (\D^\ia \cF_\ia) 
               + (\D^2 \cF_2)  
               - \half \D^\ada ( \cF_\ada -\D_\ada \cF_4 - \D_\ia \bcF_{3\,\da}) \\* &\qquad
               - 2i (\bD_\da \bR) \bcF_3^\da 
               - 2i G^\ada (\D_\ia \bcF_{3\,\da}) 
	       \bigr\}^\tri - c.c. = 0 ,
               \label{eq:sds-wz1} \\[2ex]
               \bigl\{ &
               \cF_\ada 
	       - \D_\ada \cF_4 
	       - \D_\ia \bcF_{3\,\da} \bigr\}^\tri + c.c. =0, \label{eq:sds-wz2} \\[2ex]
%
%
	       &\cF_4^\tri = \bcF_4^\tri, \label{eq:sds-wz3}
\end{align}
\end{subequations}
\vspace*{0.1ex}
\end{minipage}
}}\end{center}
which constitute the full set of consistency conditions on 
the level of abbreviations $\cF$.
The complex conjugate of \eqref{eq:ss-wz} is an additional part of this system. 

These coefficient consistency equations are the main result of this
Part. Unfortunately expanded out they fill about three pages and have been 
put into Appendix \ref{app:coeffconsist}, therefore.

\Section{Local Counterterms}
The vacuum energy functional $W$ is only determined up to the addition of
local counter terms $\delta W$, a convenient choice for which is provided by
the basis used for the anomaly, since it allows to reuse the results
from the Wess--Zumino consistency condition:
\begin{align}
  W &\equiv W + \delta W, \\
  \delta W &= \int d^8z \, E^{-1}\delta\cB \cdot \cA,
\end{align}
with $\delta\cB \cdot \cA$ analogous to \eqref{eqn:defbase}.  To
fulfil the reality requirement $\delta W = \overline{\delta W}$, it
is necessary (and sufficient) to choose the coefficients $\delta b$
from $\delta\cB$ according to $\delta \bb^{(x)} = \delta b^{(\bar x)}$
for any $x$.\footnote{In particular for coefficients of the single, real
  terms $(A)$, $(B)$, $(L)$, $(M)$, $(P)$, $(U)$, this amounts to
  taking $b^{(x)}=\bb^{(x)}$.}

Realising that
\begin{align}
  \Delta_\sigma W &= \int d^8z\, E^{-1} \sigma\, \cB \cdot \cA , \\
  \implies \quad \delta W 
    &= \Delta_{\sigma'} W \raisebox{-2ex}{$\Biggr|$}
       {\vphantom{\Biggr|}}_{\begin{subarray}{l}
          \sigma'\mapsto 1\\ 
          b^{(x)} \mapsto \delta b^{(x)} \\
          b^{(\bar x)} \mapsto \delta \bb^{(x)}
        \end{subarray}} =:  \Delta_{\sigma'} W \bigr|_\delta \;,
\end{align}
the effect of adding the local counter terms $\delta W$ to the generating
functional $W$ is seen to be
\begin{align}
  \Delta_\sigma (W + \delta W) 
     &= \Delta_\sigma (W + \Delta_{\sigma'} W \bigr|_\delta) \\
     &= \int d^8z \, E^{-1}\sigma \, \cB \cdot \cA \nonumber\\*&\quad + \int d^8z\, E^{-1} \sigma 
        \begin{aligned}[t]& \bigl\{
          \cF_{0} - \D^\ia \cF_\ia + 
          \D^2 \cF_{2} - \D^\ada \cF_\ada \\&\qquad
       +  \D^\ia \D_\ada \bcF^\da_{3} + \D_\ada\D^\ada \cF_{4} 
        \bigr\} \bigr|_\delta \;,
        \end{aligned}
\end{align}
where in the last line equation \eqref{eq:defF} has been used.

In other words, the addition of local counter terms corresponds to the
mapping
\begin{align}
  \cB \cdot \cA \mapsto \cB \cdot \cA + \begin{aligned}[t]& \bigl\{
          \cF_{0} - \D^\ia \cF_\ia + 
          \D^2 \cF_{2} - \D^\ada \cF_\ada \\&\qquad
       +  \D^\ia \D_\ada \bcF^\da_{3} + \D_\ada\D^\ada \cF_{4} 
        \bigr\} \bigr|_\delta \;.
        \end{aligned} 
\end{align}
%

\Section{S\NB-duality}
$\cN=4$ \acro{SYM} is invariant under an $\gr{SL}(2,\mathds{R})$ symmetry
that is preserved on the quantum level. Explicit calculations indicate
the symmetry is also maintained to one loop during coupling to 
gravity. Assuming that this is true to all orders, 
one might restrict the discussion of anomaly terms
to superfield expressions that are manifestly invariant under that symmetry
for the discussion of an $\cN=4$ fixed point.

The theory of modular forms easily fills an entire book \cite{Apostol:1997}, 
but the consideration here shall be restricted to $\gr{SL}(2,\mathds{R})$
invariant terms that can be build
from the basis of anomaly terms \eqref{eqn:defbase}.

In terms of the complex coupling 
$\la := \frac{4\pi}{g^2} - \frac{i\theta}{2\pi}$,
the $\gr{SL}(2,\mathds{R})$ symmetry is generated by the two transformations
\begin{align}
  \la &\mapsto \frac{1}{\la}, &
  \la &\mapsto \la + i,
\end{align}
which have this unusual form due to employing the convention of taking
the coupling constant $g^{-2}$ as the real part of $\la$.

It follows immediately that for coefficient functions $b(\la,\bla)$ in the
a\-nom\-a\-ly it holds $b=b(\la+\bla)$.

In addition one observes
\begin{align}
  \frac{1}{\la+\bla} & \mapsto \la\bla \frac{1}{\la+\bla}, \\
  \D_\ia\la          & \mapsto -\frac{1}{\la^2} \D_\ia \la, \\
  \bD_\da\bla        & \mapsto -\frac{1}{\bla^2} \bD_\da\bla, \\
  \sdualD^2\la       & \mapsto -\frac{1}{\la^2} \sdualD^2 \la, \\
  \bar\sdualD^2\bla  & \mapsto -\frac{1}{\bla^2} \bar\sdualD^2 \bla,\\
  \sdualD_\ada \D^\ia \la &\mapsto - \frac{1}{\la^2} \sdualD_\ada \D^\ia \la, 
\end{align}
where 
\begin{align}
  \sdualD^2\la  := \D^2 \la - \frac{2}{\la+\bla} (\D^\ia\la)(\D_\ia\la), \\
  \bar\sdualD^2\bla := \overline{\sdualD^2\la} = \bD^2 \bla - \frac{2}{\la+\bla} (\bD_\da\bla)(\bD^\da\bla),  
\end{align}

Therefore S-invariant expressions are given by
\begin{align}
  \frac{1}{(\la+\bla)^2} &(\sdualD^2 \la)(\bar\sdualD^2\bla), &&\sim (P),(S),(\bar S),(U)  \label{sdual1} \displaybreak[0]\\
  \frac{1}{(\la+\bla)^2} &(\D^\ia \la)(\D_\ia \bar\sdualD^2\bla), &&\sim (L),(O),(U),(\bar T)\displaybreak[0]\\
  \frac{1}{(\la+\bla)^2} &(\bD_\da \sdualD^2\la)(\bD^\da \bla), && \label{sdual2}\displaybreak[0]\\
  \frac{1}{(\la+\bla)^2} &(\D_\ada \la)(\D^\ada \bla), &&\sim (M)\displaybreak[0] \\ 
  \frac{1}{(\la+\bla)^2} &G^\ada (\D_\ia\la)(\bD_\da \bla), &&\sim (L)\displaybreak[0] \\ 
  \frac{1}{(\la+\bla)^4} &(\D^\ia\la)(\D_\ia\la)(\bD_\da\bla)(\bD^\da\bla) &&\sim (U)
\end{align}
and moreover the $\la,\bla$ independent terms $(A)$ to $(\bar D)$. 

\Section{Towards a Proof\label{sec:towards}}

For the proof of Zamolodchikov's theorem in two dimensions,
the crucial ingredient is the connection of the anomaly coefficients
to correlation functions from which the positive definite
Zamolodchikov metric was defined, see Sections~\ref{sec:locrg-ctheorem}
and \ref{sec:locrg-atheorem} in particular.

As an example of how this procedure works the consistency condition
\eqref{eq:example-wz} from the appendix shall be discussed,
\begin{align*}
     -\tfrac{i}{2} b^{(M)}_{j\bk} 
     + \beta^i b^{(T)}_{ji\bk} 
     + i b^{(L)}_{j\bk} 
     + \ihalf b^{(N)}_{ij} (\p_\bk \beta^i)   
     + \ihalf \beta^i (\p_\bk b^{(N)}_{ij}) 
     - b^{(T)}_{ij\bk} \beta^i = 0.
\end{align*}
$b^{(T)}_{ij\bk}$ is the only coefficient function that is not 
(anti\NB-)symmetric in indices of the same kind. From the expression above
it can however be 
projected out by multiplying with $\beta^j$, which leaves
\begin{align}
     \beta^j \bigl[
     b^{(M)}_{j\bk} 
     -2 b^{(L)}_{j\bk} 
     - \p_\bk (\beta^i b^{(N)}_{ij}) \bigr] = 0, \label{eq:susytrace-coef-consist}
\end{align}
In fact $b^{(N)}_{ij}$ vanishes identically as 
a consequence of the \acro{RG}
equation, which for the anomaly restricted to that coefficient reads 
\begin{align}
  \mu \frac{\p}{\p\mu} W + \beta^i \p_i W 
    &= b^{(N)}_{ij} (\D^2 \la^i )(\D^2 \la^j).     
\end{align}
Acting on it with $\frac{\delta}{\delta\la^k}\frac{\delta}{\delta\la^l}$,
gives
\begin{align}
  \mu \frac{\p}{\p\mu} \vev{\cO^k \cO^l} + \beta^i \p_i \vev{ \cO^k \cO^l }
    &= b^{(N)}_{kl} (\D^2 \delta^6(z) )(\D^2 \delta^6(z')),
\end{align}
where the left-hand side vanishes by non-renormalisation of
chiral correlation functions.
It immediately follows that $b^{(N)}_{ij} \equiv0$, which means
that equation \eqref{eq:susytrace-coef-consist} implies
\begin{align}
    \boxed{\beta^j \bbeta^\bk \bigl[ b^{(M)}_{j\bk} 
     -2 b^{(L)}_{j\bk} \bigr] = 0.} \label{locrg:quadforms}
\end{align}
This is the supersymmetric version of equation 
\eqref{eq:locrg-chi-a-g}, which reads
\begin{align*}
   \chi^g_{ij} - 2 \chi^a_{ij} =
     \lieD_\beta S_{ij} 
     - 2 \p_i \beta^k \chi^a_{kj} - \beta^k \chi^b_{kij},
\end{align*}
though from \eqref{locrg:quadforms} the right hand side
is zero when taking into account
\begin{align}
  b^{(M)} &\sim \chi^{g} - \chi^{a}, &
  b^{(L)} &\sim \chi^{g}, 
\end{align}
as will be seen from the component expansions \eqref{eq:susytrace-comp1}--\eqref{eq:susytrace-comp3}
of the next Section.
This is just as required
for a proof of the $a$-theorem, since 
$\chi^{(a)}$ can be shown to be positive definite in 
a particular scheme.
In that scheme, 
\begin{align}
  - \widehat{\chi^{a}} = \frac{x^8 S_4}{192} \vev{ \cO_i(x) \cO_j(0) },
\end{align}
where the right hand side is positive definite by unitarity. 
The set of counterterms which are needed to change to a scheme 
where $\chi^{a} = \widehat{\chi^{a}}$ were determined 
in \cite{hoehne-diplomarbeit}.

Of course the other anomaly terms might contribute 
further terms to the
simple identification between $b^{(M)}$, $b^{(L)}$ and 
$\chi^a$, $\chi^g$,
thus spoiling the success.
Actually from the whole basis for the anomaly, there is
only one term which could do so, namely $(\D^2\la)(\bD^2\bla)$,
which seems harmless since its component expansion yields only
$(\nabla^2 \la)(\nabla^2 \la^*)$.
Moreover it is expected to conspire with the $(M)$ and $(L)$
terms from the anomaly basis to form a supersymmetric version
of the ``Riegert operator'' as shall be explained now. 

\Section{Superfield Riegert Operator}
For $\cN=4$ Yang--Mills theory \cite{Osborn:2003vk} obtains a one-loop
trace anomaly that contains the operator
\begin{align}
  \frac{1}{(\la+\la^{\smash*})^2} \left(  \nabla^2 \la \nabla^2 \la^* - 2 \cG^{mn} \nabla_m \la \nabla_n \la^* - \tfrac{1}{3} \cR \nabla^m \la \nabla_m \la^* \right),
\end{align}
which basically is the Riegert operator \eqref{eq:riegert}.\footnote{The factor
  in front plus some further terms are required to make the operator
  $\gr{SL}(2,\mathds{R})$ invariant in addition.} Note that the 
bosonic Riegert operator is a direct consequence of the (bosonic) consistency
conditions for the $\cN=4$ case. It is therefore important to reproduce
the Riegert operator in the component expansion of the superfield formulation
employed here.

This result indicates an inconsistency with our result because there
does not seems to exist a superfield expression that generates this
Riegert operator in a component expansion. Therefore it cannot be
generated as part of the derived superfield trace anomaly.


Strange enough in components\marginpar{component version} a super-Weyl covariant version of this operator
is known such that the following expression \cite{Fradkin:1985am}
is invariant under super-Weyl transformations,
\begin{align}
\begin{split}
  \Lag &= e^{-1} \nabla^2 \phi^* \nabla^2 \phi - 2 (\cR_{mn} - \tfrac{1}{3} g_{mn} \cR) \nabla_m \phi^* \nabla_n \phi \\
       &\relphantom{=} - \half \bar{\chi} [ \not{\!\!D}^3 + (\cR_{mn} - \tfrac{1}{6} g_{mn} \cR ) \gamma_m D_n ] \chi \\
       &\relphantom{=} - \tfrac{3}{4} \bar\chi \gamma_m D_n \chi F_{mn} + F^* [ D^2 - \frac{1}{6} (R-\bpsi_m \cR_m )] F\\
       &\relphantom{=} + \text{(gravitino terms)} ,
\end{split}
\end{align}
with
\begin{align}
  D_m \chi &= \nabla_m \chi + \tfrac{3i}{4} \gamma_5 A_m \chi , \qquad
  D_m = (\p_m + \tfrac{3i}{2} A_m) F,
\end{align}
and $\phi,\psi,F$ the components of a chiral field of Weyl weight $0$.

Therefore one should expect a superfield version $\Delta_R^4$
of  this operator to exist such that 
\begin{align}
 \delta_\text{Weyl} \bigl[ \int d^8z E^{-1} \la \Delta_R^4 \bla \bigr] = 0,
\end{align}
with $\delta_\text{Weyl}$ indicating a super-Weyl transformation.

On\marginpar{component expansion}
the other hand one might simply use a component expansion
of all basis terms and determine the linear combination that
yields the bosonic Riegert operator \eqref{eq:riegert}
as its lowest component.

Such a component expansion can be quite involved, but fortunately
there is only a limit number of terms that can contribute.
Here the discussion shall be restricted to 
a few natural candidate terms which already produces
some interesting results.
\begin{align}
   \D^2 & \chiproj (\D^2 \la)(\bD^2\bla) \bose 
       = 256 (\nabla^2 \la)(\nabla^2 \la^*), \label{eq:susytrace-comp1} \\
   \D^2 & \chiproj  G^\ada (\D_\ia \la)(\bD_\da \bla) \bose  \label{eq:susytrace-comp2} \\
       &= 64 (\cG_{(\mu\nu)} + \quart g_{\mu\nu} \cR)
             (\nabla^\mu \la)(\nabla^\nu \la^*) 
         -\tfrac{16}{3} \cR (\nabla^\mu\la)(\nabla_\mu\la^*) 
        + \text{(imag.)}, \notag\\
   \D^2 & \chiproj (\D_\ada \la)(\D^\ada \bla) \bose \notag \\
        &= \tfrac{32}{3} \cR g_{\mu\nu} (\nabla^\mu\la)(\nabla^\nu\la^*) 
          -32 (\nabla_\mu \la)(\nabla^2 \nabla^\mu \la^*) + \text{(imag.)} \label{eq:susytrace-comp3} \\
       &= \tfrac{32}{3} \cR g_{\mu\nu} (\nabla^\mu\la)(\nabla^\nu\la^*) 
             - 32 \cR_{\mu\nu} (\nabla^\mu \la)(\nabla^\nu \la^*) 
          \nonumber\\&\relphantom{=} 
             + 32 (\nabla^2 \la)(\nabla^2 \la^*) + \text{(total deriv.)},  \nonumber
\end{align}
where the following relations have been used,
\begin{align}
  \bigl[ \nabla_\mu, \nabla_\nu \bigr] V^\rho 
     &= \cR^\rho{}_{\sigma\mu\nu}V^\sigma, \\
\begin{split}
  \nabla^2 \nabla_\mu V
     &= \nabla_\mu \nabla^2 V + \cR_{\nu\mu} \nabla^\nu V. 
\end{split}
\end{align}

First of all one should note that \eqref{eq:susytrace-comp1} can be
expressed by a linear combination of \eqref{eq:susytrace-comp2}
and \eqref{eq:susytrace-comp3} and a total derivative, which
is just the component version of \eqref{eq:candidate-relation},
\begin{align}
\begin{split}
  (\D^2\la)(\bD^2\bla) 
     &= 4 \,G_\ada (\D^\ia \la)(\bD^\da \bla) 
       + 8 (\D_\ada \la)(\D^\ada \bla) \\&\relphantom{=}
       + \text{(total derivative)}. 
\end{split} \label{eq:susytrace-candidate-relation}
\end{align}
This relation being preserved in the component expansion is a strong
indication for equations
\eqref{eq:susytrace-comp1}--\eqref{eq:susytrace-comp3} to be correct.

Up to this identity the only combination of the candidate terms
\eqref{eq:susytrace-comp1}--\eqref{eq:susytrace-comp3} that yields the
bosonic Riegert operator as its lowest component is
\begin{align}
  (\D^2\la)(\bD^2\bla) - 8 \,G_\ada (\D^\ia \la)(\bD^\da \bla).
\end{align}
This combination is not super-Weyl covariant however and
it turns out that for the anomaly basis \eqref{eqn:defbase},
there is no non-trivial super-Weyl invariant expression
that includes $(\D^2\la)(\bD^2\bla)$---or $(\D_\ada \la)(\D^\ada \bla)$
by \eqref{eq:susytrace-candidate-relation}.
\emph{In other words, there is no superfield version of the Riegert
operator for chiral fields of Weyl weight $0$}.
This is rather puzzling since the component version \emph{does} exist.
What may have gone wrong?

\Section{Discussion}
Equation \eqref{eq:susytrace-candidate-relation}
provides a rather non-trivial consistency check for the
component expansion and the Weyl variations are simple to check.
One should therefore be confident that the result of the 
previous Section is correct.

Since\marginpar{minimal \acro{SUGRA}}
the Weyl parameter in minimal supergravity is a chiral field,
it naturally also encodes superlocal $\gr{U}(1)_R$ transformations.
So perhaps one is simply requiring too much symmetry. 
Since the expressions are global $\gr{U}(1)_R$ invariant anyway,
neglecting the local symmetry corresponds to allowing terms
that contain derivatives acting on $\sigma-\bsigma$. 
Due to $\D_\ia (\sigma - \bsigma)= \D_\ia(\sigma+\bsigma)$ 
this cannot be distinguished from super-Weyl transformations. 

In\marginpar{non-minimal \acro{SUGRA}} 
non-minimal supergravity it is possible to not require
invariance under local $\gr{U}(1)_R$, and a possible 
super-Weyl covariant operator (in the conventions of \cite{Gates:1983nr})
is given by
\begin{align}
  (\mathbb{D}_\ada \la)(\mathbb{D}^\ada \bla)
\end{align}
with the Weyl covariant vector derivative for scalar chiral yields
of $\gr{U}(1)_R$ charge $y$ given by
\begin{align}
  \mathbb{D}_\ada &:= i (\bar\nabla_\da - i (\tfrac{2}{3}+y) \bar\Gamma_\da ) 
         ( \nabla_\ia + iy \Gamma_\ia ),\\
  \delta [ \mathbb{D}_\ada \la ] \!&\hphantom{:}= L \mathbb{D}_\ada \la.
\end{align}

In \marginpar{new-minimal \acro{SUGRA}}
new-minimal supergravity the $\gr{U}(1)_R$ drops from
the formulation and it is possible to give a superfield
Riegert operator for linear superfields of Weyl-weight 0 
that is covariant under the full
invariance group of the supergravity algebra \cite{Manvelyan:1995hz}
\begin{align}
  \mathbf{D}_\ada \mathbf{D}^\ada + \frac{i}{3} (\D_\ia \bT_\da + \bD_\da T_\ia) \mathbf{D}^\ada,
\end{align}
where $\mathbf{D}_\ada = \D_\ada - \frac{i}{12} (T_\ia \bD_\da + \bT_\da \D_\ia)$
is a super-Weyl covariant derivative.

The difficulties to formulate fields of arbitrary Weyl and $\gr{U}(1)_R$ 
weight in a superconformal framework are long known (see for example 
\cite{Kugo:1982cu}) and led to the introduction of a chiral compensating field.
This can be most easily illustrated taking a chiral field $\la$ as an example.
It clearly should transform under generalised super-Weyl transformations
according to 
\begin{align} 
  \la \mapsto \e^{n_+\sigma + n_-\bsigma} \la,
\end{align}
with $n_+$ a real number and $n_-=0$ in order to stay a chiral field. 
In other words the type of the field dictates a fixed relation between
its $\gr{U}(1)_R$ charge and its Weyl weight.
Therefore a single field transforming as $\Phi \mapsto \e^\sigma \Phi$
can be used to bring all other fields to a fixed Weyl and $\gr{U}(1)$ 
weight, by redefinitions of the type $\tilde{\la} = \Phi^{-n_+} \la$
for example.

A\marginpar{Weyl invariant algebra}
suitable set of invariant supergravity fields is given by
\begin{align} \label{eq:weyl-invariant-algebra}
  \DD_\ia &= \UU \D_\ia - 2 (\D^\ib \UU )M_{\ia\ib}, &
  \UU &= [\Psi^{n+1} \bar{\Psi}^{n-1}]^{-\frac{3n+1}{8n}}, \notag \\
  \bDD_\da &= \bUU \bD_\da -2 (\bD^\db \bUU) \bM_{\da\db}, \notag \\
  \DD_\ada &= \ihalf \acomm{ \DD_\ia }{ \bDD_\da }, \notag \\
  \TT_\ia &= \DD_\ia \TT , &
  \TT &= \ln \UU^4 \bUU^2,\\
  \RR &= -\frac{1}{4} ( \bD^2- 4 R) \bUU^2, &
  \WW_{\ia\ib\ic} &= \bUU^2 \UU W_{\ia\ib\ic},  \notag \\
  \GG_\ada &= \mrlap{\bUU\UU G_\ada + \half (\bDD_\da \ln \UU)(\DD_\ia \ln \UU)} \notag\\
              &\relphantom{=} \mrlap{
              + \quart \bDD_\da \DD_\ia \ln (\UU^2 \bUU^{-1})
              - \quart \DD_\ia \bDD_\da \ln (\bUU^2 \UU^{-1}), 
              } \notag
\end{align}
where $\Psi$ is a linear conformal compensator which transforms
under Weyl transformation $\vphi\mapsto \e^\sigma \vphi$ 
according to 
\begin{align}
  \Psi \mapsto \Psi' 
    &= \exp \biggl[\frac{3n-1}{3n+1} \sigma - \bsigma \biggr] \Psi,&
   \bD_\da \sigma &= 0.
\end{align}
The case $n=\frac{1}{3}$ corresponds to minimal supergravity and 
the compensator $\Phi := \bar{\Psi}$ is a chiral field.

It should be remarked that the expressions \eqref{eq:weyl-invariant-algebra}
can be easily obtained
by replacing $\sigma$ and $\bsigma$ in the Weyl transformed
objects by $-\ln \Phi$ and $-\ln \bar{\Phi}$ in a similar
way as in the bosonic case in Section~\ref{sec:riegert-review}.

One\marginpar{which compensator}
might think of taking the already known chiral compensator $\vphi^{-1}$
as the compensator $\Phi$ in \eqref{eq:weyl-invariant-algebra}.
However this use of the chiral compensator $\vphi$, which is also
a prepotential that transforms under the $\Lambda$ supergroup, 
would break invariance under that symmetry. Another interesting
possibility is the use of 
\begin{align}
  \Omega = 1 + \int d^8z' E^{-1}(z') G_{+-}(z,z'),
\end{align}
where $G_{+-}$ is the Feynman superpropagator defined by
\begin{align}
  \quart ( \D^2 - 4 \bR)_z G_{+-} (z, z') = \delta^6(z,z')
\end{align}
and $\delta^6(z,z')$ is the chiral delta distribution.

A simple consequence of the defining relation is
\begin{align}
  \bD_\da \Omega &= 0, &
  (\D^2 - 4 \bR)\Omega &= 0,
\end{align}
which implies $\Omega\mapsto \e^{-\sigma} \Omega$
under super-Weyl transformation and $\Omega$ is a suitable (though non-local)
compensator. 
For superconformal backgrounds $\Omega$ actually becomes local
and take the form
\begin{align}
  \Omega = \vphi^{-1} + O(\mathscr{H}).
\end{align}

With such a compensator the expression\marginpar{trivially Weyl invariant}
\begin{align}\label{eq:compensated-riegert}
  (\DD^2 \la)(\bDD^2 \bla) - 8 \GG_\ada (\DD^\ia \la)(\bDD^\da \bla) 
\end{align}
yields the bosonic Riegert operator and is super-Weyl invariant.
Unfortunately the latter is also true for any other expression,
so not much has been gained. In particular in the presence
of a compensator the criterion for Weyl invariance of a term is 
the absence of any functional dependence on that compensating field,
which is certainly not true for \eqref{eq:compensated-riegert}.

Another approach may be to ask what is a natural Weyl invariant
operator for an arbitrary field, such that the operator does
not coincide with the Riegert operator.
For\marginpar{linear superfield}
example
\begin{align}
  E^{-1} [(\D^2 - 4\bR)\psi][(\bD^2 - 4R)\bpsi] 
\end{align}
is invariant when $\psi \mapsto \e^{\bsigma-\sigma}\psi$.
This transformational behaviour is incompatible with $\psi$ being
a chiral field. It is possible for $\psi$ being linear, but that
assumption annihilates the operator of course.

For\marginpar{real superfield} 
a real field $V$, a Weyl invariant operator is given by
\begin{align}
  E^{-1} V \D^\ia (\bD^2 - 4 R) \D_\ia V \equiv  
  E^{-1} V \bD_\da (\D^2 - 4 \bR) \bD^\da V,
\end{align}
with additional gauge invariance $V \mapsto V + \la + \bla$,
where $\la$ and $\bla$ are chiral and anti-chiral fields 
respectively.

Since the $\cN=4$ case should also incorporate 
$\gr{SL}(2,\mathds{R})$ symmetry with invariance of
the anomaly under
\begin{align}
  \la &\mapsto \la + i, &
  \la &\mapsto \frac{1}{\la},
\end{align}
one might be tempted to use the $\gr{SL}(2,\mathds{R})$
K\"ahler form 
\begin{align*}
V = \ln \la+\bla
\end{align*}
to also include that 
symmetry. 
Of course the operator will then contain additional 
pieces acting on more than two fields. 
However those pieces which do act on only two fields
form exactly the combination \eqref{eq:susytrace-candidate-relation},
such that the Riegert operator is missing again.

It seems that there is something in the minimal supergravity
formalism that does not allow for superfield formulation of 
the Riegert operator. I strongly suspect that it is the 
$\gr{U}(1)_R$ symmetry that spoils the formulation of the 
operator by being inevitable tied to the super-Weyl transformations.



  \pdfbookmark[0]{Conclusions}{toc}
  
\begin{savequote}[\savequotewidth]
  When your work speaks for itself, don't interrupt.
  \qauthor{Henry J.\ Kaiser}
\end{savequote}

\ifBeautifulForm
\else
\Chapter{Conclusions}
\fi

For the understanding of quantum field theories, its coupling to
gravity backgrounds has proved a valuable tool. The discovery of
\adscftcorr, which realises such a coupling holographically, has
revived the interest in this idea and been a major break-through in
the understanding of strongly coupled Yang--Mills theories.  While the
original \adscft\ duality involves $\cN=4$ supersymmetric Yang--Mills
theory, it has soon been extended to less symmetric, more realistic
theories.

In\marginpar{extension of \adscft}
this work, such an extension is explored in more detail,
taking as
a starting point the $\cN=2$ supersymmetric D3/probe~D7-brane
framework of \cite{Karch:2002sh}, which is dual to $\cN=4$
supersymmetric, large $N_c$ $\gr{SU}(N_c)$ Yang--Mills theory
augmented by a small number $N_f$ of $\cN=2$ hypermultiplets in the
fundamental representation.  By holographic methods, this theory's
meson spectrum can be calculated analytically at quadratic order
\cite{Kruczenski:2003be}.

\finalize{\begin{sloppypar}}
I considered first a geometry more general than the conventional
$\AdS_5\times \mf{S}^5$ and second an instanton gauge configuration on
the D7-branes.  The general strategy was to introduce background
configurations that reproduce the conventional setting in certain
limits.  This allowed to make contact with the ordinary \adscft\
dictionary and is an important feature of this approach compared to
others in the area that is sometimes referred to as
\acro{AdS}/\acro{QCD}.
\finalize{\end{sloppypar}}

The following results were obtained:

\begin{itemize}
\item A\marginpar{chiral symmetry breaking} 
  holographic dual of spontaneous \emph{chiral symmetry}
  breaking by a bilinear quark condensate $\vev{\bpsi\psi}$ was found.
  Since such a condensate is prohibited by supersymmetry, this
  required to use a background\footnote{Here a background by Gubser and Kehagias--Sfetsos
    \cite{Gubser:1999pk,Kehagias:1999tr} was chosen.}  that completely breaks
  supersymmetry and approximates $\AdS_5 \times \mf{S}^5$ only towards
  the boundary. By standard \adscft, the boundary of the space-time is
  associated to the ultraviolet of the dual field theory, such that
  the configuration describes an $\cN=2$ theory that is relevantly
  deformed and flows to a non-supersymmetric infrared.

  I\marginpar{quark condensate} calculated the quark condensate $\vev{\bpsi\psi}$ as a function of
  the quark mass $m_q$, which gave a non-vanishing quark condensate in
  the limit $m_q\to0$; i.e.~sponetaneous chiral symmetry breaking.
  Moreover I determined the meson spectrum and demonstrated that the
  meson mode associated to the $\gr{U}(1)_A$ axial symmetry, which is
  geometrically realised as rotations, becomes massless in the $m_q\to
  0$ limit as expected for a true Goldstone\marginpar{Goldstone boson} boson. When $m_q\neq0$
  this mode becomes a pseudo-Goldstone mode, which obeys the
  Gell-Mann--Oakes--Renner relation $M_\pi^2 \sim m_q$.  In the large
  quark mass limit, the mesons lie in the supersymmetric regime such
  that their mass is degenerate and approximates the analytic results
  of the $\cN=2$ theory.

  In addition I determined the mass of highly excited scalar and
  pseudoscalar mesons, which have the interesting feature of not being
  degenerate in this setup.

\item The\marginpar{instantons on the D7}
dual description of the mixed Coulomb--Higgs branch of the
  $\cN=2$ theory was found.  The Higgs \acro{VEV} corresponds to the
  size of an instanton configuration on the supergravity side,
  establishing a link between supersymmetry and the \acro{ADHM}
  construction that was known to exist.  Such an instanton
  configuration can only exist when there are at least two flavours,
  such that a \emph{non-Abelian} Dirac--Born--Infeld action had to be
  used.  Ordering ambiguities can be avoided since a calculation to
  quadratic order is sufficient, but a crucial insight was the use of
  a singular gauge transformation to obtain the correct boundary
  behaviour consistent with the \adscft\ dictionary.

  Having overcome this major obstacle, I numerically determined the
  meson spectrum and found it to approach the analytic $\cN=2$
  spectrum in the limit of vanishing and infinite Higgs \acro{VEV},
  though in the latter case a non-trivial rearrangement was observed,
  which could be explained to arise from above singular gauge
  transformation.

\item A\marginpar{heavy-light meson} 
  geometric realisation of heavy-light mesons was developed;
  i.e.\ mesons build up from a light and heavy quark providing a
  framework for the description of B~mesons not available before.
  Since a realisation in terms of a non-Abelian D7-brane action only
  makes sense for small mass differences, a different approach has to
  be chosen.  The configuration under consideration is that of a long
  string stretched between two D7-branes with a large separation,
  where the D7-branes are arranged to correspond to a massless and a
  heavy quark respectively.
 
  I describe an effective point-particle action derived from the
  Pol\-ya\-kov action for a straight string in a semi-classical
  approximation.  After quantisation the equation of motion gives rise
  to the spectrum of mesons consisting of a massless and a heavy
  quark.  I evaluated the spectrum in the standard
  $\AdS_5\times\mf{S}^5$ background, where I could find an analytic
  formula for the numerically determined heavy-light meson masses, and
  for the non-supersymmetric backgrounds by Constable--Myers
  \cite{Constable:1999ch} and by Gubser, Kehagias--Sfetsos discussed earlier.  
  In\marginpar{B~meson} the
  former case a comparison with the experimental values of the B~meson
  mass yields a deviation of about 20\%.
\end{itemize}

The models considered in this thesis are not meant to be realistic
duals of \acro{QCD}, but instead focus on a particular aspect like
chiral symmetry breaking by a chiral quark condensate, the meson
spectrum for D3/D7 \adscft\ either non-supersymmetric deformed or with
a Higgs \acro{VEV} switched on, and the spectrum of heavy-light mesons
in several backgrounds, giving a description of B~mesons.

It\marginpar{future challenges} would be certainly interesting to
extend the techniques developed in this thesis to a more realistic
example of \acro{AdS}/\acro{QCD}.\footnote{In particular the
  heavy-light meson construction could be easily extended to other,
  more realistic models.}  Over the last years there has been steady
progress towards such a description, including string theory duals of
theories that exhibit chiral symmetry breaking
\cite{Babington:2003vm,Evans:2004ia,Evans:2005ti,Apreda:2005hj,
  Apreda:2005yz,Kruczenski:2003uq,Barbon:2004dq,Ghoroku:2004sp,
  Brevik:2005fs,Sakai:2004cn,Sakai:2005yt,Antonyan:2006vw,Bak:2004nt,
  Ghoroku:2005tf,Mateos:2006nu,Albash:2006ew,Albash:2006bs,
  Parnachev:2006dn,Aharony:2006da,Karch:2006bv}.  There are however
three major points that need to be addressed in future refinements of
\acro{AdS}/\acro{QCD}.

The\marginpar{strong coupling}
models considered here have a \acro{UV} fixed point, but they are
not asymptotically free. The weak-strong nature of the duality, which
makes \adscft\ so interesting, unfortunately means that weak coupling
in the field theory's \acro{UV} implies strong curvature towards the
boundary of the \AdS\ space, thus requiring a full string
theoretical treatment, which currently is not feasible. 
Lacking that, there are recent
attempts to circumvent the situation by introducing a \acro{UV}
cut-off in the geometry to produce phenomenological models of
\acro{QCD} dynamics \cite{Erlich:2005qh,DaRold:2005zs,
  deTeramond:2005su,Brodsky:2006uq,Boschi-Filho:2002ta,
  Boschi-Filho:2002vd,Hong:2003jm,Evans:2006dj,DaRold:2005vr,
  Shock:2006fc,Shock:2006qy,Ghoroku:2005kg,Ghoroku:2005vt,
  Ghoroku:2006cc,Karch:2006pv,Andreev:2006vy,Andreev:2006ct,
  Hambye:2005up}.

A\marginpar{backreaction}
second problematic property is the probe limit $N_f \ll N_c$, which 
corresponds to the \emph{quenched approximation} of lattice \acro{QFT}.
Additional contributions are roughly of the order $\frac{N_f}{N_c}$. 
Including the backreaction
of the D7\NB-branes on the geometry would allow the number of flavours
to be of the same order of magnitude as the number of colours.
Such backgrounds have been considered in \cite{Burrington:2004id}.


The\marginpar{separation of scales}
last important aspect is the separation of the \acro{SUSY} and
confinement scales. In the B~physics example discussed in
Section~\ref{sec:bmesons}, the B~meson is far in the supersymmetric
regime.  To change this situation one needs a background configuration
that incorporates at least two different scales.

From the recent works cited above one can read off a tendency to focus
on particular aspects of the larger problem of finding a holographic
dual of \acro{QCD} and \acro{YM} theories, an approach also to be
found in this thesis.  A challenge for the future will be to
incorporate into one model as many as possible of the insights gained
here and elsewhere since the discovery of \adscft\ duality almost ten
years ago.

\horizrule

In the second Part of this thesis 
the coupling of supersymmetric quantum field theories 
to minimal supergravity was investigated. 
Coupling a gravity background to a conformal quantum field theory 
gives rise to a conformal anomaly\marginpar{conformal anomaly}
\begin{align} \label{eq:concl-anomaly}
  \vev{ T_m{}^m } = c\, C^2 - a\, \tilde{\cR}^2 + b\, \cR^2 + f\, \Box R. \tag{$\star$}
\end{align}
In \cite{Osborn:1991gm} a space-time dependent coupling approach 
was used to calculate consistency conditions for the coefficients
in the two-dimensional anomaly providing an alternative proof 
for Zamolodchikov's $c$\NB-theorem. 
However \cite{Osborn:1991gm} did not
obtain consistency conditions sufficiently restrictive to extend the
theorem to four dimensions.

The specific project pursued here was to extend this technique to superfields
and determine the conformal anomaly for those supersymmetric field
theories whose coupling constants can be promoted to chiral fields $\la$.
A prominent example for such is given by super-Yang--Mills theories.

\noindent The steps performed in detail were:
\begin{itemize}
\item I\marginpar{basis of superfield operators}
  determined a complete ansatz for the conformal anomaly by
  finding a basis of 38 local superfield expressions of dimension 2
  and composing a linear combination with arbitrary coefficient
  functions $b(\la,\bla)$.  In the constant coupling limit, these
  coefficient functions become the superspace analogue of the
  coefficients $c$, $a$, $b$ and $f$ that appear in the bosonic
  conformal anomaly \eqref{eq:concl-anomaly}.
\item Then\marginpar{consistency conditions} 
  I calculated the Wess--Zumino consistency conditions 
  for the coefficient functions, which arise from the fact
  that Weyl transformations are Abelian.
\item Furthermore I discussed the dependence on local counterterms
  and possible consequences of S-duality in the $\cN=4$ case.
\item It\marginpar{superfield Riegert operator} 
  is noted that a superfield version of the Riegert
  operator\footnote{The Riegert operator is the unique conformally
    covariant differential operator of fourth order acting on a scalar
    field of Weyl weight 0.} is needed to make contact with an
  existing one-loop calculation \cite{Osborn:2003vk}.  Various
  approaches to the problem of finding a superfield Riegert operator
  (which is independent of the anomaly calculation presented) have
  been discussed. The conclusion is that the problem is rooted in the
  $\gr{U}(1)_R$ symmetry being built into the formalism of minimal
  supergravity in superfield formulation \emph{in a local way}, while
  on the component level the $\gr{U}(1)_R$ is only realised as a
  global symmetry.
\end{itemize}

In order to\marginpar{computer algebra?} check this assumption it would be
desirable to repeat the full calculation in a component approach.  The
sheer size of this task is daunting however: The basis for the anomaly
I found contains about 40 terms in superfield formulation plus their
complex conjugates.  As a consequence the calculation of the
Wess--Zumino consistency conditions is very involved and potentially
error prone. A component based approach will probably incorporate even
more terms and should therefore be implemented with the help of a
computer.  Unfortunately a computer based treatment of 
supergravity has a number of requirements not
satisfied by any existing computer algebra system (\acro{CAS}) today.  These
requirements are
\begin{itemize}
  \item an efficient mechanism for the representation of tensors and
        contracted indices,
  \item handling of commuting, anticommuting and non-commuting
        objects (this should include the ability to reduce a number
        of terms to a canonical basis of terms using the supergravity 
        algebra and Bianchi identities), 
  \item a way to represent non-commuting tensor valued
        functions of other objects (e.g.~for non-anticommuting 
        spinorial derivatives),
  \item making no assumption about the symmetries of the metric,
  \item allowing torsion, and
  \item no automatic expansion of compact parenthesised expressions
        into a lengthy sum of terms.
\end{itemize}
Of the existing systems, \acro{FORM} \cite{Vermaseren:2000nd} seems to be
coming the closest to these requirements since it provides a rather
low-level tensor support without restrictive internal assumptions.  Its
summarising capabilities are unsatisfactory however and may be a major
obstacle in the implementation of a computer based analysis of the trace
anomaly.

Another promising program is \textsf{Cadabra}
\cite{Peeters-AEI-2006-037, Peeters-AEI-2006-038}, which meets all of
the above requirements but is still in a development stage.

Nevertheless\marginpar{component calculation}
the next steps in a future analysis of the trace anomaly
are the implementation of a supergravity computer algebra package and
a component based analysis.  As outlined above this is a difficult
task, but the results presented in this thesis can serve as a highly
non-trivial unit test to confirm the correctness of such a package.
Then one may carry out a complete component expansion of all basis
terms and reexamine the question of whether a superfield version of
the Riegert operator does exist in minimal superfield supergravity.
This analysis can then be easily extended to non-minimal \acro{SUGRA}
and as a check one may reproduce the Riegert operator in new-minimal
\acro{SUGRA} as well.

A reimplementation of the whole calculation in a component based
approach would provide an independent source of confirmation for the
results of this thesis. \emph{If} a superfield based treatment of
minimal supergravity is consistent on the quantum level,\footnote{See
  \cite{deWit:1985bn} on why a superfield treatment of 
  minimal supergravity should be consistent and \cite{Shamir:1992ff}
  on the question of consistence of anomaly calculations in the
  presence of compensating fields.} the two calculations should
actually yield the same result, strengthening confidence in the results
presented here. Of course inconsistency would be an interesting result
in its own right.

In any case I hope to have provided a basis for 
understanding the structure of the conformal anomaly in
supersymmetric field theories coupled to supergravity.



  \begin{savequote}[\savequotewidth]
  It pays to be obvious, especially if you have a reputation for subtlety.
  \qauthor{Isaac Asimov}
\end{savequote}

\Chapter*{Acknowledgements}
\ifBeautifulForm
\else
I would like to thank my wife Olga, Johanna Erdmenger and Dieter L\"ust.
Without their support this work would not have been possible.
Many thanks also to my friend Robert Eisenreich for or despite 
his being esopucative all the time.

I am thankful to Hugh Osborn, Peter Breitenlohner, Sergei Kuzenko and
Gabriel Lopez Cardoso for helpful discussions on the subjects of supergravity,
conformal covariance and space-time dependent couplings.

Part of this thesis was supported by the \acro{DFG} Graduiertenkolleg ``The
Standard Model of Particle Physics---structure, precision tests and
extensions'' and the Max-Planck-Institut f\"ur Physik, M\"unchen.
\fi



  \appendix
  \pdfbookmark[0]{Appendix}{toc}

  \allowdisplaybreaks[4]

  \begin{savequote}[\savequotewidth]
  Genius is one per cent inspiration, ninety-nine per cent
  perspiration.  
  \qauthor{Thomas A.\ Edison}
\end{savequote}

\Chapter{Weyl Variation of the Basis\label{app:weyl-variation}}

This Chapter provides the Weyl variations of all basis terms.
The terms for $\Delta^W_\sigma$ can be extracted from those
proportional to $\sigma$, and correspondingly for the complex 
conjugated fields.

\stepcounter{equation}
\begin{align}
  E  \delta \bigl[ E^{-1} G_\ada G^\ada \bigr] 
      &= 2i G_\ada \D^\ada(\bsigma-\sigma) \tag{\theequation$A$} \\
  E \delta \bigl[ E^{-1} R\bR \bigr] 
      &= -\quart (\bD^2\bsigma)\bR - \quart(\D^2\sigma)R  \tag{\theequation$B$}\\
  E \delta \bigl[ E^{-1} R^2 \bigr] 
      &= 3(\bsigma-\sigma) R^2 - \half(\bD^2\bsigma) R  \tag{\theequation$C$}\\
  E \delta \bigl[ E^{-1} \bR^2 \bigr] 
      &= 3(\sigma-\bsigma) \bR^2 - \half(\D^2\sigma)\bR  \tag{\theequation$\bar C$}\\
  E  \delta \bigl[ E^{-1} \D^2 R \bigr] 
      &= -2 (\D^\ia\sigma)(\D_\ia R) -2 (\D^2\sigma) R
         - \quart \D^2\bD^2\bsigma  \nonumber\\
      &= -2 (\D^\ia\sigma)(\D_\ia R) -2 (\D^2\sigma) R \nonumber\\*&\quad
         +2 (\D_\ada \D^\ada \bsigma) - 2i G_\ada (\D^\ada \bsigma)  \tag{\theequation$D$} \\
  E  \delta \bigl[ E^{-1} \bD^2 \bR \bigr] 
      &= -2 (\D_\da\bsigma)(\bD^\da \bR) -2 (\bD^2\bsigma) \bR 
         - \quart \bD^2\D^2\sigma  \nonumber\\
      &= -2 (\bD_\da\bsigma)(\bD^\da \bR) - 2(\bD^2\bsigma) \bR \nonumber\\*&\quad
         +2 (\D_\ada \D^\ada \sigma) + 2i G_\ada (\D^\ada \sigma)  \tag{\theequation$\bar D$} \\
  E \delta \bigl[ E^{-1}R\D^2\la  \bigr] 
      &= -\quart (\bD^2\bsigma)(\D^2\la)+2R(\D^\ia\sigma)(\D_\ia\la)  \tag{\theequation$E$}\\
  E \delta \bigl[ E^{-1}\bR\bD^2\bla  \bigr] 
      &= -\quart (\D^2\sigma)(\bD^2\bla)+2\bR(\bD_\da\bsigma)(\bD^\da\bla)  \tag{\theequation$\bar E$}
\end{align}
\begin{align}
  E \delta \bigl[ E^{-1} R\bD^2\bla \bigr] 
      &= 3 (\bsigma-\sigma)R(\bD^2\bla) 
         - \quart (\bD^2\bsigma)(\bD^2\bla) \nonumber\\*&\quad   
         + 2R(\bD_\da\bsigma)(\bD^\da\bla)  \tag{\theequation$F$}\\
  E \delta \bigl[ E^{-1} \bR\D^2\la \bigr] 
      &= 3 (\sigma-\bsigma)\bR(\D^2\la) 
         - \quart (\D^2\sigma)(\D^2\la) \nonumber\\*&\quad   
         + 2\bR(\D^\ia\sigma)(\D_\ia\la) \tag{\theequation$\bar F$}\\
  E \delta \bigl[ E^{-1} (\D^\ia R)(\D_\ia\la) \bigr] 
      &= -2 (\D^\ia \sigma) (\D_\ia \la) R - \quart (\D^\ia\bD^2 \bsigma)(\D_\ia\la) \nonumber\\
       = -2 (\D^\ia \sigma) (\D_\ia \la) R &+ [(G^\ada - i\D^\ada)(\bD_\da\bsigma )](\D_\ia\la)  \tag{\theequation$G$} \\
  E \delta \bigl[ E^{-1} (\bD_\da \bR)(\bD^\da\bla) \bigr] 
       &= -2 (\bD_\da \bsigma) (\bD^\da \bla) \bR - \quart (\bD_\da\D^2 \sigma)(\D^\da\bla)\nonumber\\
       = -2 (\bD_\da \bsigma) (\bD^\da \bla) \bR &- [(G^\ada + i\D^\ada)(\D_\ia\sigma )](\bD_\da\bla)  \tag{\theequation$\bar G$}\\
  E \delta \bigl[ E^{-1} G_\ada \D^\ada \la \bigr] 
      &= i [\D_\ada (\bsigma-\sigma)](\D^\ada\la)   \nonumber\\*&\quad
         - \tfrac{i}{2} G_\ada (\bD^\da\bsigma)(\D^\ia\la)  \tag{\theequation$H$}\\
  E \delta \bigl[ E^{-1} G_\ada \D^\ada \bla \bigr] 
      &= i [\D_\ada (\bsigma-\sigma)](\D^\ada\bla)  \nonumber\\*&\quad
         - \tfrac{i}{2} G_\ada (\D^\ia\sigma)(\bD^\da\bla)  \tag{\theequation$\bar H$}\\
  E \delta \bigl[ E^{-1} (\D^\ada \D_\ada \la) \bigr] 
      &= (\D^\ada (\sigma+\bsigma))\D_\ada\la  
	 - R (\D^\ia\sigma)(\D_\ia\la) \notag \\*
         - \tfrac{i}{2} (\bD^\da\bsigma)&(\D_\ada \D^\ia\la) 
         + G^\ada (\bD_\da\bsigma) (\D_\ia\la) \tag{\theequation$I$} \\
  E \delta \bigl[ E^{-1} (\D^\ada \D_\ada \bla) \bigr] 
      &= (\D^\ada (\sigma+\bsigma))\D_\ada\bla 
	 - \bR (\bD_\da\bsigma)(\bD^\da\bla) \notag \\*
         - \tfrac{i}{2} (\D^\ia\sigma)&(\D_\ada \bD^\da\bla) 
         - G^\ada (\D_\ia\sigma) (\bD_\da\bla) \tag{\theequation$\bar I$} \\
  E \delta \bigl[ E^{-1} R (\D^\ia \la)(\D_\ia \la) \bigr]  
    &= -\quart (\bD^2\bsigma)(\D^\ia\la)(\D_\ia\la)  \tag{\theequation$J$}\\
  E \delta \bigl[ E^{-1} \bR (\bD_\da \bla)(\bD^\da \bla) \bigr]  
    &= -\quart (\D^2\sigma)(\bD_\da\bla)(\bD^\da\bla)  \tag{\theequation$\bar J$}\\
  E \delta \bigl[ E^{-1} \bR (\D^\ia\la)(\D_\ia\la) \bigr] 
    &= 3(\sigma - \bsigma) \bR (\D^\ia \la)(\D_\ia \la)  \nonumber\\* &\quad 
       -\quart (\D^2\sigma)(\D^\ia\la)(\D_\ia\la)  \tag{\theequation$K$}\\
  E \delta \bigl[ E^{-1} R (\bD_\da\bla)(\bD^\da\bla) \bigr] 
    &= 3(\bsigma - \sigma) R (\bD_\da \bla)(\bD^\da \bla)  \nonumber\\* &\quad 
       -\quart (\bD^2\bsigma)(\bD_\da\bla)(\bD^\da\bla)    \tag{\theequation$\bar K$}\\
  E \delta \bigl[ E^{-1} G^\ada (\D_\ia \la)(\bD_\da \bla) \bigr]  
      &= i (\D_\ada(\bsigma-\sigma))(\D^\ia\la)(\bD^\da\bla)  \tag{\theequation$L$}\\ 
  E \delta \bigl[ E^{-1} (\D_\ada \la)(\D^\ada \bla) \bigr]  
      &= -\tfrac{i}{2} (\bD_\da \bsigma)( \D_\ia \la) (\D^\ada \bla) \nonumber\\ &\quad 
         -\tfrac{i}{2} (\D^\ia\sigma)(\bD^\da \bla) (\D_\ada \la) \tag{\theequation$M$}\\
  E \delta \bigl[ E^{-1} (\D^\ada \la)(\D_\ada \la)\bigr] 
    &= -i(\bD^\da\bsigma)(\D^\ia\la)(\D_\ada\la)  \tag{\theequation$N$}\\
  E \delta \bigl[ E^{-1} (\D^\ada \bla)(\D_\ada \bla)\bigr] 
    &= -i(\D^\ia\sigma)(\bD^\da\bla)(\D_\ada\bla)  \tag{\theequation$\bar N$}\\
  E \delta \bigl[ E^{-1} (\D^\ia\la)(\D_\ada \bD^\da\bla) \bigr] 
    &= \half(\D^\ia\la) [(\D_\ada(\bsigma+\sigma))(\bD^\da\bla) \tag{\theequation$O$}
        \\*&\quad
      -i(\D_\ia\sigma)(\bD^2\bla)
      +2(\bD^\da\bsigma)(\D_\ada\bla) \bigr] \nonumber
\end{align}
\begin{align}
  E \delta \bigl[ E^{-1} (-1)(\bD^\da\bla)(\D_\ada \D^\ia\la) \bigr] 
    &= -\half(\bD^\da\bla) [(\D_\ada(\bsigma+\sigma))(\D^\ia\la) \notag\\*
      +i(\bD_\da\bsigma)(\D^2\la) &
      +2(\D^\ia\sigma)(\D_\ada\la) \bigr] \tag{\theequation$\bar O$} \\
  E \delta \bigl[ E^{-1}  (\D^2 \la)(\bD^2\bla)  \bigr] 
      &= 2 (\D^\ia\sigma)(\D_\ia\la)(\bD^2\bla) \nonumber\\*&\quad 
         + 2 (\bD_\da\bsigma)(\D^2 \la)(\bD^\da \bla)  \tag{\theequation$P$}\\
  E  \delta \bigl[ E^{-1} (\D^2\la)^2 \bigr] 
    &= 3(\sigma-\bsigma)(\D^2\la)^2 \nonumber\\*&\quad 
       + 4(\D^2\la)(\D^\ia\sigma)(\D_\ia\la)  \tag{\theequation$Q$}\\
  E  \delta \bigl[ E^{-1} (\bD^2\bla)^2 \bigr] 
    &= 3(\bsigma-\sigma)(\bD^2\bla)^2 \nonumber\\*&\quad 
       + 4(\bD^2\bla)(\bD_\da\bsigma)(\bD^\da\bla) \tag{\theequation$\bar Q$}\\
  E \delta \bigl[ E^{-1} (\D^\ia\la)(\D_\ia\la)(\D^2\la) \bigr] 
    &= 3(\sigma-\bsigma)(\D^\ia\la)(\D_\ia\la)(\D^2\la) \nonumber\\*&\quad 
       + 2 (\D^\ia\sigma)(\D_\ia\la)(\D^\ib\la)(\D_\ib\la)   \tag{\theequation$R$}\\
  E \delta \bigl[ E^{-1} (\bD_\da\bla)(\bD^\da\bla)(\bD^2\bla) \bigr] 
    &= 3(\bsigma-\sigma)(\bD_\da\bla)(\bD^\da\bla)(\bD^2\bla) \nonumber\\*&\quad 
       + 2 (\bD_\da\bsigma)(\bD^\da\bla)(\bD_\db\bla)(\bD^\db\bla)   \tag{\theequation$\bar R$} \\
  E  \delta \bigl[ E^{-1} (\D^\ia\la)(\D_\ia\la)(\bD^2\bla) \bigr] 
    &= 2 (\D^\ia\la )(\D_\ia\la)(\bD_\da\bsigma)(\bD^\da\bla)  \tag{\theequation$S$} \\
  E  \delta \bigl[ E^{-1} (\bD_\da\bla)(\D^\da\bla)(\D^2\la) \bigr] 
    &= 2 (\bD_\da\bla )(\bD^\da\bla)(\D^\ia\sigma)(\D_\ia\la)  \tag{\theequation$\bar S$} \\
  E \delta \bigl[ E^{-1} (\D_\ada\la^i)(\D^\ia\la^j)(\bD^\da\bla^\bk) \bigr] 
    &= \tfrac{i}{2} (\D^\ia \la^j)(\D_\ia\la^i)(\bD_\da\bsigma)(\bD^\da\bla^\bk)  \tag{\theequation$T$}\\
  E \delta \bigl[ E^{-1} (\D_\ada\bla^\bi)(\D^\ia\la^k)(\D^\da\bla^\bj) \bigr] 
    &= -\tfrac{i}{2} (\bD_\da \bla^\bi)(\bD^\da\bla^\bj)(\D^\ia\sigma)(\D_\ia\la^k)  \tag{\theequation$\bar T$}\displaybreak[0]\\
  E \delta \bigl[ E^{-1}  (\D^\ia\la)(\D_\ia\la)(\bD_\db\bla)&(\bD^\db\bla)\bigr] 
    = 0  \tag{\theequation$U$}\displaybreak[0] \\
  E \delta \bigl[ E^{-1} (\D^\ia\la)(\D_\ia\la)(\D^\ib\la)&(\D_\ib\la)  \bigr] \nonumber\\*
    = 3(\sigma-\bsigma)& (\D^\ia\la)(\D_\ia\la)(\D^\ib\la)(\D_\ib\la)  \tag{\theequation$V$}\displaybreak[0]\\
  E \delta \bigl[ E^{-1} (\bD_\da\bla)(\bD^\da\bla)(\bD_\db\bla)&(\bD^\db\bla)  \bigr] \nonumber\\*
    = 3(\bsigma-\sigma)& (\bD_\da\bla)(\bD^\da\bla)(\bD_\db\bla)(\bD^\db\bla)  \tag{\theequation$\bar V$}
\end{align}


  \begin{savequote}[\savequotewidth]
  Vi Victa Vis.
  \qauthor{Cicero}
\end{savequote}

\Chapter{Wess--Zumino Consistency~Condition\label{app:wzcc}}
\finalize{\vskip-3em}
\Section*{Weyl Contribution}
For the coefficients defined in 
\begin{align}
  ( \Delta^W_{\sigma} )&
     ( \Delta^W_{\sigma'} - \Delta^\beta_{\sigma'} ) W 
     = \int d^8z\, E^{-1} \sigma' \bigl\{ 
          \sigma \cC_{0} + (\D^\ia\sigma) \cC_\ia + \notag\\*& 
          (\D^2 \sigma) \cC_{2} + (\D^\ada \sigma) \cC_\ada 
     + (\D_\ada\D^\ia \sigma) \cC^\da_{3} + (\D_\ada\D^\ada\sigma) \cC_{4} 
       \bigr\},
\end{align}
one obtains
\begin{subequations}
\begin{align}
\begin{split}
   \cC_{0}  &= 
      - 3 b^{(C)}R^2 
      + 3 b^{(\bar C)}\bR^2 \\&\quad
      - \bigl[3 R(\bD^2\bla) b^{(F)} \bigr]
      + 3 \bR (\D^2\la) b^{(\bar F)} \\&\quad
      + 3 b^{(K)} \bR(\D^\ia\la)(\D_\ia \la)
      - \bigl[3 b^{(\bar K)}R(\bD_\da\bla)(\bD^\da\bla)\bigr] \\&\quad
      + 3 b^{(Q)}  (\D^2\la)^2 
      - 3 b^{(\bar Q)}(\bD^2\bla)^2 \\&\quad
      + 3(\D^\ia\la)(\D_\ia\la)(\D^2\la) b^{(R)} 
      - 3 b^{(\bar R)}(\bD_\da\bla)(\bD^\da\bla)(\bD^2\bla) \\&\quad
      + 3 b^{(V)} (\D^\ia\la)(\D_\ia\la)(\D^\beta\la)(\D_\beta\la) 
      - 3  b^{(\bar V)} (\bD_\da\bla)(\bD^\da\bla)(\bD_\db\bla)(\bD^\db\bla),
\end{split} \\[1ex]
\begin{split}
   \cC_\ia  &= 
      - 2 b^{(D)} (\D_\ia R)
      - 2 R(\D_\ia\la) b^{(E)}
      + 2 \bR (\D_\ia\la)  b^{(\bar F)}  \\&\quad
      - \bigl[2 (\D_\ia\la)R  b^{(G)} \bigr]
      - \bigl[G_\ada(\bD^\da\bla) b^{(\bar G)}\bigr] \\&\quad
      - \tfrac{i}{2} G_\ada (\bD^\da\la) b^{(\bar H)} 
      - \tfrac{i}{2} b^{(M)}  (\bD^\da\bla)(\D_\ada\la)
      - i(\bD^\da\bla)(\D_\ada\bla)  b^{(\bar N)}\\&\quad 
      - \bigl[\tfrac{i}{2}(\D_\ia\la)(\bD^2\bla)  b^{(O)}\bigr] 
      - \bigl[b^{(\bar O)} (\D_\ada \la)(\bD^\da\bla) \bigr]
      + 2 b^{(P)}(\D_\ia \la)(\bD^2\bla) \\&\quad
      + 4 b^{(Q)}(\D^2\la)(\D_\ia\la) 
      + 2(\D_\ia\la)(\D^\ib\la)(\D_\ib\la) b^{(R)} \\&\quad
      + 2 b^{(\bar S)}(\bD_\da\bla)(\bD^\da\bla)(\D_\ia\la) 
      - \tfrac{i}{2} b^{(\bar T)}(\bD_\da\bla)(\bD^\da\bla)(\D_\ia\la) ,
\end{split}\\[1ex]
\begin{split}
   \cC_{2}  &=  
      - \quart b^{(B)} R 
      - \half b^{(\bar C)}\bR  \\&\quad
      - 2 b^{(D)}R 
      - \quart (\bD^2\bla ) b^{(\bar E )} \\&\quad
      - \quart (\D^2\la) b^{(\bar F)} 
      - \quart (\bD_\da\bla)(\bD^\da\bla) b^{(\bar J)} 
      - \quart  b^{(K)} (\D^\ia\la)(\D_\ia\la),
\end{split} \\[1ex]
\begin{split}
   \cC_\ada &= 
      - 2i b^{(A)} G_\ada       
      + 2i G_\ada  b^{(\bar D)}  \\&\quad
      - i(\D_\ada\la) b^{(H)} 
      - i(\D_\ada\bla) b^{(\bar H)}  \\&\quad
      - i b^{(L)}(\D_\ia\la)(\bD_\da\bla) 
      + \bigl[ \half b^{(O)} (\D_\ia\la)(\bD_\da\bla) \bigr]
      - \bigl[ \half b^{(\bar O)} (\D_\ia\la)(\bD_\da\bla) \bigr],
\end{split}\\[1ex]
    \bcC^\da_{3} &= \bigl[ i b^{(\bar G)} (\bD^\da\bla)\bigr], \\[1ex]
    \cC_{4} &= \bigl[ 2b^{(\bar D)} \bigr].
\end{align}
\end{subequations}
For further discussion, it proves useful to sort its contents with respect to derivatives on $\la$ or $\bla$.
\begin{subequations}
\begin{align}
\begin{split}
   \cC_{0}  &= 
      - 3 b^{(C)}R^2 
      + 3 b^{(\bar C)}\bR^2 \\&\quad
      - \bigl[ 3 R(\bD^2\bla^\bi) b^{(F)}_\bi \bigr] 
      + 3 \bR (\D^2\la^i) b^{(\bar F)}_i \\&\quad
      + 3 b^{(K)}_{ij} \bR(\D^\ia\la^i)(\D_\ia \la^j)
      - \bigl[ 3 b^{(\bar K)}_{\bi\bj} R(\bD_\da\bla^\bi)(\bD^\da\bla^\bj) \bigr]  \\&\quad
      + 3 b^{(Q)}_{ij}  (\D^2\la^i)(\D^2\la^j)
      - 3 b^{(\bar Q)}_{\bi\bj} (\bD^2\bla^\bi)(\bD^2\bla^\bj) \\&\quad
      + 3 (\D^\ia\la^i)(\D_\ia\la^j)(\D^2\la^k) b^{(R)}_{ijk}
      - 3 b^{(\bar R)}_{\bi\bj\bk} (\bD_\da\bla^\bi)(\bD^\da\bla^\bj)(\bD^2\bla^\bk) \\&\quad
      + 3 b^{(V)}_{ijkl} (\D^\ia\la^i)(\D_\ia\la^j)(\D^\beta\la^k)(\D_\beta\la^l)  \\&\quad
      - 3 b^{(\bar V)}_{\bi\bj\bk\bl} (\bD_\da\bla^\bi)(\bD^\da\bla^\bj)(\bD_\db\bla^\bk)(\bD^\db\bla^\bl),
\end{split} \\[1ex]
\begin{split}
   \cC_\ia  &= 
      - 2 b^{(D)} (\D_\ia R)  
      + (\D_\ia\la^i) \biggl( - 2 R ( b^{(E)}_i + \bigl[b^{(G)}_i\bigr] +\bigl[\half b^{(I)}_i\bigr] ) + 2 \bR b^{(\bar F)}_i  \biggl) \\&\quad
      + (\bD^\da\bla^\bi) \biggl(  - \tfrac{i}{2} G_\ada  b^{(\bar H)}_\bi 
                                   - G_\ada b^{(\bar I)}_\bi 
				   - \bigl[ G_\ada b^{(\bar G)}_\bi \bigr] \biggr) \\&\quad
      + (\D_\ada\bD^\da\bla^\bi) \biggl(- \ihalf b^{(\bar I)}_\bi \biggr)
      + (\D_\ada\bla^\bi)(\bD^\da\bla^\bj) \biggl( - i b^{(\bar N)}_{\bi\bj} \biggr)\\&\quad
      + (\D_\ada\la^i)(\bD^\da\bla^\bj) \biggl( -\tfrac{i}{2} b^{(M)}_{i\bj}  - \bigl[ b^{(\bar O)}_{\bj i}\bigr]  \biggr) \\&\quad
      + (\D_\ia\la^i)(\bD^2\bla^\bj) \biggl( - \bigl[ \tfrac{i}{2} b^{(O)}_{i\bj}\bigr]  + 2 b^{(P)}_{i\bj} \biggr)\\&\quad
      + (\D^2\la^i)(\D_\ia\la^j) \biggl(  4 b^{(Q)}_{ij} \biggr) \\&\quad
      + (\D_\ia\la^i)(\D^\ib\la^j)(\D_\ib\la^k)\biggl(  2 b^{(R)}_{ijk} \biggr) \\&\quad
      + (\bD_\da\bla^\bi)(\bD^\da\bla^\bj)(\D_\ia\la^k) \biggl(
             2 b^{(\bar S)}_{\bi\bj k} 
	     - \tfrac{i}{2} b^{(\bar T)}_{\{\bi\bj\}k} \biggr) ,
\end{split}\\[1ex]
\begin{split}
   \cC_{2}  &=  
      - (\quart b^{(B)} +2 b^{(D)} ) R 
      - \half b^{(\bar C)}\bR  
      - \quart (\bD^2\bla^\bi ) b^{(\bar E )}_\bi  \\&\quad
      - \quart (\D^2\la^i) b^{(\bar F)}_i
      - \quart (\bD_\da\bla^\bi)(\bD^\da\bla^\bj) b^{(\bar J)}_{\bi\bj} 
      - \quart  b^{(K)}_{ij} (\D^\ia\la^i)(\D_\ia\la^j) ,
\end{split}\\[1ex]
\begin{split}
   \cC_\ada &=        
        2i G_\ada ( \bigl[ b^{(\bar D)} \bigr] - b^{(A)})
      + (\D_\ada\la^i) ( \bigl[b^{(I)}_i\bigr] - \bigl[ ib^{(H)}_i \bigr] ) \\&\quad
      + (\D_\ada\bla^\bi)(b^{(\bar I)}_\bi - i b^{(\bar H)}_\bi )
      + (\bigl[ \half b^{(O)}_{i\bj}\bigr] - \bigl[\half b^{(\bar O)}_{i\bj}\bigr]  - i b^{(L)}_{i\bj} ) (\D^\ia\la^i)(\bD^\da\bla^\bj) ,
\end{split}\\[1ex]
    \bcC^\da_{3} &= \bigl[ i b^{(\bar G)}_\bj (\bD^\da\bla^\bj) \bigr], \\[1ex]
    \cC_{4} &= \bigl[ 2b^{(\bar D)} \bigr].
\end{align}
\end{subequations}

\Section*{Beta Contribution}
Acting with the operator $\Delta^\beta$ on the conformal anomaly, one
first notices, that $\frac{\delta}{\delta\la^i}$ should only act on
derivatives of $\la$, since otherwise a $\sigma\sigma'$ contribution,
which vanishes from the commutator, is created.
\begin{align}
  \Delta^\beta_\sigma & (\Delta^W_{\sigma'}-\Delta^\beta_{\sigma'}) W
    = \int d^8z\, E^{-1} \sigma' \bigl\{ 
          \sigma \beta^i (\p_i \cB_{(0)}) \cA_{(0)} 
          + (\D^\ia \sigma \beta^i )\cE^i_\ia \\&
          + (\D^2 \sigma \beta^i) \cE^i_{(2)} 
          + (\D^\ada \sigma \beta^i ) \cE^i_\ada 
          + (\D_\ada \D^\ada \sigma \beta^i) \cE^i_{4} 
          + (\D_\ada \D^\da \sigma \beta^i ) \bcE^{i\da}_{3} \bigr\}, \notag
\end{align}
\begin{subequations}
\begin{align}
\begin{split}
   \cE_\ia  &= 
       \bigl[b^{(G)}_i (\D_\ia R) \bigr]
       + \bigl[2 b^{(J)}_{ij}R (\D_\ia \la^j) \bigr]
       + 2 b^{(K)}_{ij}\bar R (\D_\ia \la^j) \\&\quad
       + b^{(L)}_{i\bj} G_\ada (\bD^\da \bla^\bj) 
       + \bigl[ b^{(O)}_{i\bj}(\D_\ada\bD^\da \bla^\bj)\bigr]  \\&\quad
       + 2 b^{(R)}_{ijk}(\D_\ia\la^j)(\D^2\la^k) 
       + \bigl[ 2 b^{(S)}_{ij\bk}(\D_\ia\la^j)(\bD^2\bla^\bk)\bigr]  \\&\quad
       + b^{(T)}_{ji\bk}(\D_\ada \la^j)(\bD^\da\bla^\bk) 
       + b^{(\bar T)}_{\bj i\bk} (\D^\ada\bla^\bj)(\bD_\da\bla^\bk)  \\&\quad
       + 2 b^{(U)}_{ij\bk\bl}(\D_\ia\la^j)(\bD_\db \bla^\bk)(\bD^\db \bla^\bl) 
       + 4  b^{(V)}_{ijkl} (\D_\ia\la^j)(\D^\ib\la^k)(\D_\ib \la^l), 
\end{split}\\[1ex]
\begin{split}
   \cE_{2}  &= 
       b^{(E)}_{i}R 
       +  b^{(\bar F)}_{i}\bR 
       + b^{(P)}_{i\bj}(\bD^2\bla^\bj)  \\&\quad
       + 2 b^{(Q)}_{ij}(\D^2\la^j) 
       + b^{(R)}_{lji}(\D^\ia\la^l)(\D_\ia\la^j) 
       + b^{(\bar S)}_{\bi\bj i} (\bD_\da\bla^\bi)(\bD^\da\bla^\bj),
\end{split}\\[1ex]
\begin{split}
   \cE_\ada &=  
       \bigl[ b^{(H)}_{i}G_\ada \bigr]
       + b^{(M)}_{i\bj}(\D_\ada \bla^\bj) 
       + b^{(N)}_{ij}(\D_\ada \la^j) 
       + b^{(T)}_{ij\bk}(\D_\ia \la^j)(\bD_\da \bla^\bk),
\end{split}\\[1ex]
   \bcE^\da_{3} &=  \bigl[ b^{(\bar O)}_{\bj i} (\bD^\da \bla^\bj)\bigr], \\[1ex]
   \cE_{4} &= \bigl[ b^{(I)}_{i} \bigr].
\end{align}
\end{subequations}

\Section*{Summary}
The results of the previous two Sections can be used to
determine the $\cF$ coefficients defined by 
\begin{align}
  ( \Delta^W_{\sigma}  - \Delta^\beta_{\sigma} ) 
  &( \Delta^W_{\sigma'} - \Delta^\beta_{\sigma'} ) W \nonumber\\*&
     = \int d^8z\, E^{-1} \sigma' \bigl\{ 
          \sigma \cF_{0} + (\D^\ia\sigma) \cF_\ia + 
          (\D^2 \sigma) \cF_{2} + (\D^\ada \sigma) \cF_\ada 
        \nonumber \\*&\hphantom{=\int d^8z\, E^{-1} \sigma' \bigl\{ } 
     + (\D_\ada\D^\ia \sigma) \bcF^\da_{3} + (\D_\ada\D^\ada\sigma) \cF_{4} 
       \bigr\},
\end{align}
by expanding the Weyl and beta contribution in terms of derivatives
on $\la$ and $\bla$, keeping in mind that the $b$ coefficients and beta functions are
functions of $\la$ and $\bla$ in general, so it holds
\begin{subequations}
\begin{align}
  b &= b(\la,\bla), \\
  \D_\ia b &= (\D_\ia\la^k)(\partial_k b),\\
  \bD^\da b &= (\bD^\da \bla^\bk) (\partial_\bk b) ,\\
  \D_\ada b &= (\D_\ada \la^k )(\partial_k b) + (\D_\ada \bla^\bk) (\partial_\bk b),\\
  \bD^\da\D_\ada b &= (\bD^\da\D_\ada \la^k )(\partial_k b) 
                      + (\D_\ada \la^k )(\bD^\da\bla^\bj)(\partial_\bj\partial_k b) \nonumber\\*&\quad
                      + (\bD^\da\D_\ada \bla^\bk) (\partial_\bk b) 
                      + (\D_\ada \bla^\bk)(\bD^\da\bla^\bj) (\p_\bj\partial_\bk b), \\
  \D^\ada \D_\ada b &= 
         (\D^\ada \D_\ada \la^k) (\p_k b) 
         + (\D^\ada \D_\ada \bla^\bk)( \p_\bk b) \nonumber\\*&\relphantom{=}
	 + (\D^\ada \la^k)(\D_\ada \la^l) (\p_l \p_k b)  \nonumber\\*&\relphantom{=}
	 + 2(\D^\ada \la^k)(\D_\ada \bla^\bl)(\p_k\p_\bl b) \nonumber\\*&\relphantom{=}
	 + (\D^\ada \bla^\bk)(\D_\ada \bla^\bl)(\p_\bk \p_\bl b)
\end{align}
\end{subequations}
and similarly for $\beta^i$. This yields
\begin{subequations}
\begin{align}\label{eq:F}
\begin{split}
   \cF_{0}&=  \cC_{0} 
     + \beta^i (\p_i \cB) \cdot \cA  \\&\quad
     + (\D^\ia\la^j)(\p_j\beta^i)\cE^i_\ia  \\&\quad
     + [ (\D^2 \la^j)(\p_j\beta^i)+(\D^\ia \la^j)(\D_\ia\la^k)(\p_j\p_k \beta^i) ]\cE_2^i \\&\quad
     + [(\D^\ada \la^j)(\p_j\beta^i) + (\D^\ada \bla^\bj) (\p_\bj\beta^i) ] \cE_\ada^i \\&\quad
     + [ (\D_\ada\D^\ia \la^j)(\p_j\beta^i) 
         + (\D^\ia\la^j)(\D_\ada\la^k)(\p_k\p_j\beta^i) \\&\qquad
	 + (\D^\ia\la^j)(\D_\ada\bla^\bk)(\p_\bk\p_j\beta^i) ] \bcE_3^{\da i} \\&\quad
     + [ (\D^\ada\D_\ada \la^k)(\p_k\beta^i) 
         + (\D^\ada\D_\ada \bla^\bk)(\p_\bk\beta^i) \\&\qquad
	 + (\D^\ada \la^k)(\D_\ada\la^j)(\p_j\p_k\beta^i) 
         + 2(\D^\ada \la^k)(\D_\ada \bla^\bj) (\p_\bj\p_k \beta^i) \\&\qquad
         + (\D^\ada\bla^\bk)(\D_\ada\bla^\bj)(\p_\bj\p_\bk \beta^i)] \cE_4^i ,
 \end{split}\\[1ex]
\begin{split}
   \cF_\ia  &=  \cC_\ia 
     + \beta^i \cE_\ia^i 
     + 2 (\D_\ia\la^j)(\p_j\beta^i)\cE_2^i \\&\quad
     + [ (\D_\ada \la^k)(\p_k\beta^i) + (\D_\ada \bla^\bk)(\p_\bk\beta^i) ] \bcE_3^{\da i},
 \end{split}\\[1ex]
\begin{split}
   \cF_{2}  &=  \cC_{2} + \beta^i \cE_2^i,
 \end{split}\\[1ex]
\begin{split}
   \cF_\ada &= \cC_\ada + \beta^i \cE_\ada^i 
     + (\D_\ia\la^j)(\p_j\beta^i)\bcE_{3\;\da}^i \\&\quad
     + 2 (\D_\ada\la^k)(\p_k\beta^i)\cE_4^i 
     + 2 (\D_\ada \bla^\bk)(\p_\bk \beta^i) \cE_4^i    ,
\end{split}\\[1ex]
   \bcF^\da_{3} &= \bcC^\da_{3} + \beta^i \bcE_3^{i\da}, \\[1ex]
   \cF_{4} &= \cC_{4} + \beta^i \cE_4^i .
\end{align}
\end{subequations}
%



\begin{savequote}[\savequotewidth]
  I have made this
  longer, because I have not had the time to
  make it shorter.  
  \qauthor{Blaise Pascal, ``Lettres provinciales''}
\end{savequote}

\Chapter{Coefficient Consistency Equations\label{app:coeffconsist}}
Consistency equation \eqref{eq:sds-wz3} yields
\begin{align}
  2 \bb^{(D)} - 2 b^{(D)} + \beta^i \bb^{(\bar I)}_i - \bbeta^\bi \bb^{(\bar I)}_\bi = 0.
\end{align}
From consistency equation \eqref{eq:sds-wz2} one obtains
\begin{subequations}
\begin{align}
     &-\bb^{(A)}  + b^{(A)} = 0,\\[2ex]
\begin{split}
     &
 	      \bb^{(\bar N)}_{ij} \beta^i 
 	      + (\p_j\beta^i) \bb^{(\bar I)}_i 
 	      + 2 \p_j \bb^{(D)} 
 	      + (\p_j \bb^{(\bar I)}_i) \beta^i
 	      + b^{(I)}_j 
 	      + i b^{(H)}_j 
 	      + b^{(M)}_{j \bi} \bbeta^\bi \\&\qquad
 	      + (\p_j \bbeta^\bi) b^{(\bar I)}_\bi 
 	      + 2 (\p_j b^{(D)} )
 	      + (\p_j b^{(\bar I)}_\bi) \bbeta^\bi 
 	      - 2 b^{(G)}_j 
 	      - 2i \bbeta^\bi b^{(O)}_{j \bi}
 	       = 0,
\end{split} \\[2ex]
\begin{split}
     &
 	      b^{(\bar N)}_{\bi\bj} \beta^\bi 
 	      + (\p_\bj\bbeta^\bi) b^{(\bar I)}_\bi 
 	      + 2 \p_\bj b^{(D)} 
 	      + (\p_\bj b^{(\bar I)}_\bi) \bbeta^\bi 
 	      + \bb^{(I)}_\bj 
 	      - i \bb^{(H)}_\bj 
 	      + \bb^{(M)}_{i\bj} \beta^i \\&\qquad
 	      + (\p_\bj \beta^i) \bb^{(\bar I)}_i 
 	      + 2 (\p_\bj \bb^{(D)} )
 	      + (\p_\bj \bb^{(\bar I)}_i) \beta^i 
 	      - 2 \bb^{(G)}_\bj 
 	      + 2i \beta^i \bb^{(O)}_{\bj i}
 	      = 0 ,
\end{split} \\[2ex]
\begin{split}
              &{-} i \bb^{(L)}_{i\bj} 
	      + \bb^{(\bar T)}_{ki\bj} \beta^k
	      - i (\p_i \bb^{(G)}_\bj )
	      - (\p_i \bb^{(O)}_{\bj k}) \beta^k
	      + i b^{(L)}_{\bj i}  \\&\qquad
	      + b^{(\bar T)}_{\bk \bj i} \bbeta^\bk
	      + i (\p_\bj b^{(G)}_i )
	      - (\p_\bj b^{(O)}_{i \bk}) \bbeta^\bk = 0.
\end{split}
\end{align}
\end{subequations}
The following sets of equations have to be augmented by their
complex conjugates. From consistency condition \eqref{eq:ss-wz}
one gets
\begin{subequations}
\begin{align}&
     - 2 b^{(D)} 
     + \quart b^{(B)} 
     + b^{(A)}
     +  2 b^{(D)} 
     - \beta^i b^{(E)}_i = 0, \\[2ex]
\begin{split}&
     - 2 b^{(E)}_j 
     + 2 \beta^i b^{(K)}_{ij}
     + b^{(E)}_{i} (\p_j\beta^i) 
     + \quart (\p_j b^{(B)})  \\&\qquad
     + 2 (\p_j b^{(D)} )
     - \beta^i (\p_j b^{(E)}_i) 
     + b^{(N)}_{ij} \beta^i = 0, 
\end{split} \\[2ex]
\begin{split}&
     2 b^{(\bar F)}_j 
     + b^{(\bar F)}_{i}(\p_j\beta^i) 
     + \half (\p_j b^{(\bar C)})  \\&\qquad
     + b^{(\bar F)}_j 
     - \beta^i (\p_j b^{(\bar F)}_i) 
     + 8 \beta^i b^{(Q)}_{ij} = 0,
\end{split} \\[2ex] &
     - \tfrac{i}{2} b^{(\bar H)}_\bj 
     - b^{(\bar I)}_\bj 
     + \beta^i b^{(L)}_{i\bj} 
     + b^{(\bar E)}_\bj 
     - 4 \beta^i b^{(P)}_{i\bj}
     + (\p_\bj b^{(A)})
     - b^{(\bar I)}_\bj 
     + i b^{(\bar H)}_\bj = 0, \\[2ex]&
     - \ihalf b^{(\bar I)}_\bj 
     - i b^{(\bar E)}_\bj 
     + 4i \beta^i b^{(P)}_{i\bj} 
     + \ihalf b^{(\bar I)}_\bj 
     + \half b^{(\bar H)}_\bj 
     + \ihalf \beta^i b^{(M)}_{i\bj} = 0, \\[2ex]&   
     -\tfrac{i}{2} b^{(M)}_{j\bk} 
     + \beta^i b^{(T)}_{ji\bk} 
     + i b^{(L)}_{j\bk} 
     + \ihalf b^{(N)}_{ij} (\p_\bk \beta^i)   
     + \ihalf \beta^i (\p_\bk b^{(N)}_{ij}) 
     - b^{(T)}_{ij\bk} \beta^i = 0, \label{eq:example-wz}
 \\[2ex]
\begin{split}&
     - i b^{(\bar N)}_{\bj\bk} 
     + \beta^i b^{(\bar T)}_{\bj i\bk}
     - i b^{(\bar J)}_{\bj\bk} 
     + 2i \beta^i b^{(\bar S)}_{\bj\bk i} \\&\qquad
     + \ihalf (\p_\bk b^{(\bar I)}_\bj - i \p_\bk b^{(\bar H)}_\bj)
     + \ihalf (\p_\bk \beta^i) b^{(M)}_{i\bj}
     + \ihalf (\p_\bk b^{(M)}_{i\bj}) \beta^i = 0,
\end{split} \\[2ex] 
\begin{split}&
     4 b^{(Q)}_{kj} 
     + 2 \beta^i b^{(R)}_{ijk} 
     + 2 b^{(Q)}_{ik}(\p_j\beta^i) 
     + \quart (\p_j b^{(\bar F)}_k )\\&\qquad
     - \quart b^{(K)}_{kj} 
     - 2 \beta^i (\p_j b^{(Q)}_{ik})
     - \beta^i b^{(R)}_{jki} = 0, 
\end{split} \\[2ex]&
     2 b^{(P)}_{i\bj} 
     + b^{(P)}_{k\bj} (\p_i\beta^k) 
     + \quart (\p_i b^{(\bar E)}_\bj)
     - \beta^k (\p_i b^{(P)}_{k\bj})
     + \half b^{(L)}_{i\bj}
     + \ihalf b^{(T)}_{ki\bj}\beta^k = 0, \\[2ex]&
     2 b^{(R)}_{jkl} 
     + 4 \beta^i b^{(V)}_{ijkl} 
     + b^{(R)}_{kli}(\p_j\beta^i)
     + \quart \p_j b^{(K)}_{kl}
     - \beta^i (\p_j b^{(R)}_{kli}) = 0, \\[2ex]
\begin{split}&
     2 b^{(\bar S)}_{\bi\bj k} 
     - \tfrac{i}{2} b^{(\bar T)}_{\{\bi\bj\}k}
     + 2 \beta^l b^{(U)}_{lk\bi\bj} 
     + b^{(\bar S)}_{\bi\bj l}(\p_k\beta^l)
     + \quart \p_k b^{(\bar J)}_{\bi\bj} \\&\qquad
     - \beta^l (\p_k b^{(\bar S)}_{\bi\bj l})
     + \half (\p_\bi b^{(L)}_{k\bj})
     + \ihalf b^{(T)}_{lk\bi} (\p_\bj \beta^l)   
     + \ihalf (\p_\bj b^{(T)}_{lk\bi} ) \beta^l = 0, 
\end{split}
\end{align}
\end{subequations}
while consistency equation \eqref{eq:sds-wz1} yields
\begin{subequations}
\begin{align}
\begin{split}
  &\beta^l (\p_l \bb^{(A)}) - \bbeta^\bl (\p_\bl b^{(A)}) = 0,
\end{split}\\[2ex]
\begin{split}
  &\beta^l (\p_l \bb^{(\bar B)}) - \bbeta^\bl (\p_\bl b^{(B)}) = 0, 
\end{split}\\[2ex]
\begin{split}
  &  - 3 \bb^{(\bar C)} 
               + \beta^l (\p_l \bb^{(\bar C)})
     - 3 b^{(C)} 
               -\bbeta^\bl (\p_\bl b^{(C)})
               = 0, 
\end{split}\\[2ex]
\begin{split}
  & +2 \bb^{(\bar D)} 
               + \quart(\bb^{(D)}-\bb^{(A)}) 
               - \quart (\bb^{(B)}+2\bb^{(\bar D)}) 
               + \beta^l \bb^{(\bar G)}_l 
               + \beta^l \bb^{(\bar E)}_l \\&\quad
               - \iquart \beta^l \bb^{(\bar H)}_l
               + \beta^l (\p_l \bb^{(\bar D)})
    + \quart (b^{(D)}-b^{(A)}) 
               + \iquart \bbeta^\bl b^{(\bar H)}_\bl 
               -\bbeta^\bl (\p_\bl b^{(D)}) =0, 
\end{split}\\[2ex]
\begin{split}&
  + 2 \bb^{(\bar E)}_i 
               + 2 \bb^{(\bar G)}_i 
               + \bb^{(\bar I)}_i 
               - \p_k (\quart \bb^{(B)} 
               + 2 \bb^{(\bar D)} ) 
               - \bb^{(F)}_i 
               + 2 \beta^l \bb^{(\bar J)}_{li} 
               + \beta^l \p_i \bb^{(\bar E)}_l \\&\quad
               + 8 \beta^l \bb^{(\bar Q)}_{li} 
               +\beta^l (\p_l \bb^{(\bar E)}_i)
     + b^{(I)}_i
               + 2 b^{(E)}_i 
               - 8 \bbeta^\bl b^{(\bar P)}_{\bl i} 
               - \bbeta^\bl (\p_\bl b^{(E)}_i)
               = 0, 
\end{split}\\[2ex]
\begin{split}&
  - 3 \bb^{(\bar F)}_\bi 
               + 2 i \beta^l \bb^{(\bar O)}_{l\bi} 
               +\beta^l (\p_l \bb^{(\bar F)}_\bi)
     - b^{(F)}_\bi \\&\quad 
               + \half \p_\bi b^{(C)} 
               - 2 \bbeta^\bl b^{(\bar K)}_{\bl\bi}
               - \bbeta^\bl \p_\bi b^{(F)}_\bl 
               -\bbeta^\bl (\p_\bl b^{(F)}_\bi)
               = 0,
\end{split}\\[2ex]
\begin{split}&
  +2 \p_i \bb^{(\bar D)} 
               + 2 \bb^{(\bar E)}_i 
               + 2 \bb^{(\bar G)}_i 
               + \bb^{(\bar I)}_i 
               - \p_i(\half \bb^{(B)} + 4 \bb^{(\bar D)})
               + \beta^l \p_i \bb^{(\bar G)}_l \\&\quad
               + 2 \beta^l \bb^{(\bar J)}_{li} 
               + 2 \beta^l \p_i \bb^{(\bar E)}_l 
               +\beta^l (\p_l \bb^{(\bar G)}_i)
      +\ihalf b^{(H)}_i 
               - b^{(I)}_i 
               - b^{(G)}_i \\&\quad
               + b^{(E)}_i 
               - \bbeta^\bl b^{(L)}_{\bl i} 
               - 4 \bbeta^\bl b^{(P)}_{\bl i}
               -\bbeta^l (\p_\bl b^{(G)}_i)
               = 0,
\end{split}\\[2ex]
\begin{split}&
  - i \p_i (\bb^{(D)} - \bb^{(A)}) 
               + \half \bb^{(\bar H)}_l \p_i \beta^l 
               - \half \beta^l \p_i \bb^{(\bar H)}_l 
               + \beta^l (\p_l \bb^{(\bar H)}_i)
         - b^{(H)}_i 
               - 2i b^{(G)}_i  \\&\qquad
               - i \p_i (b^{(D)} - b^{(A)})
               - 2i b^{(E)}_i
               - 2i \bbeta^\bl b^{(L)}_{\bl i}
               - 4 \bbeta^\bl b^{(\bar O)}_{\bl i}
               + 8 i \bbeta^\bl b^{(P)}_{\bl i}\\&\qquad 
               - \half b^{(\bar H)}_\bl \p_i \bbeta^\bl 
               + \half \bbeta^\bl \p_i b^{(\bar H)}_\bl
               -\bbeta^\bl (\p_\bl b^{(H)}_i)
               = 0, 
\end{split}\\[2ex]
\begin{split}&
   -\half(\bb^{(\bar I)}_i - i \bb^{(\bar H)}_i)
               + \p_i \bb^{(D)} 
               - \half \beta^l \bb^{(\bar N)}_{li} 
               + \half \beta^l \p_i \bb^{(\bar I)}_l 
               + \half \bb^{(\bar I)}_l \p_i \beta^l
               + \beta^l (\p_l \bb^{(\bar I)}_i) \\&\quad
       -b^{(I)}_i 
               + \half (b^{(I)}_i + i b^{(H)}_i)
               - 2 b^{(E)}_i
               - \p_i b^{(D)} 
               - 2i \bbeta^\bl b^{(\bar O)}_{\bl i} 
               + 8 \bbeta^\bl b^{(P)}_{\bl i}\\&\quad
               + \half \bbeta^\bl b^{(M)}_{\bl i}
               - \half \bbeta^\bl \p_i b^{(\bar I)}_\bl 
               - \half b^{(\bar I)}_\bl \p_i \bbeta^\bl
               - \bbeta^\bl (\p_\bl b^{(I)}_i)
               =0,
\end{split}\\[2ex]
\begin{split}&
   + 2 \p_j (\bb^{(\bar E)}_i + \bb^{(\bar G)}_i + \half \bb^{(\bar I)}_i )
               - \quart \p_i \p_j \bb^{(B)} 
               - 2 \p_i \p_j \bb^{(\bar D)} 
               - 2 \p_i \bb^{(F)}_j \\&\quad
               + 2 \beta^l \p_i \bb^{(\bar J)}_{lj} 
               + \beta^l \p_i\p_j \bb^{(\bar E)}_l
               + \beta^l (\p_l \bb^{(\bar J)}_{ij})
     + 2 b^{(N)}_{ij} 
               + 2 b^{(J)}_{ij} 
               - 4 \bbeta^\bl b^{(S)}_{ij\bl}\\&\quad
               + 2i \bbeta^\bl b^{(T)}_{i\bl j} 
               - \bbeta^\bl (\p_\bl b^{(J)}_{ij})
               = 0, 
\end{split}\\[2ex]
\begin{split}&
  + 4 \bb^{(\bar K)}_{ij} 
               - 2 \p_j \bb^{(F)}_i 
               - 16 \bb^{(\bar Q)}_{ij} 
               - \half \p_i \p_j \bb^{(C)}   
               + 2 \beta^l \p_i \bb^{(\bar K)}_{lj}  
               + 8 \beta^l \bb^{(\bar R)}_{lij}\\&\quad
               + \beta^l \p_i \p_j \bb^{(F)}_l 
               + 16 \beta^l \p_i \bb^{(\bar Q)}_{lj} 
               - 4 \beta^l \bb^{(\bar R)}_{ijl}
               + \beta^l (\p_l \bb^{(\bar K)}_{ij})
     + 3 b^{(K)}_{ij} \\&\quad
               - \bbeta^\bl (\p_\bl b^{(K)}_{ij})
               = 0, 
\end{split}\\[2ex]
\begin{split}&
   \ihalf \p_i \bb^{(H)}_\bj 
               + \p_i \bb^{(I)}_\bj
               + \p_i \bb^{(G)}_\bj 
               + \bb^{(M)}_{i\bj} 
               - 2i \bb^{(O)}_{\bj i}
               + 2i \bb^{(\bar O)}_{i\bj} 
               - 8 \bb^{(P)}_{i\bj}\\&\quad
               - 2 \p_i \bb^{(E)}_\bj 
               + \beta^l \p_i \bb^{(L)}_{l\bj} 
               + 8 \beta^l \bb^{(\bar S)}_{li\bj} 
               - 2i \beta^l b^{(T)}_{il\bj} 
               + 8 \beta^l \p_i \bb^{(P)}_{l\bj}\\&\quad
               + \beta^l (\p_l \bb^{(L)}_{i\bj})
       + \ihalf \p_\bj b^{(H)}_i 
               - \p_\bj b^{(I)}_i
               - \p_\bj b^{(G)}_i
               - b^{(M)}_{i\bj} 
               - 2i b^{(O)}_{i\bj}\\&\quad
               + 2i b^{(\bar O)}_{\bj i} 
               + 8 b^{(P)}_{i\bj}
               + 2 \p_\bj b^{(E)}_i 
               - \bbeta^\bl \p_\bj b^{(L)}_{\bl i} 
               - 8 \bbeta^\bl b^{(\bar S)}_{\bl\bj i} 
               - 2i \bbeta^\bl \bb^{(T)}_{\bj\bl i}\\&\quad 
               - 8 \bbeta^\bl \p_\bj b^{(P)}_{\bl i}
               - \bbeta^\bl (\p_\bl b^{(L)}_{\bj i})
               = 0, 
\end{split}\\[2ex]
\begin{split}&
  \bb^{(M)}_{i\bj} 
               - 2i \bb^{(\bar O)}_{\bj i}
               - \half \p_i (\bb^{(I)}_\bj - i \bb^{(H)}_\bj) 
               - \half \p_\bj (\bb^{(\bar I)}_i - i \bb^{(\bar H)}_i) 
               + 2 \p_i\p_\bj \bb^{(D)}\\&\quad
               - 2 i \beta^l \bb^{(\bar T)}_{il\bj}
               + \half \bb^{(M)}_{l\bj} \p_i \beta^l
               + \half \bb^{(\bar N)}_{li} \p_\bj \beta^l 
               - \half \beta^l \p_i \bb^{(M)}_{l\bj} 
               - \half \beta^l \p_\bj \bb^{(\bar N)}_{li}\\&\quad
               + \beta^l \p_i \p_\bj \bb^{(\bar I)}_l
               + \bb^{(\bar I)}_l \p_i \p_\bj \beta^l
               + \beta^l (\p_l \bb^{(M)}_{i\bj})
     - b^{(M)}_{i\bj} 
               - 2i b^{(\bar O)}_{\bj i}\\&\quad
               + \half \p_i (b^{(I)}_\bj + i b^{(H)}_\bj) 
               + \half \p_\bj (b^{(\bar I)}_i + i b^{(\bar H)}_i) 
               - 2 \p_i\p_\bj b^{(D)}\\&\quad
               - 2 i \beta^l b^{(\bar T)}_{il\bj}
               - \half b^{(M)}_{l\bj} \p_i \beta^l
               - \half b^{(\bar N)}_{li} \p_\bj \beta^l 
               + \half \beta^l \p_i b^{(M)}_{l\bj} 
               + \half \beta^l \p_\bj b^{(\bar N)}_{li}\\&\quad
               - \beta^l \p_i \p_\bj b^{(\bar I)}_l
               - b^{(\bar I)}_l \p_i \p_\bj \beta^l
               - \beta^l (\p_l b^{(M)}_{i\bj})
               = 0, 
\end{split}\\[2ex]
\begin{split}&
  - \half \p_j (\bb^{(\bar I)}_i - i \bb^{(\bar H)}_i)
               + \p_i \p_j \bb^{(\bar D)} 
               + \half \bb^{(\bar N)}_{lj} \p_i \beta^l
               - \half \beta^l \p_i \bb^{(\bar N)}_{lj} \\&\quad
               + \half \beta^l \p_i\p_j \bb^{(\bar I)}_l 
               + \half \bb^{(\bar I)}_l \p_i \p_j \beta^l
               + \beta^l (\p_l \bb^{(\bar N)}_{ij})
     - 2 b^{(N)}_{ij} 
               + \half \p_j (b^{(I)}_i 
               + i b^{(H)}_i)  \\&\quad
               - b^{(J)}_{ij}  
               - \p_i \p_j b^{(D)}  
               - 2 i \bbeta^\bl b^{(T)}_{i\bl j}  
               + 4 \bbeta^\bl b^{(S)}_{ij\bl}  
               - \half b^{(M)}_{\bl j} \p_i \bbeta^\bl
               + \half \bbeta^\bl \p_i b^{(M)}_{\bl j} \\&\quad
               - \half \bbeta^\bl \p_i \p_j b^{(\bar I)}_\bl  
               - \half b^{(\bar I)}_\bl \p_i \p_j \bbeta^\bl
               - \bbeta^\bl (\p_\bl b^{(N)}_{ij})
                = 0, 
\end{split}\\[2ex]
\begin{split}&
  \ihalf \p_i \bb^{(I)}_\bj 
               + 2 \bb^{(\bar O)}_{i\bj} 
               + 8 i \bb^{(P)}_{i\bj}  
               - \half (\half \bb^{(\bar O)}_{i\bj} - \half \bb^{(O)}_{i\bj} - i \bb^{(L)}_{i\bj}) 
               + 2 i \p_i \bb^{(E)}_\bj \\&\quad
               + \beta^l \p_i \bb^{(\bar O)}_{l\bj}  
               - 8i \beta^l \bb^{(\bar S)}_{li\bj} 
               - 8i \beta^l \p_i b^{(P)}_{l\bj}
               - \half \beta^l \bb^{(\bar T)}_{li\bj}
               + \beta^l (\p_l \bb^{(\bar O)}_{i\bj})
     + \ihalf b^{(M)}_{\bj i} \\&\quad
               - b^{(O)}_{i\bj} 
               + \half (\half b^{(\bar O)}_{\bj i} - \half b^{(O)}_{\bj i} + i b^{(L)}_{\bj i}) 
               - \bbeta^\bl b^{(\bar T)}_{\bj\bl i}  
               +\half \bbeta^\bl b^{(\bar T)}_{\bl\bj i}
               +\bbeta^\bl (\p_\bl b^{(O)}_{i\bj})
               = 0, 
\end{split}\\[2ex]
\begin{split}&
  \ihalf \bb^{(\bar O)}_{i\bj} 
               - 2 \bb^{(P)}_{i\bj} 
               - \quart \p_i \bb^{(E)}_\bj 
               + 2 \beta^l \bb^{(\bar S)}_{li\bj} 
               + \beta^l \p_i \bb^{(P)}_{l\bj}
               +\beta^l (\p_l \bb^{(P)}_{i\bj})
               \\&\quad
     + \ihalf b^{(\bar O)}_{i\bj} 
               + 2 b^{(P)}_{i\bj} 
               + \quart \p_\bj b^{(E)}_i 
               - 2 \bbeta^\bl b^{(\bar S)}_{\bl\bj i} 
               - \bbeta^\bl \p_\bj b^{(P)}_{\bl i}
               - \bbeta^\bl (\p_\bl b^{(P)}_{i\bj})
               = 0, 
\end{split}\\[2ex]
\begin{split}&
  - \bb^{(\bar Q)}_{ij} 
               - \quart \p_i \bb^{(F)}_j 
               - \quart \bb^{(\bar K)}_{ij} 
               + 2 \beta^l \bb^{(\bar R)}_{lij} 
               + 2\beta^l \p_i \bb^{(\bar Q)}_{lj}\\&\quad
               + \beta^l \bb^{(\bar R)}_{ijl}
               +\beta^l (\p_l \bb^{(\bar Q)}_{ij})
      + 3 b^{(Q)}_{ij} 
               - \bbeta^\bl (\p_\bl b^{(Q)}_{ij})
               = 0, \\
\end{split}\\[2ex]
\begin{split}&
   + 3 \bb^{(\bar R)}_{ijk} 
               - 4 \p_i \bb^{(\bar Q)}_{jk} 
               - \quart \p_k \bb^{(\bar K)}_{ij}  
               + \half \p_i \bb^{(\bar K)}_{kj} 
               + 2 \beta^l \p_i \bb^{(\bar R)}_{ljk} \\&\quad
               + 4 \beta^l (\bb^{(\bar V)}_{lkij} + \bb^{(\bar V)}_{likj} )  
               + 2 \beta^l \p_i \p_j \bb^{(\bar Q)}_{lk}
               + \beta^l \p_k \bb^{(\bar R)}_{ijl}
               - 2 \beta^l \p_i \bb^{(\bar R)}_{jkl}  \\&\quad
               + \beta^l (\p_l \bb^{(\bar R)}_{ijk})
     + 3 b^{(R)}_{ijk} 
               -\bbeta^\bl (\p_\bl b^{(R)}_{ijk})
               = 0, 
\end{split}\\[2ex]
\begin{split}&
    \p_i (\ihalf \bb^{(\bar O)}_{j\bk} -2 \bb^{(P)}_{j\bk} ) 
               - \quart \p_i \p_j \bb^{(E)}_\bk 
               + 2 \beta^l \p_i \bb^{(\bar S)}_{lj\bk} 
               + \beta^l \p_i\p_j \bb^{(P)}_{l\bk} \\&\quad
               + \beta^l (\p_l \bb^{(\bar S)}_{ij\bk})
     +2 b^{(S)}_{ij\bk} 
               + \ihalf b^{(T)}_{ij\bk} 
               + \quart \p_i \p_j b^{(F)}_\bk 
               + \quart \p_\bk b^{(J)}_{ij}  
               - 2 \bbeta^\bl b^{(U)}_{ij\bk\bl} \\&\quad
               - \bbeta^\bl (\p_\bk b^{(S)}_{ij\bl})
               - \bbeta^\bl (\p_\bl b^{(S)}_{ij\bk})
               = 0, 
\end{split}\\[2ex]
\begin{split}&
  \ihalf \p_j \bb^{(M)}_{i\bk} 
               + \p_j \bb^{(O)}_{\bk i}
               - \half \p_i (\half \bb^{(\bar O)}_{j\bk} - \half \bb^{(O)}_{j\bk}-i\bb^{(L)}_{j\bk})
               + \beta^l \p_j \bb^{(\bar T)}_{il\bk}  \\&\quad
               + \half \bb^{(\bar T)}_{lj\bk} \p_i \beta^l
               - \half \beta^l \p_i \bb^{(\bar T)}_{lj\bk}
               +\beta^l (\p_l \bb^{(\bar T)}_{ij\bk})
     + i \p_\bk b^{(N)}_{ij} 
               + 8i b^{(S)}_{ji\bk}  \\&\quad
               - 2 b^{(T)}_{ji\bk} 
               + \half \p_i (\half b^{(\bar O)}_{\bk j} - \half b^{(O)}_{\bk j} + ib^{(L)}_{\bk j})
               + 2i \p_\bk b^{(J)}_{ij} 
               - \bbeta^\bl \p_\bk b^{(T)}_{i\bl j} \\&\quad 
               - 8i \bbeta^\bl b^{(U)}_{ij\bk\bl} 
               - 8i \bbeta^\bl \p_\bk b^{(S)}_{ij\bl}
               - \half b^{(\bar T)}_{\bl\bk j} \p_i \bbeta^\bl 
               + \half \bbeta^\bl \p_i b^{(\bar T)}_{\bl\bk j}
               - \bbeta^\bl (\p_\bl b^{(T)}_{ij\bk})
               = 0, 
\end{split}\\[2ex]
\begin{split}&
  \p_i (-2 \bb^{(S)}_{\bk\bl j} 
               + \ihalf \bb^{(T)}_{\bk\bl j}) 
               - \quart \p_i \p_j \bb^{(J)}_{\bk\bl} 
               + 2 \beta^m \p_i \bb^{(U)}_{mj\bk\bl}  
               + \beta^m \p_i\p_j \bb^{(S)}_{\bk\bl m} \\&\quad
               + \beta^m (\p_m \bb^{(U)}_{ij\bk\bl})
     -\p_\bk (-2 b^{(S)}_{ij \bl} 
               - \ihalf b^{(T)}_{ij\bl}) 
               + \quart \p_\bk \p_\bl b^{(J)}_{ij} 
               - 2 \bbeta^\bm \p_\bk b^{(U)}_{\bm\bl ij}\\&\quad
               - \bbeta^\bm \p_\bk\p_\bl b^{(S)}_{ij\bm}
               - \bbeta^\bm (\p_\bm b^{(U)}_{ij\bk\bl})
               = 0, 
\end{split}\\[2ex]
\begin{split}&
  3 \bb^{(\bar V)}_{ijkl} 
               - 2 \p_i \bb^{(\bar R)}_{jkl} 
               - \quart \p_i \p_j \bb^{(\bar K)}_{kl} 
               + \beta^m \p_i \bb^{(\bar V)}_{mjkl} \\&\quad
               + \beta^m \p_i \p_j \bb^{(\bar R)}_{klm}
               + \beta^m (\p_m \bb^{(\bar V)}_{ijkl})
     + 3 b^{(V)}_{ijkl}
               - \bbeta^\bm (\p_\bm b^{(V)}_{ijkl})
               = 0.
\end{split}
\end{align}
\end{subequations}



\Chapter{Minimal Algebra on Chiral~Fields\label{ch:chiralalgebra}}
The following follows from the superalgebra for a chiral ($\la$) or
antichiral ($\bla$) scalar superfield. Although trivial, these special cases
occur sufficiently frequent to earn explicit treatment,

\begin{subequations}
\begin{align}
  \D^2\bD^2\bla &= (8iG_\ada \D^\ada - 8 \D_\ada \D^\ada + 4 (\bD_\da \bR)\bD^\da + 8 \bR\bD^2 ) \bla,\\
  \D^\ia\D_\ada \la &= \D_\ada\D^\ia\la - 2i G_\ada\D^\ia\la, \\
  \D^\ia\D_\ada\bla &= 2i \bR \bD_\da \bla,\\
  (\D_\ia \bD^2 \bla) &= 4 (G_\ada - i\D_\ada )(\bD^\da \bla),\\
  (\D^\ia \D^2 \la)   &= 4 \bR (\D^\ia \la), \\
  (\D^2 \D_\ia \la) &= -2 \bR (\D_\ia \la), \\
  (\D^2 \bD_\da \bla) &= 4 \bR (\bD_\da \bla), \\
  \D_\ia (\D^\ib \la)(\D_\ib \la) &= - (\D_\ia \la)(\D^2 \la), \\
  (\bD^\da \D_\ada \bla) &= (\D_\ada  \bD^\da \bla) - 2 i G_\ia{}^\da (\bD_\da \bla), \\
  (\bD^\da \D_\ada \la) &= -2 i R(\D_\ia\la), \\
  (\bD_\da \D^2 \la) &= 4 (G_\ada + i \D_\ada)(\D^\ia \la), \\ 
  (\D_\ia \D_\ib \la) &= \half \veps_{\ia\ib} (\D^2 \la), \\
  (\D^\ia \D_\ada \bD^\da \bla) &= 
       -2i \D^\ada \D_\ada \bla 
       + 2i \bR \bD^2 \bla 
       + 4 G_\ada \D^\ada \bla.
\end{align}
\end{subequations}

Weyl variations for derivatives acting on chiral fields of Weyl weight $0$
are given by
\begin{subequations}
\begin{align}
  \delta' \bigl[ \la \bigr] &= 0, \\
  \delta' \bigl[ \D^\ia \la \bigr] 
     &= (\half \sigma' - \bsigma') \D^\ia\la, \\
  \delta' \bigl[ \D_\ada \la \bigr]
     &= -\half(\sigma'+\bsigma') \D_\ada \la
        - \tfrac{i}{2} (\bD_\da \bsigma') \D_\ia \la, \\
  \delta' \bigl[ \D^2 \la \bigr] 
     &= (\sigma'-2\bsigma') \D^2 \la 
        + 2 (\D^\ia \sigma') \D_\ia \la, \\
  \delta' \bigl[ \bD^2 \bla \bigr] 
     &= (\bsigma'-2\sigma') \bD^2 \bla 
        + 2 (\bD_\da \bsigma') \bD^\da \bla, \\
  \delta' \bigl[ \D_\ada \bla \bigr] 
     &= -\half(\sigma'+\bsigma')\D_\ada\bla
        -\tfrac{i}{2} (\D_\ia\sigma')\bD_\da \bla, \\
  \delta' \bigl[ \D^\ada \bD_\da \bla \bigr] 
     &= -\tfrac{3}{2}\sigma'\D^\ada \bD_\da \bla
        + \half (\D^\ada \sigma') \bD_\da \bla\nonumber\\*&\relphantom{=}
        + \tfrac{i}{2} (\D^\ia\sigma')\bD^2 \bla 
        + (\bD_\da \bsigma') \D^\ada \bla  \nonumber\\*&\relphantom{=}
        + \half (\D^\ada \bsigma') \bD_\da \bla, \\
  \delta' \bigl[ \D^\ada \D_\ia \la \bigr] 
     &= -\tfrac{3}{2} \bsigma' \D^\ada \D_\ia \la  
        + \half (\D^\ada \bsigma') \D_\ia \la\nonumber\\*&\relphantom{=}
        - \tfrac{i}{2} (\bD^\da\bsigma')\D^2 \la 
        + (\D_\ia \sigma') \D^\ada \la  \nonumber\\*&\relphantom{=}
        + \half ( \D^\ada \sigma') \D_\ia \la.
\end{align}
\end{subequations}



  \providecommand\bibtitlefont{}
  \bibliographystyle{joslac}
  \bibliography{jo-diss}

\begin{thebibliography}{100}
\expandafter\ifx\csname bibnamefont\endcsname\relax
  \def\bibnamefont#1{\textsc{#1}}\fi
\expandafter\ifx\csname bibtitlefont\endcsname\relax
  \def\bibtitlefont#1{\emph{#1}}\fi
\expandafter\ifx\csname url\endcsname\relax
  \def\url#1{\texttt{#1}}\fi
\expandafter\ifx\csname urlprefix\endcsname\relax\def\urlprefix{URL }\fi
\providecommand{\href}[2]{\url{#2}}

\bibitem{Erdmenger:2006bg}
\bibnamefont{J.~Erdmenger}, \bibnamefont{N.~Evans} and
  \bibnamefont{J.~Gro{\ss}e}.
\newblock \bibtitlefont{Heavy-light mesons from the \acro{AdS}/\acro{CFT}
  correspondence}  (2006).
\newblock
\href{http://www.arXiv.org/abs/hep-th/0605241}{hep-th/0605241}.
\newblock

\bibitem{Apreda:2006ie}
\bibnamefont{R.~Apreda}, \bibnamefont{J.~Erdmenger}, \bibnamefont{N.~Evans},
  \bibnamefont{J.~Gro{\ss}e} and \bibnamefont{Z.~Guralnik}.
\newblock \bibtitlefont{Instantons on {D7} Brane Probes and
  \acro{AdS}/\acro{CFT} with Flavour}.
\newblock Fortsch. Phys. \textbf{54}:266--274 (2006).
\newblock
\href{http://www.arXiv.org/abs/hep-th/0601130}{hep-th/0601130}.
\newblock

\bibitem{Erdmenger:2005bj}
\bibnamefont{J.~Erdmenger}, \bibnamefont{J.~Gro{\ss}e} and
  \bibnamefont{Z.~Guralnik}.
\newblock \bibtitlefont{Spectral flow on the Higgs branch and
  \acro{AdS}/\acro{CFT} duality}.
\newblock JHEP \textbf{06}:052 (2005).
\newblock
\href{http://www.arXiv.org/abs/hep-th/0502224}{hep-th/0502224}.
\newblock

\bibitem{Grosse:2005}
\bibnamefont{J.~Gro{\ss}e}.
\newblock \bibtitlefont{{MathPSfrag}: Creating Publication-Quality Labels in
  {M}athematica Plots}  (2005).
\newblock
\href{http://www.arXiv.org/abs/cs.GR/0510087}{cs.GR/0510087}.
\newblock

\bibitem{tHooft:1973jz}
\bibnamefont{G.~'t~Hooft}.
\newblock \bibtitlefont{A planar diagram theory for string interactions}.
\newblock Nucl. Phys. \textbf{B72}:461
 (1974).
\newblock

\bibitem{Gopakumar:2003ns}
\bibnamefont{R.~Gopakumar}.
\newblock \bibtitlefont{From free fields to \acro{AdS}}.
\newblock Phys. Rev. \textbf{D70}:025009 (2004).
\newblock
\href{http://www.arXiv.org/abs/hep-th/0308184}{hep-th/0308184}.
\newblock

\bibitem{Maldacena:1998re}
\bibnamefont{J.~M. Maldacena}.
\newblock \bibtitlefont{The large {$N$} limit of superconformal field theories
  and supergravity}.
\newblock Adv. Theor. Math. Phys. \textbf{2}:231--252 (1998).
\newblock
\href{http://www.arXiv.org/abs/hep-th/9711200}{hep-th/9711200}.
\newblock

\bibitem{Witten:1998qj}
\bibnamefont{E.~Witten}.
\newblock \bibtitlefont{Anti-de Sitter space and holography}.
\newblock Adv. Theor. Math. Phys. \textbf{2}:253--291 (1998).
\newblock
\href{http://www.arXiv.org/abs/hep-th/9802150}{hep-th/9802150}.
\newblock

\bibitem{Gubser:1998bc}
\bibnamefont{S.~S. Gubser}, \bibnamefont{I.~R. Klebanov} and \bibnamefont{A.~M.
  Polyakov}.
\newblock \bibtitlefont{Gauge theory correlators from non-critical string
  theory}.
\newblock Phys. Lett. \textbf{B428}:105--114 (1998).
\newblock
\href{http://www.arXiv.org/abs/hep-th/9802109}{hep-th/9802109}.
\newblock

\bibitem{Freedman:1999gp}
\bibnamefont{D.~Z. Freedman}, \bibnamefont{S.~S. Gubser},
  \bibnamefont{K.~Pilch} and \bibnamefont{N.~P. Warner}.
\newblock \bibtitlefont{Renormalization group flows from holography
  supersymmetry and a {$c$}-theorem}.
\newblock Adv. Theor. Math. Phys. \textbf{3}:363--417 (1999).
\newblock
\href{http://www.arXiv.org/abs/hep-th/9904017}{hep-th/9904017}.
\newblock

\bibitem{Lee:1998bx}
\bibnamefont{S.-M. Lee}, \bibnamefont{S.~Minwalla}, \bibnamefont{M.~Rangamani}
  and \bibnamefont{N.~Seiberg}.
\newblock \bibtitlefont{Three-point Functions of Chiral Operators in {$D = 4$},
  {$\cN = 4$} \acro{SYM} at large {$N$}}.
\newblock Adv. Theor. Math. Phys. \textbf{2}:697--718 (1998).
\newblock
\href{http://www.arXiv.org/abs/hep-th/9806074}{hep-th/9806074}.
\newblock

\bibitem{Nojiri:1998dh}
\bibnamefont{S.~Nojiri} and \bibnamefont{S.~D. Odintsov}.
\newblock \bibtitlefont{Conformal anomaly for dilaton coupled theories from
  AdS/CFT correspondence}.
\newblock Phys. Lett. \textbf{B444}:92--97 (1998).
\newblock
\href{http://www.arXiv.org/abs/hep-th/9810008}{hep-th/9810008}.
\newblock

\bibitem{Nojiri:1998yx}
\bibnamefont{S.~Nojiri} and \bibnamefont{S.~D. Odintsov}.
\newblock \bibtitlefont{Two-boundaries AdS/CFT correspondence in dilatonic
  gravity}.
\newblock Phys. Lett. \textbf{B449}:39--47 (1999).
\newblock
\href{http://www.arXiv.org/abs/hep-th/9812017}{hep-th/9812017}.
\newblock

\bibitem{Nojiri:2000kh}
\bibnamefont{S.~Nojiri}, \bibnamefont{S.~D. Odintsov} and
  \bibnamefont{S.~Ogushi}.
\newblock \bibtitlefont{Finite action in d5 gauged supergravity and dilatonic
  conformal anomaly for dual quantum field theory}.
\newblock Phys. Rev. \textbf{D62}:124002 (2000).
\newblock
\href{http://www.arXiv.org/abs/hep-th/0001122}{hep-th/0001122}.
\newblock

\bibitem{Bogolyubov:1980nc}
\bibnamefont{N.~N. Bogolyubov} and \bibnamefont{D.~V. Shirkov}.
\newblock \bibtitlefont{Introduction to the Theory of Quantized Fields}.
\newblock Intersci. Monogr. Phys. Astron. \textbf{3}:1--720
 (1959).
\newblock

\bibitem{Becchi:1974}
\bibnamefont{C.~Becchi}.
\newblock \bibtitlefont{Current Algebra {W}ard Identities in the Renormalized
  {$\sigma$} Model}.
\newblock Comm. Math. Phys. \textbf{39(4)}:329--344 (1974).
\newblock [euclid.cmp/1103860236].

\bibitem{Epstein:1975gp}
\bibnamefont{H.~Epstein} and \bibnamefont{V.~Glaser}.
\newblock \bibtitlefont{Adiabatic Limit in Perturbation Theory}  (1975).
\newblock In \bibtitlefont{Erice 1975, Proceedings, Renormalization Theory},
  Dordrecht 1976, 193-254 [CERN-TH-1344].

\bibitem{Henningson:1998gx}
\bibnamefont{M.~Henningson} and \bibnamefont{K.~Skenderis}.
\newblock \bibtitlefont{The holographic {W}eyl anomaly}.
\newblock JHEP \textbf{07}:023 (1998).
\newblock
\href{http://www.arXiv.org/abs/hep-th/9806087}{hep-th/9806087}.
\newblock

\bibitem{Akhmedov:1998vf}
\bibnamefont{E.~T. Akhmedov}.
\newblock \bibtitlefont{A Remark on the \acro{AdS}/\acro{CFT} Correspondence
  and the Renormalization Group Flow}.
\newblock Phys. Lett. \textbf{B442}:152--158 (1998).
\newblock
\href{http://www.arXiv.org/abs/hep-th/9806217}{hep-th/9806217}.
\newblock

\bibitem{Karch:1999pv}
\bibnamefont{A.~Karch}, \bibnamefont{D.~L{\"u}st} and \bibnamefont{A.~Miemiec}.
\newblock \bibtitlefont{New {$\cN = 1$} superconformal field theories and their
  supergravity description}.
\newblock Phys. Lett. \textbf{B454}:265--269 (1999).
\newblock
\href{http://www.arXiv.org/abs/hep-th/9901041}{hep-th/9901041}.
\newblock

\bibitem{Fayyazuddin:1998fb}
\bibnamefont{A.~Fayyazuddin} and \bibnamefont{M.~Spalinski}.
\newblock \bibtitlefont{Large {$N$} superconformal gauge theories and
  supergravity orientifolds}.
\newblock Nucl. Phys. \textbf{B535}:219--232 (1998).
\newblock
\href{http://www.arXiv.org/abs/hep-th/9805096}{hep-th/9805096}.
\newblock

\bibitem{Aharony:1998xz}
\bibnamefont{O.~Aharony}, \bibnamefont{A.~Fayyazuddin} and \bibnamefont{J.~M.
  Maldacena}.
\newblock \bibtitlefont{The Large {$N$} Limit of {$\cN = 2,1$} Field Theories
  from Three-Branes in {F}-theory}.
\newblock JHEP \textbf{07}:013 (1998).
\newblock
\href{http://www.arXiv.org/abs/hep-th/9806159}{hep-th/9806159}.
\newblock

\bibitem{Karch:2002sh}
\bibnamefont{A.~Karch} and \bibnamefont{E.~Katz}.
\newblock \bibtitlefont{Adding flavor to \acro{AdS}/\acro{CFT}}.
\newblock JHEP \textbf{06}:043 (2002).
\newblock
\href{http://www.arXiv.org/abs/hep-th/0205236}{hep-th/0205236}.
\newblock

\bibitem{Breitenlohner:1982jf}
\bibnamefont{P.~Breitenlohner} and \bibnamefont{D.~Z. Freedman}.
\newblock \bibtitlefont{Stability in gauged extended supergravity}.
\newblock Ann. Phys. \textbf{144}:249
 (1982).
\newblock

\bibitem{Bertolini:2001qa}
\bibnamefont{M.~Bertolini}, \bibnamefont{P.~Di~Vecchia}, \bibnamefont{M.~Frau},
  \bibnamefont{A.~Lerda} and \bibnamefont{R.~Marotta}.
\newblock \bibtitlefont{{$\cN = 2$} gauge theories on systems of fractional
  {D3}/{D7} branes}.
\newblock Nucl. Phys. \textbf{B621}:157--178 (2002).
\newblock
\href{http://www.arXiv.org/abs/hep-th/0107057}{hep-th/0107057}.
\newblock

\bibitem{Grana:2001xn}
\bibnamefont{M.~Gra{\~n}a} and \bibnamefont{J.~Polchinski}.
\newblock \bibtitlefont{Gauge / gravity duals with holomorphic dilaton}.
\newblock Phys. Rev. \textbf{D65}:126005 (2002).
\newblock
\href{http://www.arXiv.org/abs/hep-th/0106014}{hep-th/0106014}.
\newblock

\bibitem{Kruczenski:2003be}
\bibnamefont{M.~Kruczenski}, \bibnamefont{D.~Mateos}, \bibnamefont{R.~C. Myers}
  and \bibnamefont{D.~J. Winters}.
\newblock \bibtitlefont{Meson spectroscopy in \acro{AdS}/\acro{CFT} with
  flavour}.
\newblock JHEP \textbf{07}:049 (2003).
\newblock
\href{http://www.arXiv.org/abs/hep-th/0304032}{hep-th/0304032}.
\newblock

\bibitem{Myers:2006qr}
\bibnamefont{R.~C. Myers} and \bibnamefont{R.~M. Thomson}.
\newblock \bibtitlefont{Holographic mesons in various dimensions}  (2006).
\newblock
\href{http://www.arXiv.org/abs/hep-th/0605017}{hep-th/0605017}.
\newblock

\bibitem{Sakai:2003wu}
\bibnamefont{T.~Sakai} and \bibnamefont{J.~Sonnenschein}.
\newblock \bibtitlefont{Probing flavored mesons of confining gauge theories by
  supergravity}.
\newblock JHEP \textbf{09}:047 (2003).
\newblock
\href{http://www.arXiv.org/abs/hep-th/0305049}{hep-th/0305049}.
\newblock

\bibitem{Kirsch:2005uy}
\bibnamefont{I.~Kirsch} and \bibnamefont{D.~Vaman}.
\newblock \bibtitlefont{The {D3}/{D7} background and flavor dependence of Regge
  trajectories}.
\newblock Phys. Rev. \textbf{D72}:026007 (2005).
\newblock
\href{http://www.arXiv.org/abs/hep-th/0505164}{hep-th/0505164}.
\newblock

\bibitem{Arean:2006pk}
\bibnamefont{D.~Arean} and \bibnamefont{A.~V. Ramallo}.
\newblock \bibtitlefont{Open String Modes at Brane Intersections}.
\newblock JHEP \textbf{04}:037 (2006).
\newblock
\href{http://www.arXiv.org/abs/hep-th/0602174}{hep-th/0602174}.
\newblock

\bibitem{Nunez:2003cf}
\bibnamefont{C.~N{\'u}{\~n}ez}, \bibnamefont{A.~Paredes} and \bibnamefont{A.~V.
  Ramallo}.
\newblock \bibtitlefont{Flavoring the gravity dual of {$\cN = 1$} Yang-Mills
  with probes}.
\newblock JHEP \textbf{12}:024 (2003).
\newblock
\href{http://www.arXiv.org/abs/hep-th/0311201}{hep-th/0311201}.
\newblock

\bibitem{Wang:2003yc}
\bibnamefont{X.-J. Wang} and \bibnamefont{S.~Hu}.
\newblock \bibtitlefont{Intersecting branes and adding flavors to the
  Maldacena-N{\'u}{\~n}ez background}.
\newblock JHEP \textbf{09}:017 (2003).
\newblock
\href{http://www.arXiv.org/abs/hep-th/0307218}{hep-th/0307218}.
\newblock

\bibitem{Casero:2006pt}
\bibnamefont{R.~Casero}, \bibnamefont{C.~N{\'u}{\~n}ez} and
  \bibnamefont{A.~Paredes}.
\newblock \bibtitlefont{Towards the string dual of {$\cN = 1$} \acro{SQCD}-like
  theories}.
\newblock Phys. Rev. \textbf{D73}:086005 (2006).
\newblock
\href{http://www.arXiv.org/abs/hep-th/0602027}{hep-th/0602027}.
\newblock

\bibitem{Edelstein:2006kw}
\bibnamefont{J.~D. Edelstein} and \bibnamefont{R.~Portugues}.
\newblock \bibtitlefont{Gauge / string duality in confining theories}  (2006).
\newblock
\href{http://www.arXiv.org/abs/hep-th/0602021}{hep-th/0602021}.
\newblock

\bibitem{Casero:2005se}
\bibnamefont{R.~Casero}, \bibnamefont{A.~Paredes} and
  \bibnamefont{J.~Sonnenschein}.
\newblock \bibtitlefont{Fundamental matter, meson spectroscopy and non-critical
  string / gauge duality}.
\newblock JHEP \textbf{01}:127 (2006).
\newblock
\href{http://www.arXiv.org/abs/hep-th/0510110}{hep-th/0510110}.
\newblock

\bibitem{Hirayama:2006jn}
\bibnamefont{T.~Hirayama}.
\newblock \bibtitlefont{A holographic dual of \acro{CFT} with flavor on de
  Sitter space}  (2006).
\newblock
\href{http://www.arXiv.org/abs/hep-th/0602258}{hep-th/0602258}.
\newblock

\bibitem{Janik:2005zt}
\bibnamefont{R.~A. Janik} and \bibnamefont{R.~Peschanski}.
\newblock \bibtitlefont{Asymptotic perfect fluid dynamics as a consequence of
  \acro{AdS}/\acro{CFT}}.
\newblock Phys. Rev. \textbf{D73}:045013 (2006).
\newblock
\href{http://www.arXiv.org/abs/hep-th/0512162}{hep-th/0512162}.
\newblock

\bibitem{Canoura:2005uz}
\bibnamefont{F.~Canoura}, \bibnamefont{J.~D. Edelstein}, \bibnamefont{L.~A.~P.
  Zayas}, \bibnamefont{A.~V. Ramallo} and \bibnamefont{D.~Vaman}.
\newblock \bibtitlefont{Supersymmetric branes on {$\AdS_ \times Y^*(p,q)$} and
  their field theory duals}.
\newblock JHEP \textbf{03}:101 (2006).
\newblock
\href{http://www.arXiv.org/abs/hep-th/0512087}{hep-th/0512087}.
\newblock

\bibitem{Benvenuti:2005qb}
\bibnamefont{S.~Benvenuti}, \bibnamefont{M.~Mahato}, \bibnamefont{L.~A.
  Pando~Zayas} and \bibnamefont{Y.~Tachikawa}.
\newblock \bibtitlefont{The gauge / gravity theory of blown up four cycles}
  (2005).
\newblock
\href{http://www.arXiv.org/abs/hep-th/0512061}{hep-th/0512061}.
\newblock

\bibitem{Kehagias:1999tr}
\bibnamefont{A.~Kehagias} and \bibnamefont{K.~Sfetsos}.
\newblock \bibtitlefont{On running couplings in gauge theories from type-IIB
  supergravity}.
\newblock Phys. Lett. \textbf{B454}:270--276 (1999).
\newblock
\href{http://www.arXiv.org/abs/hep-th/9902125}{hep-th/9902125}.
\newblock

\bibitem{Gubser:1999pk}
\bibnamefont{S.~S. Gubser}.
\newblock \bibtitlefont{Dilaton-driven confinement}  (1999).
\newblock
\href{http://www.arXiv.org/abs/hep-th/9902155}{hep-th/9902155}.
\newblock

\bibitem{Schreiber:2004ie}
\bibnamefont{E.~Schreiber}.
\newblock \bibtitlefont{Excited Mesons and Quantization of String Endpoints}
  (2004).
\newblock
\href{http://www.arXiv.org/abs/hep-th/0403226}{hep-th/0403226}.
\newblock

\bibitem{Glozman:2004gk}
\bibnamefont{L.~Y. Glozman}.
\newblock \bibtitlefont{Chiral and {$\gr{U}(1)_A$} Restorations High in the
  Hadron Spectrum, Semiclassical Approximation and Large {$N_c$}}.
\newblock Int. J. Mod. Phys. \textbf{A21}:475--486 (2006).
\newblock
\href{http://www.arXiv.org/abs/hep-ph/0411281}{hep-ph/0411281}.
\newblock

\bibitem{Shifman:2005zn}
\bibnamefont{M.~Shifman}.
\newblock \bibtitlefont{Highly Excited Hadrons in \acro{QCD} and Beyond}
  (2005).
\newblock
\href{http://www.arXiv.org/abs/hep-ph/0507246}{hep-ph/0507246}.
\newblock

\bibitem{Arean:2007nh}
\bibnamefont{D.~Arean}, \bibnamefont{A.~V. Ramallo} and
  \bibnamefont{D.~Rodriguez-Gomez}.
\newblock \bibtitlefont{Holographic flavor on the Higgs branch}.
\newblock JHEP \textbf{05}:044 (2007).
\newblock
\href{http://www.arXiv.org/abs/hep-th/0703094}{hep-th/0703094}.
\newblock

\bibitem{Arean:2006vg}
\bibnamefont{D.~Arean}, \bibnamefont{A.~V. Ramallo} and
  \bibnamefont{D.~Rodriguez-Gomez}.
\newblock \bibtitlefont{Mesons and Higgs branch in defect theories}.
\newblock Phys. Lett. \textbf{B641}:393--400 (2006).
\newblock
\href{http://www.arXiv.org/abs/hep-th/0609010}{hep-th/0609010}.
\newblock

\bibitem{Constable:1999ch}
\bibnamefont{N.~R. Constable} and \bibnamefont{R.~C. Myers}.
\newblock \bibtitlefont{Exotic scalar states in the \acro{AdS}/\acro{CFT}
  correspondence}.
\newblock JHEP \textbf{11}:020 (1999).
\newblock
\href{http://www.arXiv.org/abs/hep-th/9905081}{hep-th/9905081}.
\newblock

\bibitem{Zamolodchikov:gt}
\bibnamefont{A.~B. Zamolodchikov}.
\newblock \bibtitlefont{`Irreversibility' of the Flux of the Renormalization
  Group in a {2-D} Field Theory}.
\newblock JETP Lett. \textbf{43}:730--732
 (1986).
\newblock

\bibitem{Anselmi:1997am}
\bibnamefont{D.~Anselmi}, \bibnamefont{D.~Z. Freedman}, \bibnamefont{M.~T.
  Grisaru} and \bibnamefont{A.~A. Johansen}.
\newblock \bibtitlefont{Nonperturbative Formulas for Central Functions of
  Supersymmetric Gauge Theories}.
\newblock Nucl. Phys. \textbf{B526}:543--571 (1998).
\newblock
\href{http://www.arXiv.org/abs/hep-th/9708042}{hep-th/9708042}.
\newblock

\bibitem{Anselmi:1997ys}
\bibnamefont{D.~Anselmi}, \bibnamefont{J.~Erlich}, \bibnamefont{D.~Z. Freedman}
  and \bibnamefont{A.~A. Johansen}.
\newblock \bibtitlefont{Positivity Constraints on Anomalies in Supersymmetric
  Gauge Theories}.
\newblock Phys. Rev. \textbf{D57}:7570--7588 (1998).
\newblock
\href{http://www.arXiv.org/abs/hep-th/9711035}{hep-th/9711035}.
\newblock

\bibitem{Cardy:1988cw}
\bibnamefont{J.~L. Cardy}.
\newblock \bibtitlefont{Is there a {$c$} Theorem in Four Dimensions?}
\newblock Phys. Lett. \textbf{B215}:749--752
 (1988).
\newblock

\bibitem{Intriligator:2003jj}
\bibnamefont{K.~Intriligator} and \bibnamefont{B.~Wecht}.
\newblock \bibtitlefont{The Exact Superconformal {R}-Symmetry Maximizes $a$}.
\newblock Nucl. Phys. \textbf{B667}:183--200 (2003).
\newblock
\href{http://www.arXiv.org/abs/hep-th/0304128}{hep-th/0304128}.
\newblock

\bibitem{Osborn:1991gm}
\bibnamefont{H.~Osborn}.
\newblock \bibtitlefont{Weyl consistency conditions and a local renormalization
  group equation for general renormalizable field theories}.
\newblock Nucl. Phys. \textbf{B363}:486--526
 (1991).
\newblock

\bibitem{Osborn:2003vk}
\bibnamefont{H.~Osborn}.
\newblock \bibtitlefont{Local couplings and {$\gr{Sl}(2,\mathbb{R})$}
  invariance for gauge theories at one loop}.
\newblock Phys. Lett. \textbf{B561}:174--182 (2003).
\newblock
\href{http://www.arXiv.org/abs/hep-th/0302119}{hep-th/0302119}.
\newblock

\bibitem{Riegert:1984kt}
\bibnamefont{R.~J. Riegert}.
\newblock \bibtitlefont{A nonlocal action for the trace anomaly}.
\newblock Phys. Lett. \textbf{B134}:56--60
 (1984).
\newblock

\bibitem{Fradkin:1985am}
\bibnamefont{E.~S. Fradkin} and \bibnamefont{A.~A. Tseytlin}.
\newblock \bibtitlefont{Conformal Supergravity}.
\newblock Phys. Rept. \textbf{119}:233--362
 (1985).
\newblock

\bibitem{Manvelyan:1995hz}
\bibnamefont{R.~P. Manvelyan}.
\newblock \bibtitlefont{Superweyl cocycle in {$d=4$} and
  super\-con\-formal-invariant operator}.
\newblock Phys. Lett. \textbf{B373}:306--308 (1996).
\newblock
\href{http://www.arXiv.org/abs/hep-th/9512045}{hep-th/9512045}.
\newblock

\bibitem{deWit:1985bn}
\bibnamefont{B.~de~Wit} and \bibnamefont{M.~T. Grisaru}.
\newblock \bibtitlefont{Compensating Fields and Anomalies}  (1985).
\newblock In 'Quantum Field Theory and Quantum Statistics: Essays in Honor of
  60th birthday of E.S. Fradkin.

\bibitem{Shamir:1992ff}
\bibnamefont{Y.~Shamir}.
\newblock \bibtitlefont{Compensating Fields and Anomalies in Supergravity}.
\newblock Nucl. Phys. \textbf{B389}:323--348 (1993).
\newblock
\href{http://www.arXiv.org/abs/hep-th/9207038}{hep-th/9207038}.
\newblock

\bibitem{Bjorken:1968dy}
\bibnamefont{J.~D. Bjorken}.
\newblock \bibtitlefont{Asymptotic Sum Rules at Infinite Momentum}.
\newblock Phys. Rev. \textbf{179}:1547--1553
 (1969).
\newblock

\bibitem{Feynman:1973xc}
\bibnamefont{R.~P. Feynman}.
\newblock \bibtitlefont{Photon-hadron Interactions}  (1972).
\newblock Reading 1972, 282p.

\bibitem{PDBook}
\bibnamefont{{S. Eidelman \normalfont{et. al.} (Particle Data Group)}}.
\newblock \bibtitlefont{Review of Particle Physics}.
\newblock Phys. Lett. B \textbf{592}:1+ (2004).

\bibitem{Sohnius:1981sn}
\bibnamefont{M.~F. Sohnius} and \bibnamefont{P.~C. West}.
\newblock \bibtitlefont{Conformal Invariance in {$\cN=4$} Supersymmetric
  {Y}ang-{M}ills Theory}.
\newblock Phys. Lett. \textbf{B100}:245
 (1981).
\newblock

\bibitem{D'Hoker:2002aw}
\bibnamefont{E.~D'Hoker} and \bibnamefont{D.~Z. Freedman}.
\newblock \bibtitlefont{Supersymmetric gauge theories and the
  \acro{AdS}/\acro{CFT} correspondence}  (2002).
\newblock
\href{http://www.arXiv.org/abs/hep-th/0201253}{hep-th/0201253}.
\newblock

\bibitem{Dall'Agata:1998wh}
\bibnamefont{G.~Dall'Agata}, \bibnamefont{K.~Lechner} and
  \bibnamefont{M.~Tonin}.
\newblock \bibtitlefont{Action for \acro{IIB} supergravity in 10 dimensions}
  (1998).
\newblock
\href{http://www.arXiv.org/abs/hep-th/9812170}{hep-th/9812170}.
\newblock

\bibitem{Dall'Agata:1998va}
\bibnamefont{G.~Dall'Agata}, \bibnamefont{K.~Lechner} and
  \bibnamefont{M.~Tonin}.
\newblock \bibtitlefont{{$D = 10$}, {$\cN = \text{\acro{IIB}}$} supergravity:
  Lorentz-invariant actions and duality}.
\newblock JHEP \textbf{07}:017 (1998).
\newblock
\href{http://www.arXiv.org/abs/hep-th/9806140}{hep-th/9806140}.
\newblock

\bibitem{Polchinski:1998rr}
\bibnamefont{J.~Polchinski}.
\newblock \bibtitlefont{String theory. Vol. 2: Superstring theory and beyond}
  (1998).
\newblock Cambridge, UK: Univ. Pr. (1998) 531 p.

\bibitem{Green:1982tk}
\bibnamefont{M.~B. Green} and \bibnamefont{J.~H. Schwarz}.
\newblock \bibtitlefont{Extended Supergravity in Ten Dimensions}.
\newblock Phys. Lett. \textbf{B122}:143
 (1983).
\newblock

\bibitem{Howe:1983sr}
\bibnamefont{P.~S. Howe} and \bibnamefont{P.~C. West}.
\newblock \bibtitlefont{The Complete {$\cN=2$}, {$d = 10$} Supergravity}.
\newblock Nucl. Phys. \textbf{B238}:181
 (1984).
\newblock

\bibitem{Schwarz:1983wa}
\bibnamefont{J.~H. Schwarz} and \bibnamefont{P.~C. West}.
\newblock \bibtitlefont{Symmetries and Transformations of Chiral {$\cN=2$} {$D
  = 10$} Supergravity}.
\newblock Phys. Lett. \textbf{B126}:301
 (1983).
\newblock

\bibitem{Schwarz:1983qr}
\bibnamefont{J.~H. Schwarz}.
\newblock \bibtitlefont{Covariant Field Equations of Chiral {$\cN=2$ $D = 10$}
  Supergravity}.
\newblock Nucl. Phys. \textbf{B226}:269
 (1983).
\newblock

\bibitem{Green:1987mn}
\bibnamefont{M.~B. Green}, \bibnamefont{J.~H. Schwarz} and
  \bibnamefont{E.~Witten}.
\newblock \bibtitlefont{Superstring Theory. Vol. 2: Loop Amplitudes, Anomalies
  and Phenomenology} (1987).
\newblock Cambridge, Uk: Univ. Pr. ( 1987) 596 P. (Cambridge Monographs On
  Mathematical Physics).

\bibitem{Polchinski:1995mt}
\bibnamefont{J.~Polchinski}.
\newblock \bibtitlefont{Dirichlet-Branes and {R}amond-{R}amond Charges}.
\newblock Phys. Rev. Lett. \textbf{75}:4724--4727 (1995).
\newblock
\href{http://www.arXiv.org/abs/hep-th/9510017}{hep-th/9510017}.
\newblock

\bibitem{Polchinski:1996fm}
\bibnamefont{J.~Polchinski}, \bibnamefont{S.~Chaudhuri} and \bibnamefont{C.~V.
  Johnson}.
\newblock \bibtitlefont{Notes on {D}-Branes}  (1996).
\newblock
\href{http://www.arXiv.org/abs/hep-th/9602052}{hep-th/9602052}.
\newblock

\bibitem{Kikkawa:1984cp}
\bibnamefont{K.~Kikkawa} and \bibnamefont{M.~Yamasaki}.
\newblock \bibtitlefont{Casimir Effects in Superstring Theories}.
\newblock Phys. Lett. \textbf{B149}:357
 (1984).
\newblock

\bibitem{Dai:1989ua}
\bibnamefont{J.~Dai}, \bibnamefont{R.~G. Leigh} and
  \bibnamefont{J.~Polchinski}.
\newblock \bibtitlefont{New Connection between String Theories}.
\newblock Mod. Phys. Lett. \textbf{A4}:2073--2083
 (1989).
\newblock

\bibitem{Giveon:1994fu}
\bibnamefont{A.~Giveon}, \bibnamefont{M.~Porrati} and
  \bibnamefont{E.~Rabinovici}.
\newblock \bibtitlefont{Target space duality in string theory}.
\newblock Phys. Rept. \textbf{244}:77--202 (1994).
\newblock
\href{http://www.arXiv.org/abs/hep-th/9401139}{hep-th/9401139}.
\newblock

\bibitem{Alvarez:1993qi}
\bibnamefont{E.~Alvarez}, \bibnamefont{L.~Alvarez-Gaume}, \bibnamefont{J.~L.~F.
  Barbon} and \bibnamefont{Y.~Lozano}.
\newblock \bibtitlefont{Some Global Aspects of Duality in String Theory}.
\newblock Nucl. Phys. \textbf{B415}:71--100 (1994).
\newblock
\href{http://www.arXiv.org/abs/hep-th/9309039}{hep-th/9309039}.
\newblock

\bibitem{Paton:1969je}
\bibnamefont{J.~E. Paton} and \bibnamefont{H.-M. Chan}.
\newblock \bibtitlefont{Generalized {V}eneziano Model with Isospin}.
\newblock Nucl. Phys. \textbf{B10}:516--520
 (1969).
\newblock

\bibitem{Leigh:1989jq}
\bibnamefont{R.~G. Leigh}.
\newblock \bibtitlefont{{D}irac-{B}orn-{I}nfeld Action from {D}irichlet Sigma
  Model}.
\newblock Mod. Phys. Lett. \textbf{A4}:2767
 (1989).
\newblock

\bibitem{Myers:1999ps}
\bibnamefont{R.~C. Myers}.
\newblock \bibtitlefont{Dielectric-branes}.
\newblock JHEP \textbf{12}:022 (1999).
\newblock
\href{http://www.arXiv.org/abs/hep-th/9910053}{hep-th/9910053}.
\newblock

\bibitem{Tseytlin:1997cs}
\bibnamefont{A.~A. Tseytlin}.
\newblock \bibtitlefont{On non-abelian generalisation of the Born-Infeld action
  in string theory}.
\newblock Nucl. Phys. \textbf{B501}:41--52 (1997).
\newblock
\href{http://www.arXiv.org/abs/hep-th/9701125}{hep-th/9701125}.
\newblock

\bibitem{Garousi:1998fg}
\bibnamefont{M.~R. Garousi} and \bibnamefont{R.~C. Myers}.
\newblock \bibtitlefont{World-volume interactions on {D}-branes}.
\newblock Nucl. Phys. \textbf{B542}:73--88 (1999).
\newblock
\href{http://www.arXiv.org/abs/hep-th/9809100}{hep-th/9809100}.
\newblock

\bibitem{Aharony:1999ti}
\bibnamefont{O.~Aharony}, \bibnamefont{S.~S. Gubser}, \bibnamefont{J.~M.
  Maldacena}, \bibnamefont{H.~Ooguri} and \bibnamefont{Y.~Oz}.
\newblock \bibtitlefont{Large {$N$} Field Theories, String Theory and Gravity}.
\newblock Phys. Rept. \textbf{323}:183--386 (2000).
\newblock
\href{http://www.arXiv.org/abs/hep-th/9905111}{hep-th/9905111}.
\newblock

\bibitem{Witten:1998zw}
\bibnamefont{E.~Witten}.
\newblock \bibtitlefont{{Anti-de~Sitter} space, thermal phase transition, and
  confinement in gauge theories}.
\newblock Adv. Theor. Math. Phys. \textbf{2}:505--532 (1998).
\newblock
\href{http://www.arXiv.org/abs/hep-th/9803131}{hep-th/9803131}.
\newblock

\bibitem{Burrington:2004id}
\bibnamefont{B.~A. Burrington}, \bibnamefont{J.~T. Liu}, \bibnamefont{L.~A.
  Pando~Zayas} and \bibnamefont{D.~Vaman}.
\newblock \bibtitlefont{Holographic duals of flavored {$\cN = 1$} super
  Yang-Mills: Beyond the probe approximation}.
\newblock JHEP \textbf{02}:022 (2005).
\newblock
\href{http://www.arXiv.org/abs/hep-th/0406207}{hep-th/0406207}.
\newblock

\bibitem{Babington:2003vm}
\bibnamefont{J.~Babington}, \bibnamefont{J.~Erdmenger}, \bibnamefont{N.~J.
  Evans}, \bibnamefont{Z.~Guralnik} and \bibnamefont{I.~Kirsch}.
\newblock \bibtitlefont{Chiral Symmetry Breaking and Pions in
  Non-supersymmetric Gauge / Gravity Duals}.
\newblock Phys. Rev. \textbf{D69}:066007 (2004).
\newblock
\href{http://www.arXiv.org/abs/hep-th/0306018}{hep-th/0306018}.
\newblock

\bibitem{Gell-Mann:1968rz}
\bibnamefont{M.~Gell-Mann}, \bibnamefont{R.~J. Oakes} and
  \bibnamefont{B.~Renner}.
\newblock \bibtitlefont{Behavior of current divergences under SU(3) x SU(3)}.
\newblock Phys. Rev. \textbf{175}:2195--2199
 (1968).
\newblock

\bibitem{Skenderis:2002vf}
\bibnamefont{K.~Skenderis} and \bibnamefont{M.~Taylor}.
\newblock \bibtitlefont{Branes in \acro{AdS} and {pp}-wave spacetimes}.
\newblock JHEP \textbf{06}:025 (2002).
\newblock
\href{http://www.arXiv.org/abs/hep-th/0204054}{hep-th/0204054}.
\newblock

\bibitem{Douglas:1995bn}
\bibnamefont{M.~R. Douglas}.
\newblock \bibtitlefont{Branes within branes}  (1995).
\newblock
\href{http://www.arXiv.org/abs/hep-th/9512077}{hep-th/9512077}.
\newblock

\bibitem{Dorey:1999pd}
\bibnamefont{N.~Dorey}, \bibnamefont{T.~J. Hollowood}, \bibnamefont{V.~V.
  Khoze}, \bibnamefont{M.~P. Mattis} and \bibnamefont{S.~Vandoren}.
\newblock \bibtitlefont{Multi-instanton calculus and the \acro{AdS}/\acro{CFT}
  correspondence in {$\cN = 4$} superconformal field theory}.
\newblock Nucl. Phys. \textbf{B552}:88--168 (1999).
\newblock
\href{http://www.arXiv.org/abs/hep-th/9901128}{hep-th/9901128}.
\newblock

\bibitem{Witten:1995gx}
\bibnamefont{E.~Witten}.
\newblock \bibtitlefont{Small Instantons in String Theory}.
\newblock Nucl. Phys. \textbf{B460}:541--559 (1996).
\newblock
\href{http://www.arXiv.org/abs/hep-th/9511030}{hep-th/9511030}.
\newblock

\bibitem{Douglas:1996uz}
\bibnamefont{M.~R. Douglas}.
\newblock \bibtitlefont{Gauge Fields and {D}-branes}.
\newblock J. Geom. Phys. \textbf{28}:255--262 (1998).
\newblock
\href{http://www.arXiv.org/abs/hep-th/9604198}{hep-th/9604198}.
\newblock

\bibitem{Belavin:1975fg}
\bibnamefont{A.~A. Belavin}, \bibnamefont{A.~M. Polyakov}, \bibnamefont{A.~S.
  Shvarts} and \bibnamefont{Y.~S. Tyupkin}.
\newblock \bibtitlefont{Pseudoparticle solutions of the {Y}ang-{M}ills
  equations}.
\newblock Phys. Lett. \textbf{B59}:85--87
 (1975).
\newblock

\bibitem{Oprisa:2005wu}
\bibnamefont{D.~Oprisa} and \bibnamefont{S.~Stieberger}.
\newblock \bibtitlefont{Six gluon open superstring disk amplitude, multiple
  hypergeometric series and Euler-Zagier sums}  (2005).
\newblock
\href{http://www.arXiv.org/abs/hep-th/0509042}{hep-th/0509042}.
\newblock

\bibitem{Guralnik:2004ve}
\bibnamefont{Z.~Guralnik}, \bibnamefont{S.~Kovacs} and \bibnamefont{B.~Kulik}.
\newblock \bibtitlefont{Holography and the Higgs branch of {$\cN = 2$}
  \acro{SYM} theories}  (2004).
\newblock
\href{http://www.arXiv.org/abs/hep-th/0405127}{hep-th/0405127}.
\newblock

\bibitem{Guralnik:2005jg}
\bibnamefont{Z.~Guralnik}, \bibnamefont{S.~Kovacs} and \bibnamefont{B.~Kulik}.
\newblock \bibtitlefont{\acro{AdS}/\acro{CFT} duality and the Higgs branch of
  {$\cN = 2$} \acro{SYM}}  (2005).
\newblock
\href{http://www.arXiv.org/abs/hep-th/0501154}{hep-th/0501154}.
\newblock

\bibitem{Babington:2003up}
\bibnamefont{J.~Babington}, \bibnamefont{J.~Erdmenger}, \bibnamefont{N.~J.
  Evans}, \bibnamefont{Z.~Guralnik} and \bibnamefont{I.~Kirsch}.
\newblock \bibtitlefont{A Gravity Dual of Chiral Symmetry Breaking}.
\newblock Fortsch. Phys. \textbf{52}:578--582 (2004).
\newblock
\href{http://www.arXiv.org/abs/hep-th/0312263}{hep-th/0312263}.
\newblock

\bibitem{Erlich:2005qh}
\bibnamefont{J.~Erlich}, \bibnamefont{E.~Katz}, \bibnamefont{D.~T. Son} and
  \bibnamefont{M.~A. Stephanov}.
\newblock \bibtitlefont{\acro{QCD} and a holographic model of hadrons}.
\newblock Phys. Rev. Lett. \textbf{95}:261602 (2005).
\newblock
\href{http://www.arXiv.org/abs/hep-ph/0501128}{hep-ph/0501128}.
\newblock

\bibitem{DaRold:2005zs}
\bibnamefont{L.~Da~Rold} and \bibnamefont{A.~Pomarol}.
\newblock \bibtitlefont{Chiral symmetry breaking from five dimensional spaces}.
\newblock Nucl. Phys. \textbf{B721}:79--97 (2005).
\newblock
\href{http://www.arXiv.org/abs/hep-ph/0501218}{hep-ph/0501218}.
\newblock

\bibitem{Evans:2006dj}
\bibnamefont{N.~Evans} and \bibnamefont{T.~Waterson}.
\newblock \bibtitlefont{Improving the infra-red of holographic descriptions of
  \acro{QCD}}  (2006).
\newblock
\href{http://www.arXiv.org/abs/hep-ph/0603249}{hep-ph/0603249}.
\newblock

\bibitem{Gates:1983nr}
\bibnamefont{S.~J. Gates}, \bibnamefont{M.~T. Grisaru}, \bibnamefont{M.~Rocek}
  and \bibnamefont{W.~Siegel}.
\newblock \bibtitlefont{Superspace, or one thousand and one lessons in
  supersymmetry}.
\newblock Front. Phys. \textbf{58}:1--548 (1983).
\newblock
\href{http://www.arXiv.org/abs/hep-th/0108200}{hep-th/0108200}.
\newblock

\bibitem{Buchbinder:1998qv}
\bibnamefont{I.~L. Buchbinder} and \bibnamefont{S.~M. Kuzenko}.
\newblock \bibtitlefont{Ideas and methods of supersymmetry and supergravity: Or
  a walk through superspace}  (1998).
\newblock Bristol, UK: IOP (1998) 656 p.

\bibitem{Bagger:1990qh}
\bibnamefont{J.~Bagger} and \bibnamefont{J.~Wess}.
\newblock \bibtitlefont{Supersymmetry and supergravity}  (1990).
\newblock JHU-TIPAC-9009.

\bibitem{Wess:1992cp}
\bibnamefont{J.~Wess} and \bibnamefont{J.~Bagger}.
\newblock \bibtitlefont{Supersymmetry and supergravity}  (1992).
\newblock Princeton, USA: Univ. Pr. (1992) 259 p.

\bibitem{Babington:2005vu}
\bibnamefont{J.~Babington} and \bibnamefont{J.~Erdmenger}.
\newblock \bibtitlefont{Space-time Dependent Couplings in {$\cN = 1$}
  \acro{SUSY} Gauge Theories: Anomalies and Central Functions}.
\newblock JHEP \textbf{06}:004 (2005).
\newblock
\href{http://www.arXiv.org/abs/hep-th/0502214}{hep-th/0502214}.
\newblock

\bibitem{Apostol:1997}
\bibnamefont{T.~M. Apostol}.
\newblock \bibtitlefont{Modular Functions and {D}irichlet Series in Number
  Theory}.
\newblock Number~41 in Graduate Texts in Mathematics. Springer-Verlag, New
  York, 2 edition (1997).

\bibitem{hoehne-diplomarbeit}
\bibnamefont{S.~H{\"o}hne}.
\newblock \bibtitlefont{Local Couplings in Quantum Field Theories}.
\newblock Master's thesis, Humboldt-Universit{\"a}t zu Berlin (2005).
\newblock Diploma Thesis.

\bibitem{Kugo:1982cu}
\bibnamefont{T.~Kugo} and \bibnamefont{S.~Uehara}.
\newblock \bibtitlefont{Conformal and {Poincar\'e} Tensor Calculi in {$\cN=1$}
  Supergravity}.
\newblock Nucl. Phys. \textbf{B226}:49
 (1983).
\newblock

\bibitem{Evans:2004ia}
\bibnamefont{N.~J. Evans} and \bibnamefont{J.~P. Shock}.
\newblock \bibtitlefont{Chiral dynamics from \acro{AdS} space}.
\newblock Phys. Rev. \textbf{D70}:046002 (2004).
\newblock
\href{http://www.arXiv.org/abs/hep-th/0403279}{hep-th/0403279}.
\newblock

\bibitem{Evans:2005ti}
\bibnamefont{N.~Evans}, \bibnamefont{J.~Shock} and \bibnamefont{T.~Waterson}.
\newblock \bibtitlefont{{D7} brane embeddings and chiral symmetry breaking}.
\newblock JHEP \textbf{03}:005 (2005).
\newblock
\href{http://www.arXiv.org/abs/hep-th/0502091}{hep-th/0502091}.
\newblock

\bibitem{Apreda:2005hj}
\bibnamefont{R.~Apreda}, \bibnamefont{J.~Erdmenger} and \bibnamefont{N.~Evans}.
\newblock \bibtitlefont{Scalar Effective Potential for {D7} Brane Probes which
  Break Chiral Symmetry}  (2005).
\newblock
\href{http://www.arXiv.org/abs/hep-th/0509219}{hep-th/0509219}.
\newblock

\bibitem{Apreda:2005yz}
\bibnamefont{R.~Apreda}, \bibnamefont{J.~Erdmenger}, \bibnamefont{N.~Evans} and
  \bibnamefont{Z.~Guralnik}.
\newblock \bibtitlefont{Strong Coupling Effective {H}iggs Potential and a First
  Order Thermal Phase Transition from \acro{AdS}/\acro{CFT} Duality}.
\newblock Phys. Rev. \textbf{D71}:126002 (2005).
\newblock
\href{http://www.arXiv.org/abs/hep-th/0504151}{hep-th/0504151}.
\newblock

\bibitem{Kruczenski:2003uq}
\bibnamefont{M.~Kruczenski}, \bibnamefont{D.~Mateos}, \bibnamefont{R.~C. Myers}
  and \bibnamefont{D.~J. Winters}.
\newblock \bibtitlefont{Towards a holographic dual of large-{$N_c$}
  \acro{QCD}}.
\newblock JHEP \textbf{05}:041 (2004).
\newblock
\href{http://www.arXiv.org/abs/hep-th/0311270}{hep-th/0311270}.
\newblock

\bibitem{Barbon:2004dq}
\bibnamefont{J.~L.~F. Barbon}, \bibnamefont{C.~Hoyos}, \bibnamefont{D.~Mateos}
  and \bibnamefont{R.~C. Myers}.
\newblock \bibtitlefont{The holographic life of the eta'}.
\newblock JHEP \textbf{10}:029 (2004).
\newblock
\href{http://www.arXiv.org/abs/hep-th/0404260}{hep-th/0404260}.
\newblock

\bibitem{Ghoroku:2004sp}
\bibnamefont{K.~Ghoroku} and \bibnamefont{M.~Yahiro}.
\newblock \bibtitlefont{Chiral symmetry breaking driven by dilaton}.
\newblock Phys. Lett. \textbf{B604}:235--241 (2004).
\newblock
\href{http://www.arXiv.org/abs/hep-th/0408040}{hep-th/0408040}.
\newblock

\bibitem{Brevik:2005fs}
\bibnamefont{I.~Brevik}, \bibnamefont{K.~Ghoroku} and
  \bibnamefont{A.~Nakamura}.
\newblock \bibtitlefont{Meson mass and confinement force driven by dilaton}.
\newblock Int. J. Mod. Phys. \textbf{D15}:57--68 (2006).
\newblock
\href{http://www.arXiv.org/abs/hep-th/0505057}{hep-th/0505057}.
\newblock

\bibitem{Sakai:2004cn}
\bibnamefont{T.~Sakai} and \bibnamefont{S.~Sugimoto}.
\newblock \bibtitlefont{Low energy hadron physics in holographic \acro{QCD}}.
\newblock Prog. Theor. Phys. \textbf{113}:843--882 (2005).
\newblock
\href{http://www.arXiv.org/abs/hep-th/0412141}{hep-th/0412141}.
\newblock

\bibitem{Sakai:2005yt}
\bibnamefont{T.~Sakai} and \bibnamefont{S.~Sugimoto}.
\newblock \bibtitlefont{More on a holographic dual of \acro{QCD}}.
\newblock Prog. Theor. Phys. \textbf{114}:1083--1118 (2006).
\newblock
\href{http://www.arXiv.org/abs/hep-th/0507073}{hep-th/0507073}.
\newblock

\bibitem{Antonyan:2006vw}
\bibnamefont{E.~Antonyan}, \bibnamefont{J.~A. Harvey}, \bibnamefont{S.~Jensen}
  and \bibnamefont{D.~Kutasov}.
\newblock \bibtitlefont{\acro{NJL} and \acro{QCD} from String Theory}  (2006).
\newblock
\href{http://www.arXiv.org/abs/hep-th/0604017}{hep-th/0604017}.
\newblock

\bibitem{Bak:2004nt}
\bibnamefont{D.~Bak} and \bibnamefont{H.-U. Yee}.
\newblock \bibtitlefont{Separation of spontaneous chiral symmetry breaking and
  confinement via \acro{AdS}/\acro{CFT} correspondence}.
\newblock Phys. Rev. \textbf{D71}:046003 (2005).
\newblock
\href{http://www.arXiv.org/abs/hep-th/0412170}{hep-th/0412170}.
\newblock

\bibitem{Ghoroku:2005tf}
\bibnamefont{K.~Ghoroku}, \bibnamefont{T.~Sakaguchi}, \bibnamefont{N.~Uekusa}
  and \bibnamefont{M.~Yahiro}.
\newblock \bibtitlefont{Flavor quark at high temperature from a holographic
  model}.
\newblock Phys. Rev. \textbf{D71}:106002 (2005).
\newblock
\href{http://www.arXiv.org/abs/hep-th/0502088}{hep-th/0502088}.
\newblock

\bibitem{Mateos:2006nu}
\bibnamefont{D.~Mateos}, \bibnamefont{R.~C. Myers} and \bibnamefont{R.~M.
  Thomson}.
\newblock \bibtitlefont{Holographic phase transitions with fundamental matter}
  (2006).
\newblock
\href{http://www.arXiv.org/abs/hep-th/0605046}{hep-th/0605046}.
\newblock

\bibitem{Albash:2006ew}
\bibnamefont{T.~Albash}, \bibnamefont{V.~Filev}, \bibnamefont{C.~V. Johnson}
  and \bibnamefont{A.~Kundu}.
\newblock \bibtitlefont{A Topology-Changing Phase Transition and the Dynamics
  of Flavour}  (2006).
\newblock
\href{http://www.arXiv.org/abs/hep-th/0605088}{hep-th/0605088}.
\newblock

\bibitem{Albash:2006bs}
\bibnamefont{T.~Albash}, \bibnamefont{V.~Filev}, \bibnamefont{C.~V. Johnson}
  and \bibnamefont{A.~Kundu}.
\newblock \bibtitlefont{Global Currents, Phase Transitions, and Chiral Symmetry
  Breaking in Large {$N_c$} Gauge Theory}  (2006).
\newblock
\href{http://www.arXiv.org/abs/hep-th/0605175}{hep-th/0605175}.
\newblock

\bibitem{Parnachev:2006dn}
\bibnamefont{A.~Parnachev} and \bibnamefont{D.~A. Sahakyan}.
\newblock \bibtitlefont{Chiral phase transition from string theory}  (2006).
\newblock
\href{http://www.arXiv.org/abs/hep-th/0604173}{hep-th/0604173}.
\newblock

\bibitem{Aharony:2006da}
\bibnamefont{O.~Aharony}, \bibnamefont{J.~Sonnenschein} and
  \bibnamefont{S.~Yankielowicz}.
\newblock \bibtitlefont{A holographic model of deconfinement and chiral
  symmetry restoration}  (2006).
\newblock
\href{http://www.arXiv.org/abs/hep-th/0604161}{hep-th/0604161}.
\newblock

\bibitem{Karch:2006bv}
\bibnamefont{A.~Karch} and \bibnamefont{A.~O'Bannon}.
\newblock \bibtitlefont{Chiral transition of {$\cN = 4$} super {Y}ang-{M}ills
  with flavor on a 3-sphere}  (2006).
\newblock
\href{http://www.arXiv.org/abs/hep-th/0605120}{hep-th/0605120}.
\newblock

\bibitem{deTeramond:2005su}
\bibnamefont{G.~F. de~Teramond} and \bibnamefont{S.~J. Brodsky}.
\newblock \bibtitlefont{The hadronic spectrum of a holographic dual of
  \acro{QCD}}.
\newblock Phys. Rev. Lett. \textbf{94}:201601 (2005).
\newblock
\href{http://www.arXiv.org/abs/hep-th/0501022}{hep-th/0501022}.
\newblock

\bibitem{Brodsky:2006uq}
\bibnamefont{S.~J. Brodsky} and \bibnamefont{G.~F. de~Teramond}.
\newblock \bibtitlefont{Hadronic spectra and light-front wavefunctions in
  holographic \acro{QCD}}.
\newblock Phys. Rev. Lett. \textbf{96}:201601 (2006).
\newblock
\href{http://www.arXiv.org/abs/hep-ph/0602252}{hep-ph/0602252}.
\newblock

\bibitem{Boschi-Filho:2002ta}
\bibnamefont{H.~Boschi-Filho} and \bibnamefont{N.~R.~F. Braga}.
\newblock \bibtitlefont{\acro{QCD} / string holographic mapping and glueball
  mass spectrum}.
\newblock Eur. Phys. J. \textbf{C32}:529--533 (2004).
\newblock
\href{http://www.arXiv.org/abs/hep-th/0209080}{hep-th/0209080}.
\newblock

\bibitem{Boschi-Filho:2002vd}
\bibnamefont{H.~Boschi-Filho} and \bibnamefont{N.~R.~F. Braga}.
\newblock \bibtitlefont{Gauge / string duality and scalar glueball mass
  ratios}.
\newblock JHEP \textbf{05}:009 (2003).
\newblock
\href{http://www.arXiv.org/abs/hep-th/0212207}{hep-th/0212207}.
\newblock

\bibitem{Hong:2003jm}
\bibnamefont{S.~Hong}, \bibnamefont{S.~Yoon} and \bibnamefont{M.~J. Strassler}.
\newblock \bibtitlefont{Quarkonium from the fifth dimension}.
\newblock JHEP \textbf{04}:046 (2004).
\newblock
\href{http://www.arXiv.org/abs/hep-th/0312071}{hep-th/0312071}.
\newblock

\bibitem{DaRold:2005vr}
\bibnamefont{L.~Da~Rold} and \bibnamefont{A.~Pomarol}.
\newblock \bibtitlefont{The scalar and pseudoscalar sector in a
  five-dimensional approach to chiral symmetry breaking}.
\newblock JHEP \textbf{01}:157 (2006).
\newblock
\href{http://www.arXiv.org/abs/hep-ph/0510268}{hep-ph/0510268}.
\newblock

\bibitem{Shock:2006fc}
\bibnamefont{J.~P. Shock}.
\newblock \bibtitlefont{Canonical coordinates and meson spectra for scalar
  deformed {$N = 4$} \acro{SYM} from the \acro{AdS}/\acro{CFT} correspondence}
  (2006).
\newblock
\href{http://www.arXiv.org/abs/hep-th/0601025}{hep-th/0601025}.
\newblock

\bibitem{Shock:2006qy}
\bibnamefont{J.~P. Shock} and \bibnamefont{F.~Wu}.
\newblock \bibtitlefont{Three flavour \acro{QCD} from the holographic
  principle}  (2006).
\newblock
\href{http://www.arXiv.org/abs/hep-ph/0603142}{hep-ph/0603142}.
\newblock

\bibitem{Ghoroku:2005kg}
\bibnamefont{K.~Ghoroku} and \bibnamefont{M.~Yahiro}.
\newblock \bibtitlefont{Holographic model for mesons at finite temperature}
  (2005).
\newblock
\href{http://www.arXiv.org/abs/hep-ph/0512289}{hep-ph/0512289}.
\newblock

\bibitem{Ghoroku:2005vt}
\bibnamefont{K.~Ghoroku}, \bibnamefont{N.~Maru}, \bibnamefont{M.~Tachibana} and
  \bibnamefont{M.~Yahiro}.
\newblock \bibtitlefont{Holographic model for hadrons in deformed {$\AdS_5$}
  background}.
\newblock Phys. Lett. \textbf{B633}:602--606 (2006).
\newblock
\href{http://www.arXiv.org/abs/hep-ph/0510334}{hep-ph/0510334}.
\newblock

\bibitem{Ghoroku:2006cc}
\bibnamefont{K.~Ghoroku}, \bibnamefont{A.~Nakamura} and
  \bibnamefont{M.~Yahiro}.
\newblock \bibtitlefont{Holographic model at finite temperature with {R}-charge
  density}  (2006).
\newblock
\href{http://www.arXiv.org/abs/hep-ph/0605026}{hep-ph/0605026}.
\newblock

\bibitem{Karch:2006pv}
\bibnamefont{A.~Karch}, \bibnamefont{E.~Katz}, \bibnamefont{D.~T. Son} and
  \bibnamefont{M.~A. Stephanov}.
\newblock \bibtitlefont{Linear confinement and \acro{AdS}/\acro{QCD}}  (2006).
\newblock
\href{http://www.arXiv.org/abs/hep-ph/0602229}{hep-ph/0602229}.
\newblock

\bibitem{Andreev:2006vy}
\bibnamefont{O.~Andreev}.
\newblock \bibtitlefont{{$1/q^2$} Corrections and Gauge / String Duality}
  (2006).
\newblock
\href{http://www.arXiv.org/abs/hep-th/0603170}{hep-th/0603170}.
\newblock

\bibitem{Andreev:2006ct}
\bibnamefont{O.~Andreev} and \bibnamefont{V.~I. Zakharov}.
\newblock \bibtitlefont{Heavy-quark Potentials and \acro{AdS}/\acro{QCD}}
  (2006).
\newblock
\href{http://www.arXiv.org/abs/hep-ph/0604204}{hep-ph/0604204}.
\newblock

\bibitem{Hambye:2005up}
\bibnamefont{T.~Hambye}, \bibnamefont{B.~Hassanain},
  \bibnamefont{J.~March-Russell} and \bibnamefont{M.~Schvellinger}.
\newblock \bibtitlefont{On the {$\Delta_I = \half$} rule in holographic
  \acro{QCD}}  (2005).
\newblock
\href{http://www.arXiv.org/abs/hep-ph/0512089}{hep-ph/0512089}.
\newblock

\bibitem{Vermaseren:2000nd}
\bibnamefont{J.~A.~M. Vermaseren}.
\newblock \bibtitlefont{New features of \acro{FORM}}  (2000).
\newblock
\href{http://www.arXiv.org/abs/math-ph/0010025}{math-ph/0010025}.
\newblock

\bibitem{Peeters-AEI-2006-037}
\bibnamefont{K.~Peeters}.
\newblock \bibtitlefont{A Field Theory Motivated Approach to Symbolic Computer
  Algebra} Preprint AEI-2006-037.

\bibitem{Peeters-AEI-2006-038}
\bibnamefont{K.~Peeters}.
\newblock \bibtitlefont{Cadabra: Reference Guide and Tutorial} Preprint
  AEI-2006-038.

\end{thebibliography}


\providecommand{\WileyBibTextsc}{}
\let\textsc\WileyBibTextsc
\providecommand{\othercit}{}
\providecommand{\jr}[1]{#1}
\providecommand{\etal}{~et~al.}

\end{document}